\documentclass[11pt]{article}
\pdfoutput=1
\usepackage{jheppub}
\usepackage{amsmath,amssymb,amsthm,graphicx,amscd,amsbsy}
\usepackage[matrix,arrow,curve]{xy}
\usepackage{hyperref}
\usepackage{color}
\usepackage[usenames,dvipsnames,svgnames,table]{xcolor}
\usepackage{multirow}
\usepackage{tabularx}
\usepackage{bm}
\usepackage{dsfont}
\usepackage{mathrsfs}
\usepackage{comment}

\usepackage{tabularx}

\def\be{\begin{eqnarray}}
\def\ee{\end{eqnarray}}

\def\lm{\limits}

\def\Tr{{\rm Tr}\,}

\def\CA{{\mathcal A}}

\def\CW{{\mathcal W}}
\def\CB{{\mathcal B}}
\def\CH{{\mathcal H}}

\def\CK{{\mathcal K}}

\def\CX{{\mathcal X}}

\def\CM{{\mathcal M}}
\def\CS{{\mathcal S}}

\def\IR{{\mathbb{R}}}
\def\IZ{{\mathds{Z}}}
\def\IN{{\mathds{N}}}
\def\IC{{\mathds{C}}}

\def\CH{\mathscr{H}}

\def\FOmega{\underline{\overline{\Omega}}}
\def\FCF{\underline{\overline{{\cal F}}}}

\def\hY{{\hat Y}}

\def\hU{{\hat \Upsilon}}
\def\hA{{\hat A}}
\def\wr{{\rm wr}}

\def\cl{{\rm cl}}

\def\sign{{\rm sign}}

\def\fa{{\mathfrak a}}
\def\fb{{\mathfrak b}}
\def\fc{{\mathfrak c}}
\def\fd{{\mathfrak d}}
\def\fh{{\mathfrak h}}
\def\fp{{\mathfrak p}}

\def\fF{{\mathfrak F}}
\def\fD{{\mathfrak D}}
\def\tSigma{{\widetilde \Sigma}}

\def\aux{\chi}
\def\bU{\overline\Upsilon}
\def\bY{\overline Y}

\def\tGamma{\tilde \Gamma^{*}}
\def\bGamma{\boldsymbol{\Gamma}^{*}}
\def\tgamma{\tilde \gamma}

\newtheorem{conj}{Conjecture}

\makeatletter
\newsavebox{\@brx}
\newcommand{\llangle}[1][]{\savebox{\@brx}{\(\m@th{#1\langle}\)}%
  \mathopen{\copy\@brx\kern-0.5\wd\@brx\usebox{\@brx}}}
\newcommand{\rrangle}[1][]{\savebox{\@brx}{\(\m@th{#1\rangle}\)}%
  \mathclose{\copy\@brx\kern-0.5\wd\@brx\usebox{\@brx}}}
\makeatother

\renewcommand{\Re}{{\rm{Re}}}
\newcommand{\Arg}{{\rm{Arg}}}

\mathchardef\mhyphen="2D

\newcounter{desccount}

\newcommand{\descref}[1]{\hyperref[#1]{#1}}

\newcommand{\CN}{{\mathcal{N}}}

\renewcommand{\Im}{{\mathrm{Im\,}}}
\newcommand{\so}{{\mathfrak{so}}}
\newcommand{\su}{{\mathfrak{su}}}

\makeatletter
\newcommand{\labitem}[2]{%
\def\@itemlabel{#1}
\item
\def\@currentlabel{#1}\label{#2}}
\makeatother

\makeatletter
\newcommand{\labitemred}[1]{%
\def\@itemlabel
\item
\def\@currentlabel{#1}\label{#1}}
\makeatother

\numberwithin{equation}{section}

\subheader{\begin{flushright}
	RUNHETC-2014-14\\
	ITEP-TH-10/14
\end{flushright}}

\title{Spectral Networks with Spin}

\abstract{The BPS spectrum of d=4 N=2 field theories in general contains
not only hyper- and vector-multipelts but also short multiplets
of particles with arbitrarily high spin. This paper extends the
method of spectral networks to give an algorithm for computing
the spin content of the BPS spectrum of d=4 N=2 field theories
of class S. The key new ingredient is an identification of the
spin of states with the writhe of paths on the Seiberg-Witten
curve. Connections to quiver representation theory and to
Chern-Simons theory are briefly discussed.}

\author[1]{Dmitry Galakhov,}
\author[2]{Pietro Longhi,}
\author[3]{Gregory W. Moore,}
\affiliation[1]
{Institute for Theoretical and Experimental Physics,\\ Moscow, Russia,}
\affiliation[1,2,3]
{NHETC and Department of Physics and Astronomy, Rutgers University,\\
Piscataway, NJ 08855--0849, USA}

\emailAdd{galakhov@physics.rutgers.edu}
\emailAdd{longhi@physics.rutgers.edu}
\emailAdd{gmoore@physics.rutgers.edu}

\date{\today}

\begin{document}

\maketitle

\section{Introduction and summary}

This paper is about the BPS spectrum of a class of field theories with  $d=4$  ${\cal N}=2$ Poincar\'e supersymmetry known as 
class $\CS$. Recently, there has been some progress in finding effective algorithms for determining the BPS spectrum of these 
theories.  One such algorithm, known as the method of spectral networks  \cite{GMN5,GMN6,WWC}, is based on the geometry of the Seiberg-Witten 
curve, presented as a branched cover of another, ``ultraviolet'' curve. 
Thus far, the spectral network technique has only been used to extract information on the BPS index - a signed sum over 
the BPS Hilbert space at fixed charge. In this paper the method is refined to give an algorithm to compute the spin content, 
or more properly, the protected spin character, of the space of BPS states at a given charge. 

Let us put this result into some context. 
The study of BPS states and of their relations to several areas of mathematics has sparked considerable interest in recent years.  For some recent reviews see \cite{FelixKlein, Cecotti, Denef:2007vg, KS,Kontsevich:2009xt, Pioline:2011gf}.
Class $\CS$ theories descend from twisted compactifications of the $(0,2)$ six-dimensional theory, (the ``$\CS$'' is for ``six''). 
They are characterized by an ADE-type Lie algebra $\mathfrak{g}$, a punctured Riemann surface $C$, and certain data $D$ at the punctures characterizing codimension two defects. 
They are denoted $\CS[\mathfrak{g}, C, D]$. The investigation of the BPS spectrum in these theories has led to a number of interesting connections with the mathematics of Hitchin systems, 
integrable field theories, and cluster algebras and cluster varieties.

A characteristic feature of class $\CS$ theories is the existence of a quantum moduli space of vacua, with a Coulomb branch $\CB=\oplus_{k=1}^{K-1}H^{0}(C,K^{\otimes d_{k}})$ 
where the gauge symmetry is spontaneously broken to a $(K-1)$-dimensional maximal torus.  (In the $A_{K-1}$ case $(d_{1},\dots,d_{K-1})=(2,\dots,K)$). 
At a generic point $u\in\CB$ the IR dynamics admits a locally valid Lagrangian description \cite{SW}. 
The Coulomb branch is divided into disjoint chambers by marginal stability walls: real codimension one loci where the BPS spectrum jumps discontinuously. The BPS Hilbert spaces on two sides of a wall are related by wall-crossing formulae
 \cite{KS,Kontsevich:2009xt, Denef:2007vg, GMN1, susygalaxy,Pioline:2011gf,dimofte-gukov-soibelman}.
Therefore, in principle, since the Coulomb branch is connected, a knowledge of the BPS spectrum at some point  allows one  to recover the spectrum everywhere else by wall-crossing. 
But life is not so simple: walls can be dense, and the problem of determining the spectrum at any point at all can be challenging. 
One technique to study the BPS spectrum at a generic point $u\in\CB$ is   based on spectral networks \cite{GMN5,GMN6,WWC}. We briefly review it in section \ref{sec:Review}. This framework employs the input data of the theory $\CS[\mathfrak{g},C,D]$ and provides a description of the BPS spectrum at any point $u\in\CB$. While the range of applicability of this technique is rather large, the information it provides about BPS states is, in a sense, somewhat limited: in its current status of development the only information that can be extracted about BPS states of the 4d gauge theory is the BPS index. As emphasized by recent work \cite{WWC}, BPS spectra can exhibit a rather rich structure which is missed by the BPS index alone. A more refined quantity such as the \emph{Protected Spin Character}
(PSC) provides instead a richer description, capturing in particular the spin content of BPS states. It is therefore highly desirable to develop a framework allowing for a systematic investigation of refinements of the BPS index such as the PSC. 
Developing such a framework is precisely the aim of the present paper: our main result is a proposal for extracting PSC data from spectral networks, thus generalizing the BPS index formula of \cite{GMN5}. More precisely, we propose a method for computing the spin of both \emph{framed} and \emph{vanilla} BPS states. (The terminology comes from \cite{GMN3,GMN4,GMN5}.) The framework we propose does not follow from a first principles derivation, but relies on some conjectures, for which we provide some tests.

We will argue below that spectral networks actually contain much more information than hitherto utilized. 
In section \ref{sec:conjectures} we formulate precise conjectures explaining where such extra information sits within the network data, and how it encodes spin degeneracies.
A key ingredient is the refinement of the classification of soliton paths induced by \emph{regular homotopy}. After introducing a suitable formal algebra associated with this refinement, in section \ref{sec:FPT} we provide the related generalization of the formal parallel transport of \cite{GMN5}. This involves establishing a refined version of the \emph{detour rules}, whose physical interpretation explains the wall-crossing of framed BPS states. The refinement by regular homotopy allows one to associate to each path $\fa$ an integer known as the \emph{writhe} $\wr(\fa)$, consisting of a certain signed sum over self-intersections. We identify the writhe with the spin of a framed BPS state, while its charge is given by the canonical projection to relative homology. In the same way as framed degeneracies are good probes to study vanilla BPS indices, the framed PSCs obtained in this way serve the same purpose for computing vanilla PSCs. 

Important consistency checks come from the halo picture of framed wall-crossing \cite{GMN3, susygalaxy}, which was crucial in linking jumps of PSCs at walls of marginal stability and the motivic Kontsevich-Soibelman wall-crossing formula \cite{KS,Kontsevich:2009xt}.
The main idea here is to associate a path $\wp$ on the ultraviolet curve $C$ with a supersymmetric interface between surface defect theories \cite{GMN4}. 
We find that the halo picture easily emerges within our proposal if we restrict to a certain type of susy interface. We provide a criterion that distinguishes this special class and call them \emph{halo-saturated interfaces}. Physically, their crucial feature is that their wall-crossing behavior mimics that of line defects \cite{GMN3}.
The wall-crossing behavior of more generic interfaces is one issue which remains only partially understood, in particular it would be desirable to shed light on the halo interpretation of the framed wall crossing of generic interfaces. In section \ref{sec:gen_interf} we study a particular example and find some apparent tension with the halo picture. However, by taking into account a refinement of the homology on $\Sigma$ induced by the presence of the interface, we eventually find a reconciliation with the halo interpretation. 
A systematic understanding of how the halo picture fits with our conjectures for generic interfaces is left as an interesting and important open problem for the future. 

We would like to mention another curious conjecture, even though it is not central to the
main development of the paper. Only certain states in a vanilla multiplet will bind to
a generic interface \cite{GMN4,GMN5,2d4d}. This suggests that each state within the vanilla
multiplet can be associated with a \emph{subnetwork} of the critical network $\CW_{c}$, and that the halos forming around the interface depend on how the latter\footnote{More properly, the relative homology class associated to it.} intersects the various subnetworks. Towards the end of section \ref{sec:gen_interf} we mention this hypothesis when discussing contributions from ``phantom'' halos to the $\CK$-wall jumps of framed PSCs, while we defer a more detailed study to appendix \ref{app:L-r}, where supporting evidence is also offered.

We leave a proof of our conjectures to future work. In the present note we concentrate on how they are realized in various examples, and on their consequences. The results are in perfect agreement with other approaches, such as results derived 
from  motivic wall-crossing (see for example \cite{WWC}) or from quiver techniques.  In particular, we consider the rich playground provided by the wild BPS states investigated in \cite{WWC}. These wild BPS states
  typically furnish high-dimensional and highly reducible representations of the group of spatial rotations. In a  wild chamber  of the Coulomb branch one finds BPS multiplets of arbitrarily high spin.
In this phase of the IR theory the number of BPS states grows exponentially with the mass, a surprising fact for a gauge theory \cite{WWC,Kol:1998zb,Kol:2000tw}. We will apply our techniques both to the \emph{herd networks}, which describe a particular type of wild state, as well as to a new type of wild critical network which is a close cousin of the herds.
Wild spectral networks have been associated with algebraic equations for generating functions of BPS indices \cite{WWC,GP,KS,Kontsevich:2009xt}. 
For instance, it was found (see \cite[eq.(1.1)]{WWC}) that herd networks  encode an algebraic equation familiar in the context of the tropical vertex group. 
By exploiting our construction of the formal parallel transport, we derive a  deformation  of that equation
\be\label{eq:fnceqn}
P(z,y)=1+z\prod\lm_{s=-(m-2)}^{m-2}P(zy^{2s},y)^{m-1-|s|} \,,
\ee
which is of a \emph{functional} nature. We check that (\ref{eq:fnceqn}) correctly describes the generating function of PSCs, and discuss its consistency with quiver representation theory (in particular with Kac's theorem \cite{WWC,Reinike} and Poincar\'e polynomial stabilization \cite{Reineke}).

Finally, since the use of formal variables and the introduction of the writhe might seem artificial to some readers,  in \S\ref{sec:chern-simons} we propose a framework in which all these crucial ingredients arise naturally. A quantization of the moduli space of flat abelian connections on the Seiberg-Witten curve naturally yields an operator algebra resembling that of our formal variables. From a slightly different viewpoint, our formal variables may be thought of as Wilson operators of a certain abelian Chern-Simons theory. From this perspective both the refined classification of paths (which are singular knots in our case) by regular homotopy and the role of the writhe are no surprise at all (see e.g. \cite{dunne}). 
We do not develop the relation of our story to Chern-Simons theory in detail, rather we limit ourselves to some preliminary remarks. However we do expect an interpretation of our refined construction of the formal parallel transport as a \emph{map between observables} of two distinguished Chern-Simons theories. We hope to return to this point in the future.

\section{Protected Spin Characters from writhe}\label{sec:conjectures}
\subsection{Review of framed BPS states}\label{sec:Review}
%
Framed BPS states, introduced in \cite{GMN3}, appear in the context of four-dimensional $\CN=2$ gauge theories with the insertion of certain line defects.
In the Coulomb phase 
of the gauge theory, one may consider the effect of inserting a line defect $L_{\zeta}$ preserving a $osp(4^{*}|2)_{\zeta}$ sub-superalgebra (for notations see \cite{GMN3}). The preserved supercharges depend on the choice of $\zeta\in\IC^{*}$, and the surviving algebra develops a new type of BPS bound
\be
	E+\Re(Z/\zeta)\geq 0\,.
\ee
States in the Hilbert space $\CH_{u,L,\zeta}$ which saturate this bound are the framed BPS states, the subspace spanned by these is denoted $\CH^{BPS}_{u,L,\zeta}$. The introduction of the line defect also modifies the usual ``vanilla'' grading of the BPS Hilbert space to
\be
	\CH^{BPS}_{u,L,\zeta} = \bigoplus_{\gamma\in\Gamma_{L}}\CH_{u,L,\zeta,\gamma}
\ee
where $\Gamma_{L}$ is a torsor for the vanilla lattice gauge lattice $\Gamma$, and there is an integral-valued pairing $\langle\gamma_{L},\gamma\rangle\in\IZ$ defined for any $\gamma_{L}\in\Gamma_{L},\, \gamma\in\Gamma$.

The framed BPS bound determines a new type of marginal stability wall, termed BPS walls:
\be
	\widehat W(\gamma)=\{(u,\zeta)\,|\,Z_{\gamma}(u)/\zeta\in\IR_{-}\}\,\subset\,\CB\times\IC^{*}\,,
\ee
where $Z_{\gamma}$ denotes the central charge of a populated vanilla BPS state of charge $\gamma$.
Near these loci some framed BPS states look like halo BPS particles (with charge $\gamma_{h}\in \Gamma$) bound to a non-dynamical ``core charge'' $\gamma_{c}\in\Gamma_{L}$. Given a choice of moduli $(u,\zeta)$ where a framed BPS state of charge $\gamma_{c}+\gamma_{h}$ is stable, its energy is
\be
	E = -\Re(Z_{\gamma_{c}+\gamma_{h}}(u)/\zeta) \leq  -\Re(Z_{\gamma_{c}}(u)/\zeta)+|Z_{\gamma_{h}}(u)|\,
\ee
the inequality saturates at BPS walls, where boundstates become marginally stable. 
At BPS walls, these states mix with the continuum of vanilla BPS states, whose BPS bound is $E \geq |Z_{\gamma_{h}}(u)|$, the Fock space of framed states therefore gains or loses a factor, this is the halo picture of the framed wall-crossing phenomenon. Part of its importance stems from the fact that it underlies a physical derivation of the Kontsevich-Soibelman wall-crossing formula, and of its motivic counterpart \cite{KS,Kontsevich:2009xt,GMN3,susygalaxy,pitp}.

As suggested by the halo picture recalled above, framed BPS states furnish representations of $\so(3)$ of spatial rotations as well as of $\su(2)_{R}$.
The \emph{framed protected spin character} (PSC) is defined as
\be
	\FOmega(L_{\zeta},u,\gamma;y):=\Tr_{\CH_{u,L,{\zeta},\gamma}}y^{2J_{3}} (-y)^{2I_{3}}
\ee
where $J_{3},I_{3}$ are Cartan generators of $\so(3)$ and $\su(2)_{R}$.
Similarly, the \emph{vanilla PSC} is defined as
\be\label{eq:PSC}
	\Omega(u,\gamma;y):=\Tr_{\mathfrak{h}_{\gamma}}\,y^{2J_{3}} (-y)^{2I_{3}} = \sum_{m\in\IZ}a_{m}(\gamma)\, (-y)^{m}
\ee
where $\mathfrak{h}_{\gamma}$ is the Clifford vacuum of the BPS Hilbert space\footnote{see e.g. \cite{WWC,pitp} for more details.}, $\gamma\in \Gamma$ and the last equality defines the integers $a_{m}(\gamma)\in\IZ$.

It is useful to consider a generating function of framed BPS degeneracies
\be
\begin{split}
	F(u,L,\zeta,\gamma_{c};y) & = \sum_{\gamma_{h}\in\Gamma}\FOmega(u,L_{\zeta},\gamma_{c}+\gamma_{h};y) \, X_{\gamma_{c}+\gamma_{h}} \\
	& = \Tr_{\FCF_{\gamma_{c}}(u,L,\zeta)}\, y^{2 J_{3}}(-y)^{2 I_{3}}\, e^{Q} \,,
\end{split}
\ee
where $X_{\gamma}$ are formal variables realizing the group algebra of $\Gamma$ acting on $\Gamma_{L}$, namely
\be
	X_{\gamma_{c}}X_{\gamma_{h}}=X_{\gamma_{h}}X_{\gamma_{c}}=X_{\gamma_{c}+\gamma_{h}}\qquad X_{\gamma_{h}} X_{\gamma_{h}'}=X_{\gamma_{h}+\gamma_{h}'}\qquad \forall \gamma_{c}\in\Gamma_{L};\ \gamma_{h},\gamma_{h}'\in\Gamma\,.
\ee
We denoted by $\FCF_{\gamma_{c}}{(u,L,\zeta)}$ the Fock space of framed BPS states of core charge $\gamma_{c}$, while $Q$ is a linear operator on this Fock space which evaluates to $\log X_{\gamma}$ on a state of charge $\gamma$. The fact that $F$ is expressed as a trace over the Fock space of framed states, together with the halo-creation/decay mechanism explained in \cite[\S 3.4]{GMN3}, entail that across a BPS wall
\be\label{eq:physical-spin-formula}
	F^{\pm} = F^{\mp}\,\prod_{\gamma_{h}}\prod_{m\in\IZ}\ \prod_{m'=- 2J_{\gamma_{c},\gamma_{h}}}^{2J_{\gamma_{c},\gamma_{h}}}\ \Big(1 + (-y)^{m}\, y^{m'} X_{\gamma_{h}} \Big)^{a_{m}(\gamma_{h})}
\ee
where $2J_{\gamma_{c},\gamma_{h}}+1=|\langle\gamma_{c},\gamma_{h}\rangle|$ is the dimension of the $\so(3)$ irrep accounting for the ``orbital'' degrees of freedom of the halo.
It is worth noting that $a_{m}(\gamma)\geq 0\ \forall m$ if the vanilla BPS state in question is a fermion, while $a_{m}(\gamma)\leq 0\ \forall m$ for bosons\footnote{The Clifford vacua $\mathfrak{h}_{\gamma}=(j,j_{R})$ of bosons have $j$ half-integer, while fermions have integer $j$. An interpretation for this shift can be found in \cite{Denef:2002ru,Denef:2007vg}.}.

\medskip

Having reviewed the definition of BPS states, we shall now review how their counting goes. There are different approaches to this problem: on the one hand these states admit a semiclassical description (see \cite{Tong,MRV3}), on the other hand the six-dimensional engineering of line defects \cite{GMN3,GMN4,GMN5} has been successfully exploited to derive general expressions for generating functions in class $\CS$ theories of the $A$ type. In the rest of this paper we will introduce and study a generalization of the second approach, we therefore end this section by reviewing this technique.

A class $\CS$ theory of the $A_{k}$ type is defined by a punctured Riemann surface $C$ together with some data $D$ at the punctures \cite{Klemm:1996bj,Witten:1997sc,Gaiotto:2009we,GMN2}, we will sometimes refer to $A_{k},C,D$ as the ``UV data'' of the theory. These objects define a classical integrable system (the Hitchin system) $\CM_{H}$ together with a fibration (the Hitchin fibration) by Lagrangian tori over $\CB_{H}$ (the Hitchin base). In the context of gauge theories, $\CB_{H}\equiv \CB$ is identified with the Coulomb branch of the four-dimensional theory in question. At each $u\in\CB$, the spectral curve of the Hitchin system $\Sigma_{u}\subset T^{*}C$ is identified with the Seiberg-Witten curve which captures the low-energy dynamics of the gauge theory. The tautological 1-form $\lambda$ in $T^{*}C$ plays the role of the SW differential.
The canonical projection $\pi:\Sigma_{u}\to C$ defines a ramified covering of $C$. To this covering is associated a one-parameter family of spectral networks $\CW(u,\vartheta)$ \cite{GMN5}. Loosely speaking, a spectral network is a collection of oriented paths -- termed \emph{streets}  -- on $C$ carrying certain \emph{soliton data}; both the shape of the streets and the soliton data are determined by a set of rules; it will be important in the following that such paths can be lifted to $\Sigma$ in a way dictated by the data they carry. We will not provide a review of spectral networks, we refer the reader to the original paper \cite{GMN5} or to \cite{FelixKlein,WWC} for self-contained presentations.

Spectral networks are useful for several reasons. From the mathematical viewpoint they establish a local isomorphism between moduli spaces of flat connections known as the \emph{nonabelianization map}. From a physical point of view they give a means to compute BPS spectra of various types, including as the ``vanilla'' and ``framed'' spectra.

Given a network with its soliton data, the counting of framed BPS degeneracies is relatively simple. Given two surface defects $\mathbb{S}_{z},\, \mathbb{S}_{z'}$  \cite{Alday:2009fs,surf-op2,Gaiotto:2009fs,GMN4,GMN5} localized at $x^{1}=x^{2}=0$ in spacetime, a UV susy interface interpolating between them is associated to a relative homotopy class\footnote{Actually, $\wp$ is a relative homotopy class on the unit tangent bundle of $C$, as explained in \cite{GMN5}. For simplicity we suppress this detail for the moment.} $\wp\subset C$. At fixed values of $(u,\vartheta)$ the framed degeneracies for the interface $L_{\wp}$ are determined by the combinatorics of ``detours''
\be
	F(\wp,u,\vartheta) = \sum_{i,j'}\sum_{\pi\in\Gamma_{ij'}(z,z')} \FOmega(L_{\wp},u,\vartheta,\pi) \, X_{\pi}
\ee
where we extended the formal $X$ variables to take values in the \emph{homology path algebra} (see \cite[\S 2]{WWC}). The first sum runs over all pairs of sheets of the covering $\Sigma$ (a choice of local trivialization is understood), and the second sum is determined by the soliton data on the streets of $\CW$ crossed by $\wp$. Each $\pi$ is associated with a \emph{2d-4d framed BPS state} (see \cite[\S 4]{GMN5} also for the notation $\Gamma_{ij'}(z,z')$), the $\FOmega$ are the corresponding framed degeneracies.

As the parameter $\vartheta\in \IR/2\pi\IZ$ varies, $\CW(u,\vartheta)$ undergoes a smooth evolution, except at special values of $\vartheta$ for which the topology of the network jumps. These jumps occur precisely when $\vartheta=\Arg(Z_{\gamma}(u))$, where $\gamma$ is the charge of some populated (vanilla) BPS state. At a generic point $u\in \CB$ this singles out a one-dimensional sublattice $\Gamma_{c}\subset\Gamma$. This phenomenon is key to extracting the kind of BPS degeneracies of interest to us, and occurs precisely at the BPS walls (also termed $\CK$-walls in \cite{GMN5}). \\

\begin{figure}[h!]
\begin{center}
\includegraphics[width=0.30\textwidth]{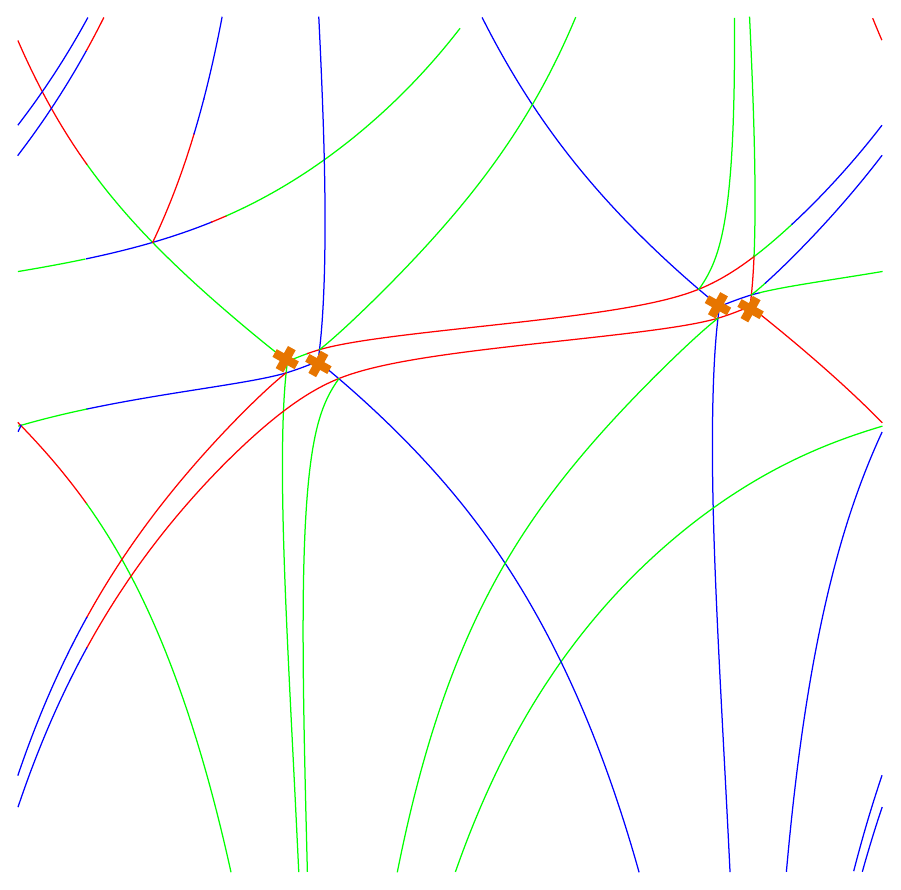}\hfill\includegraphics[width=0.30\textwidth]{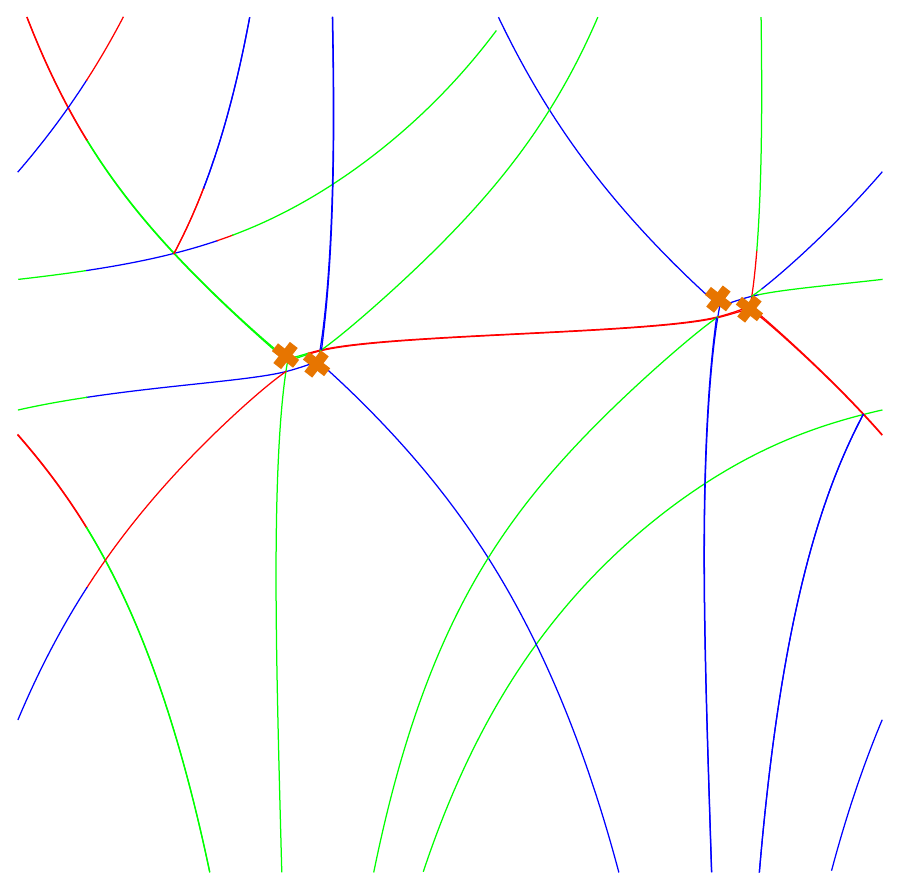}\hfill\includegraphics[width=0.30\textwidth]{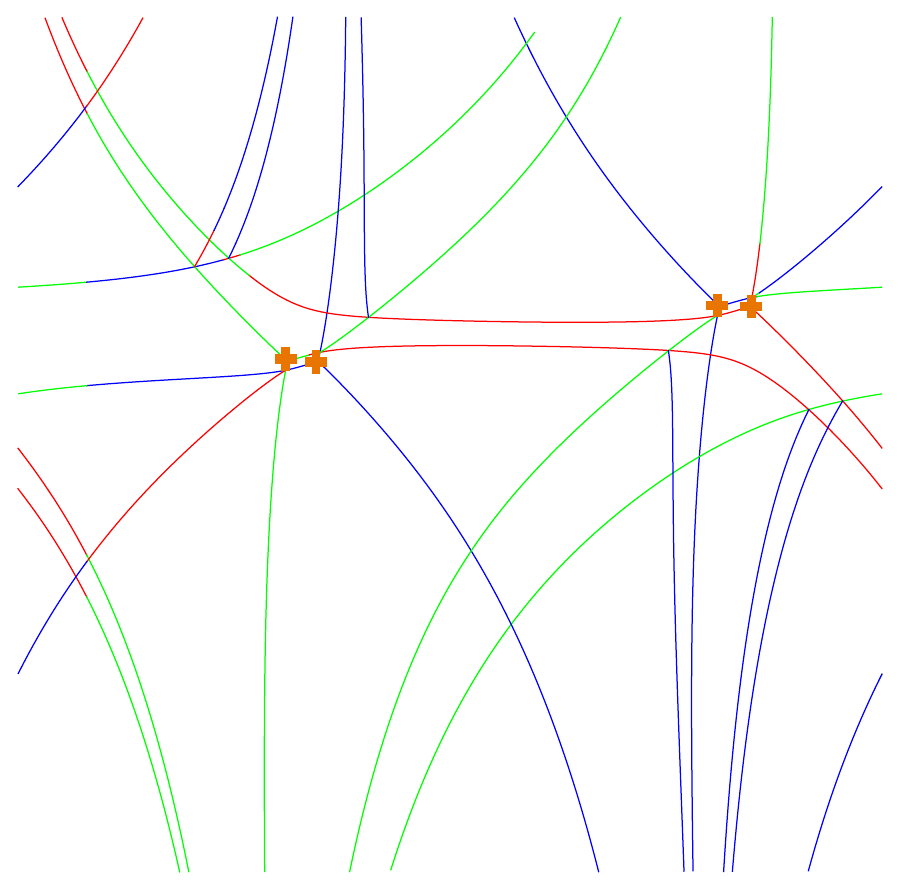}
\caption{The $\CK$-wall jump from a hypermultiplet: different colors represent streets carrying solitons of different types ($ij,jk,ki$-types).}
\label{fig:flip}
\end{center}
\end{figure}

More specifically, at a critical value $\vartheta=\vartheta_{c}$ a part of the network becomes ``degenerate,'' literally two or more streets of type $(ij)$ and $(ji)$ overlap completely, we call this part of the network $\CW_{c}\subset\CW$. The soliton data and the topology of the subnetwork $\CW_{c}$  define a discrete set (possibly infinite) of closed paths on $\Sigma$, usually indicated by $\{L(\gamma)\}_{\gamma\in\Gamma_{c}}$ whose homology class is
\be\label{eq:line-defect-property}
	[L(\gamma)]=\Omega(\gamma)\cdot \gamma\,.
\ee
The generating functions of framed degeneracies jump across $\CK$-walls, in a way that is captured by a universal substitution rule of the twisted formal variables $Y_{a}$\footnote{The $Y$ are related to the $X$ by a choice of quadratic refinement of the charge lattice(s). They obey a twisted algebra, e.g. $Y_{\gamma}Y_{\gamma'}=(-1)^{\langle\gamma,\gamma'\rangle}Y_{\gamma+\gamma'}$. For more details we refer the reader to \cite[\S 2]{WWC}.}
\be\label{eq:K-wall-classical}
	Y_{a} \quad\mapsto\quad Y_{a}\prod_{\gamma\in\Gamma_{c}}(1-Y_{\gamma})^{\langle a,L(\gamma)\rangle}\,.
\ee
The vanilla BPS indices control the \emph{change} of framed BPS indices with the variation of $\vartheta$ across a $\CK$-wall. Examples illustrating this mechanism can be found in \cite{GMN5,WWC}.

\subsection{Framed spin and writhe}\label{sec:framed-spin}

\subsubsection{Statement of the conjecture}\label{sec:framed-PSC-conjecture}
As reviewed above, for any UV susy interface $L_{\wp}$ the corresponding 2d-4d framed BPS states at $(u,\vartheta)$ are associated with relative homology classes of detours $\pi\in\Gamma_{ij'}(z,z')$.

We now introduce a refinement of the classification of paths that will be of central importance for the rest of the paper. Let $I=[0,1]$ be the unit interval parametrized by $t$, and consider an immersion $f:I\to X$ into a Riemann surface $X$, namely a smooth map such that $f_{*}:T_{t}I\to T_{f(t)}X$ is injective (i.e. the path never has zero velocity). A regular homotopy between two immersions is a homotopy through immersions. For \emph{open} paths we additionally require that the tangent directions at endpoints be constant throughout the homotopy.
This defines an equivalence relation: a \emph{regular homotopy class} is an equivalence class. 

From now on we will be assuming that all self-intersections of paths are transverse. Choose $\wp$ to be any \emph{regular homotopy class} on $C^{*}$ with endpoints $z,z'$. 
We define the following spaces
\be
	C^{*} := C \setminus \{z,z'\}\,, \qquad \Sigma^{*} := \Sigma \setminus \pi^{-1}(\{z,z'\})\,.
\ee
The detours of $\wp$ can be likewise classified by regular homotopy classes on $\Sigma^{*}$, because the network contains more information than just relative homology for soliton paths (we will further clarify this point below). We will adopt gothic letters such as $\fp$ to denote regular homotopy classes of detours on $\Sigma^{*}$, the refinement just introduced allows us to perform the assignment
\be
	\fp \quad\mapsto \quad \wr(\fp)
\ee
where $\wr$ is the \emph{writhe} of $\fp$, defined as a signed sum of $\pm1$ over transverse self-intersections of $\fp$. In the parametrization associated with regular homotopy $\fp:[0,1]\to\Sigma^{*}$, a self-intersection is a point where $\fp(t_{1})=\fp(t_{2})$. For $t_{2}>t_{1}$ the associated sign is positive if the tangent vector $\dot\fp(t_{1})\in T_{\fp(t_{1})}\Sigma^{*}$ points clockwise from $\dot\fp(t_{2})$ (in the short-way around); the sign is negative otherwise.
Let $\bGamma_{ij'}(z,z')$ be the set of regular homotopy classes of paths $\fp$ on $\Sigma^{*}$ which begin at $z^{(i)}$ and end at ${z'}^{(j')}$. There is a natural map
\be
	\beta:\, \bGamma_{ij'}(z,z') \, \longrightarrow \tGamma_{ij'}(z,z')
\ee
which identifies different regular homotopy classes $\fp$ belonging to the same relative homology class $\pi$ on the unit tangent bundle of $\Sigma^{*}$, which we denote $\tSigma^{*}$. 
%
%
Correspondingly, we also define $\tGamma:=H_{1}(\tSigma^{*},\IZ)$, then $\tGamma_{ij'}(z,z')$ is a torsor for this lattice\footnote{To avoid possible confusion, let us note that if $\fp',\fp$ differ by $n$ \emph{counter-clockwise} \emph{contractible curls} (defined below in \S \ref{sec:twisted-variables}) then $\wr(\fp')=\wr(\fp)-n$ while $\beta(\fp')=\beta(\fp)+n H$, according to our conventions.}.
%
%
%
%
\noindent We are now ready to state our conjecture on the spin of framed BPS states.

\begin{conj}\label{conj:framed-spin}
If a framed 2d-4d BPS state for an interface determined by $\wp$ is represented by a regular homotopy class $\fp$ then its spin $2J_{3}$ is the writhe of $\fp$.
\end{conj}

Recall that the surface defects choose a direction in space and $J_3$ is
defined as the spin around this direction.
Therefore, given a spectral network $\CW$ and an interface $\wp$, we define
\be\label{eq:framed-spin-conj}
\begin{split}\label{eq:framed-degeneracies}
	& \FOmega(L_{\wp},u,\vartheta,\fp)\,:=\, \#\text{ of detours in class }\fp\,, \\
	& \FOmega(L_{\wp},u,\vartheta,\pi;y) := \sum_{\fp\,|\,\beta(\fp)=\pi} y^{\wr(\fp)} \, \FOmega(L_{\wp},u,\vartheta,\fp)\,,
\end{split}
\ee
together with
\be\label{eq:framed-spin-functional}
	F(\wp,u,\vartheta;y) := \sum_{i,j'}\sum_{\pi\in\tGamma_{ij'}(z,z')} \FOmega(L_{\wp},u,\vartheta,\pi;y) \, \hY_{\pi}\,,
\ee
where $\hY$ are formal variables associated with relative homology classes $\pi\in H_{1}^{rel}(\tSigma^{*},\tSigma\setminus\tSigma^{*})$ as well as with homology classes in $\tgamma\in H_{1}^{}(\tSigma^{*},\IZ)$ subject to the relations
%
%
\be
	\hY_{\pi}\hY_{\tgamma}=y^{\langle\pi,\tgamma\rangle}\,\hY_{\pi+\tgamma} \qquad \hY_{\tgamma}\hY_{\pi}=y^{\langle\tgamma,\pi\rangle}\,\hY_{\pi+\tgamma} \qquad \hY_{\tgamma}\hY_{\tgamma'}=y^{\langle\tgamma,\tgamma'\rangle}\,\hY_{\tgamma+\tgamma'}\,,
\ee
as well as
\be
	\hY_{\pi+n H} = (-y)^{n} \hY_{\pi}\qquad \hY_{\tgamma+n H} = (-y)^{n} \hY_{\tgamma} \,,
\ee
where $H$ is the generator of $H_{1}(\tSigma^{*},\IZ)$ corresponding to a cycle wrapping the fiber, going counter-clockwise (this implies a choice of orientation on $\Sigma^{*}$).


Although not obvious from these definitions, we will prove below in section \ref{subsec:F-rhom-inv} that $F(\wp,u,\vartheta;y)$ only depends on the regular homotopy class of $\wp$ on $C^{*}$.

\subsection{Vanilla Protected Spin Characters from Spectral Networks}

The main goal of this paper is to propose (vanilla/framed) PSC formulas based on spectral network data. Our approach will be to identify susy interfaces whose framed wall-crossing is described by formula (\ref{eq:physical-spin-formula}). By considering the ``classical limit'' $y\to -1$,  it is clear that this formula won't hold for generic interfaces, we therefore need to focus on a specific subset of interfaces, which we now define.

\subsubsection{A special class of susy interfaces}\label{subsubsec:spec-lin-int}
To understand the motivations behind the definition to come, it is instructive to dissect and compare formulae (\ref{eq:physical-spin-formula}) and (\ref{eq:K-wall-classical}).

To make a meaningful comparison, we shall take $F^{-}=X_{\gamma_{c}}$ meaning that there are no halo states with core charge $\gamma_{c}$ at $\vartheta^{-}$ (here $\vartheta=\Arg(\zeta)$). The classical limit of (\ref{eq:physical-spin-formula}) reads
\be
		F^{+} = X_{\gamma_{c}}\,\prod_{\gamma_{h}} \Big(1 - (-1)^{\langle\gamma_{c},\gamma_{h}\rangle} X_{\gamma_{h}} \Big)^{|\langle\gamma_{c},\gamma_{h}\rangle|\Omega(\gamma_{h})}\,,
\ee
switching to twisted variables \cite{GMN3,WWC} the above reads
\be
		F^{+} = Y_{\gamma_{c}}\,\prod_{\gamma_{h}} (1 - Y_{\gamma_{h}} )^{|\langle\gamma_{c},\gamma_{h}\rangle|\Omega(\gamma_{h})}\,.	
\ee
Now taking $[a]=\gamma_{c}$ to be a closed cycle with a basepoint on $\Sigma$, we find the expected match between the two formulae, since (\ref{eq:line-defect-property}) ensures that
\be\label{eq:defect-pairing-property}
	\langle \gamma_{c},L(\gamma)\rangle  \, = \, \langle\gamma_{c},\gamma\rangle \,\cdot \Omega(\gamma).
\ee
On the other hand, for general $a$ there is no relation between $\langle a,L(\gamma)\rangle$ and $\Omega(\gamma)$.

As remarked in \cite[\S 6.4]{GMN5}, this reflects the fact that $L(\gamma)$ contains more information than the charge of 4d BPS degeneracies, such as how they are arranged on $C$ (the charge is a homology class, while $L(\gamma)$ are exact paths).
In appendix \ref{app:L-r} we suggest a physical interpretation of this phenomenon in terms of halos formed by single states of $\mathfrak{h}_{\gamma}$.

\medskip

Now consider a critical subnetwork $\CW_{c}\subset\CW(u,\vartheta_{c})$. In the British resolution ($\vartheta=\vartheta_{c}-\epsilon$) each two-way street $p\in\CW_{c}$ carries two soliton data sets. While in \cite{GMN5} soliton data is classified by relative homology classes, there is much more information available in the network. The requirement of homotopy invariance regulates the propagation of soliton paths across streets of the networks, in a way described by the \emph{six-way joint rules} \cite[app.A]{GMN5} \cite{WWC}. Keeping track of the joints involved in the propagation of a soliton path, it is therefore possible to associate to each soliton an oriented curve made of lifts of streets%
\footnote{Actually, homotopy invariance is employed in \cite{GMN5} to establish a 2d wall-crossing formula for solitons classified by \emph{relative homology classes} on $\tSigma$. However we will show below in section \ref{sec:joint-rules} that the \emph{same} set of 6-way rules -- now applied to regular homotopy classes of soliton paths -- follows from studying regular-homotopy invariance of a certain formal parallel transport.}.
We consider it up to regular homotopy and refer to this refined object as a \emph{soliton path}, while we preserve the terminology \emph{soliton classes} for the relative homology classes.

Let $p$ be an $ij/ji$-type two way street of $\CW_{c}$, one may join any soliton path from the $ij$-type soliton data set with any soliton path from the $ji$-type data set to make a closed path. We denote by $\Pi(p)$ the set of all combinations of soliton paths from the two data sets of $p$, classified by regular homotopy (as closed paths, i.e. without a basepoint specified) on $\Sigma^{*}$. A generic element $\ell \in \Pi$ will thus be a class of closed oriented curves on $\Sigma$. By genericity its homology class will belong to the sublattice associated with the $\CK$-wall
\be
	[\ell] \in \Gamma_{c}\,.
\ee
We also define
\be
	\Pi(\CW_{c}) := \bigcup_{p\in\CW_{c}}  \Pi(p)\,.
\ee

For any UV susy interface $L_{\wp}$, we may consider the lifts of $\{\wp^{(i)}\}=\pi^{-1}(\wp)\subset\Sigma$. We will say that $\wp$ is \emph{halo-saturated} if at least one of its lifts satisfies
\be\label{eq:saturation-def}
\begin{split}
	(I) \ & \quad \langle \wp^{(i)} , \ell \rangle \neq 0 \\
	(II) \ & \quad \frac{\langle \wp^{(i)} , \ell \rangle}{\langle \wp^{(i)} , \ell' \rangle}  = \frac{[\ell]}{[\ell']}
\end{split}\qquad\qquad \forall \ell,\ell'\in\Pi(\CW_{c})\,.
\ee
This is our special class of susy interfaces, their essential feature is that in a neighborhood of the $\CK$-wall of interest, their $4d$ framed wall-crossing is the \emph{same} as that of a suitable line \emph{defect} (the $2d$ framed wall-crossing may be different though). 

For a halo-saturated interface $\wp$, and halo charge $\gamma_{h}\in\Gamma$, choose any $\ell\in\Pi(\CW_{c})$ such that $[\ell]=\gamma_{h}$ then we define
\be
	J_{\wp^{(i)},\gamma_{h}} := \frac{1}{2}\left(|\langle \wp^{(i)},\ell \rangle|  -1  \right)\,.
\ee

\subsubsection{The vanilla PSC formula}\label{sec:VanillaPSC}
Let $\wp$ be a halo-saturated susy interface for $\CW_{c}$, with $\wp^{(i)}$ being the lift satisfying (\ref{eq:saturation-def}). Note that $\wp^{(i)}$ provides a trivialization for the torsor $\tGamma_{ij'}(z,z')$ (hence an isomorphism with $\tGamma$). 
%
%
In particular, we may single out a certain sub-torsor $\tGamma_{c,\wp^{(i)}}\simeq [\wp^{(i)}]+\tGamma_{c} \subset \tGamma_{ij'}(z,z')$, where $\tGamma_{c}\subset\tGamma$ is the critical sublattice\footnote{Let $\tilde\Gamma_{c}$ be the preimage of $\Gamma_{c}$ under the natural map $\tilde\Gamma\to\Gamma$. There is also a natural map $\tGamma\to\tilde\Gamma$ obtained by filling the circle fibers above $\Sigma\setminus\Sigma^{*}$, then $\tGamma_{c}$ is the preimage of $\tilde\Gamma_{c}$. } %
at $\vartheta_{c}$.
Considering the related restriction\footnote{An explicit example will be provided below: the first line of (\ref{eq:pureSU2-full-spin}) contains the full partition function (hence being the $ii$-component of the counterpart of (\ref{eq:framed-spin-functional})), while the LHS of (\ref{eq:su2-expected}) is the corresponding restriction to the sub-torsor determined by $\wp^{(i)}$ (the counterpart of (\ref{eq:restricted-framed-spin-functional})). The remaining terms in (\ref{eq:pureSU2-full-spin}) don't appear in (\ref{eq:su2-expected}) since they clearly do not belong to the sub-torsor: their homology classes are not of the form $[\wp^{(i)}]+\tgamma_{c}$ with $\tgamma_{c}\in\tGamma_{c}$.} of the partition function of framed BPS states (\ref{eq:framed-spin-functional}):
\be\label{eq:restricted-framed-spin-functional}
	F_{\wp^{(i)}}(\wp,u,\vartheta;y) := \sum_{\pi\in\tGamma_{c,\wp^{(i)}}} \FOmega(L_{\wp},u,\vartheta,\pi;y) \, \hY_{\pi}\,,
\ee
we can formulate our second conjecture.

\begin{conj}
As $\vartheta$ varies across the $\CK$-wall there exist integers $\{a_{m}(\tgamma_{h})\}_{m\in\IZ}$ such that
%
\be\label{eq:PSC-conjecture}
	F_{\wp^{(i)}}(\wp,u,\vartheta_{c}^{+};y) = F_{\wp^{(i)}}(\wp,u,\vartheta_{c}^{-};y) \,\left[\prod_{\tilde\gamma_{h}}\prod_{m\in\IZ}\ \Phi_{ n(\tgamma_{h}) } \Big( (-y)^{m}\,\hY_{\tilde\gamma_{h}} \Big)^{a_{m}(\tilde\gamma_{h})}\right]^{\pm1}\,,
\ee
%
moreover the $a_{m}(\tgamma_{h})$ only depend on $\gamma_{h}$, and they are precisely the Laurent coefficients of $\Omega(u,\gamma_{h};y)$\footnote{See (\ref{eq:PSC}).}.
\end{conj}
\noindent The $\Phi_{n}(\xi)$ are finite-type dilogarithms
\be
	\Phi_{n}(\xi) := \prod_{s=1}^{|n|}(1+y^{-{\rm sgn}(n)\, (2s-1)}\xi)\,,
\ee
and 
\be
	n(\tgamma_{h})=2J_{\wp^{(i)},\tilde\gamma_{h}}+1\,.
\ee
The sign is determined by the direction in which the $\CK$-wall is crossed: a framed BPS state of halo charge $\gamma_{h}$ is stable on the side where the Denef radius $\langle \gamma_{h},\wp^{(i)}\rangle\,/\,2\, \Im (Z_{\gamma_{h}}/e^{i\vartheta})$ is positive, the sign is therefore positive when going from the unstable side to the stable one and vice versa\footnote{further  details can be found in \cite{GMN3}.}.

\medskip

The practical value of this conjecture comes from taking (\ref{eq:framed-spin-functional}) into account at the same time. The latter allows to compute $F(\wp,u,\vartheta_{c}^{\pm};y)$, while (\ref{eq:PSC-conjecture}) states how to extract the $a_{m}(\gamma)$ (Laurent coefficients of the PSC). In Section \ref{sec:applications} we will provide supporting evidence for these conjectures.

\subsubsection{Framed spin wall-crossing of generic interfaces}\label{sec:generic-interfaces}
Halo-saturated interfaces are just a special class of susy interfaces, it is natural to ask whether we can say something about the framed wall-crossing of more generic choices. 
Our conjecture \ref{conj:framed-spin} offers a partial answer to this: the $2J_{3}$ eigenvalue of a framed BPS state is still identified with the writhe of the corresponding detour. The conjecture doesn't restrict to halo-saturated interfaces.

A crucial property of our special class of interfaces is that it allows to extract the \emph{vanilla} PSC of states associated with halo particles. Generic interfaces instead are not guaranteed to capture this information, this fact is unrelated to the counting of spin and was evident already in the classical story \cite{GMN4,GMN5,FelixKlein,2d4d}. The simplest example of what could go wrong is provided by a ``bare'' interface: tuning the moduli $(u,\zeta)$ to vary within a sufficiently small region near a $\CK$-wall, we may choose an interface which doesn't intersect with the network for any value of the moduli; certainly as the $\CK$-wall is crossed, this interface wouldn't capture information of vanilla PSC's, because it lacks halos of any sort.
While simple, this example points to an essential difference between interfaces and defects: the pairing $\langle \gamma_{c},\gamma_{h}\rangle$ between an (infrared) defect of charge $\gamma_{c}$ and a halo particle of charge $\gamma_{h}$ is a topological quantity, it can't be smoothly deformed to zero; on the contrary the intersection pairing $\langle a,\gamma_{h}\rangle$ between a halo charge and an (infrared) interface $a$ is well-defined on the respective homology classes only \emph{after} the endpoints of $a$ are deleted from $\Sigma$. More concretely, let us look back at equation (\ref{eq:K-wall-classical}), which applies both to IR line defects and interfaces. The wall-crossing of an IR defect of charge $\gamma_{0}\in\Gamma$ will be governed by $\langle \gamma_{0},L(\gamma)\rangle= \Omega(\gamma)\, \langle \gamma_{0},\gamma\rangle$ for $\gamma\in\Gamma_{c}$. On the other hand for an interface $\langle a,L(\gamma)\rangle$ cannot be cast into the form $\Omega(\gamma)\langle a,\gamma\rangle$, precisely because the latter pairing is not well-defined. We will come back to generic interfaces in section \ref{sec:gen_interf}, where we analyze in some detail explicit examples.

\section{Formal parallel transport} \label{sec:FPT}

In this section we describe the construction of a formal parallel transport on the UV curve $C$, employing the data of a flat abelian connection on $\Sigma$ and a spectral network. The discussion parallels closely that of \cite{GMN5}: the transport along a path $\wp$ on $C$ gets corrected  by ``detours'' corresponding to the soliton data on streets crossed by $\wp$; the novelty will consist of keeping track of a suitable refinement of the soliton data.

After defining the formal parallel transport, we show that it enjoys twisted homotopy invariance, thus reproducing the transport by a flat non-abelian connection on $C$. As already noticed in \cite{GMN5}, homotopy invariance is tightly connected to pure 2d wall-crossing, in our context this will lead to a refined version of the 2d WCF.

With respect to the PSC conjectures formulated above, this section's purpose is two-fold. First, we will provide a precise definition
of the generating function of framed PSC's, in terms of detours. Second, we will derive the generalization of the six-way joint rules of \cite[app.A]{GMN5} on which the definition of soliton paths relies.

\subsection{Twisted formal variables}\label{sec:twisted-variables}
Let $C,\Sigma,\CW$ be a triplet consisting of a punctured Riemann surface $C$, a ramified $K$-fold covering $\pi:\Sigma\to C$ and a spectral network subordinate to the covering.
For convenience we will sometimes label the sheets of $\Sigma$, implicitly employing a trivialization of the covering. We will restrict $\CW$ to WKB-type spectral networks, although everything should carry over in a straightforward way to general spectral networks (as defined in \cite[\S 9.1]{GMN5}).
We define
\be
	C^*=C\setminus\{z_1,\ldots,z_n\}\,,\qquad \Sigma^{*} = \Sigma \setminus \pi^{-1}\left(\{z_1,\ldots,z_n\}\right)
\ee
where $\{z_1,\ldots,z_n\}$ is a collection of points (away from the branching locus) with $n\geq 2$.

A \emph{path} on $\Sigma^*$ (or $C^{*}$) will be understood as a \emph{regular homotopy class} of curves on $\Sigma^*$ (resp. $C^{*}$).
%
%
%
%
We will say that two paths $\fa,\fb$ are \emph{composable} into $\fa\fb$ if $\rm{end}{(\fa)}=\rm{beg}{(\fb)}$ and the corresponding tangent directions are equal at that point.

\begin{figure}[h!]
\begin{center}
\includegraphics[width=0.45\textwidth]{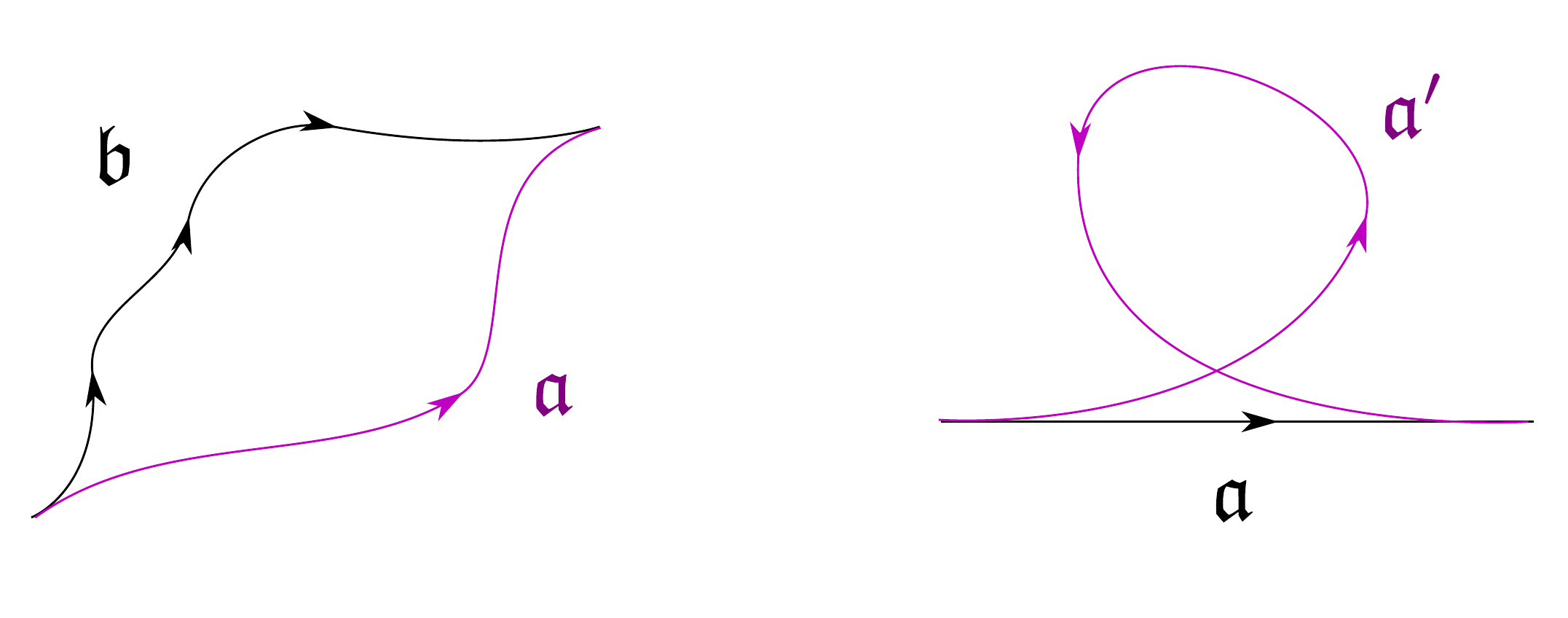}
\caption{On the left: $\fa,\fb$ are regular-homotopic. On the right: $\fa,\fa'$ differ by a contractible curl. Thus $\hU_{\fa}=\hU_{\fb}$ and $\hU_{\fa}=-\hU_{\fa'}$}
\label{fig:path-basics}
\end{center}
\end{figure}

\noindent To each path $\fa$ we associate a formal variable $\hU_{\fa}$, then we consider the unital noncommutative algebra %
%
%
over the ring $\IZ$ generated by the $\hat \Upsilon_{\frak{a}} $ and subject to the following relations:
\begin{enumerate}
\item If $\fa,\fb$ are \emph{regular-homotopic} (see figure \ref{fig:path-basics})  then
\be\label{eq:abelian-reg-hom}
	\hU_{\fa}=\hU_{\fb}
\ee
\item\label{it:concatenation} The product rule
\be\label{eq:product-rule}
	\hU_{\fa}\hU_{\fb}=\left\{ \begin{array}{ll} \hU_{\fa\fb} & \text{if $\fa,\fb$ are composable}\\ 0 &  \text{otherwise}\end{array}\right.
\ee %
\item Two paths $\fa$ and $\fa'$, such that the natural pushforwards of $\beta(\fa)$ and $\beta(\fa')$ to $H_{1}(\Sigma^{*})$ coincide, are said to differ by a \emph{contractible curl} if there exists a regular homotopy which takes $\fa\to\fa'$ except for a sub-interval of the domain $[t,t']\subset[0,1]$, where they differ by a curl (see figure \ref{fig:path-basics}). Contractible curls can be oriented clockwise or counterclockwise, for paths differing by a contractible curl
\be
\label{eq:abelian-curl}
	\hU_{\fa}=-\hU_{\fa'}\,.
\ee
\end{enumerate}
%
%

\subsection{Definition of $\fF(\wp)$: detours}
Let $\wp$ be any path (in the sense specified above) on $C^*$, subject to the condition that
\be
	\{{\rm beg}(\wp),{\rm end}(\wp)\}\,\subset\,C\setminus C^{*}\,.
\ee
We associate a formal parallel transport $\fF(\wp)$, according to the following rules.

When $\wp\cap\CW=\emptyset$
\be
	\fF(\wp) = \fD(\wp):=\sum_{i=1}^{K}\hU_{\wp^{(i)}}\,,
\ee
where $\wp^{(i)}$ are the lifts of $\wp$.

\begin{figure}[h!]
\begin{center}
\includegraphics[width=0.85\textwidth]{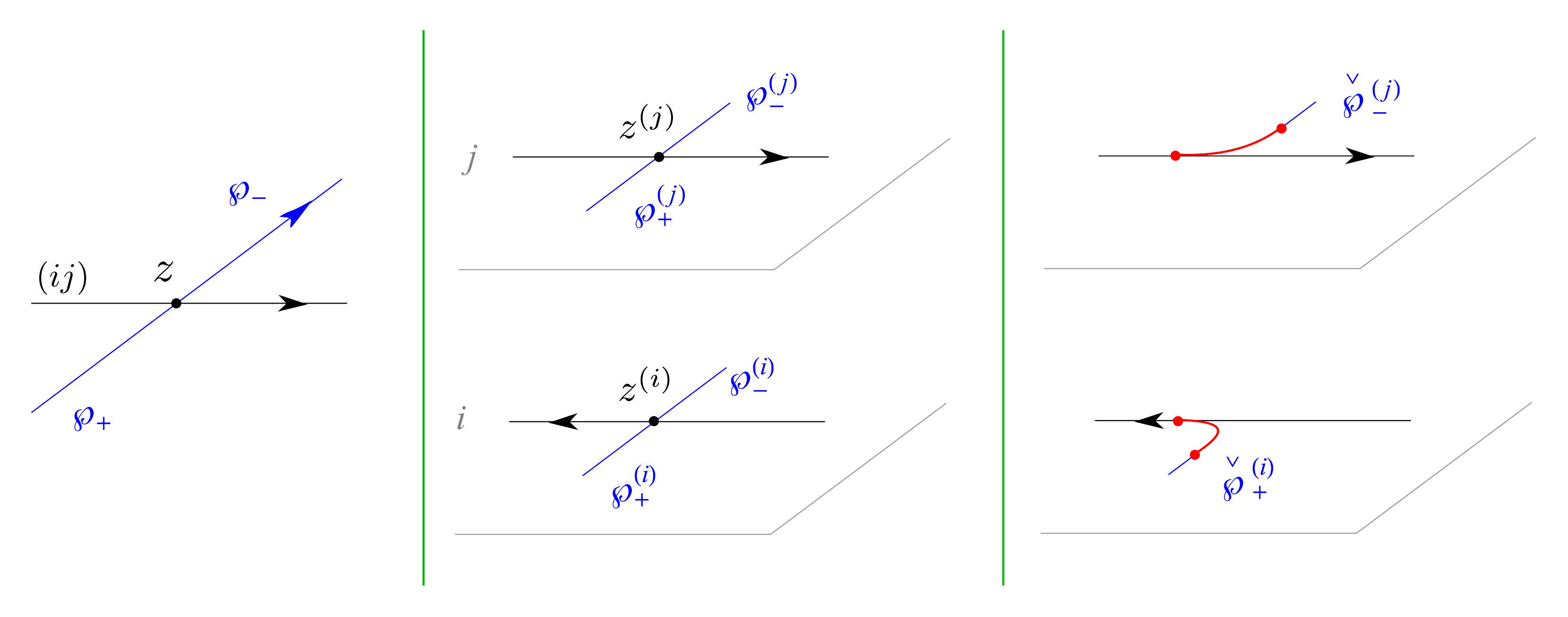}
\caption{Splitting and deforming $\pi^{-1}(\wp)$.}
\label{fig:detour-definition}
\end{center}
\end{figure}

On the other hand, when $\wp$ intersects $\CW$ at some point $z$ on a one-way $ij$ street $p$, it picks up contributions from soliton paths supported on $p$, and we have
\be\label{eq:detour-hU}
	\fF(\wp) := \fD(\wp) + \sum_{\fa\in\bGamma_{ij}(p)}\mu_{r}(\fa,p)\,\hU_{\check\wp_{+}^{(i)}\,\fa\,\check\wp_{-}^{(j)}}\,.
\ee
The sum runs over all regular homotopy classes with endpoints on the lift of $p$, the $\mu_{r}(\fa,p)$ are the refined soliton degeneracies: they are integers associated with soliton paths in each regular homotopy class $\fa\in\bGamma_{ij}(z,z)$ and they are constant along $p$. The $\mu_{r}$ are uniquely determined by rules that will be presently discussed. In analogy with \cite[\S 3.5]{GMN5}, there is a relation between $\fa,\fa'$ which differ by a contractible curl
\be\label{eq:soliton-sign}
	\mu_{r}(\fa,p)=-\mu_{r}(\fa',p)\,.
\ee
The $\check\wp_{\pm}^{(\alpha)},\, \alpha=i,j$ are defined by splitting $\wp$ at $z$ into $\wp_{\pm}$ and considering a deformation of the lifts that matches the initial/final tangent directions of the soliton path $\fa$ on sheets $i,j$ of $\Sigma^*$; this is illustrated in figure \ref{fig:detour-definition}.
%
%
%
%
%
%

Before moving on, let us introduce a convenient piece of notation: in order to deal with transports crossing several streets, it will sometime be convenient to rewrite (\ref{eq:detour-hU}) as
\be
	\fF(\wp)\, = \,\fD(\wp_{+})\,\left( 1+ \sum_{\fa}\mu_{r}(\fa,p)\,\hU_{\fa}  \right)\,\fD(\wp_{-})\,.
\ee

\subsection{Twisted homotopy invariance}\label{sec:twisted-homotopy}
We now study the constraints of twisted homotopy invariance for the formal parallel transport. More precisely, for any path $\wp$ on $C^*$, we require $F(\wp)$ to depend only on the regular homotopy class of $\wp$. Similarly to the classical case \cite{GMN5}, this requirement will induce constraints on the refined soliton content of the network. In fact, the whole analysis we will carry out is very close to that of \cite{GMN5}, the only difference is that instead of relative homology classes on the circle bundle $\tSigma$ (resp. $\widetilde C$), we work with regular homotopy classes on $\Sigma^{*}$ (resp. $C^{*}$).

\subsubsection{Contractible curl}
\begin{figure}[h!]
\begin{center}
\includegraphics[width=.20\textwidth]{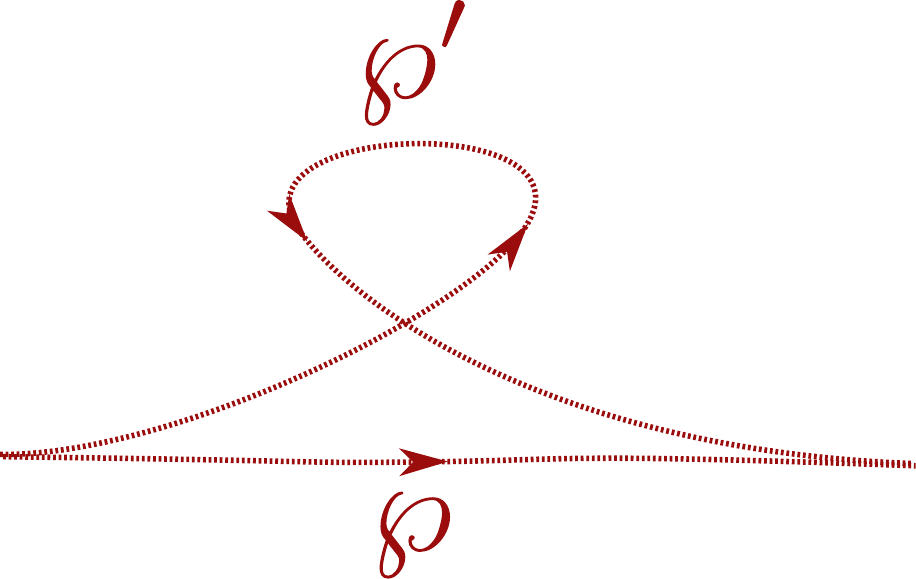}
\caption{Two paths in the same relative homology class on $\Sigma$, which are not regular-homotopic.}
\label{fig:curl}
\end{center}
\end{figure}

Before we get to actual twisted homotopy invariance, let us briefly illustrate the meaning of ``twisting''. For the paths depicted in figure \ref{fig:curl}, we have
\be
\begin{split}
	& \fF(\wp) = \fD(\wp) = \sum_{i}\hU_{\wp^{(i)}}\,, \qquad  \fF(\wp') = \fD(\wp') = \sum_{i}\hU_{{\wp'}^{(i)}}
\end{split}
\ee
Where $\wp^{(i)},\, {\wp'}^{(i)}$ are regular homotopy classes on $\Sigma$, corresponding to the lifts of $\wp,\, \wp'$. Since $\hU_{{\wp'}^{(i)}} = - \hU_{{\wp}^{(i)}}$, the formal transports are simply related as
\be
	\fF(\wp) = -\fF(\wp')\,.
\ee

\subsubsection{Homotopy across streets}
\begin{figure}[h!]
\begin{center}
\includegraphics[width=.5\textwidth]{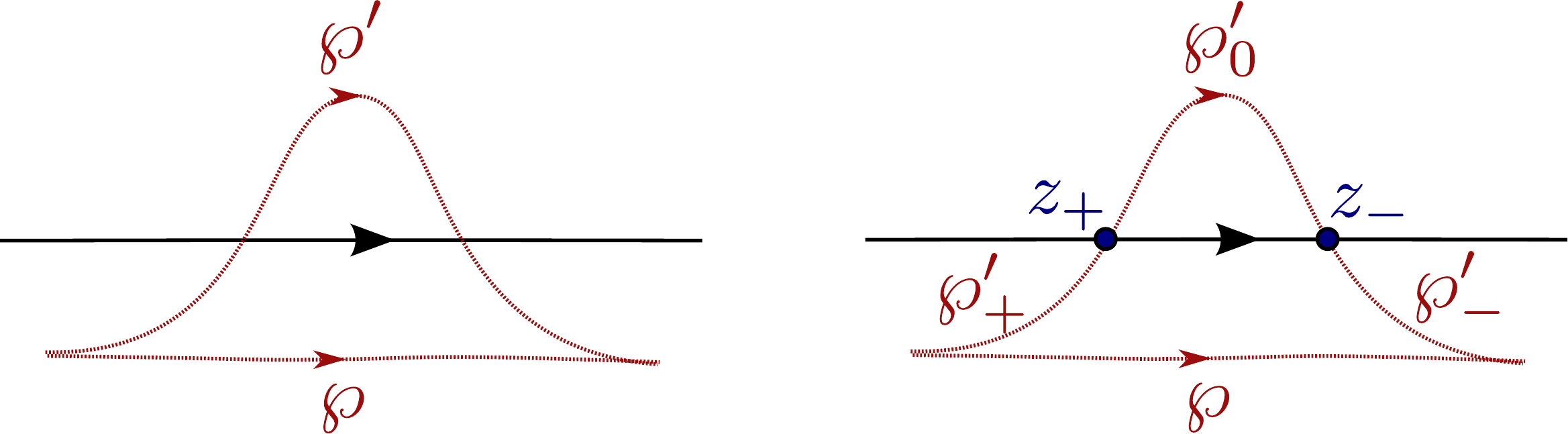}
\caption{Paths differing by a regular homotopy across a street of the network.}
\label{fig:bump}
\end{center}
\end{figure}

The simplest homotopy requirement to take into account is the one shown in figure \ref{fig:bump}, where a path $\wp$  is homotoped to $\wp'$ across a one-way street $p$ of $ij$ type.

\noindent The transports read
\be
\begin{split}
	\fF(\wp) & = \fD(\wp) \\
	\fF(\wp') & = \fD(\wp'_{+}) %
	\Big( 1+ \sum_{\fa}\mu_{r}(\fa,p) \, \hU_{\fa_{z_{+}}}\Big) %
	\fD(\wp'_{0})%
	\Big( 1+\sum_{\fb} \mu_{r}(\fb,p) \, \hU_{\fb_{z_{-}}}\Big) %
	\fD(\wp'_{-}) \\
	& = \fD(\wp') + \sum_{\fa}\mu_{r}(\fa,p) \,%
	\Big(  \hU_{{\wp'}_{+}^{(i)}}   \hU_{\fa_{z_{+}}}   \hU_{{\wp'}_{0}^{(j)}}  \hU_{{\wp'}_{-}^{(j)}}   %
	+  \hU_{{\wp'}_{+}^{(i)}}  \hU_{{\wp'}_{0}^{(i)}}   \hU_{\fa_{z_{-}}}   \hU_{{\wp'}_{-}^{(j)}}  \Big) \\
	& = \fD(\wp')
\end{split}
\ee
where, in the last step, we made use of (\ref{eq:product-rule}) and (\ref{eq:abelian-curl}), given that ${{\wp'}_{+}^{(i)}}\fa_{z_{+}}{\wp'}_{0}^{(j)}{\wp'}_{-}^{(j)}$ and ${{\wp'}_{+}^{(i)}}  {{\wp'}_{0}^{(i)}}   {\fa_{z_{-}}}   {{\wp'}_{-}^{(j)}}$ differ precisely by a contractible curl. Since $\fD(\wp)\equiv \fD(\wp')$ by virtue of (\ref{eq:abelian-reg-hom}), this establishes invariance of the formal transport.

\subsubsection{Branch Point}\label{sec:branch-pt-hom}
\begin{figure}[h!]
\begin{center}
\includegraphics[width=.33\textwidth]{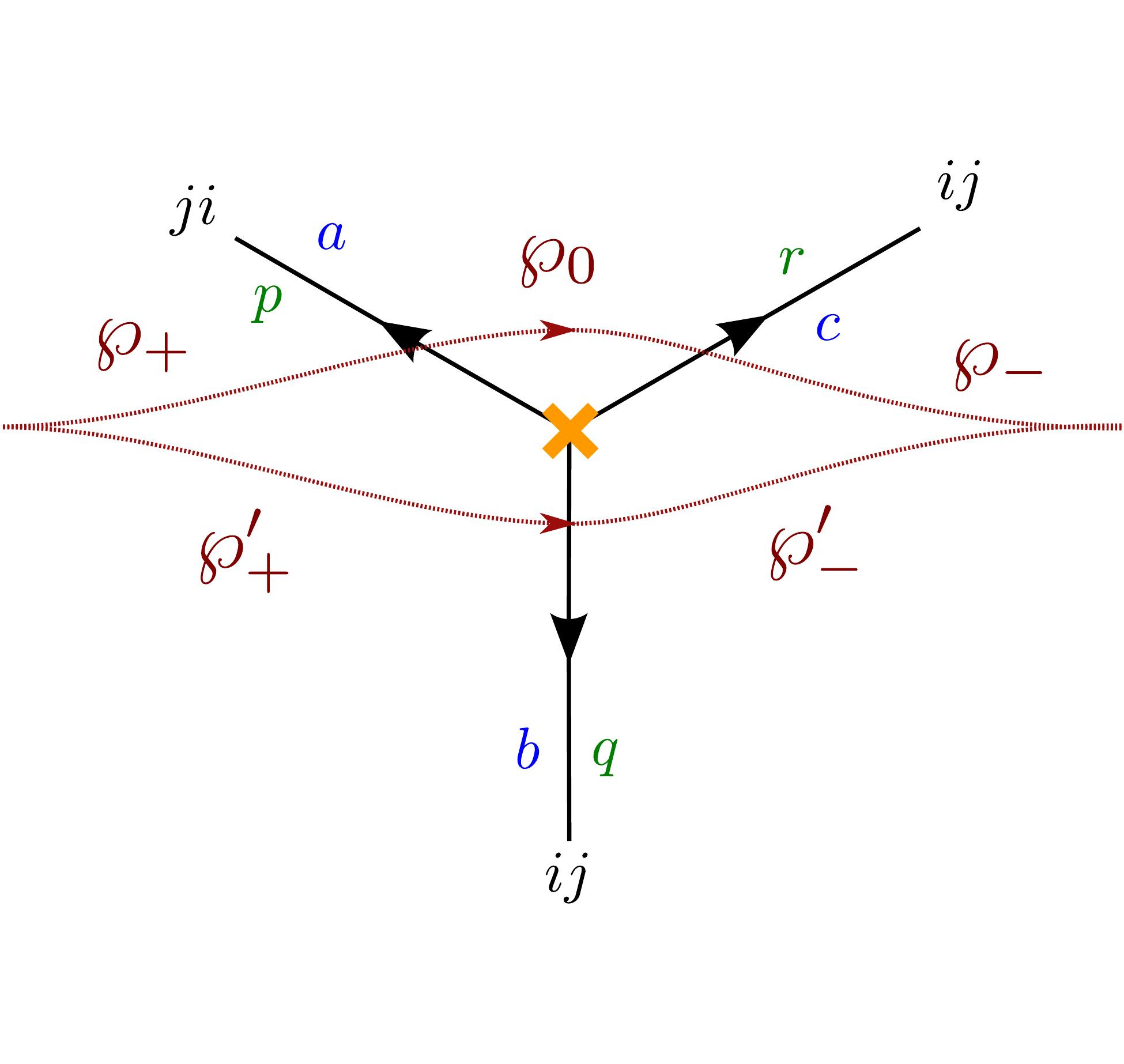}
\includegraphics[width=.65\textwidth]{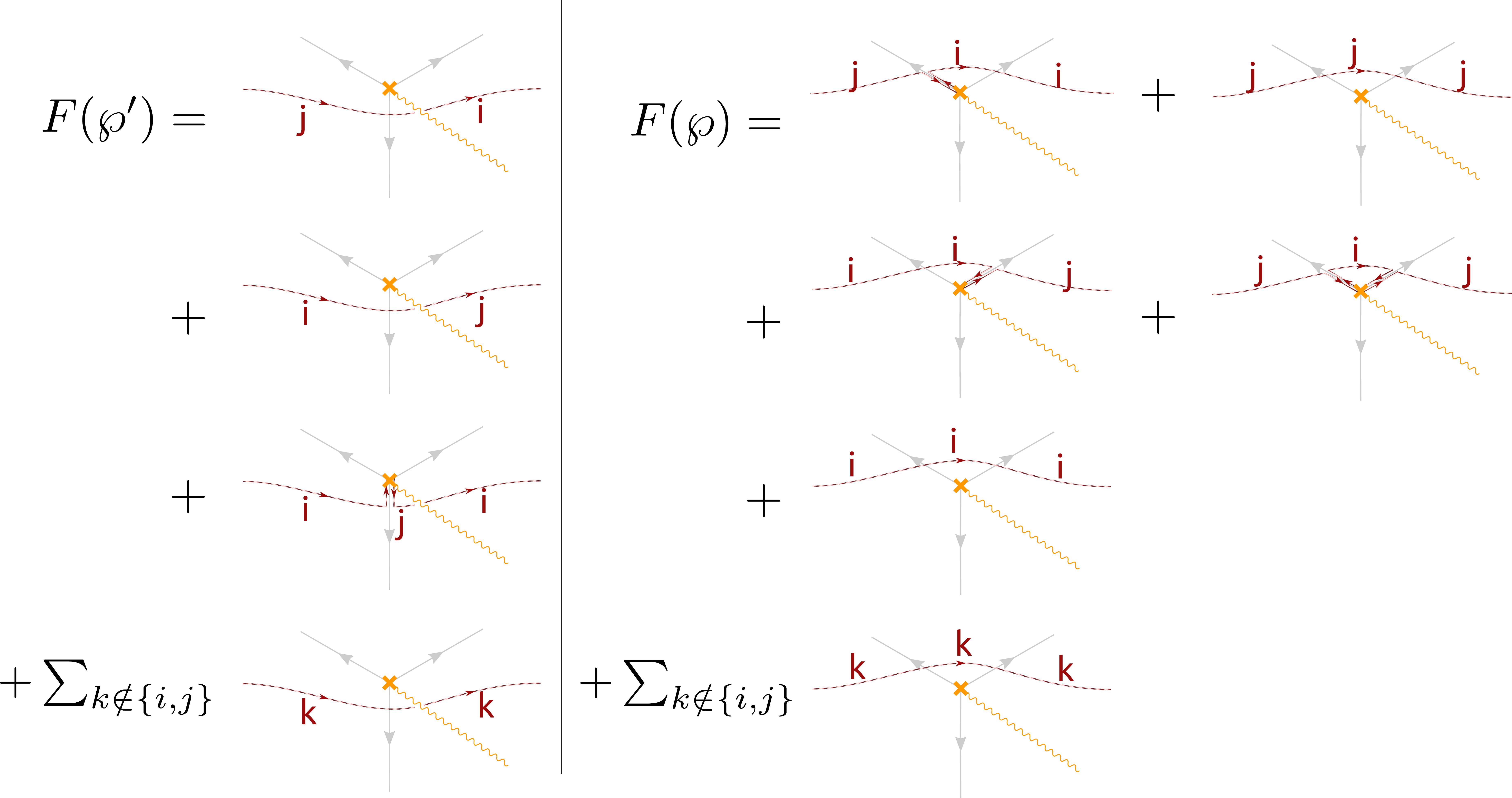}
\caption{Regularly-homotopic paths across a branch point. Indicated in green are the street labels, and in blue the simpleton path labels.}
\label{fig:branch-point}
\end{center}
\end{figure}

Homotopy invariance across branch points will provide some nontrivial constraints for \emph{simpleton} degeneracies, just as in \cite{GMN5}. Considering two paths on $C^*$ as depicted in figure \ref{fig:branch-point}, we study their transports component-wise.\\

\noindent Starting with the  $ji$ component%
\footnote{Notice that $\wp'$ crosses the $ij$ branch cut, as shown on the right frame of figure \ref{fig:branch-point}.}%
, we have
\be
\begin{split}
	\fF(\wp')_{ji}  & = \hU_{{\wp'}^{(ji)}} \\
	\fF(\wp)_{ji}  & = \mu_{r}(\fa,p) \, \hU_{{\wp}_{+}^{(j)}} \hU_{\fa} \hU_{\wp_{0}^{(i)}} \hU_{\wp_{-}^{(i)}}\\
\end{split}
\ee
since ${{\wp}_{+}^{(j)}} {\fa} {\wp_{0}^{(i)}} {\wp_{-}^{(i)}} $ is regular homotopic to ${\wp'}^{(ji)}$, this gives
\be\label{eq:simpleton-mu}
	\mu_{r}(\fa,p)=1 \,.
\ee
A similar computation for the $ij$ component reads
\be
\begin{split}
	\fF(\wp')_{ij}  & = \hU_{{\wp'}^{(ij)}} \\
	\fF(\wp)_{ij}  & = \mu_{r}(\fc,r) \, \hU_{{\wp}_{+}^{(i)}}  \hU_{\wp_{0}^{(i)}} \hU_{\fc} \hU_{\wp_{-}^{(j)}}\\
\end{split}
\ee
once again, noting that ${{\wp}_{+}^{(i)}} {\wp_{0}^{(i)}}  \fc {\wp_{-}^{(j)}} $ is regular homotopic to ${\wp'}^{(ij)}$ yields
\be
	\mu_{r}(\fc,r) = 1\,.
\ee
Repeating with the $ii$ components of the transports
\be
\begin{split}
	\fF(\wp')_{ii}  & = \mu_{r}(\fb,q) \, \hU_{{\wp'}_{+}^{(i)}}   \hU_{\fb} \hU_{{\wp'}_{-}^{(ji)}}\\
	\fF(\wp)_{ii}  & = \hU_{{\wp}^{(i)}} \\
\end{split}
\ee
again, we find
\be
	\mu_{r}(\fb,q) = 1\,.
\ee

\noindent Employing the results obtained so far, we can also evaluate the $jj$ components
\be
\begin{split}
	\fF(\wp')_{jj}  & = 0 \\
	\fF(\wp)_{jj}  & = \hU_{{\wp}^{(j)}} + \mu_{r}(\fa,p)\mu_{r}(\fc,r) \, \hU_{{\wp}_{+}^{(j)}} \hU_{\fa} \hU_{\wp_{0}^{(i)}} \hU_{\fc} \hU_{\wp_{-}^{(j)}}\\
	& = \hU_{{\wp}^{(j)}} + \hU_{{\wp}_{+}^{(j)}} \hU_{\fa} \hU_{\wp_{0}^{(i)}} \hU_{\fc} \hU_{\wp_{-}^{(j)}} = 0
\end{split}
\ee
where we used the fact that ${\wp}^{(j)}$ and ${{\wp}_{+}^{(j)}} {\fa} {\wp_{0}^{(i)}} {\fc} {\wp_{-}^{(j)}}$ differ exactly by a contractible curl.

\noindent Finally, for the $k\ell$  components ($k,\ell\neq i,j$) we have
\be
\begin{split}
	\fF(\wp')_{k\ell}  & = \fF(\wp)_{k\ell}
\end{split}
\ee
trivially.

\subsubsection{Joints}\label{sec:joint-hom}
\begin{figure}[h!]
\begin{center}
\includegraphics[width=.4\textwidth]{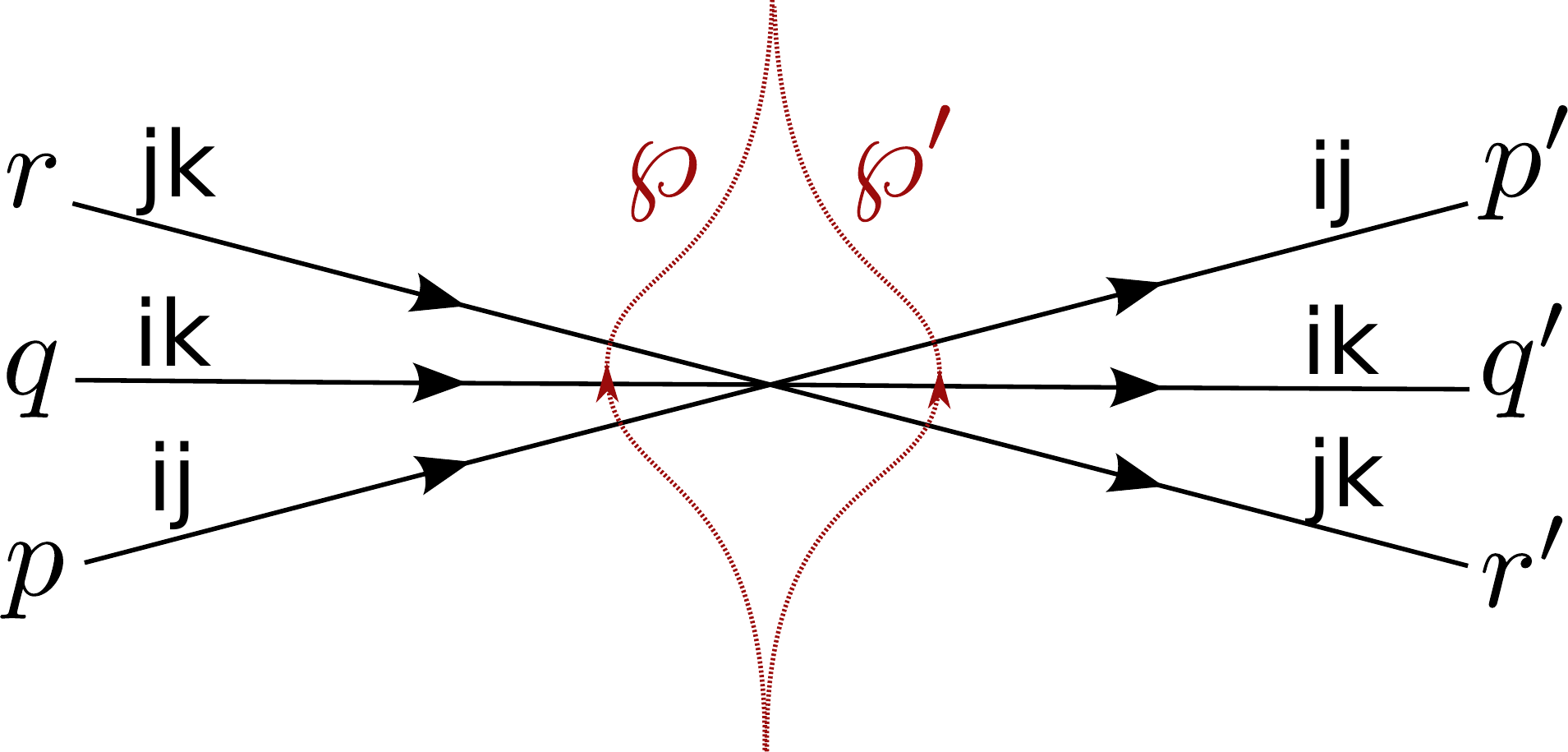}
\caption{Two paths $\wp,\wp'$ within the same restricted regular homotopy class, across a branch point.}
\label{fig:joint}
\end{center}
\end{figure}

Finally, let us examine homotopy invariance across joints of the network, as depicted in figure \ref{fig:joint}.\footnote{One should also examine joints of streets of types $ij$ and $k\ell$, which do not involve the birth/death of new solitons. The analysis is straightforward and exactly parallel to that of \cite{GMN5}, we omit it here and refer the reader to \S 5.2 of the reference.}

\noindent The formal transports are computed by means of the detour rules, and read
\be
\begin{split}
	\fF(\wp) & = \fD(\wp_{+}) %
	\Big( 1+\sum_{\fa}\mu_{r}(\fa,p) \, \hU_{\fa}\Big) %
	\Big( 1+\sum_{\fb}\mu_{r}(\fb,q) \, \hU_{\fb}\Big) %
	\Big( 1+\sum_{\fc}\mu_{r}(\fc,r) \, \hU_{\fc}\Big) %
	\fD(\wp_{-}) \\
	\fF(\wp') & = \fD(\wp'_{+}) %
	\Big( 1+\sum_{\fc}\mu_{r}(\fc,r') \, \hU_{\fc}\Big) %
	\Big( 1+\sum_{\fb}\mu_{r}(\fb,q') \, \hU_{\fb}\Big) %
	\Big( 1+\sum_{\fa}\mu_{r}(\fa,p') \, \hU_{\fa}\Big) %
	\fD(\wp_{-})
\end{split}
\ee
where it is understood that the detours involve suitable deformations as illustrated in figure \ref{fig:detour-definition}.

Setting $F(\wp)=F(\wp')$, the $ij,jk,ik$ components are respectively
\be
\begin{split}
	& \sum_{\fa}\mu_{r}(\fa,p)\,\hU_{\wp_{+}^{(i)}} \hU_{\fa} \hU_{\wp_{-}^{(j)}}  = \sum_{\fa}\mu_{r}(\fa,p') \, \hU_{{\wp'}_{+}^{(i)}} \hU_{\fa}  \hU_{{\wp'}_{-}^{(j)}} \\
	& \sum_{\fc}\mu_{r}(\fc,r)\,\hU_{\wp_{+}^{(j)}} \hU_{\fc} \hU_{\wp_{-}^{(k)}}  = \sum_{\fc}\mu_{r}(\fc,r') \, \hU_{{\wp'}_{+}^{(j)}} \hU_{\fc}  \hU_{{\wp'}_{-}^{(k)}} \\
	&  \hU_{\wp_{+}^{(i)}} \Big( \sum_{\fa,\fc}\mu_{r}(\fa,p)\mu_{r}(\fc,r)\, \hU_{\fa}\hU_{\fc} + \sum_{\fb}\mu_{r}(\fb,q) \hU_{\fb} \Big) \hU_{\wp_{-}^{(k)}} = \sum_{\fb}\mu_{r}(\fb,q') \, \hU_{{\wp'}_{+}^{(i)}} \hU_{\fb}  \hU_{{\wp'}_{-}^{(k)}}
\end{split}
\ee
from which the following 2d wall-crossing formula follows
\be\label{eq:CV-WCF}
\begin{split}
	&    \mu_{r}(\fa,p') = \mu_{r}(\fa,p) \\
	&    \mu_{r}(\fc,r') = \mu_{r}(\fc,r) \\
	& \mu_{r}(\fb,q') = \mu_{r}(\fb,q) + \sum_{\fa,\fb|\fc}\mu_{r}(\fa,p) \mu_{r}(\fc,r)
\end{split}
\ee
where the last sum runs over $\fa,\fc$ whose concatenation $\fa\fc$ is regular-homotopic to $\fb$, so that $\hU_{\fb} = \hU_{\fa} \hU_{\fc}$\footnote{More precisely, the correct statement is that one has to concatenate $\fa,\,\fc$ by gluing an extra small arc between them to match endpoint tangents. Similarly, in order to compare
$\fa\fc$ to $\fb$ one must further add small arcs at the endpoints of $\fa\fc$, in order to match the initial and final directions of $\fb$. These modifications are inessential here, since we adopt, by definition of the detour rules, paths with all the suitable modifications, and eventually we actually compare $\hU_{\wp_{+}^{(i)}  \fa\fc \wp_{-}^{(k)} }$ to $\hU_{\wp_{+}^{(i)}  \fa\fb \wp_{-}^{(k)} }$. Although irrelevant in this context, this issue was dealt with in Appendix B of \cite{WWC}.}. This
concludes the study of homotopy invariance of the formal parallel transport.

\subsection{Invariance of $F(\wp)$ under regular homotopy}\label{subsec:F-rhom-inv}
In the previous section we established the invariance of $\fF(\wp)$ under regular homotopy, for $\wp\subset C$ with $\{{\rm beg}(\wp),{\rm end}(\wp)\}\subset C\setminus C^{*}$.

Let us now choose $C^{*}$ (resp. $\Sigma^{*}$) as in section \ref{sec:framed-PSC-conjecture}, i.e. the set of auxiliary punctures now only includes the endpoints $z,z'$ of $\wp$ (resp. $\pi^{-1}\{z,z'\}\subset\Sigma$). Using the detour rules, the formal parallel transport can be written in the generic form
\be\label{eq:q-Darboux}
	\fF(\wp) = \sum_{ij'}\sum_{\fp\in{\bGamma}_{ij'}^{(r)}(z,z')}\,\FOmega(L_{\wp},u,\vartheta,\fp)\, \hU_{\fp}
\ee
where the sum is over all regular homotopy classes $\fp$ of detours of $\wp$ on $C^{*}$, and the coefficients of the series are defined by this expression.
According to (\ref{eq:soliton-sign}) and in analogy to \cite[\S 3.5]{GMN5}, these degeneracies obey
\be
	\FOmega(L_{\wp},u,\vartheta,\fa)= - \FOmega(L_{\wp},u,\vartheta,\fa')
\ee
for $\fa,\fa'$ differing by a contractible curl.

We take this as the \emph{definition} of the refined framed degeneracies introduced in (\ref{eq:framed-degeneracies}).

\medskip


Since $\fF(\wp)$ involves exclusively paths $\fp\in\bGamma(z,z'):=\sqcup_{ij'}\bGamma_{ij'}(z,z')$, we may associate to each of them its own writhe $y^{\wr(\fp)}$. Then we can 
consider a \emph{linear map} (it is not an algebra map!)
\be\label{eq:rho}
	\rho (\hU_{\fp}):=y^{\wr(\fp)}\hY_{\beta(\fp)}\,,
\ee
in \S\ref{sec:chern-simons} below we will propose some physical intuition for this map.
For convenience we adopt the following definition
\be
	\FOmega(L_{\wp},u,\vartheta,\fp;y):=\FOmega(L_{\wp},u,\vartheta,\fp)\, y^{\wr(\fp)}\,.
\ee

Collecting regular homotopy classes $\fp$ on $\Sigma^{*}$ that all belong to the preimage of a relative homology class $\pi$ on $\tSigma^{*}$, the formal parallel transport maps to
\be
\begin{split}
	\rho\Big(\fF(\wp)\Big) & = \sum_{ij'}\sum_{\pi\in\tGamma_{ij'}(z,z')}\sum_{\fp\,|\,\beta(\fp)=\pi}\,\FOmega(L_{\wp},u,\vartheta,\fp;y)\,\hY_{\pi} \\
	& = \sum_{ij'}\sum_{\pi\in\tGamma_{ij'}(z,z')}\,\FOmega(L_{\wp},u,\vartheta,\pi;y)\,\hY_{\pi}\,.
\end{split}
\ee

Since $\fF(\wp)$ is a (twisted) invariant of regular homotopy of $\wp$, 
%
%
defining
\be
	F(\wp) := \rho\Big(\fF(\wp)\Big)
\ee
establishes the twisted regular homotopy invariance claimed below (\ref{eq:framed-spin-functional}).

\subsection{Joint rules for two-way streets}\label{sec:joint-rules}
\begin{figure}[h!]
	\begin{center}
		 \includegraphics[width=0.4\textwidth]{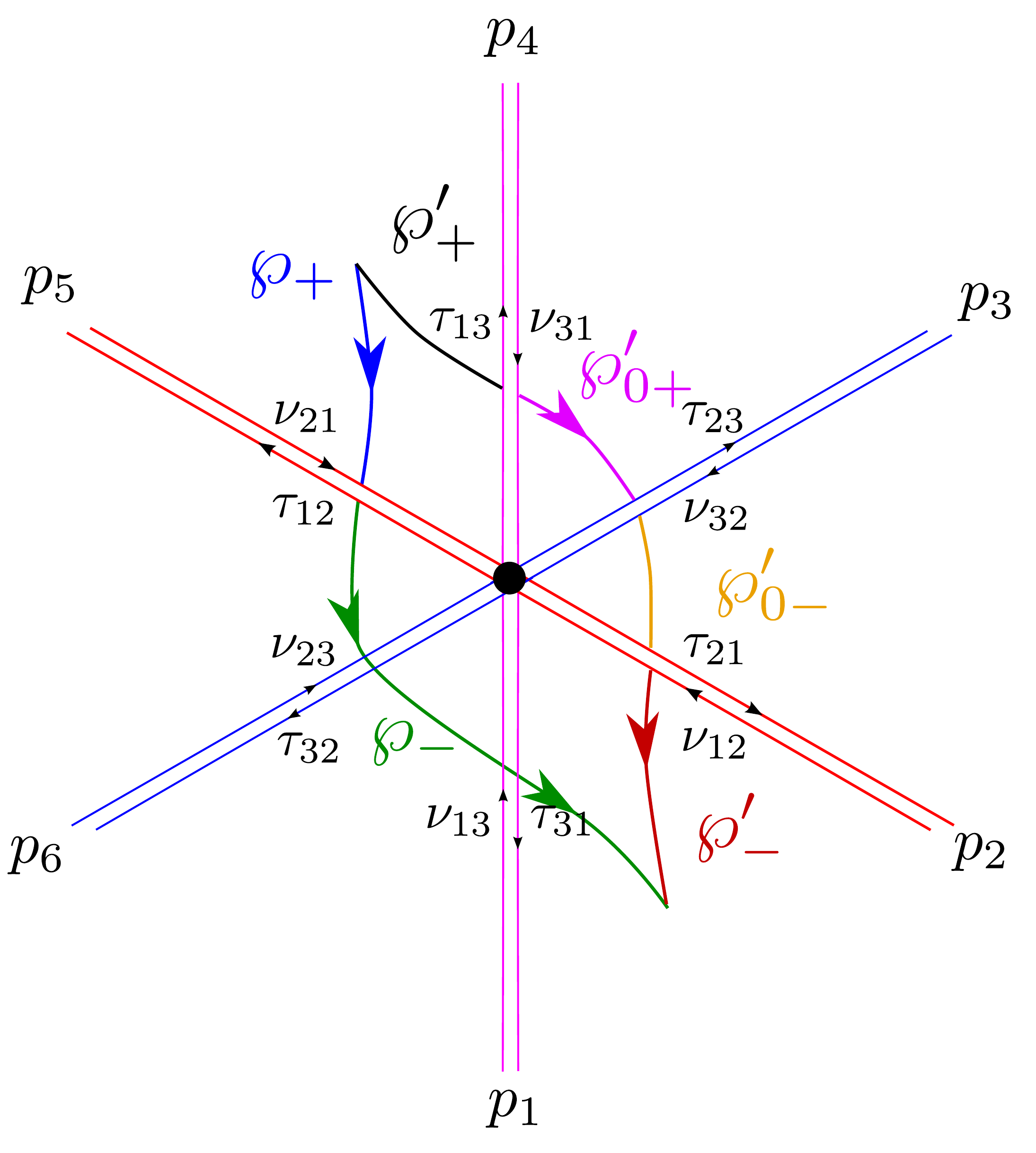}
		\caption{A six-way joint of 2-way streets, in the British resolution.  Each 2-way street has two ``lanes'' (one-way streets), one of type $ij$ and another of type $ji$ and $i,j\in\{1,2,3\}$ according to the labels shown next to each street. Each lane carries its own soliton content, indicated next to it.}
		\label{fig:six-way}
	\end{center}
\end{figure}
As in \cite{GMN5,WWC}, the key to computing vanilla BPS spectra are certain equations relating the soliton content on the 2-way streets meeting at a joint. This subsection is devoted to presenting the corresponding refined version. Considering the example shown in figure \ref{fig:six-way}, there are six two-way streets, each one carrying two soliton sets. The soliton data is encoded into the generating functions denoted $\tau,\nu$, one for each street $p$ of type $ij$:
\be
		\tau_{ij} = \sum_{\fa}\mu_{r}(\fa,p)\hU_{\fa}\qquad 	\nu_{ij} = \sum_{\fb}\mu_{r}(\fb,p)\hU_{\fb}
\ee
where $\fa,\fb$ are $ij,\,ji$ solitons supported on $p$.

Choosing paths $\wp,\wp'$ as shown, invariance of the formal parallel transport entails, in particular
\be
	 \fF_{12}(\wp) = \fF_{12}(\wp')
\ee
where, explicitly we have
\be
\begin{split}
	& \fF_{12}(\wp) = \hU_{\wp_{+}^{(1)}}\,\tau_{12}\,\hU_{\wp_{-}^{(2)}} \\
	& \fF_{12}(\wp') = \hU_{{\wp'}_{+}^{(1)}}\,\tau_{13}\,\hU_{{\wp'}_{0_{+}}^{(3)}}\,\nu_{32}\, \hU_{{\wp'}_{0_{-}}^{(2)}}\hU_{{\wp'}_{-}^{(2)}} %
	+ \hU_{{\wp'}_{+}^{(1)}}\,\hU_{{\wp'}_{0_{+}}^{(1)}}\,\hU_{{\wp'}_{0_{-}}^{(1)}}\,\nu_{12}\, \hU_{{\wp'}_{-}^{(2)}}
\end{split}
\ee
To lighten notation, we will write the constraint of homotopy invariance simply in the form%
\footnote{\label{foot:eta}As noted in appendix B of \cite{WWC}, this expression is incomplete. It should involve a certain formal variable, called $\eta$ in the reference, to account for small arcs that need to be added to match tangent directions of solitons of $\tau_{13}$ with those of $\nu_{32}$ and their composition with the solitons of $\nu_{12}$. In our context, we suppress the $\eta$ because later on, when computing generating functions for 4d BPS states, we will be actually always working with homotopy invariance of auxiliary paths $\wp,\wp'$ and such $\eta$ is subsumed in the rules for deforming detour paths.}
\be
	\tau_{12} = \nu_{12} + \tau_{13} \nu_{32}\,.
\ee

Similar, appropriate choices of auxiliary paths $\wp,\wp'$ allow to recover the desired joint soliton rules
\begin{equation}\label{eq:6way}
	\begin{array}{lr}
	\begin{aligned}
	\tau_{12} &= \nu_{12} +  \tau_{13} \nu_{32}, \\
	\tau_{23} &= \nu_{23} +  \tau_{21} \nu_{13}, \\
	\tau_{31} &= \nu_{31} +  \tau_{32} \nu_{21},
	\end{aligned} &
	 \begin{aligned}
	\tau_{21} &= \nu_{21} +  \nu_{23} \tau_{31}, \\
	\tau_{32} &= \nu_{32} +  \nu_{31} \tau_{12}, \\
	\tau_{13} &= \nu_{13} +  \nu_{12} \tau_{23}.
	\end{aligned}
\end{array}
\end{equation}
these look exactly the same as the rules in \cite{GMN5,WWC}, with the only difference that we are working with regular homotopy classes on $\Sigma^{*}$.

\subsubsection{Definition of soliton paths}
In section \ref{subsubsec:spec-lin-int} we defined halo-saturated susy interfaces based on the notion of \emph{soliton paths}, in this section provide more detail about the latter.

Let $p\in\CW_{c}$ be a two-way street of $ij$-type, it may be thought of as a pair of one-way streets $p_{ij}$ and $p_{ji}$. To determine the soliton paths going through street $p$ we proceed as follows. $p_{ij}$ has an orientation, let us denote $J[p,ij]$ the joint from which it flows out, similarly $J[p,ji]$ is the joint associated with $p_{ji}$. At $J[p,ij]$ we may consider the rules (\ref{eq:6way}), expanding them in terms of \emph{incoming} soliton generating functions, for example
\be
	\tau_{12} = \nu_{12} + \nu_{13} \nu_{32} + \nu_{12}\nu_{23}\nu_{32} + \nu_{12}\nu_{21}\nu_{13}\nu_{32} + \nu_{12}\nu_{23}\nu_{31}\nu_{13}\nu_{32}+\dots
\ee
where $\nu_{\alpha\beta}=0$ whenever the corresponding street isn't carrying solitons.

The lift $p_{ij,\Sigma}=\pi^{-1}(p_{ij})$ contains two components, $p_{ij}^{(i)}, p_{ij}^{(j)}$. We start constructing paths by \emph{concatenating} the lifts of streets involved in such sums, in the order dictated by the above formulae. For example, if $p=p_{5}$ from figure \ref{fig:six-way} we would consider several paths:
\be
\begin{split}
	& p_{5,12}^{(1)} \,\cdot\, p_{2,12}^{(1)}\,  (\dots)\, p_{2,12}^{(2)} \,\cdot\, p_{5,12}^{(2)}\\
	& p_{5,12}^{(1)} \,\cdot\, p_{1,13}^{(1)}\,  (\dots)\, p_{1,13}^{(3)}\,\cdot\, p_{3,32}^{(3)}\,  (\dots)\, p_{3,32}^{(2)} \,\cdot\, p_{5,12}^{(2)}\\
	& \dots
\end{split}
\ee
where $(\dots)$ are placeholders, which will be filled upon iteration of this procedure: namely taking into consideration the junctions at the \emph{other} ends of the streets involved (e.g. $J[p_{2},12]$ in the first line, $J[p_{1},13]$ and $J[p_{3},32]$ in the second line, and so on).
Iterating this procedure, one eventually reaches two-way streets terminating on branch-points.
If the branch-point in question sources only \emph{one} two-way street, then the $(\dots)$ are simply dropped for that street. If there is more than one two-way street ending on the branch-point, one must take into account further detours, as explained e.g. in \cite[app.A]{GMN5}. The procedure involved is a straightforward generalization of the one for joints, we skip its description.

Thus we have constructed (possibly infinite) sets of \emph{open} soliton paths, associated with $p_{ij}$ and $p_{ji}$. Joining them pairwise produces the \emph{closed} soliton paths employed in section \ref{subsubsec:spec-lin-int}.

\section{Applications and examples}\label{sec:applications}
\subsection{Vectormultiplet in $SU(2)$ SYM}
The simplest nontrivial example is the spectral network of the vectormultiplet of charge $\gamma$ in the weak coupling regime of $SU(2)$ SYM \cite{GMN2}.
\begin{figure}[h!]
\begin{center}
\includegraphics[width=0.45\textwidth]{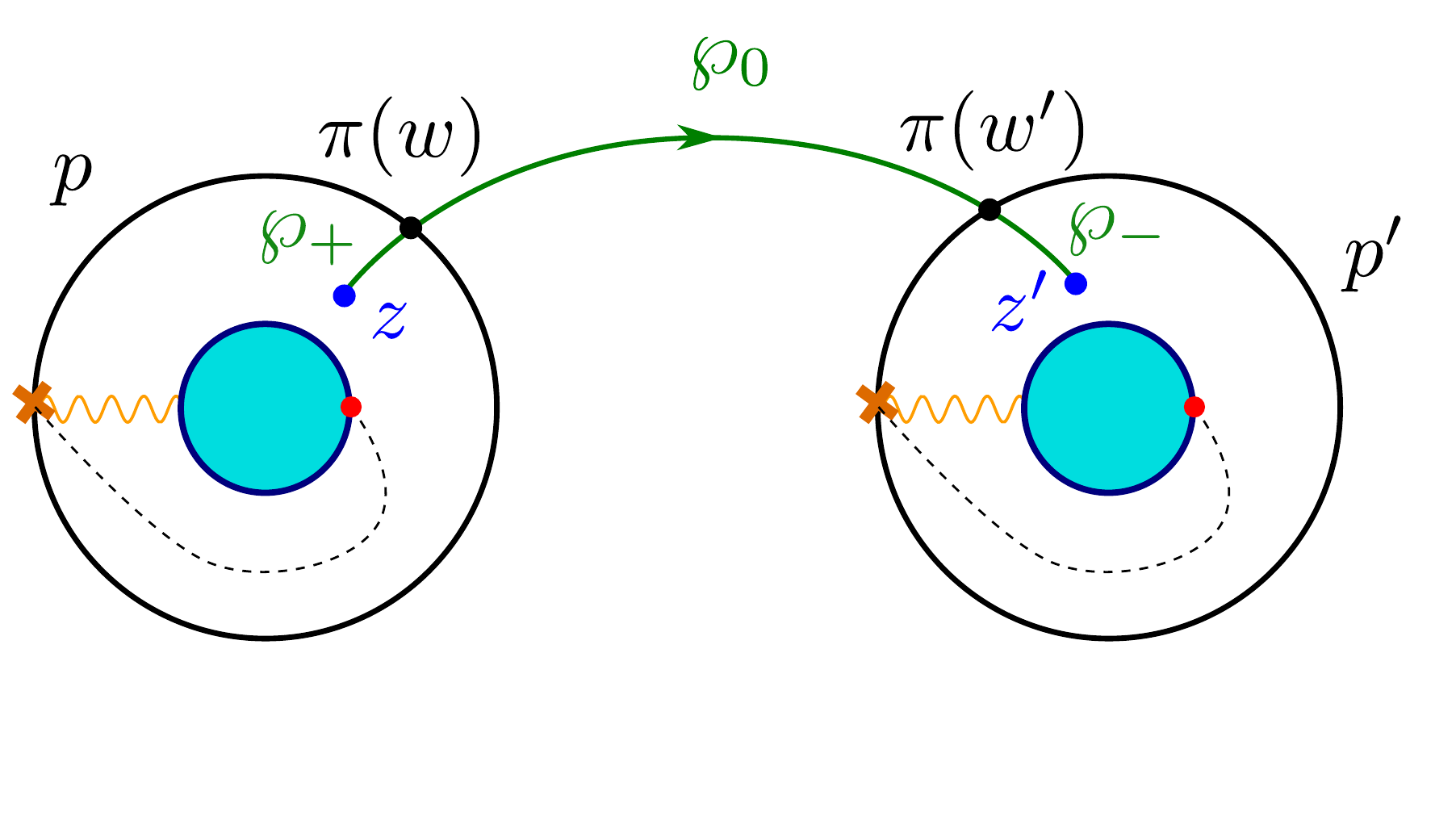}\includegraphics[width=0.24\textwidth]{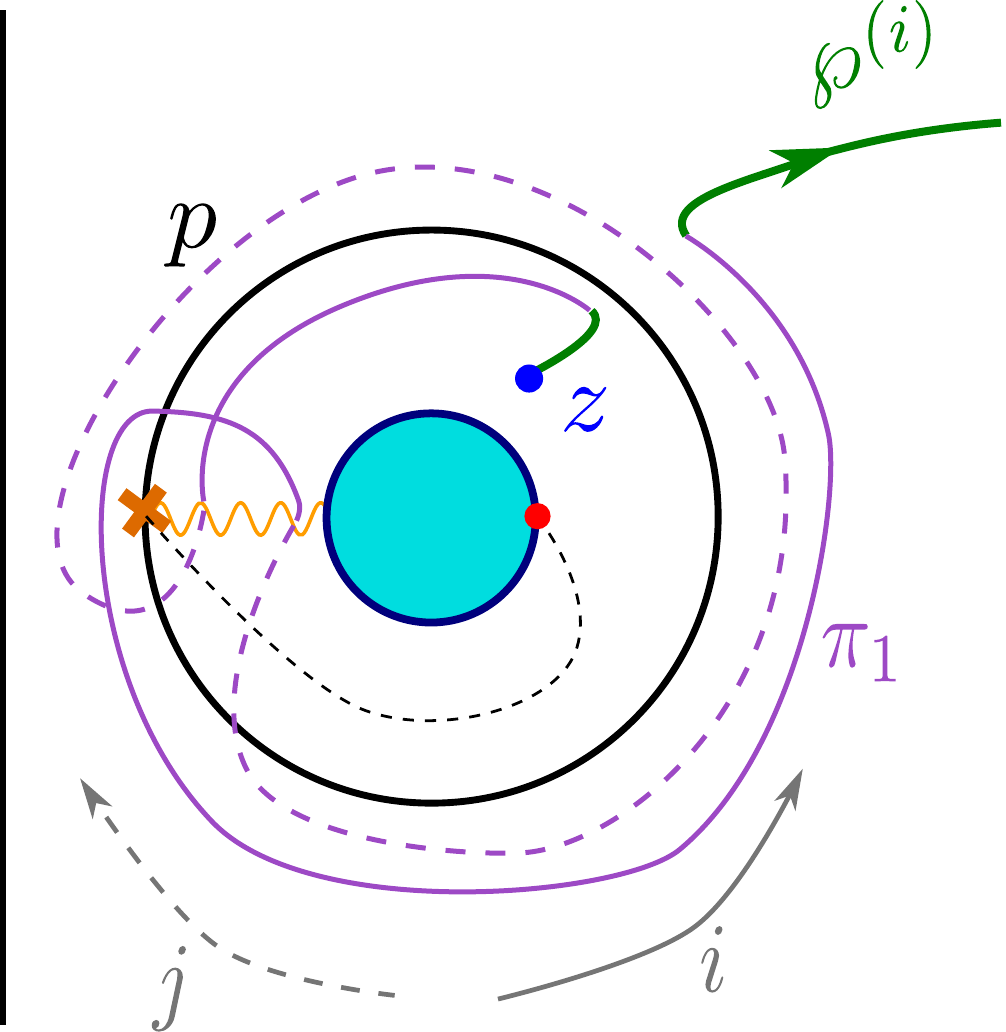}
\caption{On the left, the spectral network at the critical phase; $C$ is a cylinder, two way streets are solid lines, branch points and cuts are in orange. $z,z'\in C$ label punctures associated with UV surface defects, a choice of halo-saturated interface is shown in green. Red dots mark singularities of the WKB flow (see \cite{GMN2}). On the right: an example of a detour $\pi_{1}$; the WKB flow is indicated in grey for the two sheets of $\Sigma$.}
\label{fig:su2}
\end{center}
\end{figure}
In order to choose a halo-saturated interface, we need first to construct $\Pi(\CW_{c})$. The critical sub-network $\CW_{c}$ is depicted with solid black lines in figure \ref{fig:su2}; applying the detour rules to the branch-point of street $p$ we find
\begin{figure}[h!]
\begin{center}
\includegraphics[width=0.24\textwidth]{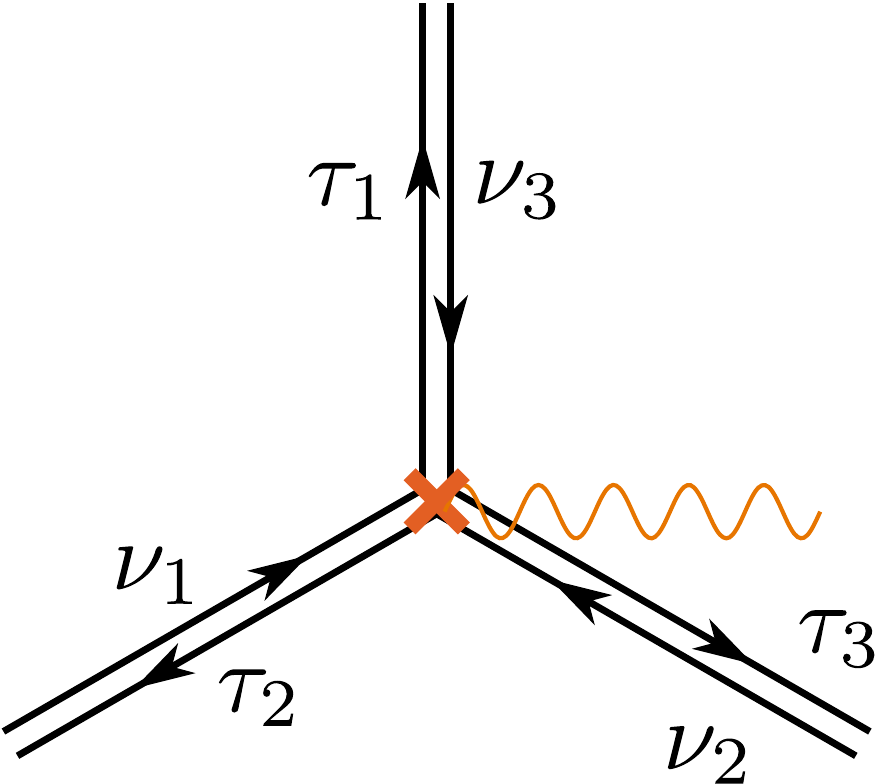}
\caption{The branch point of the two way street $p$ from figure \ref{fig:su2}.}
\label{fig:su2-BP}
\end{center}
\end{figure}
\be\label{eq:vm-solitons}
\begin{split}
	& \nu_{3} = \tau_{2}\, X_{\tgamma}\qquad \nu_{1} = \tau_{1}\, X_{\tgamma} \qquad \nu_{2}=0\\
	& \tau_{2} = X_{a_{2}}\qquad \tau_{1} = X_{a_{1}}+ \tau_{1}\, X_{\tgamma}
\end{split}
\ee
where $a_{1}$ is an $ij$-type soliton (it runs from sheet $i$ to sheet $j$), while $a_{2}$ is of type $ji$. We used $\tau_{n}=X_{a_{n}}+\nu_{n}$ \cite[app.A]{GMN5} with $a_{n}$ denoting the \emph{simpleton} paths sourcing from the branch point, and $\tgamma\in \tilde\Gamma$ is the tangent lift of $\gamma$ (which is the ``critical'' charge $\gamma_{c}$ corresponding to the $\CK$-wall, although we will be avoiding such notation to avoid confusion with the ``core'' charge of a halo boundstate).
Therefore, let us define
\be
	Q(p) = 1+\tau_{1}\nu_{3} =1+ X_{a_{1}}\, (1-X_{\tgamma})^{-1}\, X_{a_{2}}\, X_{\tgamma}^{} = (1-X_{\tgamma})^{-1}
\ee
where we used\footnote{The fact that $X_{a_{1}}X_{a_{2}}=1$ may not be obvious at first glance. This is a technical identity that reflects the choice of concatenating $a_{1}$ with $a_{2}$ the short way around, thus not going around the street $p$, thus giving a contractible cycle. This occurrence is dictated by the fact that we chose to indicate explicitly the parallel transport of solitons around $p$ by factors of $X_{\tgamma}$ in (\ref{eq:vm-solitons}), accordingly it would be wrong to write $X_{a_{1}}X_{a_{2}}=X_{\tgamma}$ since it would introduce extra powers of $X_{\tgamma}$.} $X_{a_{1}}X_{\tgamma}=X_{\tgamma}X_{a_{1}}$ and $X_{a_{1}}X_{a_{2}} = 1$ (the closure map ${\text{cl}}(a_{1}a_{2})=0\in \tGamma$ is understood, see \cite{WWC}); this expression for $Q(p)$ agrees with the expected one \cite{GMN2,GMN5}. A similar computation for $p'$ reveals the same contribution. Keeping track of \emph{soliton paths} we find that
\be
\begin{split}
	& \Pi(p) = \{\ell_{n}\}_{n=1}^{\infty} \qquad \ell_{1} = \mathring p_{\Sigma} \qquad \ell_{n} = \ell_{n-1}\,\ell_{1}\,,\\
	& \Pi(p') = \{\ell'_{n}\}_{n=1}^{\infty} \qquad \ell'_{1} = \mathring p'_{\Sigma} \qquad \ell'_{n} = \ell'_{n-1}\,\ell'_{1}\,,
\end{split}
\ee
where $\mathring  p_{\Sigma}$ is the regular homotopy class of the lift of street $p$ to $\Sigma^{*}$, where it is understood in this example that the endpoints of the two components of the lifts are pairwise glued together (at the ramification point, see figure \ref{fig:su2}) into a single closed path. The notation $\ell_{n-1}\ell_{1}$ needs some further clarification: these are regular homotopy classes of closed curves, hence composition is ambiguous. Denote $p_{\Sigma}^{(i)},p_{\Sigma}^{(j)}$ the two components of $\pi^{-1}(p)$. Then to construct $\ell_{2}$ one takes two copies of $p_{\Sigma}$: ${p'}_{\Sigma},{p''}_{\Sigma}$ and glues ${p'}_{\Sigma}^{(i)}{p'}_{\Sigma}^{(j)}{p''}_{\Sigma}^{(i)}{p''}_{\Sigma}^{(j)}$, then gluing the endpoint of ${p''}^{(j)}_{\Sigma}$ with the starting point of ${p'}^{(i)}_{\Sigma}$ gives an actual closed path, $\ell_{2}$ is the corresponding regular homotopy class. The construction generalizes straightforwardly to $\ell_{n-1}\ell_{1}$.

Noting that $[\ell_{n}]=[\ell'_{n}]=n\tgamma$, we deduce immediately that $\wp$ as depicted in figure \ref{fig:su2} satisfies both conditions (\ref{eq:saturation-def}).

Choose a trivialization of the cover such that $\tau_{1}$ carries contributions from $ij$-solitons (this together with the WKB flow fixes all other sheet labels), compatibly with the right frame of fig. \ref{fig:su2}. Then studying the detours of $\wp$ we find
\be
\begin{split}
	F_{ii}(\wp,\vartheta_{c}^{-}) & = X_{\wp^{(i)}} + X_{\wp_{+}^{(i)}}\Big(1+\sum_{n=1}^{\infty}X_{a_{1}+n\tgamma}\Big)X_{\wp_{0}^{(j)}}\Big(1+\sum_{n=1}^{\infty}X_{a'_{1}+n\tgamma}\Big)X_{\wp_{-}^{(i)}} \\
	F_{ii}(\wp,\vartheta_{c}^{+}) & = X_{\wp_{+}^{(i)}}\,Q(p)\,X_{\wp_{0}^{(i)}}\,Q(p')\,X_{\wp_{-}^{(i)}} + X_{\wp_{+}^{(i)}}\Big(1+\sum_{n=1}^{\infty}X_{a_{1}+n\tgamma}\Big)X_{\wp_{0}^{(j)}}\Big(1+\sum_{n=1}^{\infty}X_{a'_{1}+n\tgamma}\Big)X_{\wp_{-}^{(i)}}
\end{split}
\ee
where the second term of both expressions corresponds to $ij$ detours on street $p$ composed with $ji$ detours on street $p'$, while the first term in the second expression counts $ii$ detours on both $p$ and $p'$. We took into account that all 2d soliton degeneracies $\mu_{\pm}(a_{1}+n\tgamma), \mu_{\pm}(a'_{1}+n\tgamma)$ are $1$ in this example. The contribution from halos of core charge $\wp^{(i)}$ undergoes the jump
\be
	X_{\wp^{(i)}}\quad \mapsto \quad X_{\wp^{(i)}} (1-X_{\tgamma})^{-2}
\ee
in agreement with $\langle\wp^{(i)},L(\gamma)\rangle=\langle\wp^{(i)}, -p_{\Sigma} - p'_{\Sigma}\rangle = -2$.

Now we take into account the writhe: first note that the writhe of $\ell_{n}, \ell'_{n}$ with respect to the detour points $w,w'$ (see figure \ref{fig:su2}) is
\be
	\wr(\ell_{n},w) = -n\qquad \wr(\ell'_{n},w) = n\,,
\ee
as clarified by the right frame of figure \ref{fig:su2}. Therefore we find
\be\label{eq:pureSU2-full-spin}
\begin{split}
	F_{ii}(\wp,\vartheta_{c}^{-};y) & = \hY_{\wp^{(i)}} + \sum_{n,n'=0}^{\infty}y^{-n+n'}\,\hY_{\wp_{+}^{(i)} a_{1} \wp_{0}^{(j)}a'_{1}\wp_{-}^{(i)}+(n+n')\tgamma} \\
	F_{ii}(\wp,\vartheta_{c}^{+};y) & = \hY_{\wp_{}^{(i)}} + \sum_{n=1}^{\infty} (y^{-n}+y^{n})\,\hY_{\wp_{}^{(i)}+n\tgamma}+   \sum_{n,n'=1}^{\infty} y^{-n+n'}\,\hY_{\wp_{}^{(i)}+(n+n')\tgamma} \\
	& +  \sum_{n,n'=0}^{\infty}y^{-n+n'}\,\hY_{\wp_{+}^{(i)} a_{1} \wp_{0}^{(j)}a'_{1}\wp_{-}^{(i)}+(n+n')\tgamma}
\end{split}
\ee
This agrees with the expected jump for $\hY_{\wp^{(i)}}$, indeed according to our conjecture we expect
\be\label{eq:su2-expected}
\begin{split}
	F_{\wp^{(i)}}(\wp,\vartheta_{c}^{-};y) = \hY_{\wp^{(i)}} \quad \mapsto \quad 
	& \hY_{\wp^{(i)}}\, \big(1 - \, \hY_{\tgamma} \big)^{-1}\, \big(1 -y^{-2}\, \hY_{\tgamma} \big)^{-1} \\
	= \quad & \hY_{\wp^{(i)}} \, \Phi_{1}((-y)^{-1}\,\hY_{\tgamma})^{-1}\, \Phi_{1}((-y)^{}\,\hY_{\tgamma})^{-1} \\
	= \quad & F_{\wp^{(i)}}(\wp,\vartheta_{c}^{+};y)
\end{split}
\ee
where $J_{\wp^{(i)},\tgamma}=0$ in this setup, together with $a_{m}(\gamma)=-\delta_{m,\pm1}$.

\subsection{The $3$ - herd}
The next nontrivial example is provided by a class of critical networks known as $k$-herds \cite{WWC}. The case $k=1$ is trivial, while the 2-herd is just another network for the vectormultiplet studied above. The first interesting case is therefore $k=3$, we focus on this although our analysis can be straightforwardly extended to higher integer $k$.

\begin{figure}[h!]
\begin{center}
\includegraphics[width=0.64\textwidth]{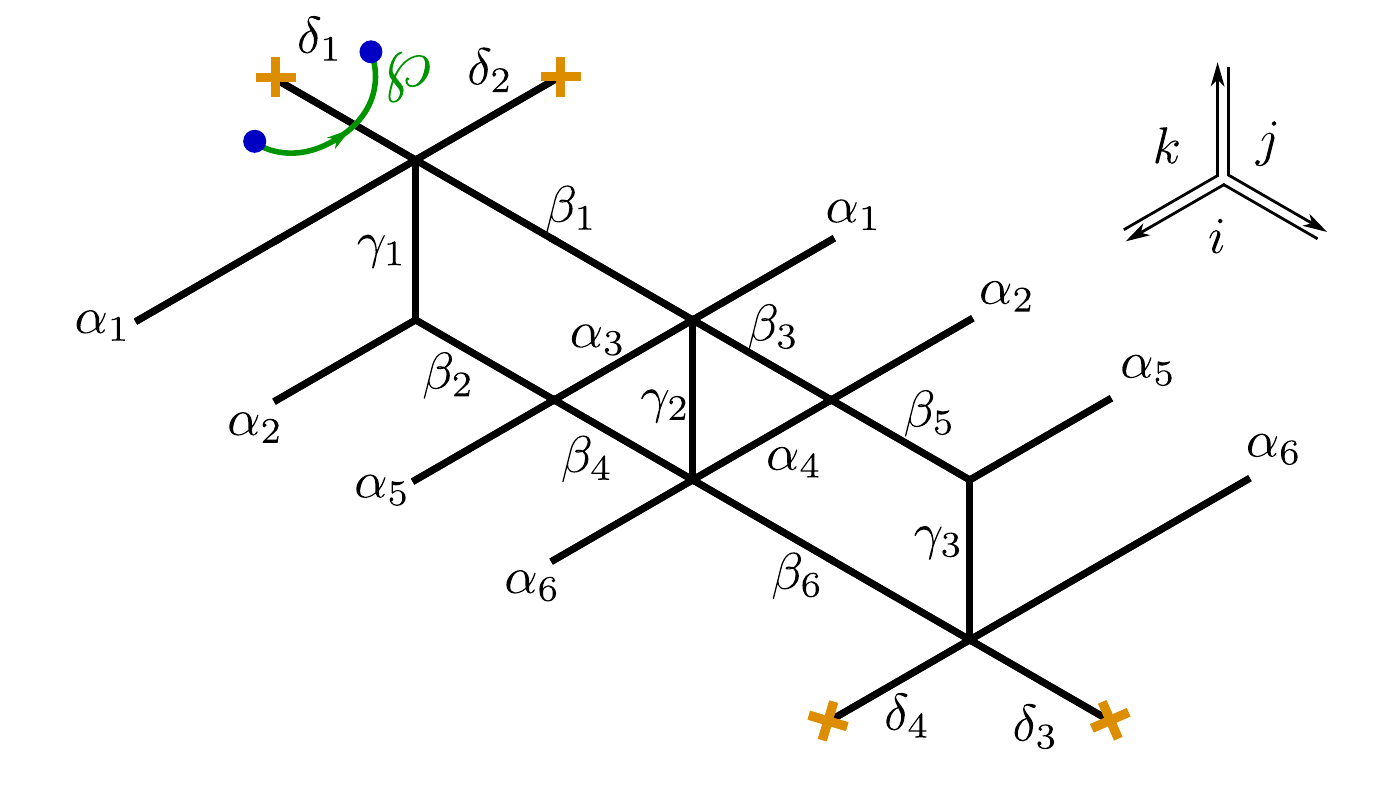}
\caption{The street map of the $3$-herd, on the cylinder $C$ which has been cut. The network streets are glued according to their labeling. The schematic direction of WKB flow of the three types of streets is displayed on the upper right, for example the branch-point on the upper-left of the herd sits at the end of an $ij$-type street and is therefore an $ij$-type branchpoint.}
\label{fig:3herd}
\end{center}
\end{figure}

The soliton content of the 2-way streets of the 3-herd has been studied in great detail in \cite{WWC}. Let $\tgamma$ be the generator of the critical sublattice corresponding to the $\CK$-wall. Recall that it may be constructed from $\CW_{c}$ as the homology class of a weighted sum of lifts of the streets of the network, where the weights are dictated by the soliton data. Rather than describing precisely the set $\Pi(\CW_{c})$ it will be sufficient for us to note (see in particular \S C.6.2 of the reference) that, for any street $p$ and any two soliton paths $\fa,\fb$ (of $ij$/$ji$ types respectively) supported on $p$, $\ell={\rm cl}(\fa\fb)$ is characterized by
\be
	[\ell] = n\tgamma\quad\Leftrightarrow\quad \ell \ni \{\delta_{1,\Sigma},\delta_{2,\Sigma},\delta_{3,\Sigma},\delta_{4,\Sigma}\} \quad n \text{ times},
\ee
where inclusion of $\delta_{i,\Sigma}$ stands for the fact that the solitons run through the $i$-th ramification point $n$ times. Street names refer to figure \ref{fig:3herd}, and $\gamma$ is the generator of the critical sublattice (with orientation fixed by $\vartheta_{c}$).

With this information at hand, we can make a simple choice of a halo-saturated $\wp$, displayed in figure \ref{fig:3herd}. Other choices are clearly possible. The machinery of section \ref{sec:FPT} establishes relations between generating functions for different halo saturated choices of $\wp$: different choices of $\wp$ within the same regular homotopy class on $C^{*}$ are equivalent.

By direct inspection of soliton paths involved in detours of $\wp^{(i)}$, one finds the following framed generating functions
\be\label{eq:3-herd-functionals}
\begin{split}
	F_{ii}(\wp,\vartheta_{c}^{-};y) & = \hY_{\wp^{(i)}} \\
	F_{ii}(\wp,\vartheta_{c}^{+};y) & = \hY_{\wp^{(i)}}  +  \big(y^{-2}+1+y^{2}\big) \hY_{\wp^{(i)}+\tgamma}\\
	& +\big(y^{-6}+2y^{-4}+3 y^{-2}+ 3 +3 y^2+2 y^4+y^{6}\big) \hY_{\wp^{(i)}+2\tgamma} \\
	& + \big(y^{-12}+2 y^{-10}+5 y^{-8}+8 y^{-6}+11 y^{-4}+12 y^{-2}+13 \\
	& +12 y^{2}+11 y^{4}+ 8 y^{6}+5 y^{8}+ 2 y^{10}+ y^{12}\big) \hY_{\wp^{(i)}+3\tgamma} + \cdots \\
	& = \hY_{\wp^{(i)}} \, \prod_{m\in [1]}\Phi_{1}((-y)^{m}\, \hY_{\tgamma})\, \prod_{m'\in[5/2]}\Phi_{2}((-y)^{m'}\, \hY_{2\tgamma})^{-1} \\
	& \times  \prod_{m''\in[3]\oplus [5]}\Phi_{3}((-y)^{m''}\, \hY_{3\tgamma})^{} \quad \cdots
\end{split}
\ee
where the notation $m\in [k]$ stands for $m\in\{-2k,-2k+2,\dots,2k-2,2k\}$.

Due to the simplicity of $\wp$, we have $F_{ii}(\wp,\vartheta_{c}^{\pm};y) \equiv F_{\wp^{(i)}}(\wp,\vartheta_{c}^{\pm};y)$ (cf. (\ref{eq:restricted-framed-spin-functional})), we thus find agreement -- up to terms of order $X_{3\tgamma}$ -- with the conjectured pattern 
of (\ref{eq:PSC-conjecture}):
\be
\begin{split}
	F_{\wp^{(i)}}(\wp,\vartheta_{c}^{-};y) = \hY_{\wp^{(i)}} \quad \mapsto \quad & \hY_{\wp^{(i)}}\, \prod_{n=1}^{\infty}\,\prod_{m\in \IZ} \Phi_{n}((-y)^{m} \, \hY_{n\tgamma})^{a_{m}(n\tgamma)} \\
	= \quad & F_{\wp^{(i)}}(\wp,\vartheta_{c}^{+};y)\,.
\end{split}
\ee
Moreover we recover the structure
\be\label{eq:3-herd-cliff}
	\fh_{\gamma}=[1]\,,\qquad \fh_{2\gamma}=\left[{5\over 2}\right]\,, \qquad \fh_{3\gamma} = [3]\oplus [5]\,,
\ee
as irreps of $\so(3)$, in agreement with \cite[app.A]{WWC}.

\medskip

In section \ref{sec:herd-psc} below, we will provide a derivation of the generating function employed above, obtained by a careful analysis of the soliton paths involved. In fact, we will provide such data for $k$-herds of any value of $k$. Adopting the same kind of $\wp$ as in our example, the above analysis extends straightforwardly to higher $k$ and $n$, allowing for a direct comparison with \cite[app.A]{WWC}, this provides further checks of the conjectures.

\subsection{The $3\,\mhyphen\,(2,3)$ - herd}\label{subsec:off-diag-herd}
We now move to a more complicated example, introducing a whole new type of critical network. It was shown in \cite{WWC} that in higher rank gauge theories there can be {wild walls} on the Coulomb branch. These are marginal stability walls $MS(\gamma,\gamma')$ with $|\langle\gamma,\gamma'\rangle|>2$ across which {wild BPS states} are created/lost. Wild BPS states are particularly interesting for us, because their Clifford vacua $\fh_{a\gamma+b\gamma'}$ typically consist of large and highly reducible representations of $\so(3)$, providing rich examples for testing our conjectures.

The critical networks of wild BPS states remained largely unexplored insofar. Except for states of charge $n(\gamma+\gamma')$ whose networks -- in \emph{some} regions of the Coulomb branch -- are known to be $k$-herds, no other cases have previously been studied. It is well known that all states of charges $a\gamma+b\gamma'$ for
\be
	\frac{k-\sqrt{k^2-4}}{2} < {\frac{a}{b}} <\frac{k+\sqrt{k^2-4}}{2}
\ee
are wild.

\medskip

\begin{figure}[h!]
\begin{center}
\includegraphics[width=0.88\textwidth]{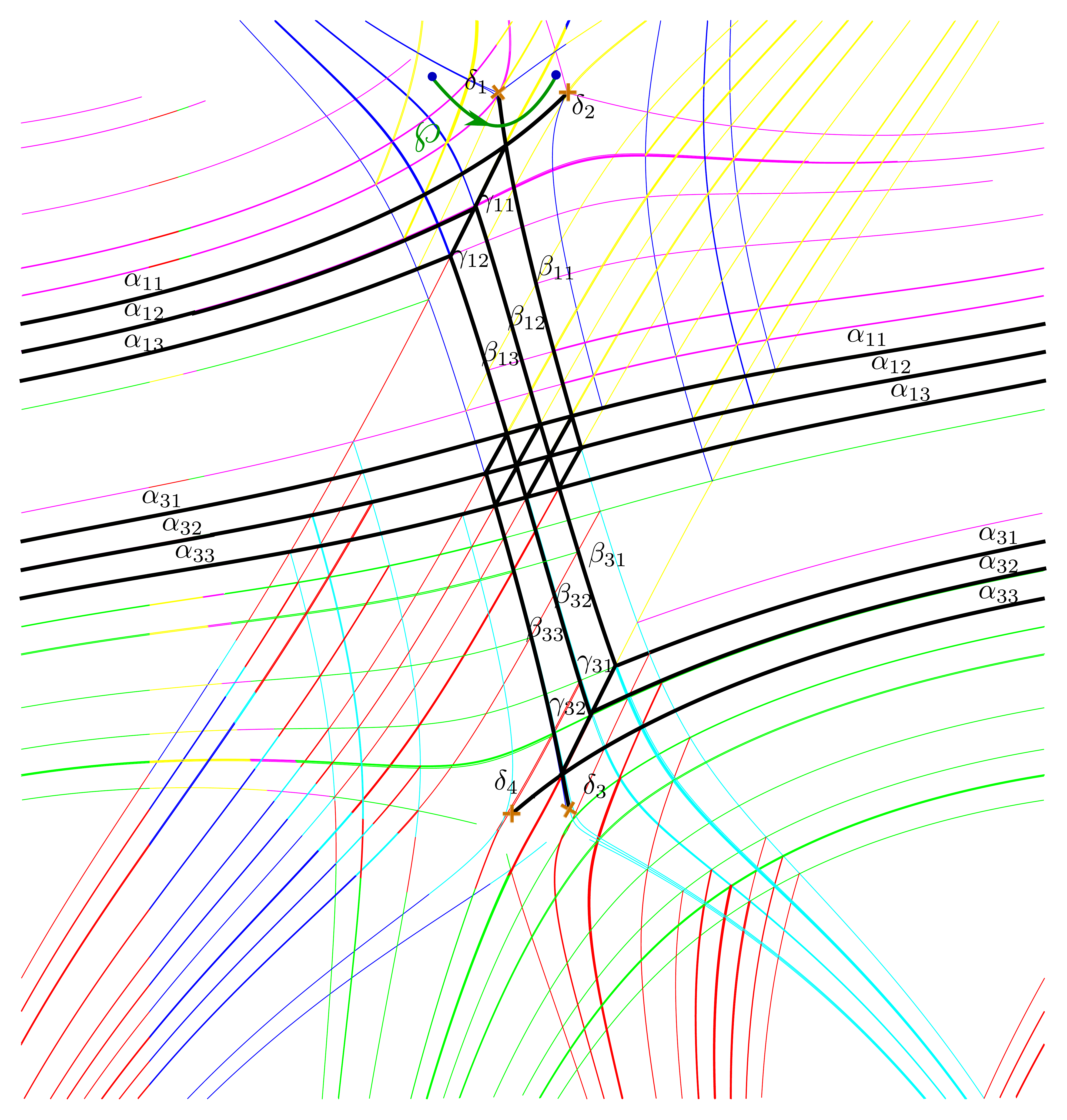}
\caption{The street map of the $3\,\mhyphen\,(2,3)$-herd, on the cylinder $C$. Different colors denote streets carrying solitons of different types, such as $ij,jk,ki$-type solitons. Two-way streets are marked in black. Recall there is an identification of the far left and far right endpoints of $\alpha_{11}$ in the figure, and so forth.}
\label{fig:3-23herd}
\end{center}
\end{figure}

We will now fix $k=3$ and $a=2, b=3$. This kind of state appears in one of the wild chambers of $SU(3)$ SYM: choosing the same point on the Coulomb branch as in \cite[\S 3.4]{WWC}, and tuning to $\vartheta = 5.22181$, the critical network of figure \ref{fig:3-23herd} appears.

By direct inspection of the soliton paths associated with each two-way street, one finds that
\be\label{eq:3-23-solitons}
	[\ell] = n\tgamma\quad\Leftrightarrow\quad \ell \ni \{2 \times \delta_{1,\Sigma},3 \times \delta_{2,\Sigma},2 \times \delta_{3,\Sigma},3 \times \delta_{4,\Sigma}\} \quad n \text{ times}.
\ee
with street names referring to figure \ref{fig:3-23herd}, and $\gamma$ being the generator of the critical sublattice (with orientation fixed by $\vartheta_{c}$). This structure could have been expected on homological grounds, being a mild generalization of the $3$-herd case.

Choosing $\wp$ as in fig \ref{fig:3-23herd} satisfies the halo-saturation condition. By direct inspection, the corresponding framed generating functions are
\be
\begin{split}
	F_{ii}(\wp,\vartheta_{c}^{-};y) & = \hY_{\wp^{(i)}} \\
	F_{ii}(\wp,\vartheta_{c}^{+};y) & = \hY_{\wp^{(i)}}  +   \left(y^{-7}+2 y^{-5}+4 y^{-3}+6 y^{-1}+ 6 y + 4 y^{3} + 2 y^{5} + y^{7}\right)\hY_{\wp^{(i)}+\tgamma} + \cdots \\
	& = \hY_{\wp^{(i)}} \,  \prod_{m\in[1]\oplus [1]\oplus [3]}\Phi_{2}((-y)^{m}\, \hY_{\tgamma})^{} \quad\times \  \Big(\cdots\Big)
\end{split}	
\ee
Due to the simplicity of $\wp$, we have $F_{ii}(\wp,\vartheta_{c}^{\pm};y) \equiv F_{\wp^{(i)}}(\wp,\vartheta_{c}^{\pm};y)$, we thus find agreement -- up to terms of order $X_{\tgamma}$ -- with the conjectured pattern\footnote{Note that, given any soliton path $\ell\in\Pi(\CW_{c})$,  $[\ell]=n\tgamma$ implies that $\langle\wp^{(i)},\ell\rangle = 2n$. Therefore $J_{\wp^{(i)},n\tgamma}=(2n-1)/2$ and the orbital $m'$ runs over $2n$ different values, thus reproducing correctly the subscript of the dilogarithms. The factor of $2$ comes from (\ref{eq:3-23-solitons}), had we chosen $\wp$ to cross $\delta_{2}$ or $\delta_{4}$, the corresponding factor would be $3$ instead of $2$.}:
\be
\begin{split}
	F_{\wp^{(i)}}(\wp,\vartheta_{c}^{-};y) = \hY_{\wp^{(i)}} \quad \mapsto \quad & \hY_{\wp^{(i)}}\, \prod_{n=1}^{\infty}\,\prod_{m\in \IZ} \Phi_{n}((-y)^{m} \,  \hY_{n\tgamma})^{a_{m}(n\gamma)} \\
	= \quad & F_{\wp^{(i)}}(\wp,\vartheta_{c}^{+};y)\,.
\end{split}
\ee
Moreover we recover the structure
\be
	\fh_{\gamma} = [1]\oplus[1]\oplus[3]\,,
\ee
as irreps of $\so(3)$, in agreement with \cite[app.A]{WWC}.

\medskip

The above analysis can in principle be extended to other values of $k,a,b$. In appendix \ref{app:off-diag-herds} we sketch the structure a large class of critical networks, which we call \emph{off-diagonal herds}\footnote{T.~Mainiero has
independently come to the picture of the off-diagonal herds and is currently
studying them.}.

\subsection{Generic interfaces and halos} \label{sec:gen_interf}
We now come back to generic interfaces, as mentioned above in \S\ref{sec:generic-interfaces}, our conjectures on the spin of framed BPS states naturally extend to these. There is a simple reason for studying generic interfaces: on the one hand they generically won't capture enough information to compute vanilla PSCs, but on the other hand their wall-crossing is of a more generic type, and studying it allows one to gain further insight into the implications of our conjectures.

In particular, the framed wall-crossing of IR line defects can be understood from a physical viewpoint in terms of a \emph{halo picture} \cite{GMN3}. The fact that some framed BPS states arrange into halos is particularly important for computing (framed/vanilla) PSCs because halos furnish representations of the group of spatial rotations. Thus the halos naturally encode the spin content of framed BPS states, for this reason the halo picture played a crucial role in establishing a physical derivation of the motivic KS wall-crossing formula. Given the importance and the success of this picture, it is of particular significance to check whether predictions based on our conjectures are compatible with it.

To make the question sharper, note that in the case of generic interfaces $\so(3)$ is broken to a Cartan subalgebra by the surface defects, which are stretched --say-- along the $x^{3}$ axis, thus we cannot expect the same type of halos that appeared in the case of line defects. So what kind of halo picture can we expect? The breaking of the rotational symmetry will induce a distinction among the states of a vanilla multiplet according to their $J_{3}$ eigenvalue. We may then expect to have halos of vanilla BPS states selectively binding to the interface depending on the $J_{3}$ quantum number. Before sharpening the question further, let us illustrate the latter statement with a simple example.

\begin{figure}[h!]
\begin{center}
\includegraphics[width=0.45\textwidth]{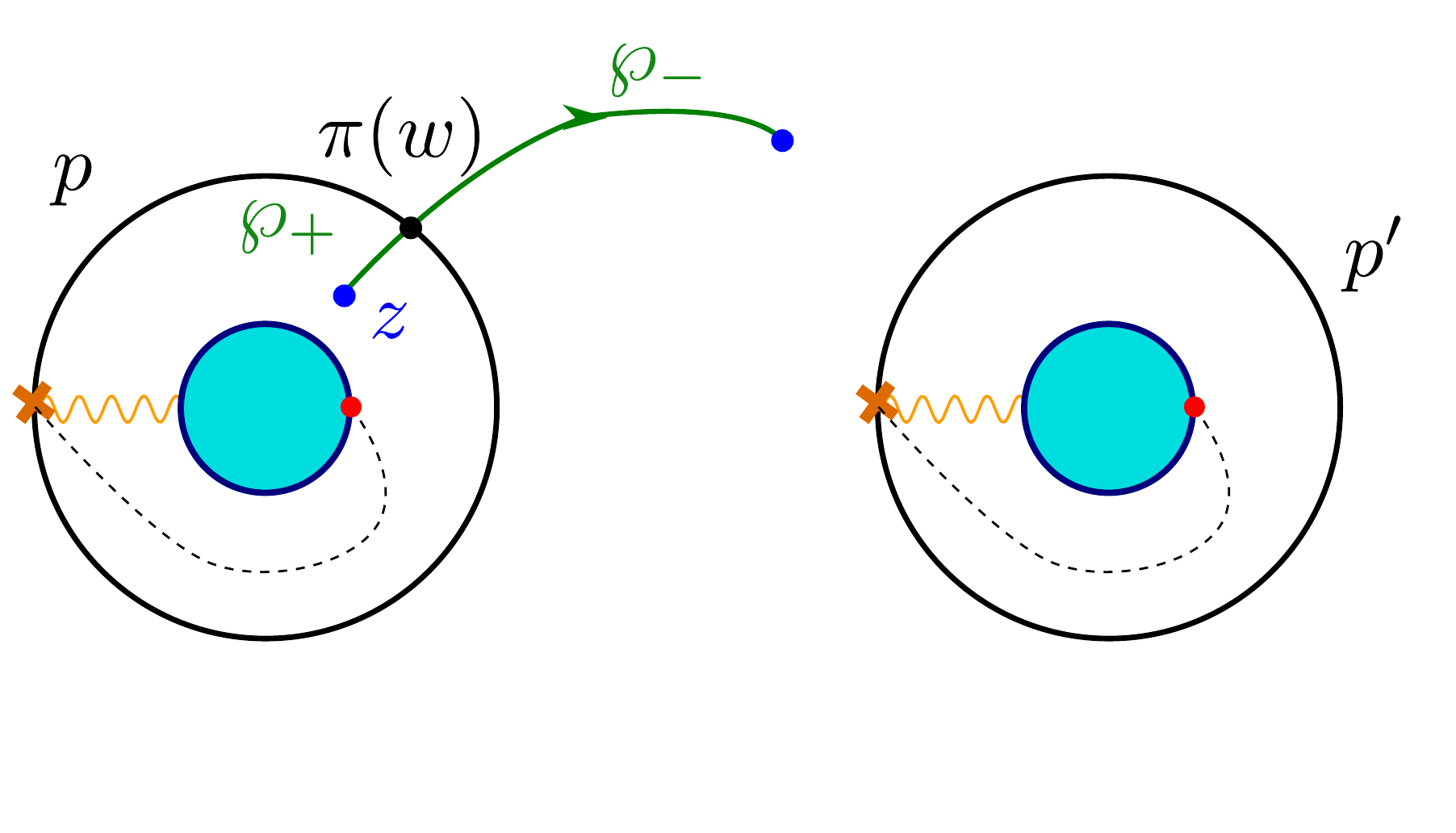}
\caption{The SU(2) vectormultiplet critical network, with a different choice of susy interface $\wp$. This choice is not halo-saturated.}
\label{fig:su2-variant}
\end{center}
\end{figure}

Consider a variant of the pure $SU(2)$ interface encountered above, as shown in figure \ref{fig:su2-variant}.

The removal of $\pi^{-1}({\rm end}(\wp))$ from $\Sigma$ now distinguishes between closed cycles coming from lifts of $p$ and those from lifts of $p'$. The sub-lattice of critical gauge charges generated by these lifts thus gets \emph{resolved} with respect to the case of a halo saturated interface and is now two dimensional. We denote by $\tgamma,\tgamma'$ the generators associated with $p$ and $p'$ respectively\footnote{A one-dimensional sub-lattice obviously has two possible generators, however the choice of $\vartheta_{c}$ canonically lifts the degeneracy.}. It follows easily from the above analysis that we now have

\be\label{eq:su2-generic-interface}
\begin{split}
	F_{ii}(\wp,\vartheta_{c}^{-};y) & = \hY_{\wp^{(i)}} \\
	F_{ii}(\wp,\vartheta_{c}^{+};y) & = \hY_{\wp_{}^{(i)}} + \sum_{n=1}^{\infty} y^{-n}\,\hY_{\wp_{}^{(i)}+n\tgamma}  
	= \hY_{\wp_{}^{(i)}} \, \Phi_{1}((-y)^{-1}\, \hY_{\tgamma})^{-1}\,,
\end{split}
\ee
comparing with (\ref{eq:su2-expected}) one realizes that the interface binds not to the whole vanilla multiplet, but only a ``partial'' halo is formed, as if the interface is binding only to vanilla states with $2J_{3}=-1$. 
Introducing the quantum-dilogarithms
\be\label{eq:q-dilog}
\Phi(\xi) := \prod_{k=1}^{\infty} (1+y^{2k-1}\xi)^{-1}\,,
\ee
the above can be recast into the suggestive form
\be\label{eq:su2-suggestive}
\begin{split}
	F_{ii}(\wp,\vartheta_{c}^{+};y) & = \Phi((-y)^{-1}\hY_{\tgamma})^{a_{-1}(\gamma)}  F_{ii}(\wp,\vartheta_{c}^{-};y)  \,   \Phi((-y)^{-1}  \,  \hY_{\tgamma})^{-a_{-1}(\gamma)}\\
	& = {\cal O}\, F_{ii}(\wp,\vartheta_{c}^{-};y)  \,   {\cal O}^{-1}\\
	{\cal O} & = \Phi((-y)^{+1}\hY_{\tgamma'})^{a_{+1}(\gamma)} \Phi((-y)^{-1}\hY_{\tgamma})^{a_{-1}(\gamma)} 
\end{split}
\ee
with $a_{\pm1}(\gamma) = -1$ (cf. (\ref{eq:su2-expected})) and where in the first line we used identity (\ref{eq:finite-quantum}) as well as the equivalence $\Phi_{n}(z) = \Phi_{-n}(y^{-2n}z)$. In the second line we used the fact that $\langle\wp^{(i)},\tgamma'\rangle=0$ hence $\hY_{\tgamma'}\hY_{\wp^{(i)}}=\hY_{\wp^{(i)}}\hY_{\tgamma'}$.

We will say that \emph{the framed wall-crossing of a generic interface is compatible with the halo picture if the $\CK$-wall jump of the generating function of framed PSCs can be expressed as a conjugation by quantum dilogarithms}\footnote{The choice of this criterion, as opposed to just demanding some jump of the form (\ref{eq:su2-generic-interface}), is more stringent. It is relatively easy to find an expression for a $\CK$-wall jump as a product of finite-type dilogs, but not all finite-type dilogs correspond to conjugation by quantum dilogs.}.

\begin{figure}[h!]
\begin{center}
\includegraphics[width=0.64\textwidth]{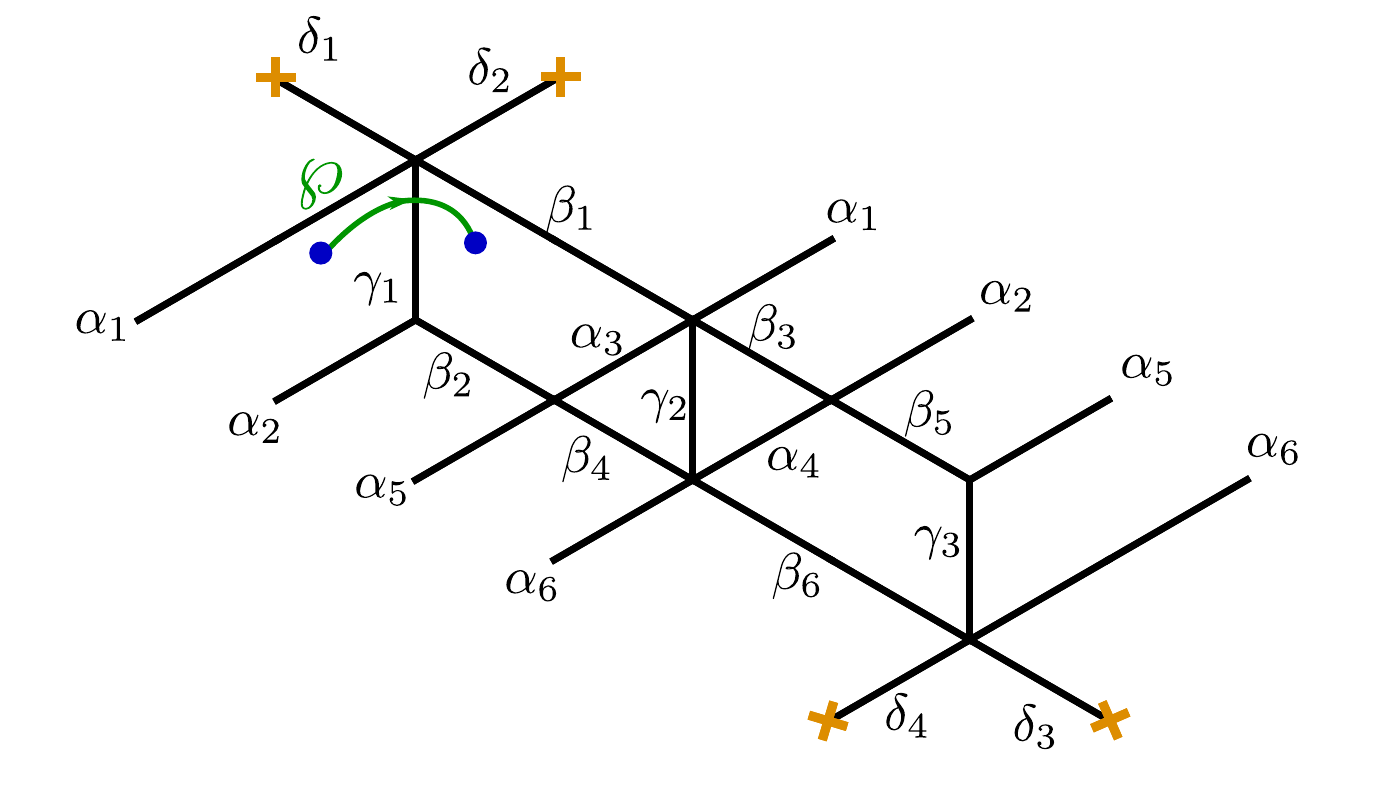}
\caption{The $3$-herd critical network, with a different choice of $\wp$. This choice is not halo-saturated.}
\label{fig:3herd-variant}
\end{center}
\end{figure}

As the next example will illustrate, it is not at all obvious that this criterion will be satisfied in general. Let us consider a different choice of interface for the $3$-herd, as shown in figure 
\ref{fig:3herd-variant}.

After removing endpoints of $\wp$ from $C$ and their lifts from $\Sigma$, there are two basic refined homology classes that we need to consider. They obey
\be
	\langle\wp^{(j)},\tgamma_{1}\rangle=-1\qquad \langle\wp^{(j)},\tgamma_{2}\rangle=0 \qquad \tgamma_{1}-\tgamma_{2}=\tgamma_{f}
\ee
where $\tgamma_{f}$ 
corresponds to a small cycle circling ${\rm end}(\wp^{(j)})$ clockwise. $\tgamma_{f}$ is in the annihilator of $\langle\,,\,\rangle$ when restricted to $\tilde\Gamma^{*}$. We will refer to it as a ``technical flavor charge''.

By direct inspection we find the following detour generating functions
\be\label{eq:3-herd-variant}
\begin{split}
	F_{jj}(\wp,\vartheta_{c}^{-};y) & = \hY_{\wp^{(j)}} \\
	F_{jj}(\wp,\vartheta_{c}^{+};y) & = \hY_{\wp^{(j)}}  +  y^{2}\, \hY_{\wp^{(j)}+\tgamma_{1}}+ (y+ y^5)\hY_{\wp^{(j)}+\tgamma_{1}+\tgamma_{2}}+2 y^4\hY_{\wp^{(j)}+2\tgamma_{1}} \\
	& + 5  y^6 \hY_{\wp^{(j)}+3\tgamma_{1}}+(5 y^7  +5  y^3) \hY_{\wp^{(j)}+2\tgamma_{1}+\tgamma_{2}}\\
	& + (y^{10} +2  y^8 + y^4 +y^{-2}+2 )\hY_{\wp^{(j)}+\tgamma_{1}+2\tgamma_{2}}+\cdots
\end{split}
\ee
this jump of $F(\wp)$ presents a little puzzle: as explained in appendix \ref{app:halo-factorization} it cannot be immediately expressed as conjugation by quantum dilogarithms. This would seem to indicate some tension between our conjectures and the halo picture, for the case of generic interfaces.

However, by introducing a technical assumption on certain ``flavor charges'' associated with the endpoints of the interface, we found that the above expression \emph{can} be massaged into a factorizable form. We will presently provide the details of this computation. While it is not clear to us what the generalization to generic interfaces should be, we expect one to exist.
To recover the halo picture, we start with the identity
\be\label{eq:bad}
\begin{split}
	\hY_{\tgamma_{1}}=\hY_{\tgamma_{2}}\hY_{\tgamma_{f}}=\hY_{\tgamma_{f}}\hY_{\tgamma_{2}}\ \Rightarrow\ \hY_{\wp^{(j)}+n_{1}\tgamma_{1}+n_{2}\tgamma_{2}} &=y^{-n_{2}} \hY_{\wp^{(j)}+(n_{1}+n_{2})\tgamma_{1}}\hY_{-n_{2}\tgamma_{f}}\\
	& =y^{n_{2}}\hY_{-n_{2}\tgamma_{f}} \hY_{\wp^{(j)}+(n_{1}+n_{2})\tgamma_{1}}
\end{split}
\ee
to turn the above into
\be
\begin{split}
	F_{jj}(\wp,\vartheta_{c}^{+};y) & = \hY_{\wp^{(j)}}  +  y^{2}\, \hY_{\wp^{(j)}+\tgamma_{1}}+ (y^{2}+ y^6)\hY_{-\tgamma_{f}}\hY_{\wp^{(j)}+2\tgamma_{1}}+2 y^4\hY_{\wp^{(j)}+2\tgamma_{1}} \\
	& + 5  y^6 \hY_{\wp^{(j)}+3\tgamma_{1}}+(5 y^8  +5  y^4) \hY_{-\tgamma_{f}}\hY_{\wp^{(j)}+3\tgamma_{1}}\\
	& + (y^{12} +2  y^{10} + y^6 +1+2y^{2} )\hY_{-2\tgamma_{f}}\hY_{\wp^{(j)}+3\tgamma_{1}}+\cdots
\end{split}
\ee
then (this is our technical assumption\footnote{Recall that $\tgamma_{f}$ is a ``technical'' flavor charge, arising from the removal of endpoints of the interface from $\Sigma$. It is therefore natural to assume that formal variables --which should be related to holonomies of a flat connection on a line bundle over $\Sigma$-- should resemble trivial holonomy around this cycle.} ) taking $\hY_{\tgamma_{f}}\to 1$
\be\label{eq:q-dilog-factorized}
\begin{split}
	F_{jj}(\wp,\vartheta_{c}^{+};y) 
	& = \hY_{\wp^{(j)}} \Big( 1 +  y^{3}\, \hY_{\tgamma_{1}}+ (y^{4}+2 y^{6} + y^8)\, \hY_{2\tgamma_{1}} \\
	& + (y^{15} +2  y^{13}  +5 y^{11} + 6  y^{9} +5  y^{7} +2y^{5}  + y^{3} )\, \hY_{\wp^{(j)}+3\tgamma_{1}} +\cdots\Big) \\
	& =  {\cal O}   \,\hY_{\wp^{(j)}}\,    {\cal O}^{-1}
\end{split}
\ee
with 
\be\label{eq:3-herd-dilogs}
\begin{split}
	{\cal O} & = \Phi_{}((-y)^{2}  \hY_{\tgamma_{1}})\, \Phi_{}((-y)^{3}  \hY_{2\tgamma_{1}})^{-1} \, \Phi_{}((-y)^{5}  \hY_{2\tgamma_{1}})^{-1} \\
	& \times \Phi \left((-y)^{10}  \hY_{3\tgamma_1}\right) \Phi \left((-y)^8 \hY_{3\tgamma_1}\right) \Phi \left((-y)^6  \hY_{3\tgamma_1}\right){}^2 \\
	& \times \Phi \left((-y)^4  \hY_{3\tgamma_1}\right) \Phi \left((-y)^2  \hY_{3\tgamma_1}\right)\, \cdots \\
	& = \prod_{n>0}\prod_{m\in\IZ}\Phi((-y)^{m}\,\hY_{n\tgamma_{1}})^{c_{n,m}}\,.
\end{split}	
\ee
All values of $m$ appearing in the factorization are compatible with the vanilla spin content (\ref{eq:3-herd-cliff}), moreover the exponents satisfy
\be
	0\leq \frac{c_{n,m}}{a_{m}(n\tgamma_{1})}\,\leq \, 1
\ee 
for $a_{m}(n\tgamma_{1})$ defined by (\ref{eq:PSC-conjecture}), this is compatible with the interpretation that each dilog is the contribution to the Framed Fock space by $|c_{n,m}|$ vanilla oscillators of corresponding charge and $2J_{3}$ eivgenvalue. Hence we recover the picture that the generic interface interacts unevenly with different states within a vanilla multiplet, as in the previous $SU(2)$ example. Only some of the vanilla states bind to the interface as the $\CK$-wall is crossed, while another part of the vanilla multiplet does not. 

It is worth mentioning that, based on the observations and the conjecture of appendix \ref{app:L-r}, it should be possible to enhance (\ref{eq:3-herd-dilogs}) with dilog factors corresponding to other states in the vanilla multiplet as well, in the same fashion as in (\ref{eq:su2-suggestive}). This would reinforce the picture of a generic interface interacting selectively with vanilla states depending on their $J_{3}$ quantum number: the ``phantom'' quantum dilogs would be those of states that do not couple to the interface. For \emph{halo-saturated} interfaces on the other hand all states of a multiplet contribute to the jump, there are no ``phantoms'', hence the choice of terminology. We will not pursue the study of generic interfaces further, although it would certainly be interesting to gain a systematic understanding of these phenomena.

\medskip



To sum up, we have given a sharp criterion to determine whether our conjectures are compatible with the halo picture, based on whether the $\CK$-wall jump of the related generating function of framed states can be expressed in terms of conjugation by quantum dilogarithms. However we do not have a general proof that this is always the case, and we have seen that it takes some care to check that the halo picture is recovered even in simple examples. It would be good to clarify these matters further.

\section{$m$-herds}\label{sec:herds}

Herds, already encountered above, are a class of critical networks occurring in higher rank gauge theories \cite{WWC}. As reviewed in section \ref{subsec:off-diag-herd}, these theories have wild chambers on their moduli space of vacua, where BPS particles of charges $\gamma,\gamma'$ with $\langle\gamma,\gamma'\rangle>2$ can form stable {wild} BPS boundstates. Herd networks correspond to ``slope 1'' boundstates, i.e. states with charge of the form $n(\gamma+\gamma')$ with $n\geq 1$.

In this section we study the refined soliton content of herds, relying on equations (\ref{eq:6way}). From the refined soliton data, vanilla PSCs of wild BPS states can be extracted. The main result is a \emph{functional equation}  for the generating function of PSCs. Our analysis applies to $m$-herds for any positive integer $m$.

\subsection{The horse and other preliminaries}\label{sec:Qinside}

\begin{figure}[h!]
\begin{center}
\includegraphics[width=0.4\textwidth]{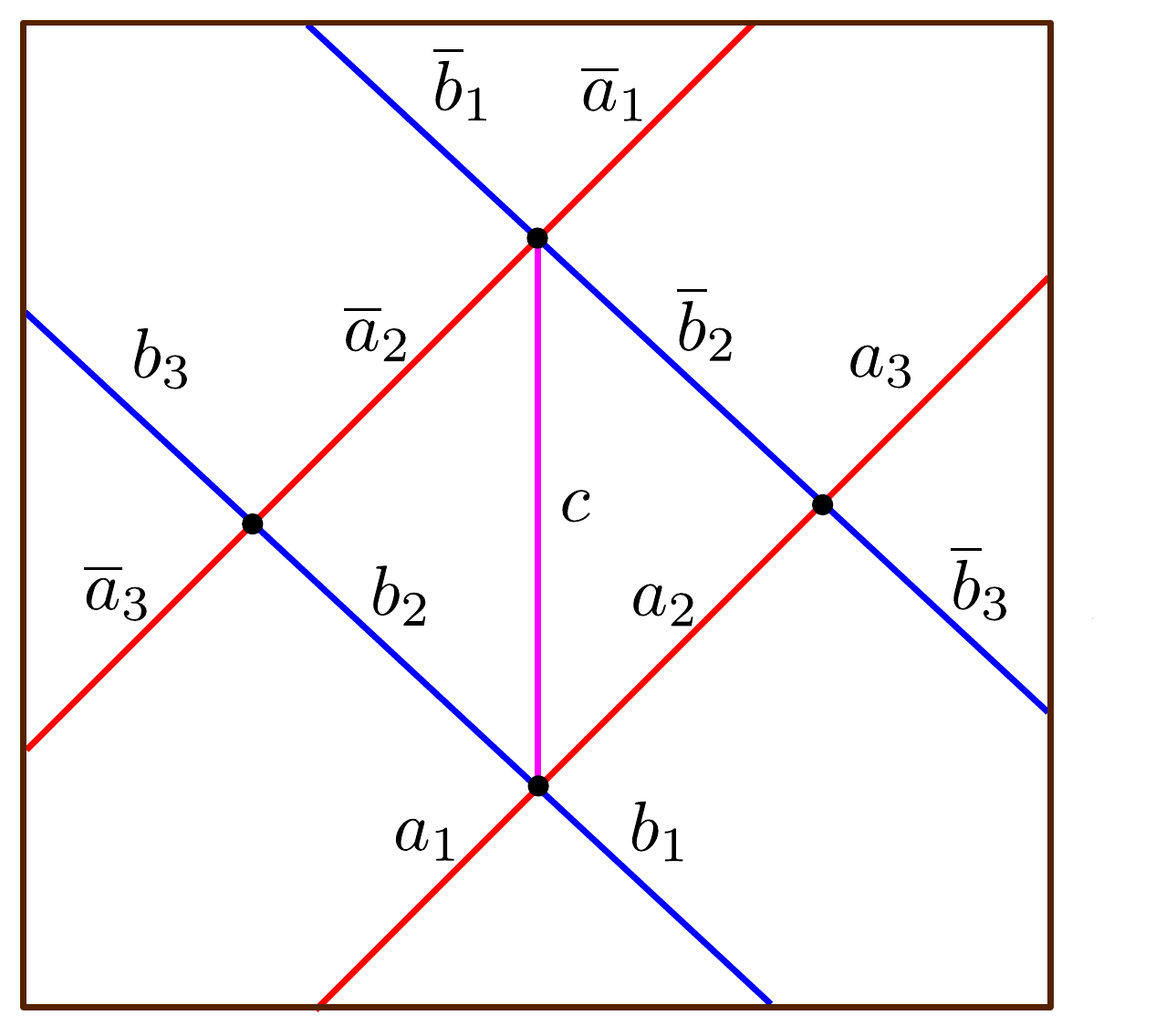}
\end{center}
\label{fig:horse}
\caption{The horse. All streets are assumed to be 2-way (in some cases certain streets are actually one-way, but this case is automatically handled by our setup) and directions correspond to conventions explained in the text. We stress that $p,\bar p$ are distinct streets and are not identified: while the bar stresses the symmetry of the equations, we do not impose particular boundary conditions.}
\end{figure}

$m$-herds are constructed by gluing together $m$ copies of an elementary subnetwork called \emph{the horse} (a.k.a. the $1$-herd, with suitable boundary conditions \cite{WWC}), shown in figure \ref{fig:horse}. We therefore start by studying the soliton content of the horse, and then move on to $m>1$.

The lift of this kind of network involves 3 sheets of the cover $\pi:\Sigma\to C$, say $i,j,k$; there are then three types of two-way streets: $ij, \,jk, \,ik$ marked by blue, red and purple colors on the figure respectively.

Recall that each two-way street can be thought of as a pair of one-way streets flowing in opposite directions. Therefore to each two-way street $p$ we associate a refined soliton generating function $D_{p}$  (resp. $U_{p}$) for the one-way street flowing downwards (resp. upwards). We fix conventions such that one-way streets of types $ij$, $jk$, $ik$ flow upwards (they carry solitons of the $U$-type), while $ji$, $kj$, $ki$ flow downwards (they carry solitons of the $D$-type).

Without loss of generality we choose the British resolution, then applying the 6-way joint rules (\ref{eq:6way}) to all four joints, we find the following set of identities
\be
\begin{split}
D_{\bar b_2}=D_{\bar b_1},\quad D_c=D_{\bar a_1}D_{\bar b_1},\quad D_{\bar a_2}=D_{\bar a_1}\hat Q(\bar b_2)\\
U_{b_2}=U_{b_1},\quad U_c=U_{b_1}U_{a_1},\quad U_{a_2}=\hat Q(b_2)U_{a_1}\\
D_{\bar b_3}=D_{\bar b_2},\quad U_{a_3}=\hat Q(\bar b_3) U_{a_2}\\
U_{b_3}=U_{b_2},\quad D_{\bar a_3}=D_{\bar a_2}\hat Q(b_3)
\end{split} \label{eq:horse}
\ee

To each street we may associate two generating functions
\be
\hat Q^{(\alpha)}(p):= 1+D_p U_p\qquad \hat Q^{(\beta)}(p):=1+U_p D_p
\ee
where $\alpha = j,k,k$ and $\beta=i,j,i$ respectively for the three types of streets. In (\ref{eq:horse}) we suppressed the superscripts, but it is understood that the suitable choice of $\hat Q^{(\star)}$ appears: this is determined by compatibility of concatenation of paths with those within the $D,U$ they multiply.

In the same way as equations (\ref{eq:6way}) were derived from homotopy invariance of off-diagonal terms of the formal parallel transport, there is a corresponding set of equations descending from homotopy invariance of diagonal terms (the story is closely parallel to \cite[eq.s (6.18)-(6.19)]{GMN5}). These may be cast into the form of a ``conservation law'' for different streets coming into one joint, for example referring to figure \ref{fig:six-way} we have for the sheet-${1}$ component
\be\label{eq:Qhi}
\hU_{\wp^{(1)}_{+}}\hat Q^{(1)}(p_1) \hU_{\wp^{(1)}_{0}}  \hat Q^{(1)}(p_2)\hU_{\wp^{(1)}_{-}}=\hU_{{\wp'}^{(1)}_{+}}\hat Q^{(1)}(p_5)\hU_{{\wp'}^{(1)}_{0}}\hat Q^{(1)}(p_4)\hU_{{\wp'}^{(1)}_{-}}
\ee
where $\wp,\wp'$ are here understood to be broken apart into pieces compatibly with the necessary concatenations. Analogous expressions hold for other streets and sheets combinations.

To keep track properly of the writhe of detours, it is more convenient to express the above rule with a richer notation.
Consider a path $\chi$ with endpoints on $\Sigma\setminus\Sigma^{*}$, intersecting $\pi^{-1}(\CW_{c})$ somewhere. An example is provided in fig.\ref{fig:3herd}, where the path $\chi$ may be taken to be $\wp^{(i)}$. Let $p$ be one of the streets whose lift is crossed by $\chi$, the intersection splits $\chi$ into two pieces denoted $\chi_\pm$.
Associated with $\chi$ we can construct a ``corrected'' detour generating function 
$Q_{\chi}(p)$ defined by the following relation
\be
\hY_{\beta(\chi)}Q_{\aux}(p,y,z):=\rho\left(\hU_{\chi_+}\hat Q(p)\hU_{\chi_-}\right)
\ee
where $z=y \hY_{\gamma_c}$ and $\rho$ was defined in (\ref{eq:rho}).
Where we implicitly made use of the fact that all detours' homology classes can be decomposed as $\pi=\beta(\chi)+n\gamma_c$. 
As will be evident in the following, the ``correction'' by $\chi$ consists of extra units of writhe induced by possible intersections of $\chi_{\pm}$ with the soliton detours to which they concatenate.

Moreover, it is easy to show that the ``conservation rule'' (\ref{eq:Qhi}) carries over through the map $\rho$:
\be\label{eq:Q-homotopy}
Q_{\aux}(p_1,y,z)Q_{\aux}(p_2,y,z)=Q_{\aux}(p_4,y,z)Q_{\aux}(p_5,y,z)
\ee
in fact, choosing the auxiliary paths as in fig. \ref{fig:qref}, multiplying both sides of (\ref{eq:Qhi}) by $\hU_{\chi_+}$ and $\hU_{\chi_-}$ from
the left and from the right respectively, accounting for the regular homotopy classes\footnote{More precisely, the $\delta_{p}$ are open regular homotopy classes on $\Sigma^{*}$ consisting of concatenations of $ij$ solitons with $ji$ solitons supported on $p$.} of detours $\delta_p$ from each street $p$, and applying the morphism $\rho$, we find
\be
\sum\lm_{\delta_{p_1},\delta_{p_2}}y^{\wr(\chi_+\delta_{p_1}\delta_{p_2}\chi_-)}Y_{\beta(\chi)+\beta(\delta_{p_1})+\beta(\delta_{p_2})}=
\sum\lm_{\delta_{p_4},\delta_{p_5}}y^{\wr(\chi_+\delta_{p_5}\delta_{p_4}\chi_-)}Y_{\beta(\chi)+\beta(\delta_{p_5})+\beta(\delta_{p_4})}
\ee
Noting that the mutual intersections of the detours paths $\delta_{p_i}$ all vanish, it is easy to see that both sides factorize into
\be
\begin{split}
& Y_{\beta(\chi)}\left(\sum\lm_{\delta_{p_1}}y^{\wr(\chi_+\delta_{p_1}\chi_-)}y^{\langle [\chi],[\delta_{p_1}] \rangle}Y_{\beta(\delta_{p_1})}\right)
\left(\sum\lm_{\delta_{p_2}}y^{\wr(\chi_+\delta_{p_2}\chi_-)}y^{\langle [\chi],[\delta_{p_2}] \rangle}Y_{\beta(\delta_{p_2})}\right)=\\
& =Y_{\beta(\chi)}\left(\sum\lm_{\delta_{p_4}}y^{\wr(\chi_+\delta_{p_4}\chi_-)}y^{\langle [\chi], [\delta_{p_4}] \rangle}Y_{\beta(\delta_{p_4})}\right)
\left(\sum\lm_{\delta_{p_5}}y^{\wr(\chi_+\delta_{p_5}\chi_-)}y^{\langle [\chi], [\delta_{p_5}] \rangle}Y_{\beta(\delta_{p_5})}\right)
\end{split}
\ee
establishing eq.\ref{eq:Q-homotopy}.

The above derivation keeps holding if we start moving the point where the path $\aux$ is connected to the street $p$, while preserving the homotopy class of the detours. 
In this way we can simultaneously uniquely assign generating functions to each street whose lift to the $i$-th sheet is contained in a contractible chart on $\Sigma^*$.

In the following we will omit the subscript $\aux$, leaving understood that we will always be working with such ``corrected'' generating functions.

\subsection{Herds}

\begin{figure}[h!]
\begin{center}
\includegraphics[width=0.8\textwidth]{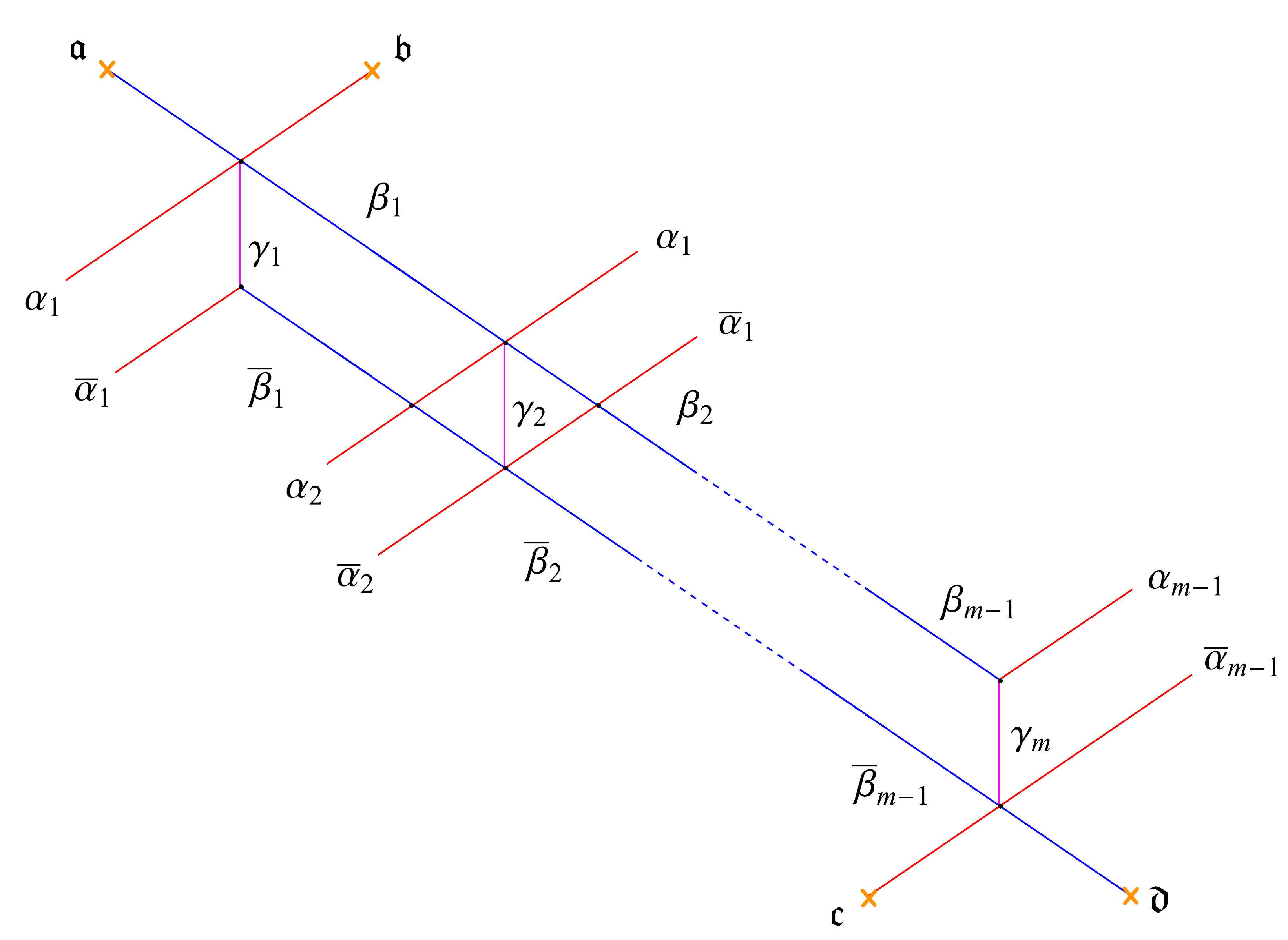}
\end{center}
\caption{The $m$-herd, the streets are glued together according to labels. Typically the herd ``wraps'' a tube of $C$, in the picture the tube has been cut along a spiral and opened up.  \label{fig:m-herd}}
\end{figure}

An $m$-herd is a critical network consisting of a sequence of horses glued together, see for example figure \ref{fig:m-herd}. The outer legs of each horse are either glued to external legs of neighboring horses, 
or terminate on branching points, as displayed. Just like horses, herds lift to a triple of sheets $i,j,k$, we adopt the same sheet labels and color conventions for the streets of herds as we used for the horse. Thus, for example, branch points $\fa$ and $\fd$ are of type $(ij)$, while  $\fb$, $\fc$ are of type $(jk)$.

Denoting the formal variables of the four simpletons by $\hU_{\fa}$, ..., $\hU_{\fd}$ (cf. figure \ref{fig:m-herd}), these soliton paths can be propagated through the $m$-herd by using the rules derived in the previous section together with gluing rules
\be
\begin{split}
D_{\alpha_1}=\hU_{\fb}\hat Q(\beta_1),\qquad&  D_{\alpha_n}=D_{\alpha_{n-1}}\hat Q(\beta_n)\hat Q(\bar \beta_{n-1})\,,\\
U_{\bar\alpha_{m-1}}=\hat Q(\bar\beta_{m-1}) \hU_{\fc},\qquad & U_{\bar\alpha_{n-1}}=\hat Q(\bar\beta_{n-1})\hat Q(\beta_n) U_{\bar\alpha_n}\,,\\
D_{\beta_n}=\hU_{\fa},\qquad & U_{\bar\beta_n}=\hU_{\fd}\,.
\end{split}
\ee
These are obtained by applying (\ref{eq:horse}) to the network, for example 
\be
	D_{\alpha_{2}} = D_{\alpha_{1}} \, \hat Q(\bar\beta_{2})\,\hat Q(\beta_{1})
\ee
is derived from
\be
	D_{\bar a_{3}} = D_{\bar a _{2}}\,\hat Q(b_{3}) = \Big( D_{\bar a_{1}}\, \hat Q(\bar b_{2}) \Big)\,\hat Q(b_{3}) \,.
\ee

Thus we find the following expression for the generating function of a generic \emph{vertical} 2-way street $\gamma_n$
\be
\begin{split}
\hat Q(\gamma_n)=1+D_{\alpha_{n-1}}D_{\beta_{n-1}}U_{\bar\beta_{n}}U_{\bar\alpha_{n}}=\\
=1+\hU_{\fb}\hat Q(\beta_1)\left(\prod\lm_{j=2}^{n-1}\hat Q(\beta_j)\hat Q(\bar\beta_{j-1})\right)\hU_{\fa}\hU_{\fd}\left(\prod\lm_{j=n}^{m-2}\hat Q(\bar\beta_j)\hat Q(\beta_{j+1})\right)\hat Q(\bar\beta_{m-1})\hU_{\fc}\,.
\end{split}
\ee

It is easily seen that each $\hat Q$ generating function is a formal power series in a single word, then we consider assigning an {\emph{algebraic generating function}} to each $\hat Q$, as follows. For example,
\be
\begin{split}
\hat Q(\beta)=\sum\lm_{n\in\IN} \omega_n(y,\beta) \left(\hU_{\fa}\hU_{\fd}\hU_{\fc}\hU_{\fb}\right)^n,\quad Q(\beta,y,z)=\sum\lm_{n\in\IN} \omega_n(y,\beta) z^n\\
\hat Q(\bar\beta)=\sum\lm_{n\in\IN} \omega_n(y,\bar\beta) \left(\hU_{\fc}\hU_{\fb}\hU_{\fa}\hU_{\fd}\right)^n,\quad Q(\bar\beta,y,z)=\sum\lm_{n\in\IN} \omega_n(y,\bar\beta) z^n
\end{split}
\ee
where $\omega_n$ are Laurent polynomials in $y$ arising after casting the $\hat Q$ into this form by means of (\ref{eq:product-rule}) and (\ref{eq:KS-rem}).
It is important to note that the two words made of simpleton variables (in the expressions for $\beta,\bar\beta$ respectively) are {\emph{ different}}.
Moreover, in constructing these functions we hid the involvement of necessary transition functions which actually extend the simpleton paths across the herd (see \cite[app.C]{WWC}). We fix a prescription for the transport of soliton paths as follows: the transport must be carried out along streets of the same soliton type (for example to join $\fa,\fd$ we can continue them across the $\beta$-type streets) plus any one of the vertical streets of $\gamma$-type.

In particular, the generating functions of vertical $\gamma$-streets $\hat Q(\gamma_n)$ are formal series in the single word $\hU_{\fb}\hU_{\fa}\hU_{\fd}\hU_{\fc}$. Performing the due substitutions of the generating functions $\hat Q(\beta),\hat Q(\bar\beta)$ into the expression for $\hat Q(\gamma_n)$ we end up with an expression in which different words are {\emph{scrambled}}. To make some order, we employ a trick explained in appendix \ref{app:dirtytrick}, manipulating the expressions of $\hat Q(\gamma_{n})$ by means of the equivalence
\be
\begin{split}
\big(\hU_{\fa}\hU_{\fd}\big) \big( \hU_{\fc}\hU_{\fb}  \big)  \dot= \, y^{2\left\langle\cl(\fa\fd),\cl(\fc\fb)\right\rangle}\, \big( \hU_{\fc}\hU_{\fb}  \big)  \big(\hU_{\fa}\hU_{\fd}\big)\,,
\end{split}
\ee
the symbol $\dot=$ means that both sides have the same image under $\rho$.

For an $m$-herd, we have simply $\left\langle\cl(\fa\fd),\cl(\fc\fb)\right\rangle=m$.
Taking this into account, we may rewrite the equations for the formal series $\hat Q$ in terms of algebraic ones which include corrections from generic auxiliary paths $\aux$
\be\label{eq:aux}
\begin{split}
&Q_{\aux}(\gamma_n,y,z)=1+z\, y^{\wr\left(\aux_+\fb\fa\gamma_n\fd\fc\aux_-\right)}\times\\
&\times \left(\prod\lm_{j=1}^{n-1}Q_{\aux}(\beta_j,y,z)\right)  \left(\prod\lm_{j=n+1}^{m-1}Q_{\aux}(\beta_j,y,zy^{-2m})\right) \left(\prod\lm_{j=n}^{m-1}Q_{\aux}(\bar\beta_j,y,z)\right)\left(\prod\lm_{j=1}^{n-2}Q_{\aux}(\bar\beta_j,y,zy^{2m})\right)
\end{split}
\ee
where in $\fb\fa\gamma_n\fd\fc$ it is understood that simpletons are propagated through the network in the way explained above, and the extra powers of $y$ in the arguments of $Q$'s account for due reorderings.
Here the path $\aux$ intersects the 2-way street $\gamma_n$ on sheet $i$ and the factor $\gamma_n$ inside $y^{\wr\left(\aux_+\fb\fa\gamma_n\fd\fc\aux_-\right)}$ means a lift of this 2-way street to another sheet. 

Switching to ``universal'' generating functions, all corresponding to a specific path $\chi=\wp^{(i)}$ as drawn in fig.\ref{fig:3herd}, gives simply
\be
\wr\left(\aux_+\fb\fa\gamma_n\fd\fc\aux_-\right)=2n-1-m
\ee
Applying homotopy invariance (\ref{eq:Q-homotopy}) thus yields
\be
\begin{split}
& \qquad \qquad\qquad Q(\fa,y,z)=Q(\fd,y,z)=\prod\lm_{j=1}^{m-1}Q(\gamma_j,y,z),\\
& Q(\beta_n,y,z)=\prod\lm_{j=n+1}^{m-1}Q(\gamma_j,y,z),\quad\qquad Q(\bar\beta_n,y,z)=\prod\lm_{j=1}^{n-1}Q(\bar\beta_j,y,z)
 \end{split}
\ee

After substitution of the ansatz
\be
Q(\gamma_n,y,z)=P(z y^{2n-1-m},y)
\ee
into (\ref{eq:aux}) all the equations in this system turns into the same single equation with a parameter $z$ shifted by powers of $y$ in different ways. This defining equation reads
\be\label{eq:P}
P(z,y)=1+z\prod\lm_{s=-(m-2)}^{m-2}P(zy^{2s},y)^{m-1-|s|} \,.
\ee
This is a \emph{functional} equation for  power series in $z$, with Laurent polynomials in $y$ as their coefficients.

In the limit $y^2\rightarrow 1$ all terms in the product on the r.h.s. become the same, then powers just sum up to $(m-1)^2$, properly reproducing the {\emph{algebraic}} herd equation found in \cite[eq.(1.2)]{WWC}. It therefore generalizes Prop.3.1 of \cite{Reineke2}  and the defining equation of (1.4) of \cite{GP} to the ``refined'' case.

Given a solution to the {{ functional}} equation (\ref{eq:P}), generating functions on other 2-way streets follow simply by
\be\label{eq:P_genf}
\begin{split}
& Q(\fa,y,z)=Q(\fd,y,z)=\prod\lm_{s=-\frac{m-1}{2}}^{\frac{m-1}{2}} P(zy^{2s},y),\\
& Q(\gamma_n,y,z)=P(z y^{2n-1-m},y)
\end{split}
\ee
where the product is assumed to be taken either over integers or over half-integers.

Finally, we should note that, due to choice of (commutative) variables in this section, there is a controlled shift in powers of $y$ as compared\footnote{More precisely, the extra $y^{n}$ factor is omitted in (\ref{eq:PSC-conjecture}) because it is reproduced by $\hY_{\wp^{(i)}}\hY_{n\tgamma_{h}}=y^{n}\,\hY_{\wp^{(i)}+n\tgamma_{h}}$.} to the expected factorization (\ref{eq:PSC-conjecture}):
\be\label{eq:Q-fact}
	Q(\fa,y,z) = \prod_{\tgamma_{h}}\prod_{m\in\IZ}\ \Phi_{n(\tgamma_{h})} \Big( (-y)^{m}y^{n(\tgamma_{h})} z^{n(\tgamma_{h})} \Big)^{a_{m}(\tilde\gamma_{h})}\,.
\ee

\subsection{Herd PSC generating functions}\label{sec:herd-psc}

To conclude our discussion of herds, we examine some explicit solutions to the functional equation (\ref{eq:P}).

\paragraph{2-herd:}

Eq.(\ref{eq:P}) is algebraic in this case and can be solved explicitly
\be
P(z,y)=(1-z)^{-1}\,.
\ee
Thus
\be
Q(\fa,y,z)=Q(\fb,y,z)=Q(\fc,y,z)=Q(\fd,y,z)=(1- z y)(1-z y^{-1})\,,
\ee
corresponding to the expected vectormultiplet
\be
\Omega(\gamma_c,y)=y+y^{-1}, \quad \Omega(n\gamma_c,y)=0, \; n\geq 2
\ee

\paragraph{3-herd:}
$m=3$ provides the first non-trivial example, since in this case (\ref{eq:P}) is no longer algebraic. Nevertheless one can study its solutions perturbatively, introducing the series
\be\label{eq:P-series}
P(z,y)=1+\sum\lm_{n=1}^{\infty} \omega^{(m)}_n(y)z^n\,.
\ee
We find the following perturbative solution
\be\label{eq:P-3-herd}
\begin{split}
P(z,y)&=1+z+\left(y^{-2}+2+y^2\right)z^2+\left(y^{-6}+2 y^{-4}+5 y^{-2}+6+5 y^{2}+2 y^{4}+y^{6}\right)z^3+\\
&+\left(y^{-12}+2 y^{-10}+5 y^{-8}+10 y^{-6}+16 y^{-4}+23 y^{-2}+26+23 y^{2}+16 y^{4}+10 y^{6}\right.\\
& +\, \left.5 y^{8}+2 y^{10}+y^{12}\right) z^4+O(z^5)\\
&=\Phi_1(y z)\Phi_2\left((-y)^{-1} y^2 z^2\right)^{-1} \Phi_2\left((-y) y^2 z^2\right)^{-1}\Phi_3\left((-y)^{-4} y^3 z^3\right)\Phi_3\left((-y)^{-2} y^3 z^3\right)  \\
&\times \Phi_3\left( y^3 z^3\right)^2 \Phi_3\left( (-y)^2 y^3 z^3\right)\Phi_3\left((-y)^4 y^3 z^3\right)\Phi_4\left((-y)^{-9} y^4 z^4\right)^{-1}\Phi_4\left((-y)^{-7} y^4 z^4\right)^{-1} \\
&\times \Phi_4\left((-y)^{-5} y^4 z^4\right)^{-3}\Phi_4\left((-y)^{-3} y^4 z^4\right)^{-4}\Phi_4\left((-y)^{-1} y^4 z^4\right)^{-5}\Phi_4\left((-y) y^4 z^4\right)^{-5} \\
&\times \Phi_4\left((-y)^3 y^4 z^4\right)^{-4}\Phi_4\left((-y)^{5} y^4 z^4\right)^{-3}\Phi_4\left((-y)^{7} y^4 z^4\right)^{-1}\Phi_4\left((-y)^{9} y^4 z^4\right)^{-1}(1+O(z^5))\,.
\end{split}
\ee
Relations (\ref{eq:P_genf}) and (\ref{eq:Q-fact}) allow to extract the corresponding PSCs: denoting  $\chi_s(y)=(y^{2s+1}-y^{-(2s+1)})/(y-y^{-1})$
\be
\begin{split}
\Omega(\gamma_c,y)
	&=\chi_{1}(y)\\
\Omega(2\gamma_c,y)
	&=\chi_{\frac{5}{2}}(y)\\
\Omega(3\gamma_c,y)
	&=\chi_{3}(y)+\chi_{5}(y)\\
\Omega(4\gamma_c,y)
& =\chi_{\frac{5}{2}}(y) + 2 \chi_{\frac{9}{2}}(y) + \chi_{\frac{11}{2}}(y) +2 \chi_{\frac{13}{2}}(y) + \chi_{\frac{17}{2}}(y)
\end{split}
\ee
as anticipated in (\ref{eq:3-herd-cliff}). These results agree in fact with the ones derived by means of the \emph{motivic Kontsevich-Sobeilman wall-crossing formula} \cite[Appendix A.2]{WWC}.

\section{Extra remarks}\label{sec:extra}



\subsection{Kac's theorem and Poincar\'e polynomial stabilization}\label{sec:quiver}

\paragraph{Kac's theorem.}

As discussed in \cite[\S 8.2]{WWC}, Kac's theorem (see e.g. \cite{Reinike}) implies a charge-dependent bound on the highest-spin irreps in the Clifford vacua of BPS states. The highest admissible spin is related to the dimensionality of the corresponding quiver variety. In the case of interest to us, $m$-Kronecker quivers, the maximal spin for a state of charge $(n,n)$ is
\be\label{eq:j_max}
2J_{\rm max}^{(\rm quiver)}(n)=(m-2)n^2+1
\ee
Recall that Laurent decomposition of the PSC reads
\be
\Omega_n (y)=\sum\lm_{m=-2J_{\rm max}}^{2J_{\rm max}}a_m(n)\,(-y)^m\,,
\ee
also note that the highest power of $y$ for the $z^{k}$ term of the generating function comes from
\be
\Phi_k((-y)^{2J_{\rm max}(k)}y^k z^k)\sim z^k y^{k-1+2J_{\rm max}(k)}+\ldots\,.
\ee
Then let us study the maximal power of $y$ for the $z^{k}$ term of $m$-herd generating functions, as predicted by equation (\ref{eq:P}).
To do so, we consider the series expansion (\ref{eq:P-series}), where coefficients $\omega_k$ are Laurent polynomials in $y$. For an $m$-herd, the first two read
\be
	\omega^{(m)}_1(y)=1,\qquad\qquad \omega^{(m)}_2(y)=\frac{y^{2(m-1)}+y^{-2(m-1)}-2}{\left(y+y^{-1}-2\right)\left(y+y^{-1}+2\right)}\,.
\ee
Equation (\ref{eq:P}) implies a recursion relation for the coefficients of (\ref{eq:P-series}). The contribution to a particular Taylor coefficient in front of $z$ can be represented as a sum over partitions $t_{s,j}$. We label non-negative integers $t_{s,j}$ by a pair of integers $(s,j)$; $s$ corresponds to a contribution of a term with a shift controlled by $s$ in (\ref{eq:P}), while $j$ distinguishes formally between the terms with the same $s$ gathered into powers in (\ref{eq:P}). We sum over all possible values of $t_{s,j}$ inserting a Kronecker symbol, so that only a few contribute. The recursion relation reads
\be\label{eq:rec_omega}
\omega^{(m)}_k(y)=\sum\lm_{t_{s,j}=0}^{\infty}y^{2\sum\lm_{s,j}s t_{s,j}}\left(\prod\lm_{s=-(m-2)}^{m-2}\prod\lm_{j=1}^{m-1-|s|}\omega^{(m)}_{t_{s,j}}(y)\right)\delta_{k-1,\sum\lm_{s,j}t_{s,j}}
\ee
The highest power of $y$ is contributed by $t_{m-2,1}=k-1$ with all the others $t$'s set to zero, therefore we may recast the above as a recursion relation for the the maximal power $\alpha_k$ for $y$ in $\omega^{(m)}_k(y)$, together with a boundary condition:
\be
\alpha_k=\alpha_{k-1}+2(m-2)(k-1),\qquad \alpha_1=0\,,
\ee
which is solved by
\be
\alpha_k=(m-2)k(k-1)\,.
\ee
Since $Q$ is related to $P$ by (\ref{eq:P_genf}),
the highest power of $y$ in the coefficient of $z^{k}$ is $\alpha_k+(m-1)k$. Hence, finally, the highest spin for the $(n,n)$ state reads
\be
	2J_{\rm max}^{(\rm herds)}(k)+k-1=\alpha_k+(m-1)k\,.
\ee
This entails a beautiful agreement of our formula (\ref{eq:P}) with previously known results from quiver representation theory
\be
	2J_{\rm max}^{(\rm herds)}(n)=2J_{\rm max}^{(\rm quiver)}(n)=(m-2)n^2+1\,.
\ee

\paragraph{Poincar\'e polynomial stabilization.}

The relation between quiver representation theory and BPS state counting extends to Poincar\'e polynomials.
In our particular example the representation of the Kronecker quiver with $m$ arrows and a dimensional vector $(n,n)$ is a collection of $m$ elements of ${\rm End}(\IC^{n})$ \cite{Reineke,CoHA}:
\be
R=\mathop{\rm Hom}\left(\IC^n,\IC^n\right)^{\oplus m}
\ee
It has a natural action of the gauge group $G=GL(n,\IC)\times GL(n,\IC)$. The BPS states are associated with $G$-equivariant cohomologies of the quiver representations.

The relation between the Poincar\'e polynomial and the PSC reads
\footnote{
It would be more precise to call quantity $\chi^{(m)}_n(y)$ a $\chi_y$-genus, though if the moduli space is smooth it can be identified with the Poincar\'e polynomial (see the discussion in \cite[section 2.5]{Engineering}).
}
\be
\begin{split}
\chi^{(m)}_n(y):= 
&\sum\lm_k \beta_{n,k}^{(m)}y^{2k}\\
=&y^{2J_{\rm max}(n)}\Omega^{(m)}(y,n\gamma_c) 
\end{split}\ee
where $\beta_{n,k}^{(m)}$ ($k=0,\dots,{\rm dim}\,\CM$) are corresponding suitably defined\footnote{
The BPS indices for generalized $m$-herds are not simple
 Euler-characteristics (of stable or semi-stable moduli).
 The reason is that the contributions to $\Omega(n \gamma)$ for $n>1$ involve
 contributions from threshold bound states, or, in  the language of quivers,
 from semi-stable representations of the Kronecker quiver.

 The failure can be seen most drastically for the $m$-herd: where the Euler
 characteristic $\chi(n)$ for the moduli space of stable representations of
 the Kronecker $m$-quiver, with dimension  vector $(n,n)$, vanishes for $n>1$ (see
 the proof of the $m$-herd functional equation in \cite{Reineke2}). See also discussion in \cite[s.6.5, s.7]{Reineke}.
 
 We thank T.~Mainiero for this valuable remark.} 
 Betti numbers of the moduli space of representations, and $\Omega^{(m)}(y,n\gamma_c)$ denotes the PSC of a BPS state of charge $(n,n)$, with $m$ being the charge pairing of elementary constituents.

Explicit computations \cite{WWC} of the Betti numbers suggest that they \emph{stabilize}: there is a well-defined limit
\be
\lim\lm_{n\longrightarrow\infty} \beta_{n,k}^{(m)}=\beta^{(m)}_{\infty,k}\,,
\ee
which can be recast as a limit for a polynomial
\be
\lim\lm_{n\longrightarrow\infty}\chi_n^{(m)}(y)=\chi_{\infty}^{(m)}(y)\,.
\ee
Moreover, by direct inspection, this limit turns out to be independent of $m$: $\chi_{\infty}^{(m)}(y)=\chi(y)$ for all $m$; this observation implies \emph{another} interesting limit
\be
\lim\lm_{m\longrightarrow\infty}\chi_n^{(m)}(y)=\chi_{n}^{(\infty)}(y)\,.
\ee
It turns out that these limiting polynomials are known. In fact they correspond to the Poincar\'e polynomials of the classifying space $B\left((GL_{n}\times GL_{n})/\IC^*\right)$ where $\IC^{*}$ is the subgroup of elements $(\lambda\,\mathbb{I},\lambda^{-1}\,\mathbb{I})$ \cite{Reineke}.

Numerical experiments indicate that Betti numbers satisfy an interesting inequality
$\beta_{n,k}^{(m)}\leq\beta_{n,k}^{(m+1)}$, implying in turn
\be
	\beta_{\infty,k}^{(m)}\leq \beta_{\infty,k}^{(\infty)}
\ee
though it remains unclear why Betti numbers saturate this bound for every $m\geq 3$.

It is interesting to investigate how this convergence interplays with the equation for the generating function (\ref{eq:P}). As a preliminary remark, notice that the expansion of the dilog product in the generating function allows one to relate coefficients in the formal series to the PSC
\be
	\omega_n^{(m)}(y)=y^{\alpha_n-2J_{\rm max}(n)}\frac{1-y^{2n}}{1-y^2}\Omega^{(m)}(y,n\gamma_c)\left(1+O(y^{(m-1)n})\right)\,.
\ee
and by $O(y^p)$ we denote a formal series in $y$, starting with a term of degree $p$. It is simple to observe this relation since $\Phi_n(\xi)=1+y^{1-2n}\frac{1-y^{2n}}{1-y^2}\xi+O(\xi)$
\be
\prod\lm_{k\in \IZ}\Phi_n \left( (-y)^k y^n z^n \right)^{a_k(\tilde\gamma_h)} =1+y^{1-n}\frac{1-y^{2n}}{1-y^2}\left(\sum\lm_{k\in\IZ}a_k(\tilde\gamma_h)(-y)^k\right)z^n+O(z^{n+1})
\ee
and corrections from lower dilogarithms can be estimated by lowest values of the powers of $y$ they bring in.

Introducing the series
\be
	\tilde\chi_n^{(m)}(y):=y^{-(m-2)n(n-1)}\omega_n^{(m)}(y)\,,
\ee
we can focus on its stabilization since (assuming $|y|<1$)
\be
\begin{split}
	&\lim\lm_{n\rightarrow\infty}\tilde\chi_n^{(m)}(y)=(1-y^2)^{-1}\chi_{\infty}^{(m)}(y)\,,\\
	&  \lim\lm_{m\rightarrow\infty}\tilde\chi_n^{(m)}(y)=\frac{1-y^{2n}}{1-y^2}\chi_n^{(\infty)}(y)\,.
\end{split}
\ee

Performing the substitution $\omega^{(m)}_n(y)\mapsto \tilde\chi^{(m)}_n(y) y^{(m-2)n(n-1)}$, $s\mapsto s-(m-2)$ into (\ref{eq:rec_omega}) we arrive at the following recursion relation
\be\label{eq:rec_chi}
\begin{split}
&\tilde\chi^{(m)}_k(y)=\sum\lm_{t_{s,j}=0}^{\infty}y^{2\sum\lm_{s,j}s t_{s,j}+{(m-2)\sum\lm_{(s,j)\neq(s',j')}t_{s,j}t_{s',j'}}}\times\\
&\times\left(\prod\lm_{s=0}^{2(m-2)}\prod\lm_{j=1}^{m-1-|s-(m-2)|}\tilde\chi^{(m)}_{t_{s,j}}(y)\right)\delta_{k-1,\sum\lm_{s,j}t_{s,j}}
\end{split}
\ee
where the second summation in the power of $y$ goes over different pairs of indices. In the limit $m\rightarrow\infty$, precisely that summation causes a \emph{localization} (assuming $|y|<1$ and noticing that the power is non-negative) on partitions of $k-1$ satisfying $\sum\lm_{(s,j)\neq(s',j')}t_{s,j}t_{s',j'}=0$, these are partitions consisting of just one $t_{s,j}=k-1$ with all the others being zero. Thus we are eventually left with a summation over positions $(s,j)$
\be
\tilde \chi^{(\infty)}_k (y)=\sum\lm_{s=0}^{\infty}(1+s)y^{2s(k-1)}\tilde \chi^{(\infty)}_{k-1} (y)\,.
\ee
This reproduces the result from quiver representation theory
\be
	\chi(y,B\left((GL_{n}\times GL_{n})/\IC^*\right))=\frac{1-y^2}{(1-y^2)^{2n}}\tilde\chi^{(\infty)}_n(y)=\frac{1-y^2}{\prod\lm_{j=1}^n (1-y^{2j})^2}\,,
\ee
the corresponding limiting Poincar\'e series reads
\be
	\chi(y)=\frac{1-y^2}{\prod\lm_{j=1}^{\infty} (1-y^{2j})^2}\,.
\ee

\subsection{Chern-Simons, formal variables and the writhe}\label{sec:chern-simons}
In this section we propose a different perspective on the formal variables introduced in \S\ref{sec:twisted-variables}, together with a natural explanation for the appearance of the writhe and of the map $\rho$ introduced in (\ref{eq:rho}), two prominent characters of our story.

The formal variables $\hU$ employed above have a natural interpretation in terms of a quantized twisted flat connection. Before turning to the twisted connection, let us consider a classical flat abelian $\IC$-valued connection on $\Sigma$, subject to certain boundary conditions at punctures. We take the logarithm of the holonomy to be fixed to $\mathfrak{m}_s$ at the puncture $z_s$. Let $\CX_{\gamma}$ be coordinates on the moduli space $\CM_{ab}\simeq Hom(\pi_{1}(\Sigma),\IC^{\times})$ with fixed choices of $\mathfrak{m}$, obeying
\be\label{eq:classical-Poisson}
\begin{split}
	& \{\CX_\gamma,\CX_{\gamma'}\}=\langle\gamma,\gamma'\rangle\CX_\gamma\CX_{\gamma'} \,,\qquad %
	\CX_\gamma\CX_{\gamma'}=\CX_{\gamma+\gamma'}\,.
\end{split}
\ee
These coordinates are holonomies
%
%
\be
	\CX_{\gamma} = \exp\oint_{\gamma}\CA^{ab}\,,
\ee
where $\CA^{ab}$ is required to have canonical structure
%
%
\be\label{eq:poisson-alg}
	\{  \CA^{ab}_\mu(w),  \CA^{ab}_\nu(w')   \} = \frac{1}{k}\,\epsilon_{\mu\nu}\,\delta^{(2)}(w-w')\,,
\ee
where $w,w'$ are local coordinates on $\Sigma$ and we have used $k=1$ in (\ref{eq:poisson-alg}). $\epsilon_{\mu\nu}$ is the Levi-Civita symbol normalized to $\epsilon_{12}=1$. Given a flat connection with this Poisson bracket, its transports indeed obey (\ref{eq:classical-Poisson}). This also coincides with the algebra of Darboux coordinates of \cite{GMN1} (cf eq. (2.3) of \cite{GMN2}).

Notice that the canonical structure of this flat connection coincides with the equal-time Poisson bracket of a Chern-Simons gauge field on $\Sigma$, with noncompact gauge group $\IC^{\times}$. In the spirit of this observation, it is easy to see that promoting the Poisson bracket to a commutator
\be\label{eq:basic-comm}
	[\hA_{\mu}(w),\hA_{\nu}(w')]=2 \log y \,\epsilon_{\mu\nu}\delta^{(2)}(w-w'),
\ee
produces corresponding ``quantum'' noncommutative holonomies obeying precisely the algebra of our $y$-twisted formal variables
\be\label{eq:closed-coordinates}
	 \bY_{\gamma} = \exp  \oint_{\gamma}\hA \, , \qquad\qquad \bY_{\gamma}\bY_{\gamma'}=y^{\langle\gamma,\gamma'\rangle}\bY_{\gamma+\gamma'}\,.
\ee
Honest gauge invariant holonomies should be path-ordered, however if a closed path does not self-intersect, then path-ordering has no effect since the commutator (\ref{eq:basic-comm}) only contributes to transverse (self-)intersections. On the other hand, if the path does contain self-intersections, the path-ordered transport will depend on a choice of basepoint $p\in\gamma$
\be
	\bU_{\gamma_{p}} = P \exp\oint_{p}^{p}\hat A\,.
\ee

Closed self-intersecting curves on surfaces are also known as \emph{singular knots}, Wilson lines associated to singular knots on the plane in abelian Chern-Simons theory were studied in \cite{dunne}, where it was shown that the algebra of $\bU_{\gamma_{p}}$ matches that of $y^{\wr(\gamma_{p})}\,\bY_{\gamma}$, this motivates (\ref{eq:rho}), and offers a \emph{natural explanation for the appearance of the writhe as a consequence of path-ordering} of quantum holonomies. In particular this relation reveals that the $\bY$ also enjoy gauge invariance, being proportional to the $\bU$ up to a constant. There is an analogous story for open paths\footnote{Although open Wilson lines aren't gauge invariant, they are gauge covariant and this is enough to ensure that their algebra is gauge invariant.}.

In the proofs of twisted homotopy invariance of \S\ref{sec:twisted-homotopy}, it was crucial to deal with a \emph{twisted} flat connection, concretely we repeatedly used the fact that holonomy around a contractible cycle\footnote{For contractible curls winding counter-clockwise} equals $-y$, resulting in (\ref{eq:abelian-curl}). At the classical level, one way to construct such a connection is to consider the unit circle bundle $\tSigma\to\Sigma$ with a flat $U(1)$ connection having fixed holonomy equal to $-1$ around the circle fiber; then to each path on $\Sigma$ one associates the transport of this connection along the \emph{tangent framing lift} of the path to $\tSigma$. To the best of our knowledge, quantum twisted flat connections have not been discussed in the literature. A reasonable approach to quantizing a twisted flat connection is to leave the holonomy on the fiber fixed to a constant, while quantizing the holonomies on $\Sigma$ in a way consistent with the symplectic structure. Alternatively, using the data of a spin structure we can identity the moduli space of twisted flat connections with the moduli space of ordinary flat connections and quantize the latter. Either way we produce transports obeying the twisted algebra of our formal variables $\hU$.

The above discussion of quantum flat connections is only meant to provide an heuristic motivation for the definition of formal variables in section \S\ref{sec:twisted-variables}. In particular, it ignores the important subtleties associated with the quantization of Chern-Simons connections with noncompact gauge group.
A more thorough investigation of how our formal variables can be modeled on quantum holonomies of a Chern-Simons connection should clearly be possible, given a number of works available in the literature on noncompact Chern-Simons (see e.g. \cite{Witten-CS-1,Bar-Natan-Witten,dimofte,Witten:2010cx}). We leave this for future work.


We expect that quantum Chern-Simons theory will provide an interesting
perspective on the key formula, equation (\ref{eq:q-Darboux}). We recall from
\cite[\S 10]{GMN5} that given the data of a spectral network one
can construct a ``nonabelianization map,''  taking a flat $\IC^*$-connection
$\nabla^{\rm ab}$ on $\Sigma$ to a flat $GL(K,\IC)$-connection $\nabla^{\rm nonab}$
on $C$. The key formula defining this map expresses the parallel transport of
$\nabla^{\rm nonab}$ along a path $\wp$ on $C$ in terms of a sum of parallel transports
by $\nabla^{\rm ab}$ on $\Sigma$, weighted by framed BPS degeneracies.
(See, for example, equation (16.17) of \cite{FelixKlein}.) In the quantum
setting, $P\exp\int_{\wp} \nabla^{\rm nonab}$ and $P \exp \int_{\gamma_{ij'}}  \nabla^{\rm ab}$
become quantum operators on Chern-Simons theory Hilbert spaces $\CH^{\rm nonab}(C)$ and $\CH^{\rm ab}(\Sigma)$,
respectively. We conjecture that there is an isomorphism between
these Hilbert spaces $\phi:\CH^{\rm nonab}(C)\rightarrow\CH^{\rm ab}(\Sigma) $ allowing us to interpret equation (\ref{eq:q-Darboux}) as a quantum version of the
nonabelianization map\footnote{Related considerations have appeared in \cite{Cecotti:2011iy}   and \cite{3dNetworks}.}:
\be
\phi\left(\Tr P\exp\oint_{\wp} \nabla^{\rm nonab}\right)\phi^{-1}=\sum\lm_{\fa}\FOmega(\wp,\fa;y) \exp \oint_{\fa}  \nabla^{\rm ab}
\ee 
We stress that this is a conjecture, motivated by the present paper, and further
work is needed to make precise sense of the formula. 
We hope to return to this topic and make these ideas more precise
in future work.

\medskip

An interpretation of $y$-twisted formal variables in terms of deformation quantization of the above Poisson brackets was already suggested in \cite[\S 6.2]{GMN3}. The relation of BPS states of class $\CS$ theories to Chern-Simons Wilson lines was already pointed out in \cite{CS-classS1,CS-classS2}. In those works Chern-Simons theory appeared when considering compactifications of M5 branes in certain backgrounds, via the duality of Chern-Simons theory to open topological strings (see also \cite{top-str1,top-str2}). Although we didn't find a straightforward connection to our setup, we take such results as supporting evidence that our formal variables can be related to quantum parallel transports.

We expect there will also be very interesting further connections with non-compact WZW models and Toda theories \cite{KZtoBPZ1,KZtoBPZ2}, 
using the theory of Verlinde operators  \cite{Alday:2009fs,surf-op2} and $\beta$-ensembles.  See  \cite{Triality} for a recent review 
of the current state of the art. Closely related to this is the theory of check operators 
\cite{check,GMM} which should provide new perspectives on the quantum version of the Darboux expansion 
alluded to above.

\section*{Acknowledgements}

We thank Emanuel Diaconescu, Tudor Dimofte, Davide Gaiotto, Tom Mainiero and Andy Neitzke 
for useful discussions and  correspondence.
The work of DG, PL, and GM is supported by the DOE under grants
SC0010008,
ARRA-SC0003883,
DE-SC0007897. 
The work of GM is partly supported by the NSF Focused Research Group award DMS-1160591.
GM thanks the Aspen Center for Physics for hospitality while completing this work. The ACP is
supported in part by the National Science Foundation under Grant No. PHYS-1066293.
The work of DG is partly supported by RFBR 13-02-00457, 14-01-31395-mol-a, NSh-1500.2014.2.

\appendix
\section{Generating function detailed calculation}\label{sec:GF_det}
In this Appendix we present a simple technique allowing one to calculate the writhe effectively for soliton paths encoded by certain graphs on branched spectral covers, and show its application to a direct computation of several first terms in expansions like (\ref{eq:P-3-herd}).

\subsection{Singular writhe technique}
Schematically, the spectral network may be though of as a graph. We would like to adapt the usual notion of writhe for smooth curves to the singular\footnote{Here \emph{singular} does not denote the presence of self-intersections as, for instance, in the math literature on knots. We refer instead to discontinuities of the tangent vector of solitons paths, these occur in correspondence of (lifts of) joints.} curves arising in this setting. In order to compute the writhe it will be necessary to keep track of the order in which the path runs (in different transverse directions) through a self-intersection. For this purpose, we will resolve paths pictorially by drawing under/over-crossing. The path-ordering convention is that what runs below runs first.

\begin{figure}[h!]
\begin{center}
\includegraphics[width=0.4\textwidth]{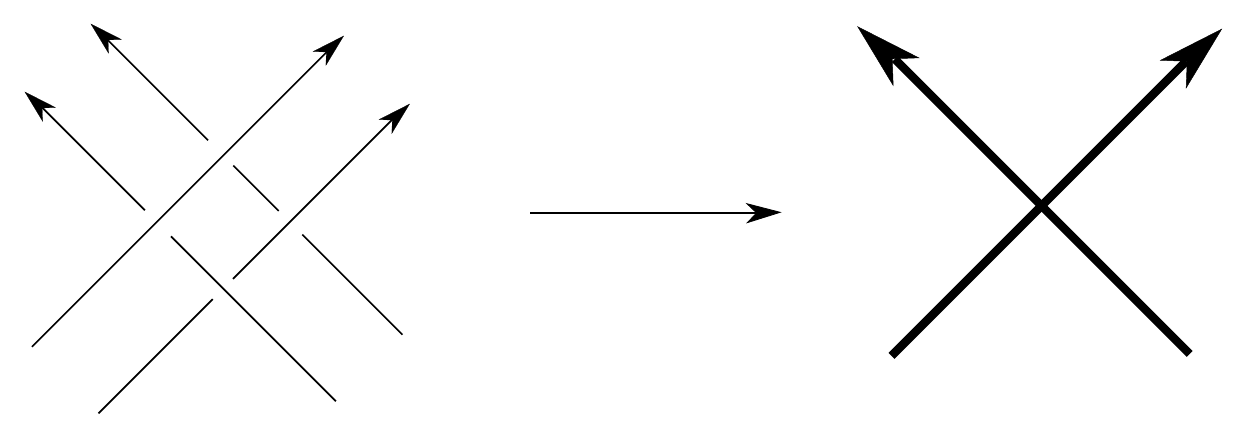}
\end{center}
\caption{Resolved intersection \label{fig:th-int}}
\end{figure}

For example, consider a self-intersection through which a single path runs multiple times as shown in figure \ref{fig:th-int}). 
We can slightly resolve the critical angle $\vartheta_c$ to get the picture on the left hand side, then it is simple to notice that the four segments give contributions to the writhe by pairs, so it is enough to sum these contributions pairwise for the thick intersections. The contribution for this particular case is $-2$.

\begin{figure}[h!]
\begin{center}
\includegraphics[width=0.8\textwidth]{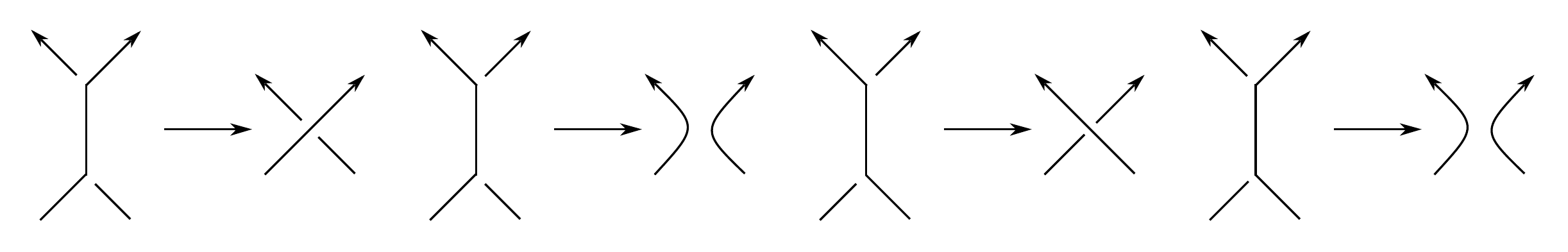}
\end{center}
\caption{\label{fig:sing-wr} On the singular writhe calculation. At junctions, soliton paths may intersect somewhat ambiguously, with ``half-integer'' units of intersections to be taken into account. Upon summation the contributions from \emph{all junctions} involved, intersection numbers can eventually be correctly computed and found to be integers. From left to right we show the addition of half-integer intersection numbers at pairs of junctions:
$-\frac{1}{2}-\frac{1}{2}=-1$, $-\frac{1}{2}+\frac{1}{2}=0$, $+\frac{1}{2}+\frac{1}{2}=+1$, $+\frac{1}{2}-\frac{1}{2}=0$}
\end{figure}

Another interesting possible issue are ``half-intersections". They occur when two lines going first parallel split and go in different directions, this happens due to splitting of intersections as depicted in fig.\ref{fig:sing-wr}. From these pictures it is clear that we can assign half-contributions to half-intersections assuming that these half-contributions will be summed up to an integer result.

Thus to compute the ``singular writhe" we choose a base point, in each self-intersection lines come in with some tangent vectors $\vec v_k$ at corresponding ``times" $t_k$, so, finally the formula for the writhe reads
\be
\begin{split}
\wr(\gamma)=&-\sum\lm_{i\in{\rm int's}} \sum\lm_{k,m}\sign(t_k^{(i)}-t_m^{(i)})\sign \left[\vec v_k^{(i)},\vec v_m^{(i)}\right]+\\
&-\frac{1}{2}\sum\lm_{i\in\frac{1}{2}{\rm int's}} \sum\lm_{k,m}\sign(t_k^{(i)}-t_m^{(i)})\sign \left[\vec v_k^{(i)},\vec v_m^{(i)}\right]
\end{split}
\ee

\subsection{Diagram rules}
In the next subsection we present results for detour writhe calculations in the cases of 2-herds and 3-herds. We schematize the corresponding detours by diagrams denoting resolutions of paths in 6-way joints.

In this way we can reformulate the 6-way joint rules in a pictorial form. As an example, consider
\be
\tau_{ki}(p_{S})=
\begin{picture}(80,80)(0,18)
\includegraphics[scale=0.2]{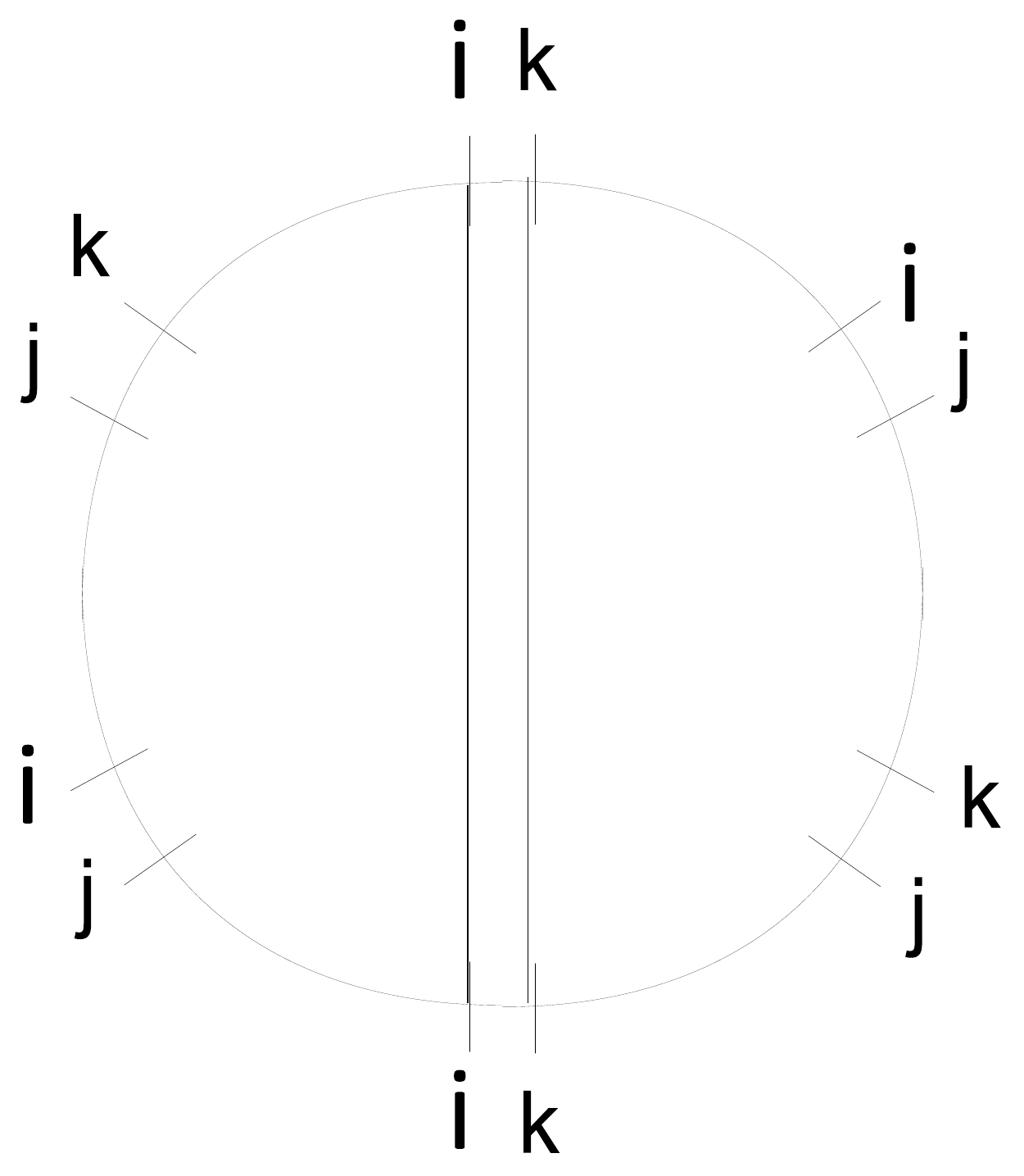}
\end{picture}+
\begin{picture}(80,80)(0,18)
\includegraphics[scale=0.2]{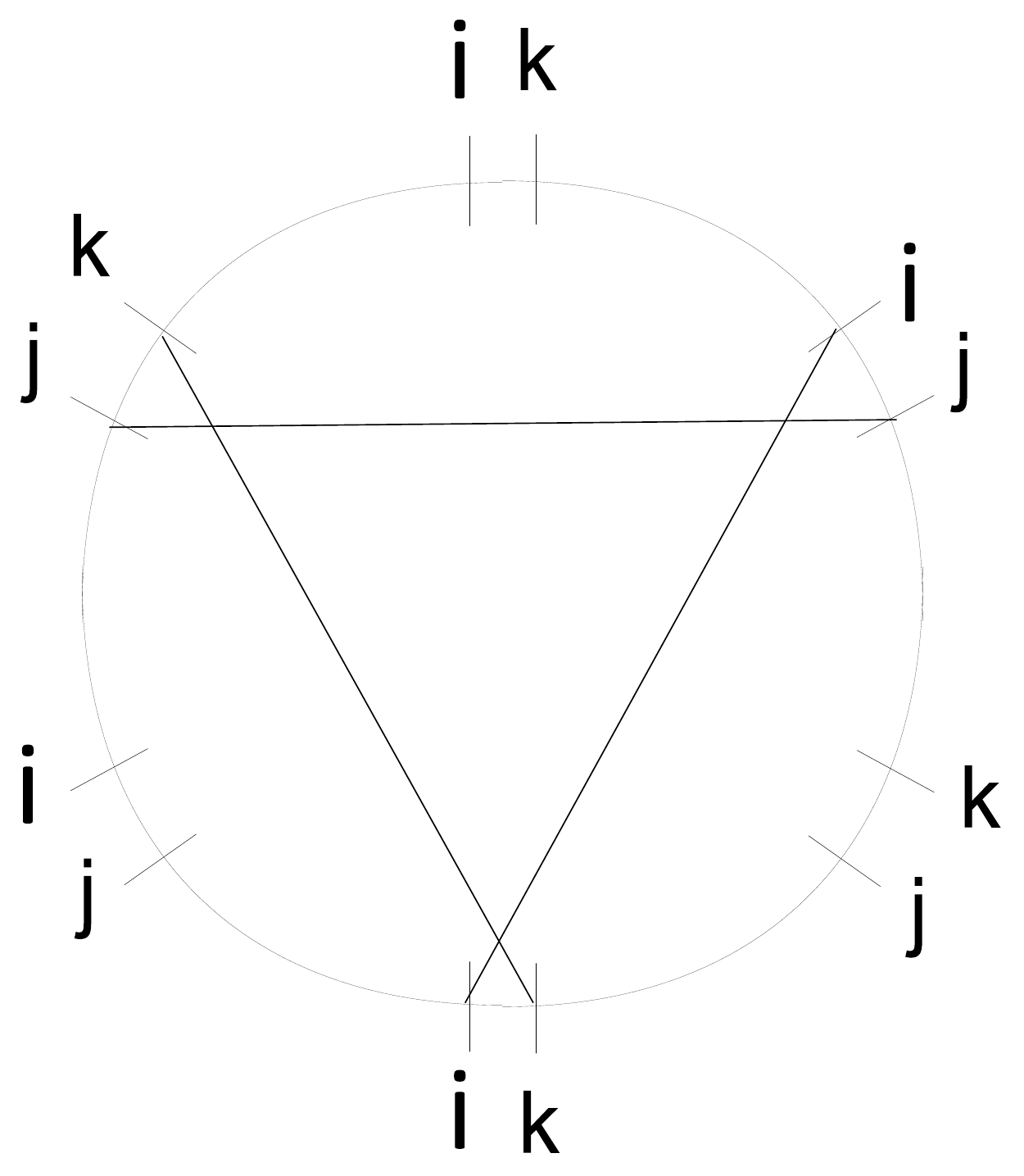}
\end{picture}+ \begin{picture}(80,80)(0,18)
\includegraphics[scale=0.2]{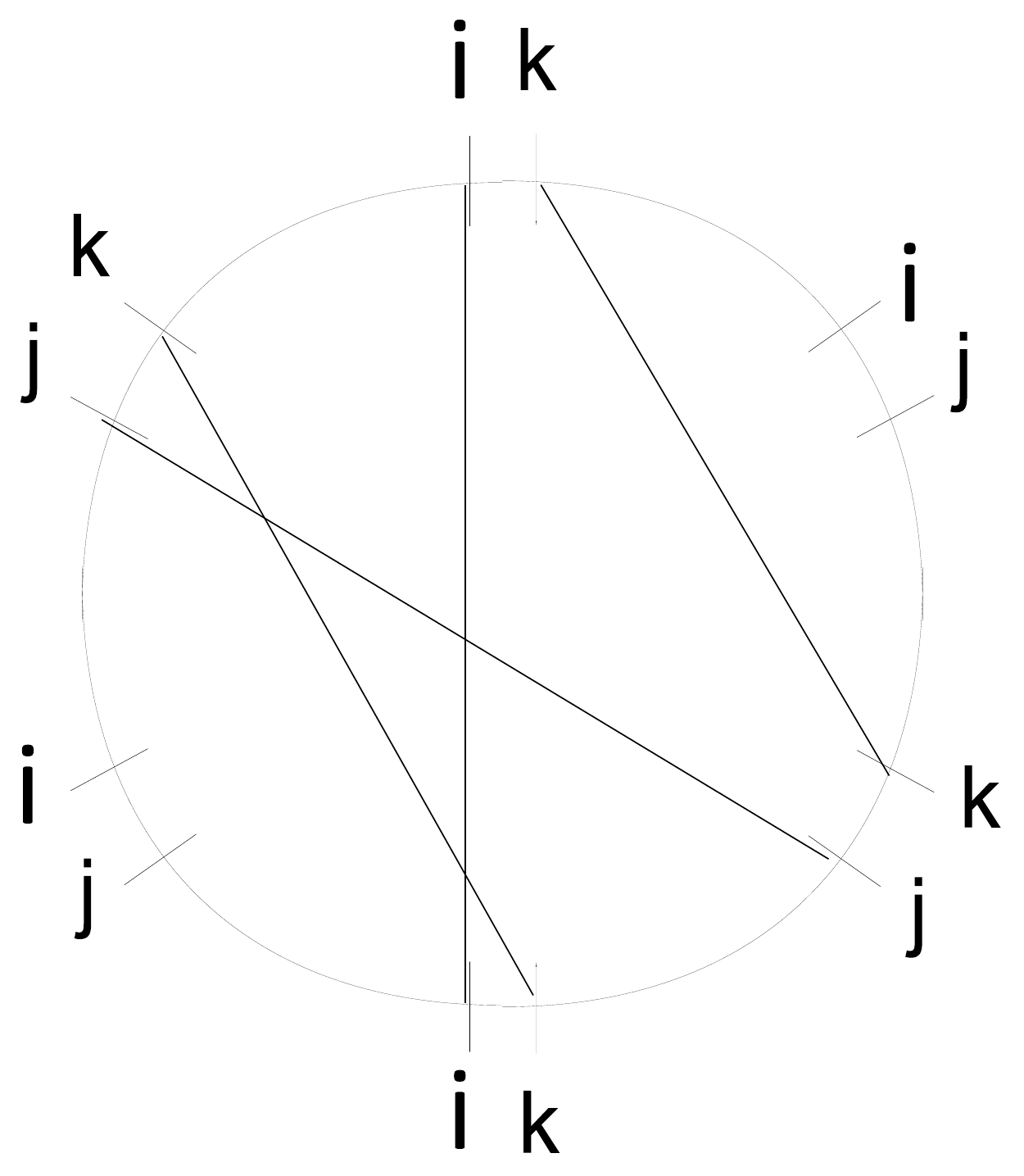}
\end{picture}+\begin{picture}(80,80)(0,18)
\includegraphics[scale=0.2]{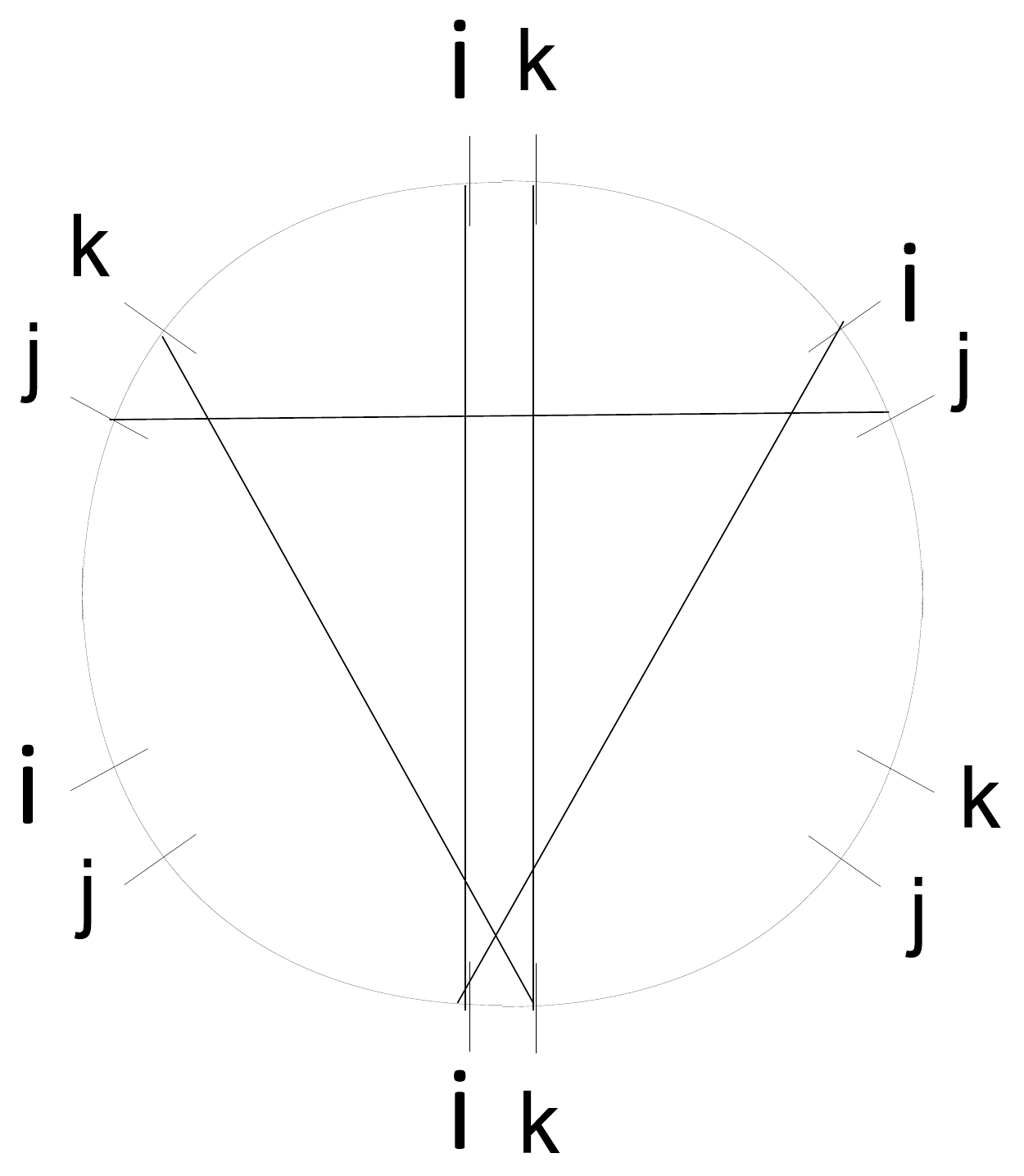}
\end{picture}+\ldots
\ee
here we describe the outgoing soliton generating function for the street $p_{S}$ ($S$ is for \emph{South}) attached to the bottom gate of the joint: solitons of type $ki$ start from this gate on sheet $k$ going upwards and end up on sheet $i$ going downwards to return back to the bottom gate. What happens in between is described by following the lines connecting the various gates. For example, in the first term we simply have straight connections to the upper gate, this corresponds to solitons contributed from $\nu_{ki}(p_{N})$ (the street at the northern gate). In the second term   we have solitons starting on sheet $k$ on the south gate, propagating on sheet $k$ to the NW gate, then propagating through the network, then coming back to the NW gate on sheet $j$, then propagating to the NE gate on sheet $j$ and going once more through the network and coming back to the NE gate on sheet $i$ and finally propagating back to the S gate, and on $p_{S}$ on sheet $i$. Solitons in the second term are those encoded into $\nu_{kj}(p_{NW})\nu_{ji}(p_{NE})$. Further terms bear analogous interpretations.

We present whole detour diagrams, calculate corresponding writhes and restore the generating function $Q(\fd,y,z)$ (eq.(\ref{eq:P_genf})). The results will be:
\be
\begin{split}
& Q^{(\rm 2-herd)}(\fd,y,z)=1+(y+y^{-1})z+(y^2+1+y^{-2})z^2+O(z^3)\\
& Q^{(\rm 3-herd)}(\fd,y,z)=1+(y^2+1+y^{-2})z+(y^6+2y^4+3y^2+3+3y^{-2}+2y^{-4}+y^{-6})z^2+O(z^3)
\end{split}
\ee

\begin{figure}[h!]
\begin{center}
\includegraphics[scale=0.42]{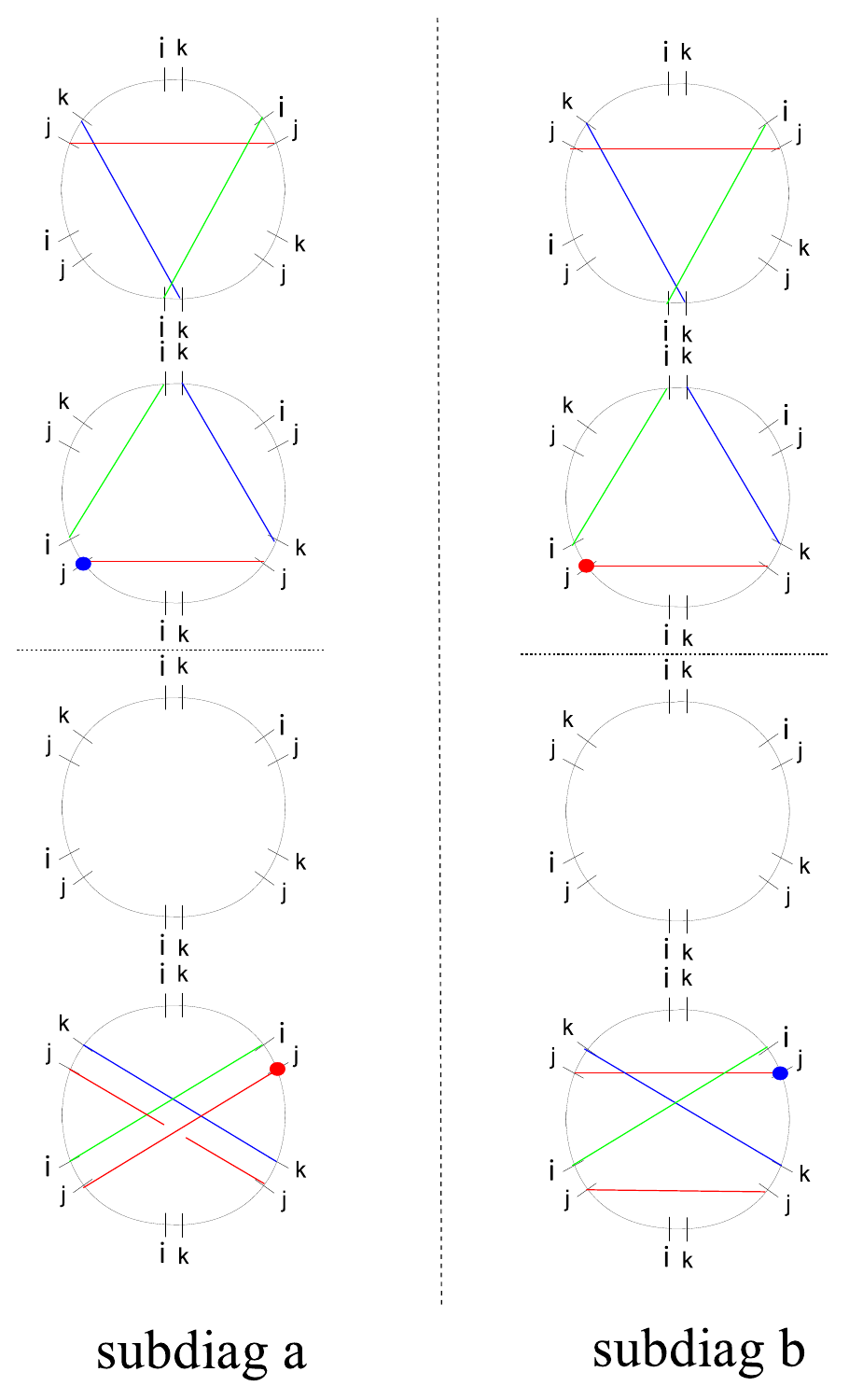}\hfill\includegraphics[scale=0.42]{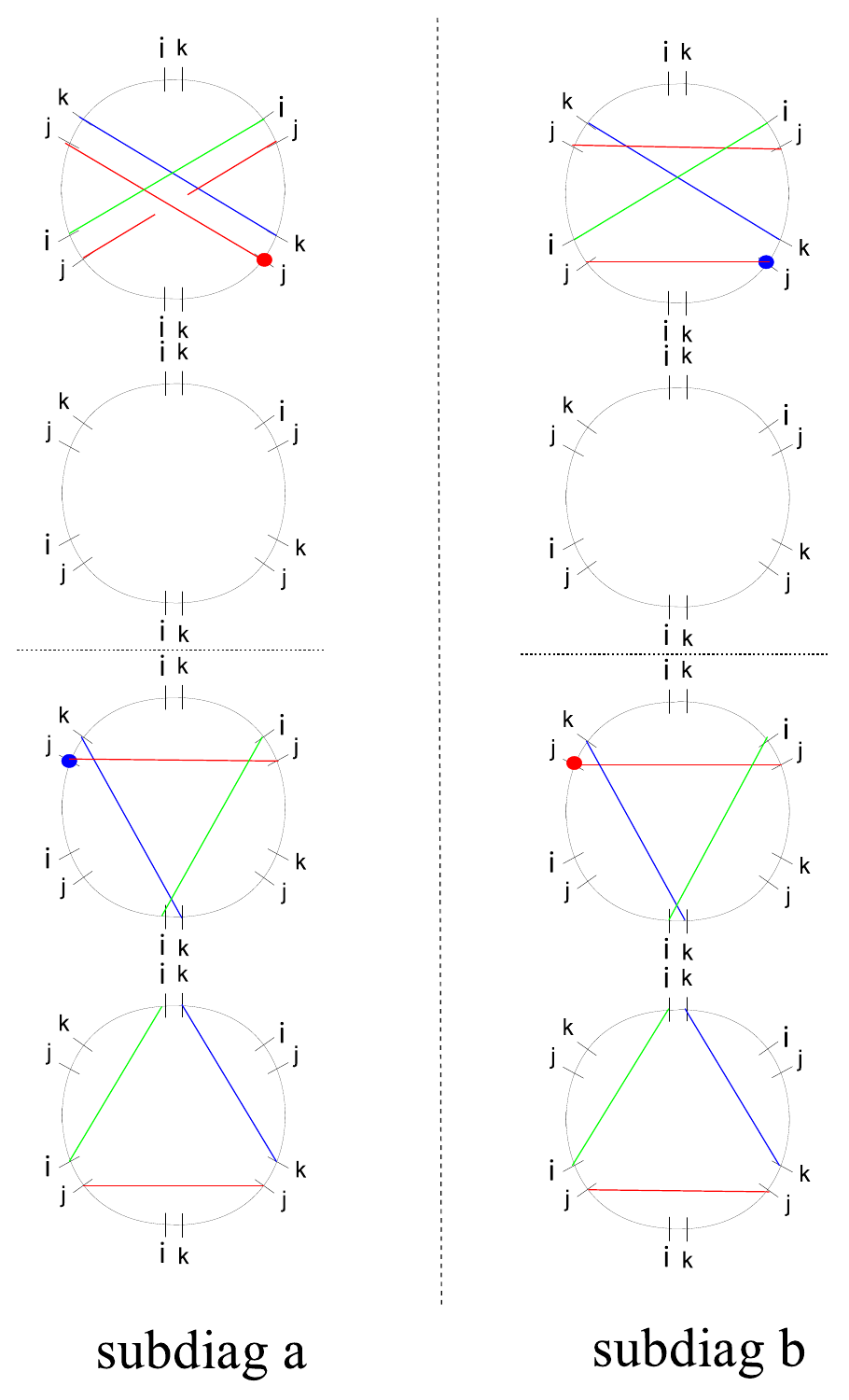}\hfill\includegraphics[scale=0.42]{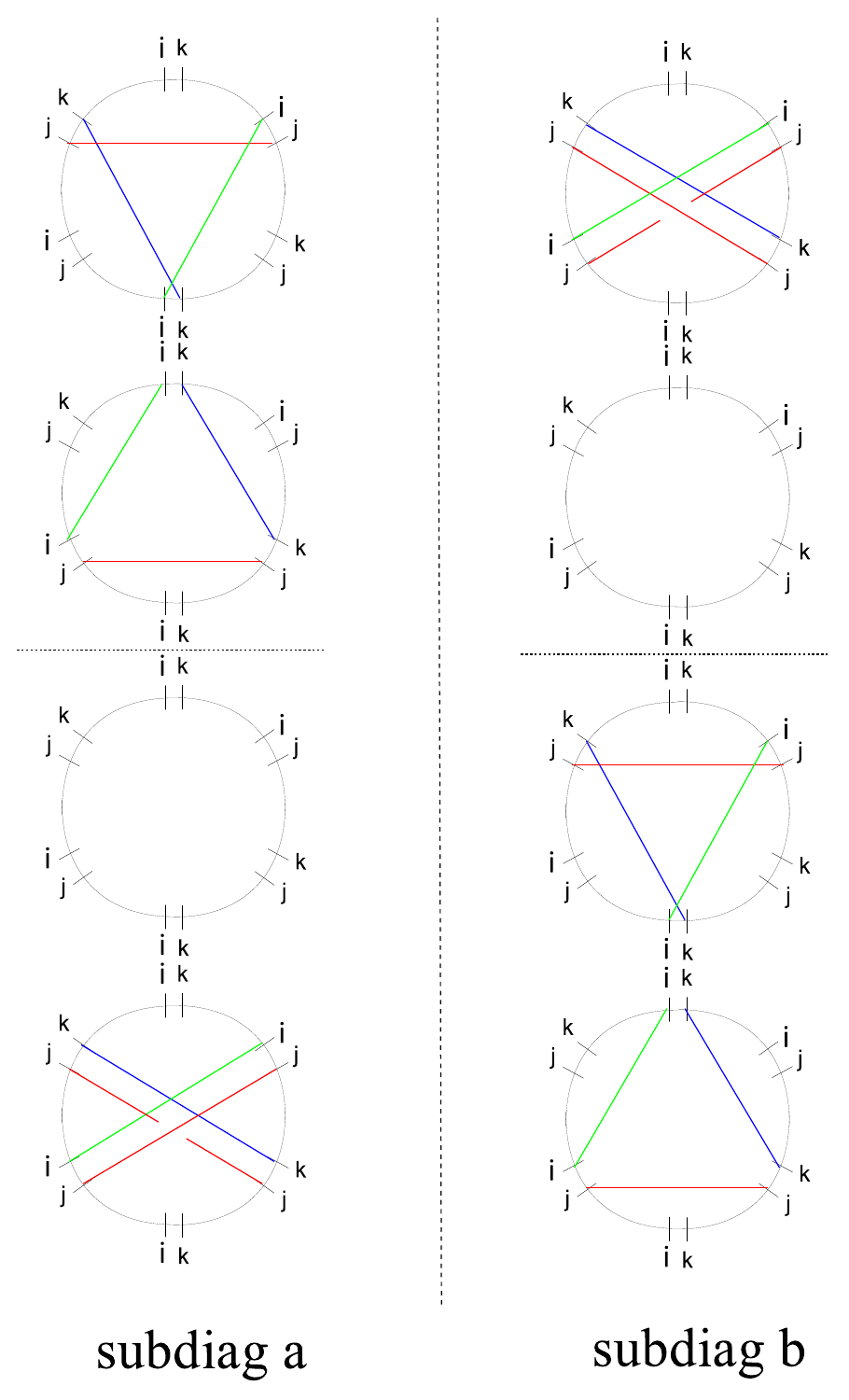}
\caption{Diagrams of order two in the 2-herd. Since each path goes twice around the herd, we split the path into two and represented each piece separately, the dashed vertical line separates two pieces of the same path, the thick dot indicates where one piece joins the other.}
\label{fig:h2-order2}
\end{center}
\end{figure}

\textbf{Diagram rules:}
\begin{itemize}
  \item {\color{green} Green} lanes go along $i$th sheet, {\color{red} red} lanes go along $j$th sheet, {\color{blue} blue} lanes go along $k$th sheet
  \item {\color{green} Green} and {\color{blue} blue} lanes go from right to left, {\color{red}red} lanes go from left to right
  \item We calculate the generating function for $\fd$ branching point street, thus we always start computing writhe from $\fd$ branching point from $k$th sheet.
  \item For $n$-th order contributions (meaning that the detour's homology class is $n$ times the generator of the critical lattice), in order to keep things tidy, we split the picture into $n$ layers. To reconstruct the path one has to glue the layers back together, the endpoints of a piece of path drawn on a layer are marked by thick {\color{red}red} and {\color{blue}blue} dots. The {\color{red}red} dot is where the jump to the next layer begins, the {\color{blue} blue} dot is where the jump from the previous layer lands. For example, see figure \ref{fig:h2-order2}. On those diagrams where dots are missing layers are glued in the point $\fd$.
\end{itemize}

\subsection{2-herd diagrams}

\subsubsection{Diagrams for order one}
\begin{figure}[h!]
\begin{center}
\includegraphics[scale=0.42]{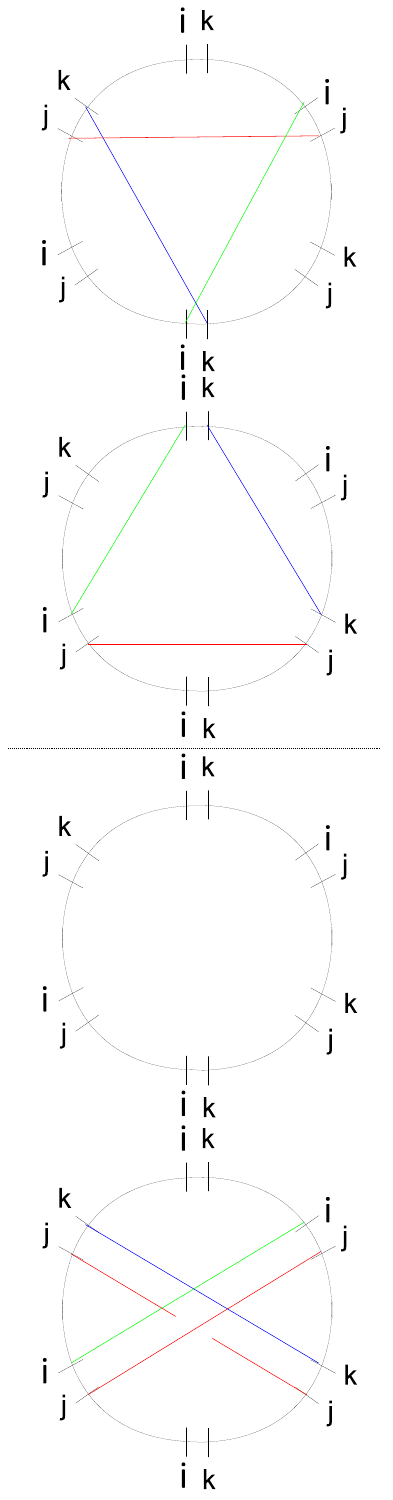}\hspace{30pt}\includegraphics[scale=0.42]{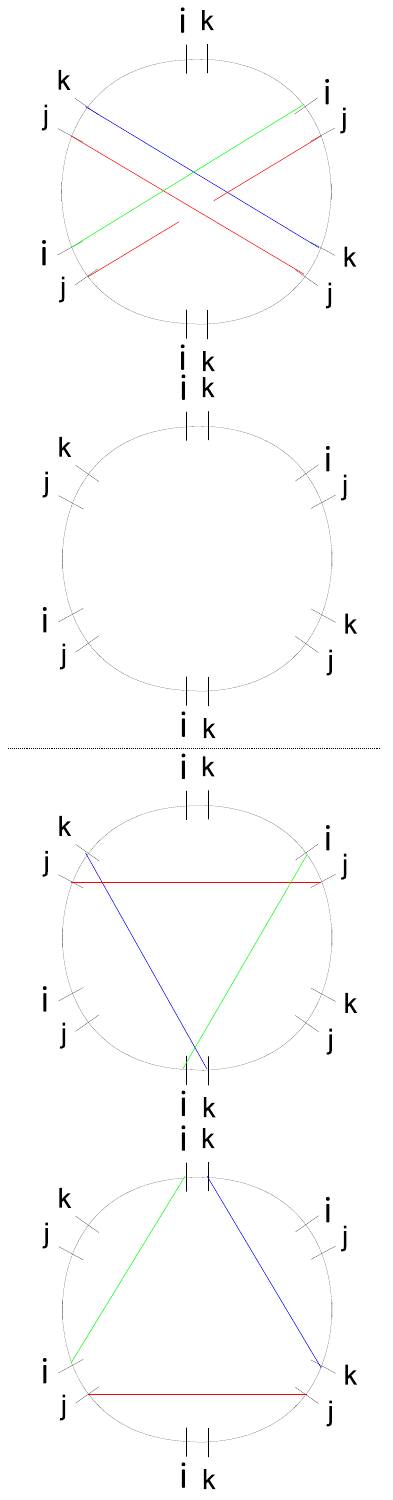}
\caption{The two diagrams for order one in the 2-herd}
\label{fig:2herd-ord1}
\end{center}
\end{figure}

\begin{figure}[h!]
\begin{center}
\includegraphics[width=0.55\textwidth]{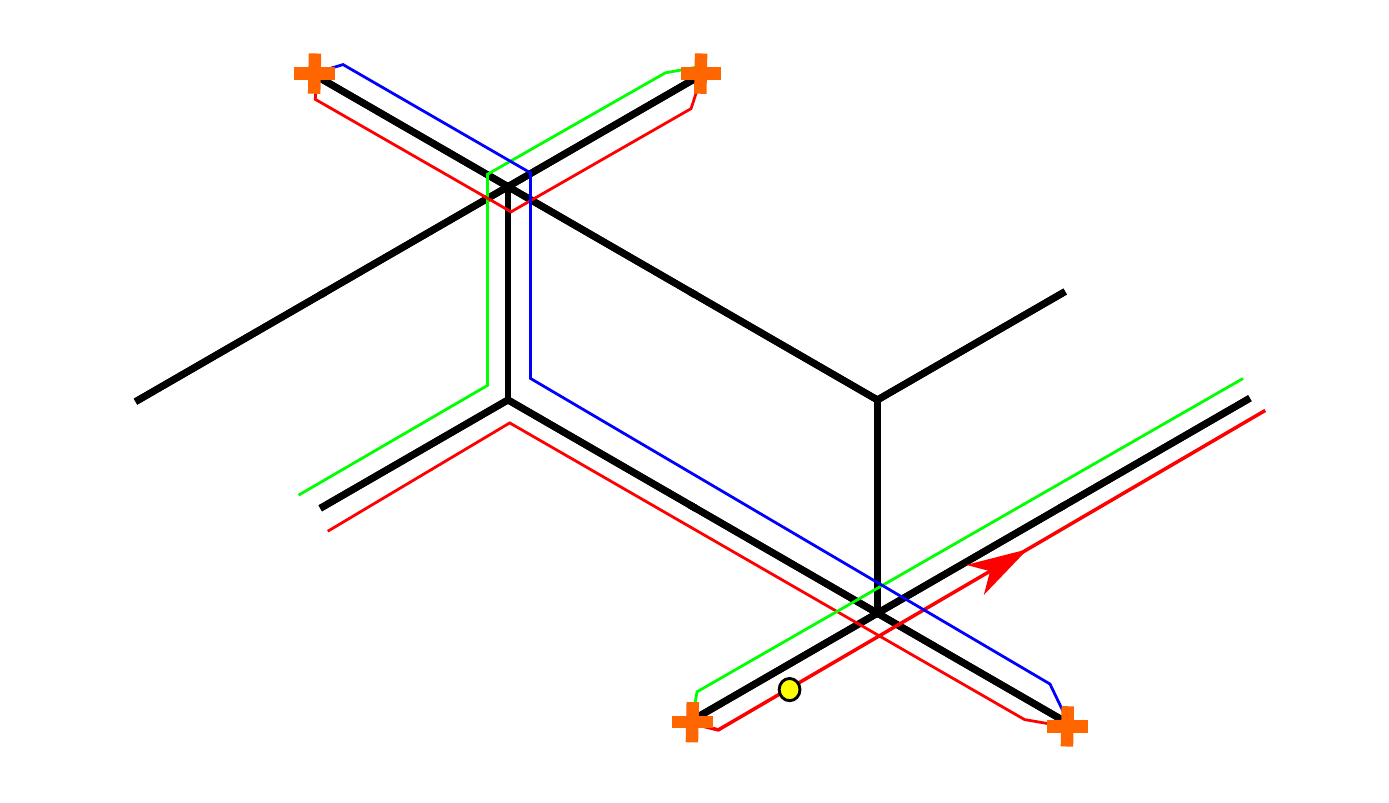}
\caption{The path corresponding to the diagram on the left of figure \ref{fig:2herd-ord1}. Recall that the endpoints of streets on the far left are identified with endpoints of streets on the far right, as the herd wraps a tube of the Riemann surface $C$. The starting point of the detour is indicated by a yellow dot, there is only one self-intersection at the lowest joint, where two red lines (both run on sheet $i$) cross each other. The overall writhe of this detour is therefore $-1$.}
\label{fig:full-path}
\end{center}
\end{figure}

We now introduce a diagrammatic representation of detours. We draw the joints of a critical network as circles and mark the six gates of each joint. The two herd has four joints: two for each horse, these are shown in each column of figure \ref{fig:2herd-ord1}: focusing on the left column, the upper two joints correspond to the joints of the upper horse, while the lower two joints are the joints of the lower horse (see fig. \ref{fig:full-path}), the horizontal dashed line separating them denotes the distinction of joints of a horse from those of the other. The topology of the network determines how the different gates are mutually connected by two-way streets, or whether they attach to streets ending on branch points. Each column describes a detour path on $\Sigma$, the path is constructed out of the segments shown in the figure (each color corresponds to a sheet of $\Sigma$) as well as of lifts of streets attached to the gates on which segments end.
All paths are conventionally taken with a basepoint on sheet $i$ on the terminal 2-way streets on the SW branch-point of the herd, they are constructed starting from the basepoint and following segments through joints, and connecting streets from one joint to the next one. 
As an example, in figure \ref{fig:full-path} we reproduce in full detail the path described by the diagram on the left of fig. \ref{fig:2herd-ord1}.

\begin{tabular}{|c|cc|}
  \hline
  diag $\#$ &1&2\\
  \hline
  contribution &$y^{-1}$&$y$\\
  \hline
\end{tabular}
\subsubsection{Diagrams for order two}
\begin{tabular}{|c|ccc|}
  \hline
  diag $\#$ &1&2&3\\
  \hline
  contribution &$y^{-2}$&$y^2$& 1\\
  \hline
\end{tabular}
\subsection{3-herd diagrams}
\subsubsection{Diagrams for order one}
\begin{figure}[h!]
\begin{center}
\hfill\includegraphics[scale=0.42]{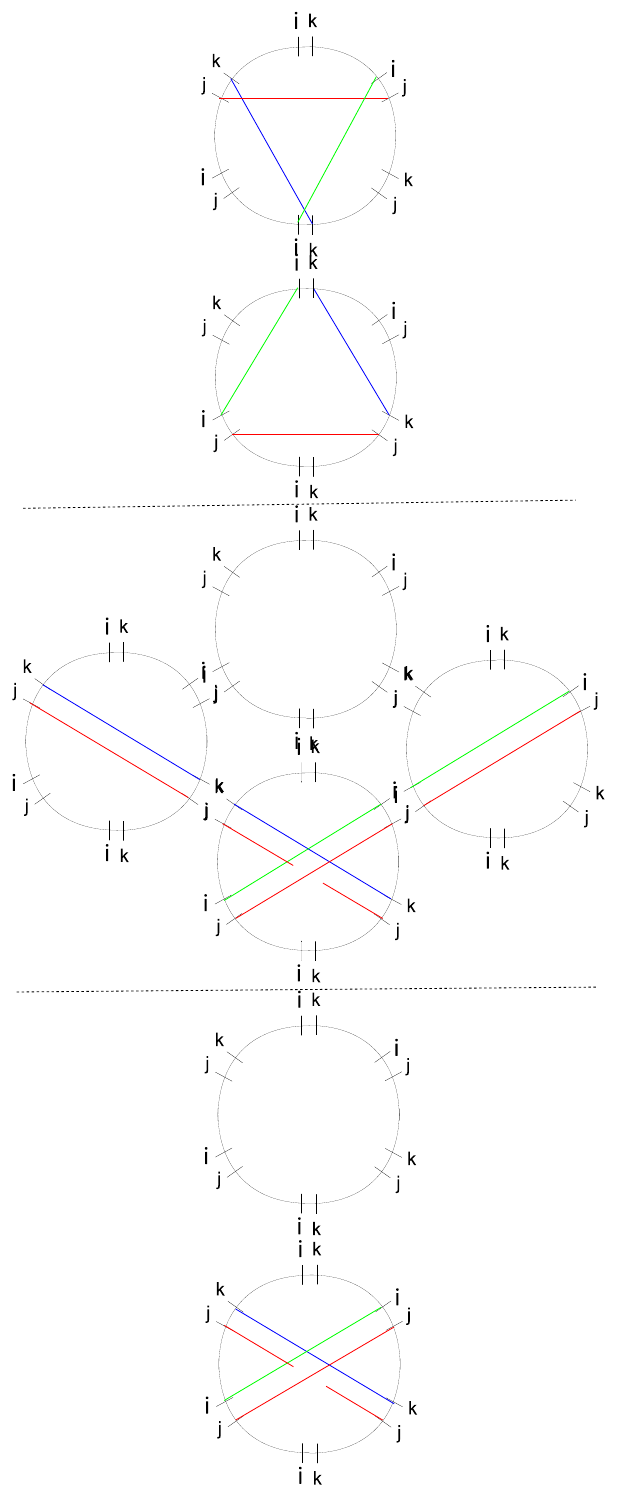}\hfill\includegraphics[scale=0.42]{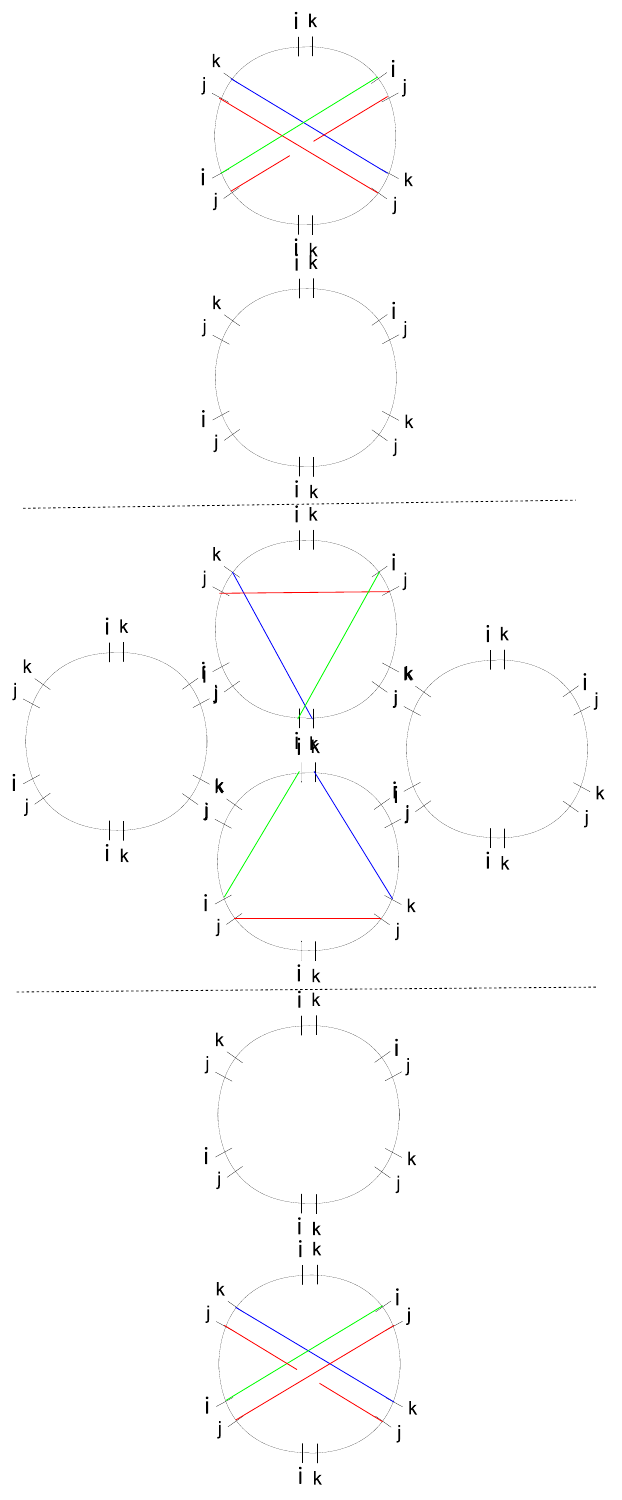}\hfill\includegraphics[scale=0.42]{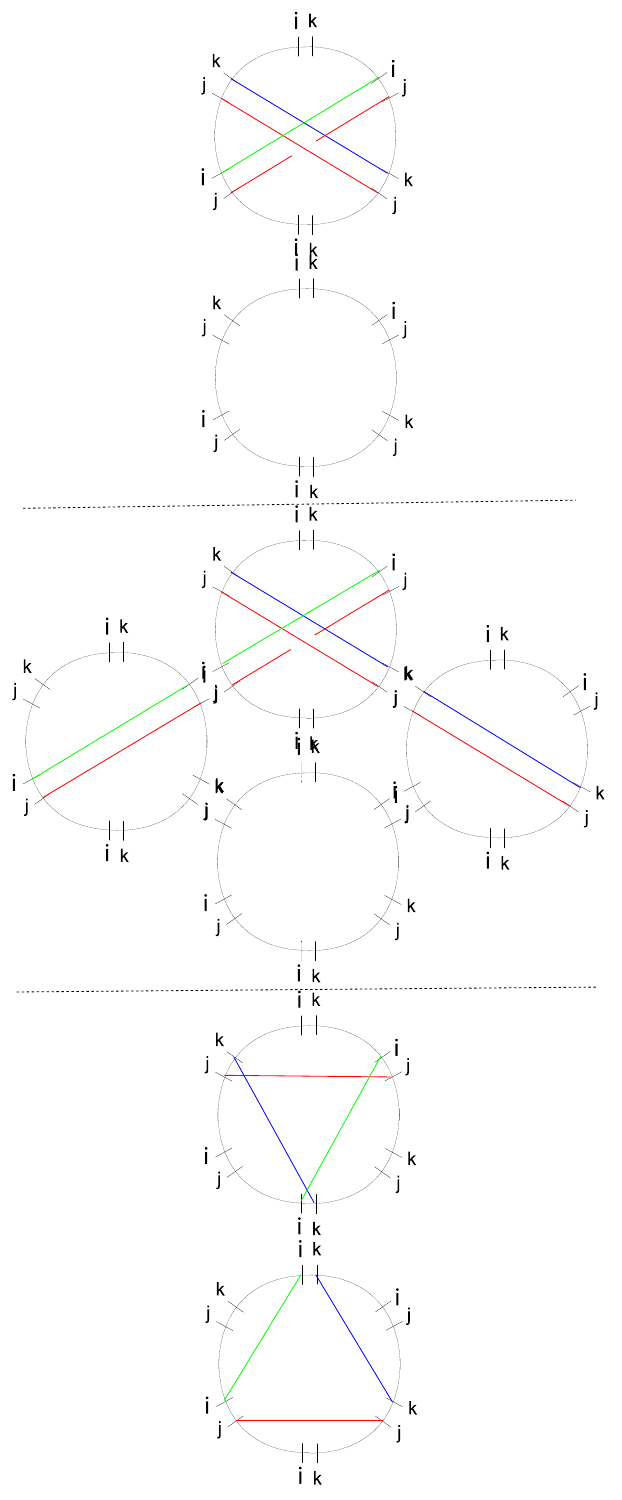}\hfill\hfill
\caption{Diagrams for the order one in 3-herd}
\end{center}
\end{figure}
\begin{tabular}{|c|ccc|}
  \hline
  diag $\#$ &1&2&3\\
  \hline
  contribution &$y^{-2}$& 1& $y^2$\\
  \hline
\end{tabular}

\subsubsection{Diagrams for order two}

\begin{figure}[h!]
\begin{center}
\includegraphics[scale=0.42]{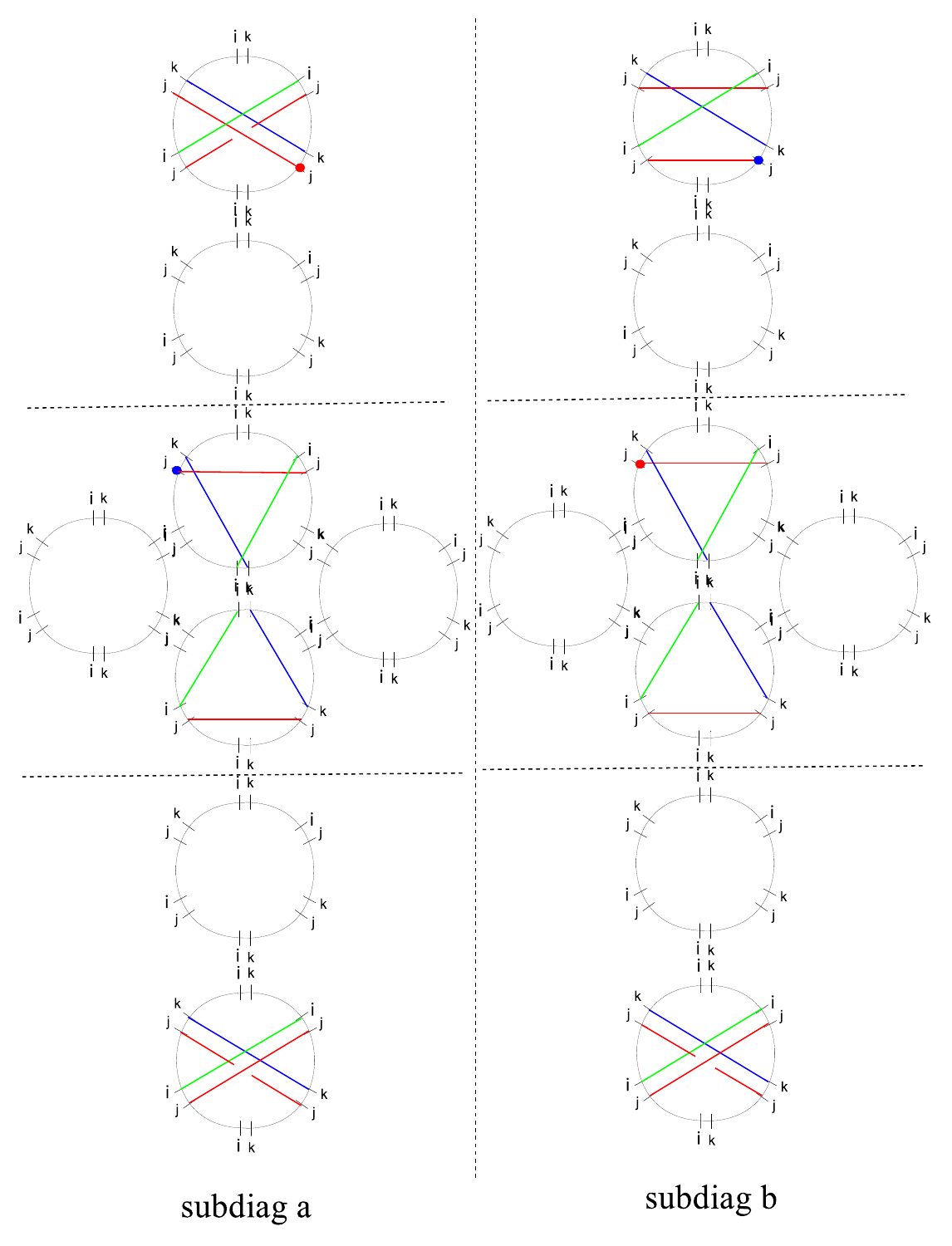}\includegraphics[scale=0.42]{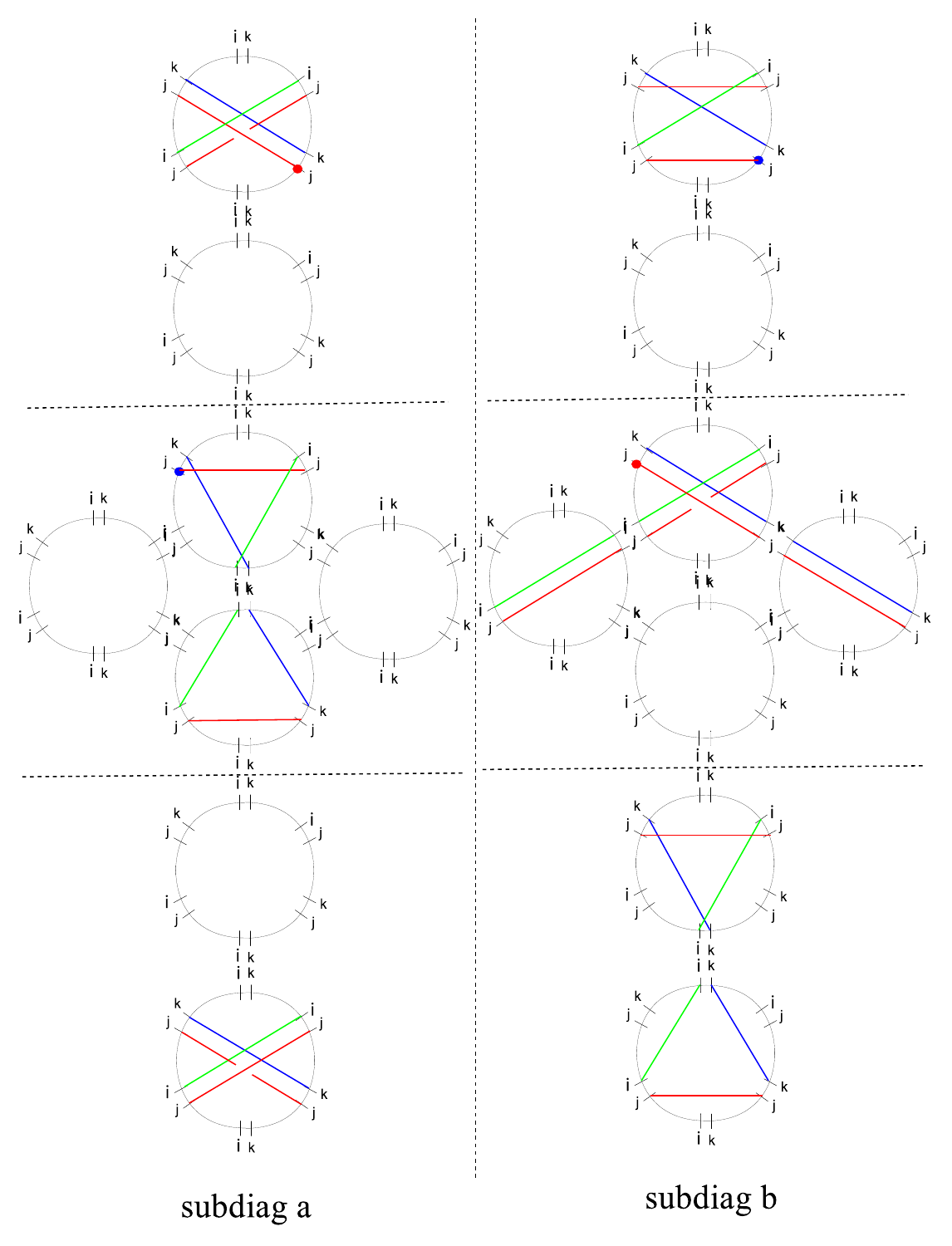}\includegraphics[scale=0.42]{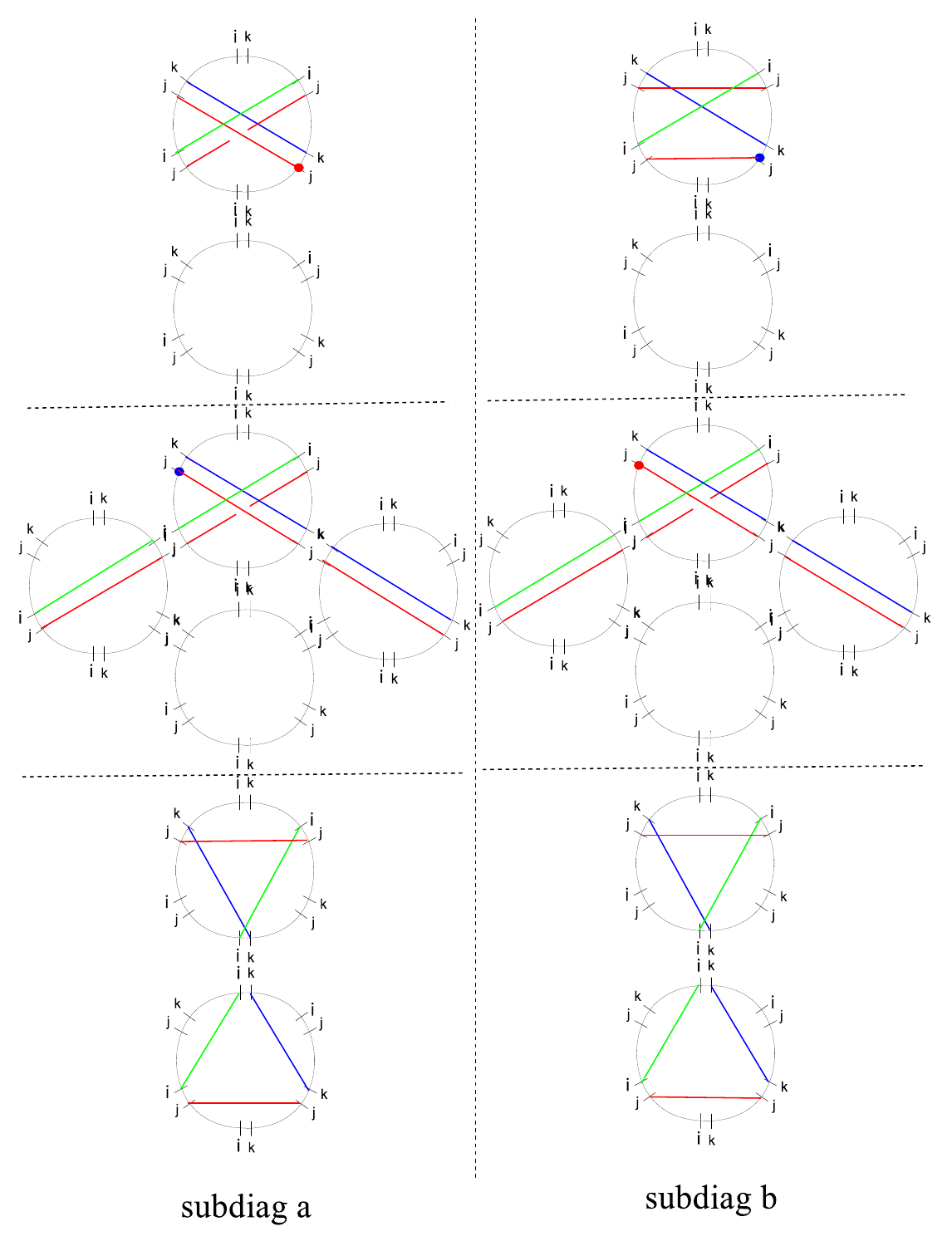}
\includegraphics[scale=0.42]{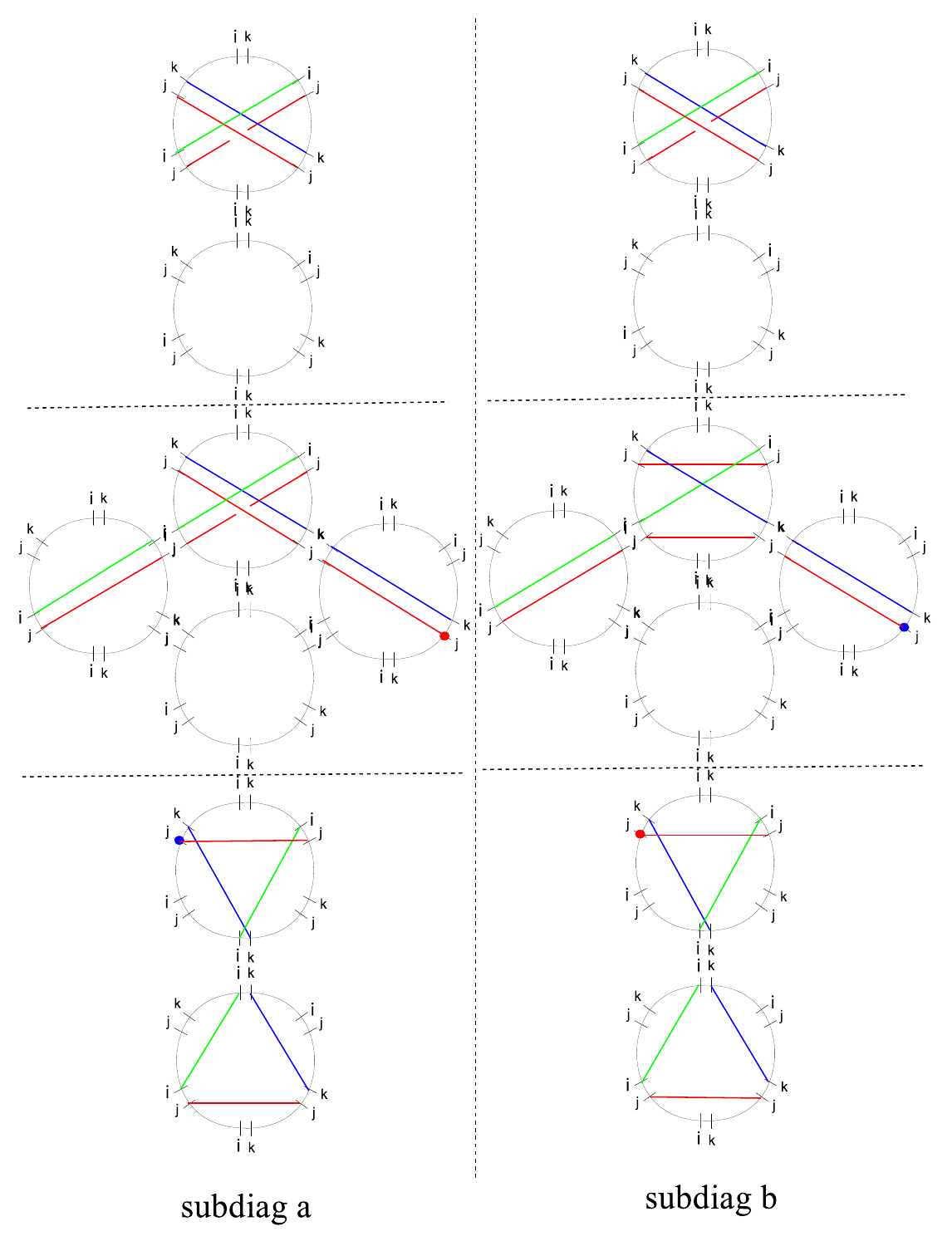}\includegraphics[scale=0.42]{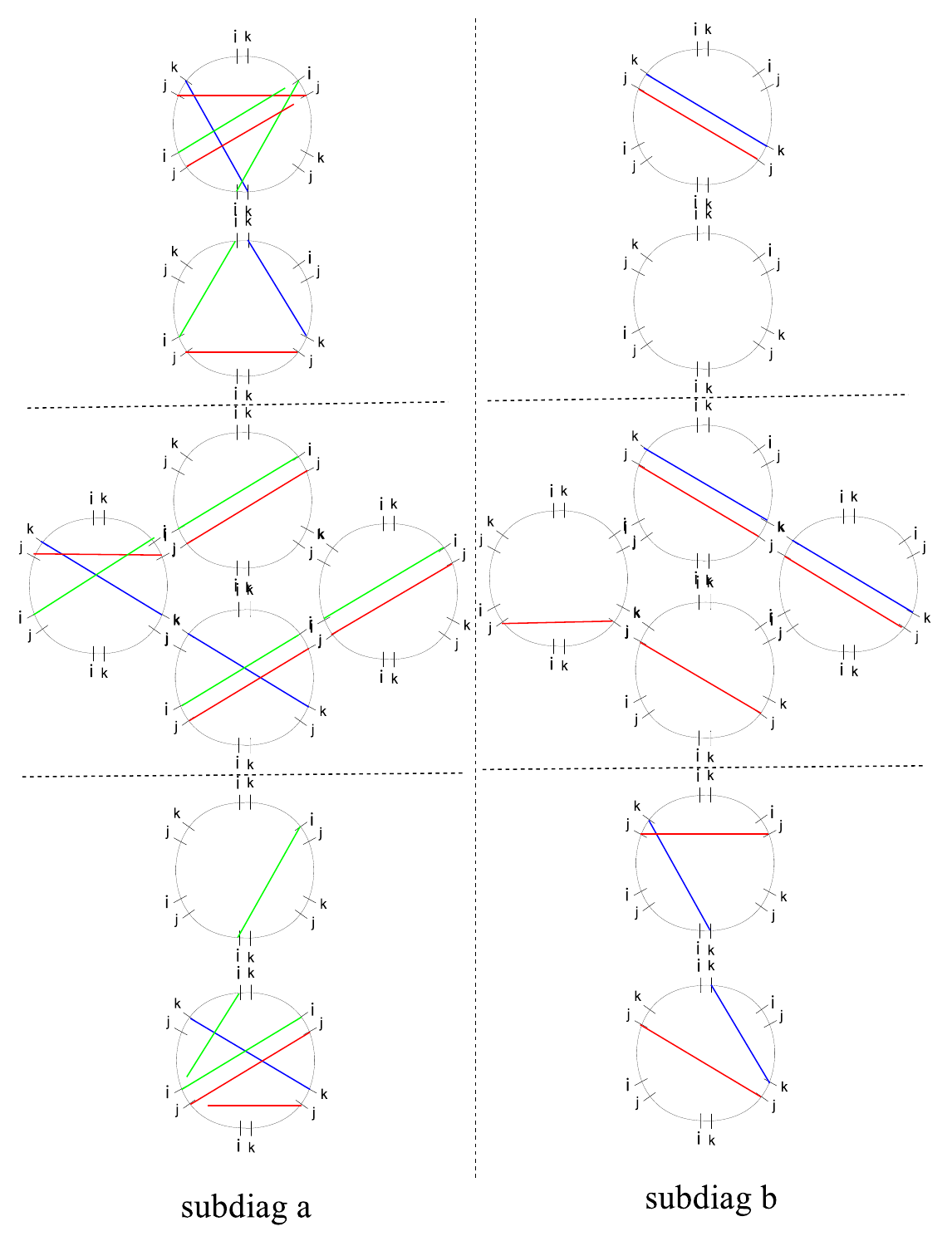}\includegraphics[scale=0.42]{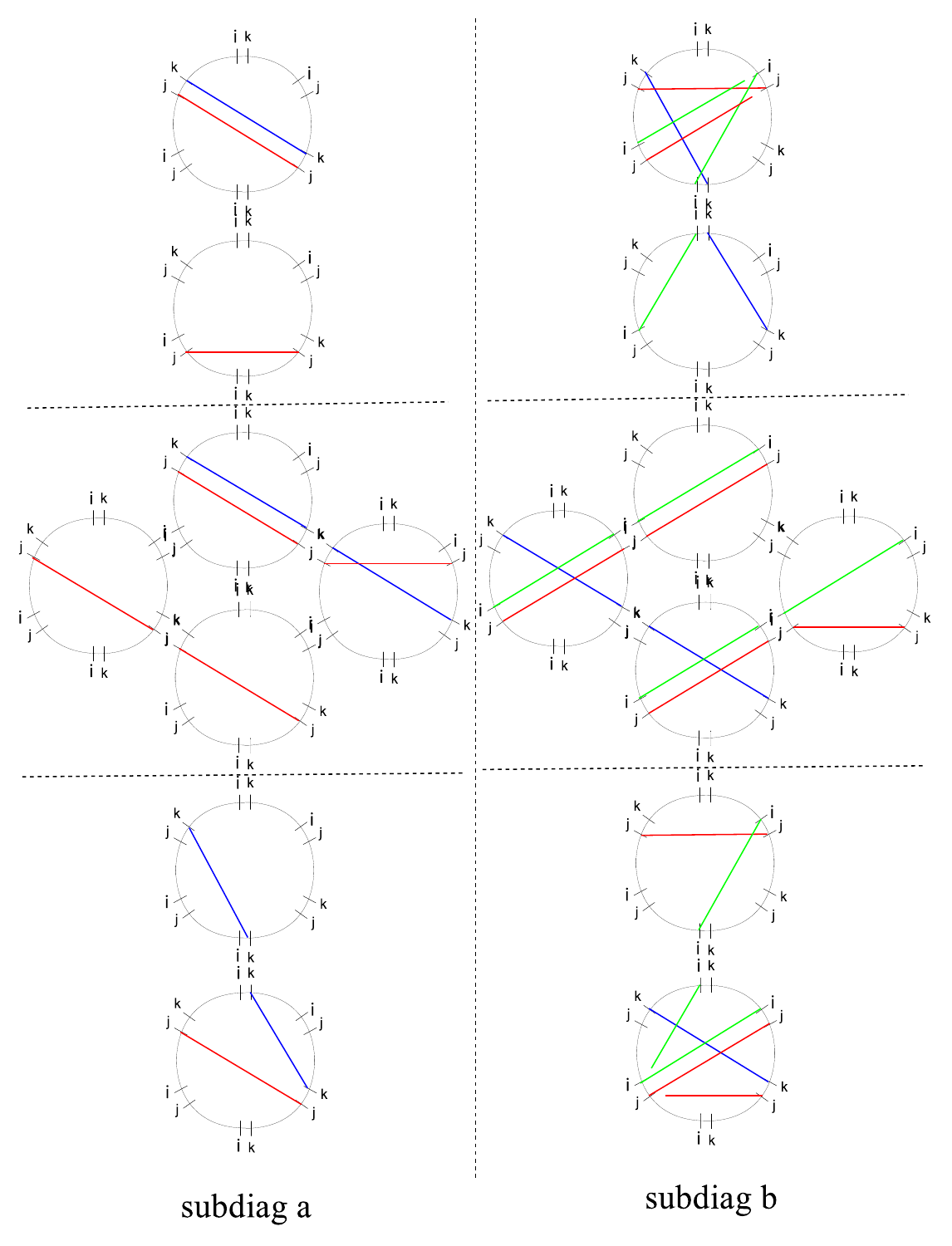}
\includegraphics[scale=0.42]{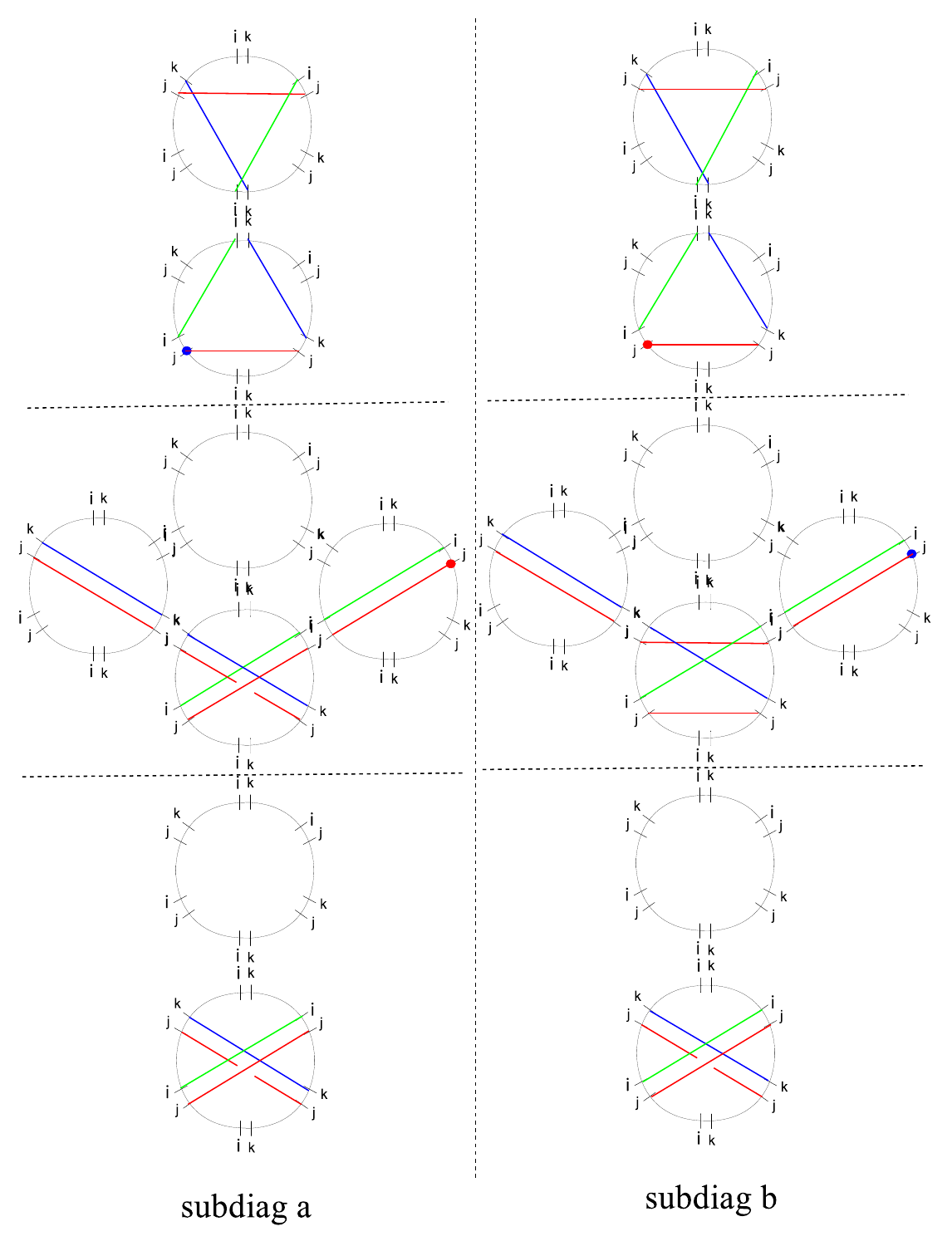}\includegraphics[scale=0.42]{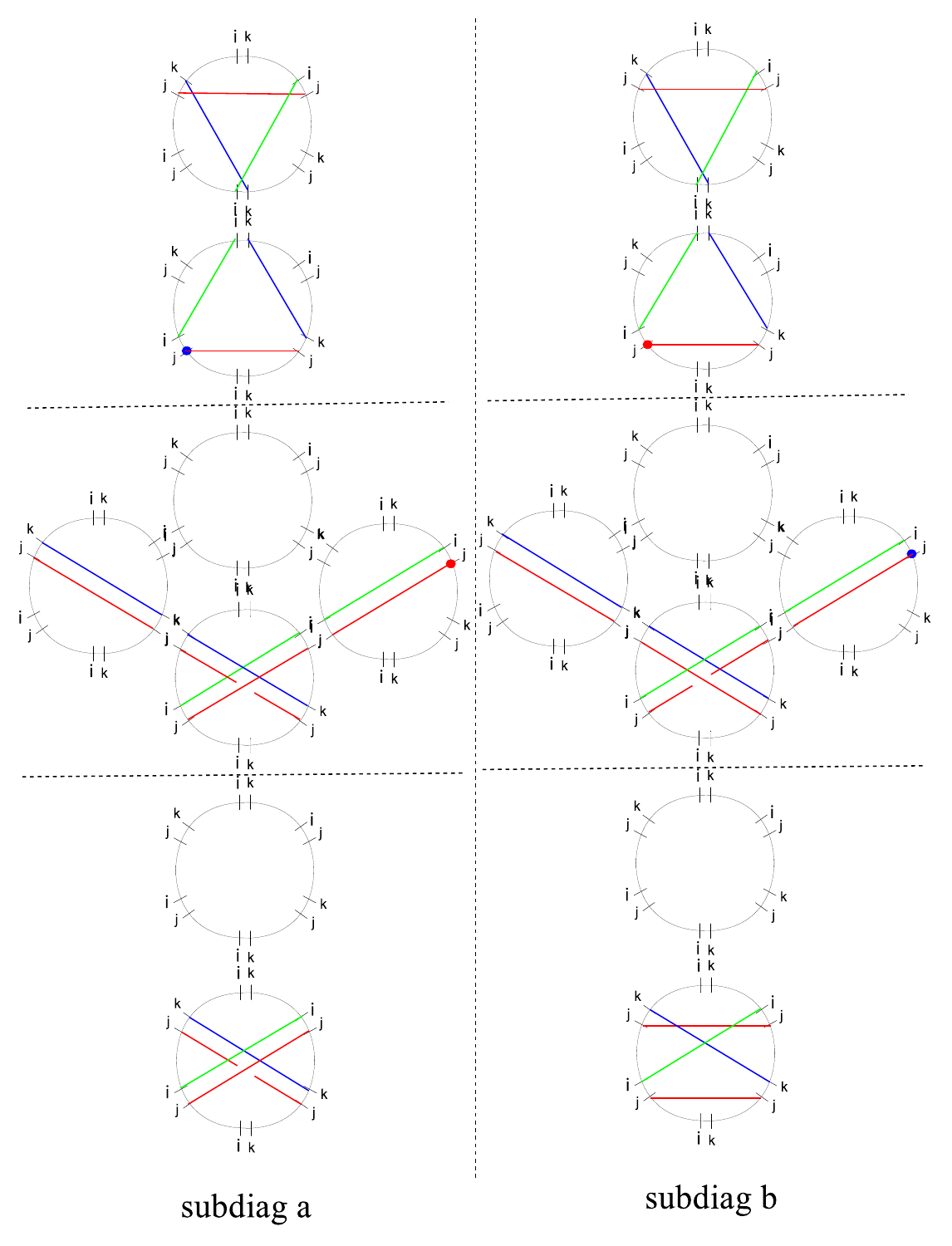}
\caption{Diagrams $\#$1-8 for the order two in 3-herd}
\end{center}
\end{figure}

\begin{figure}[h!]
\begin{center}
\includegraphics[scale=0.42]{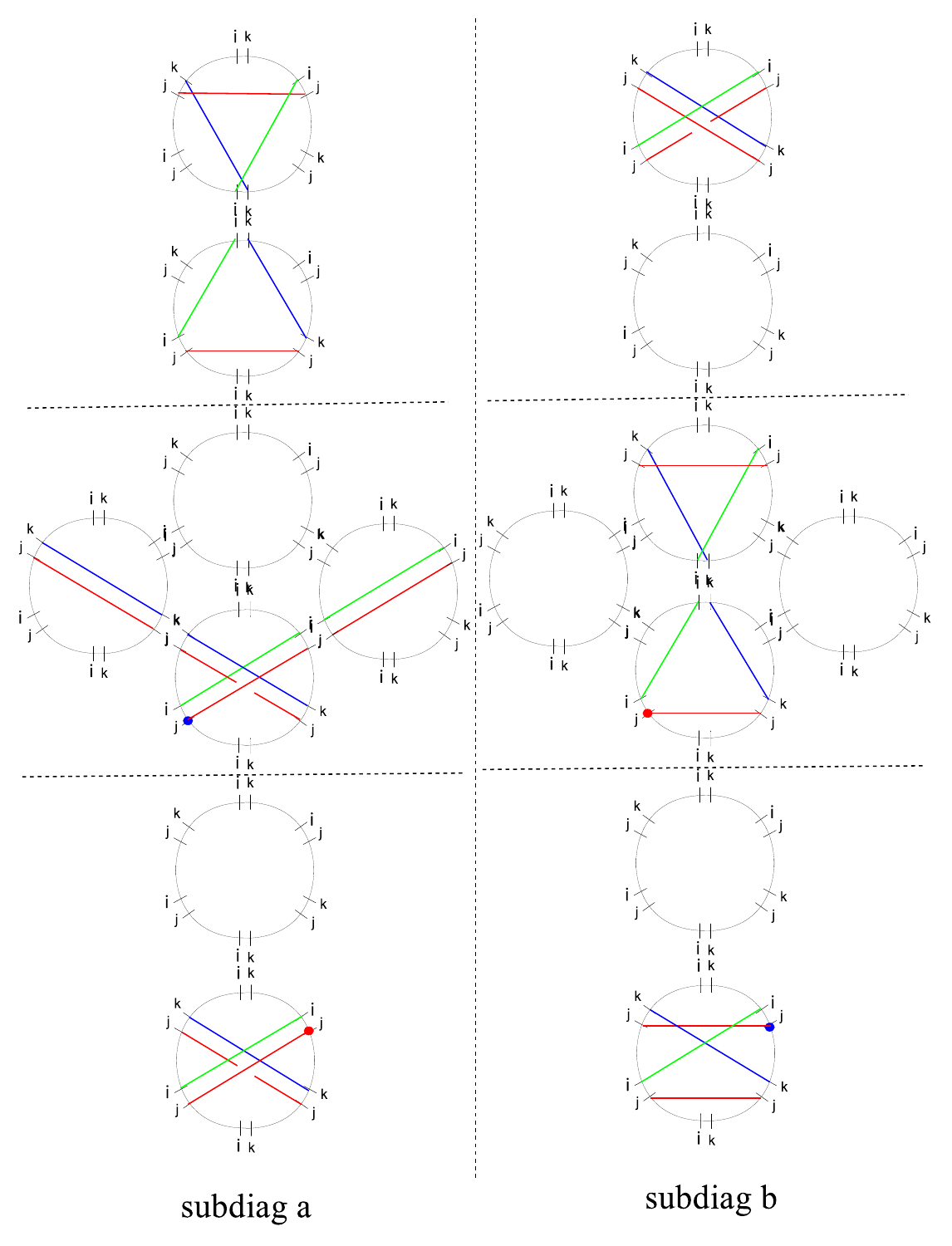}\includegraphics[scale=0.42]{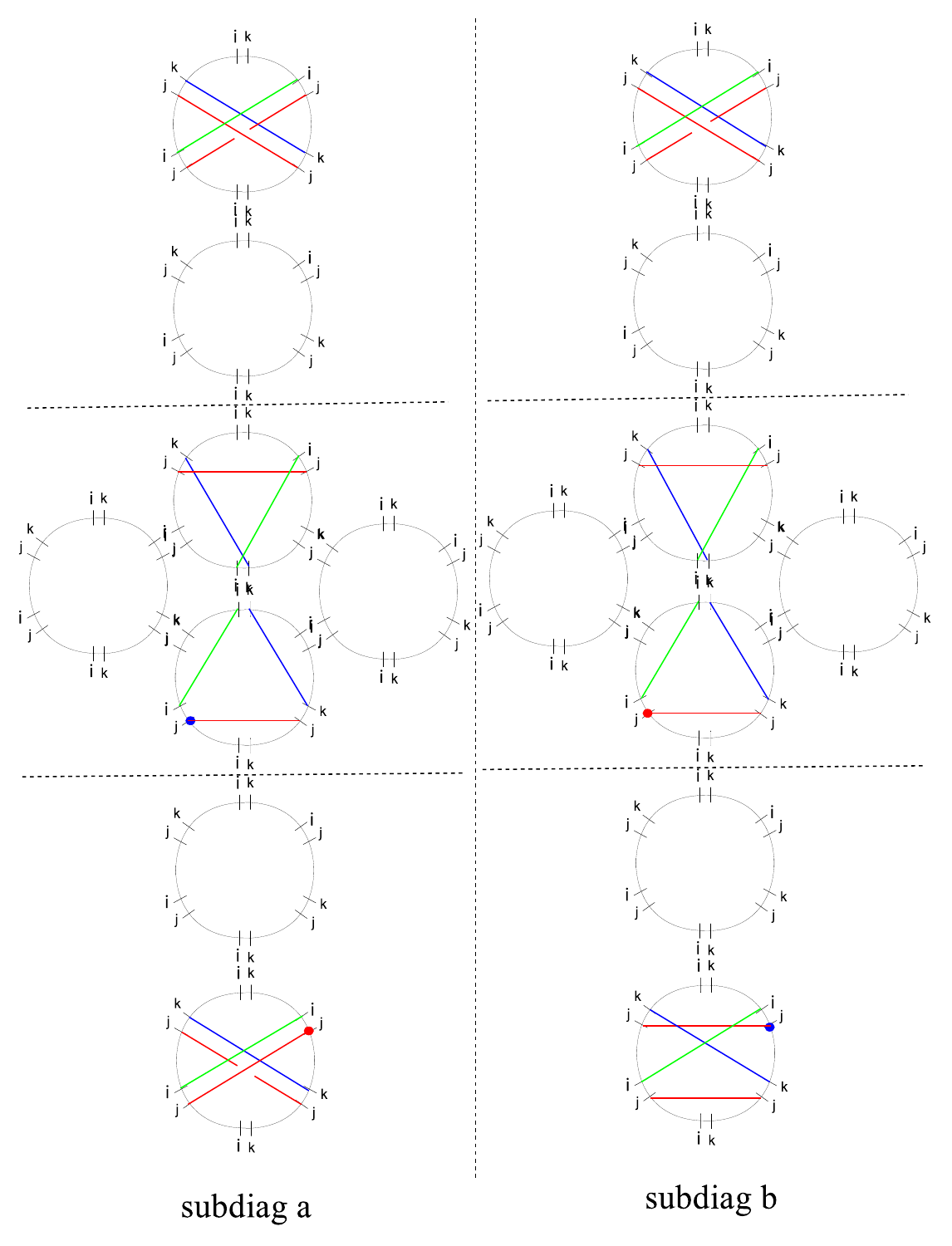}\includegraphics[scale=0.42]{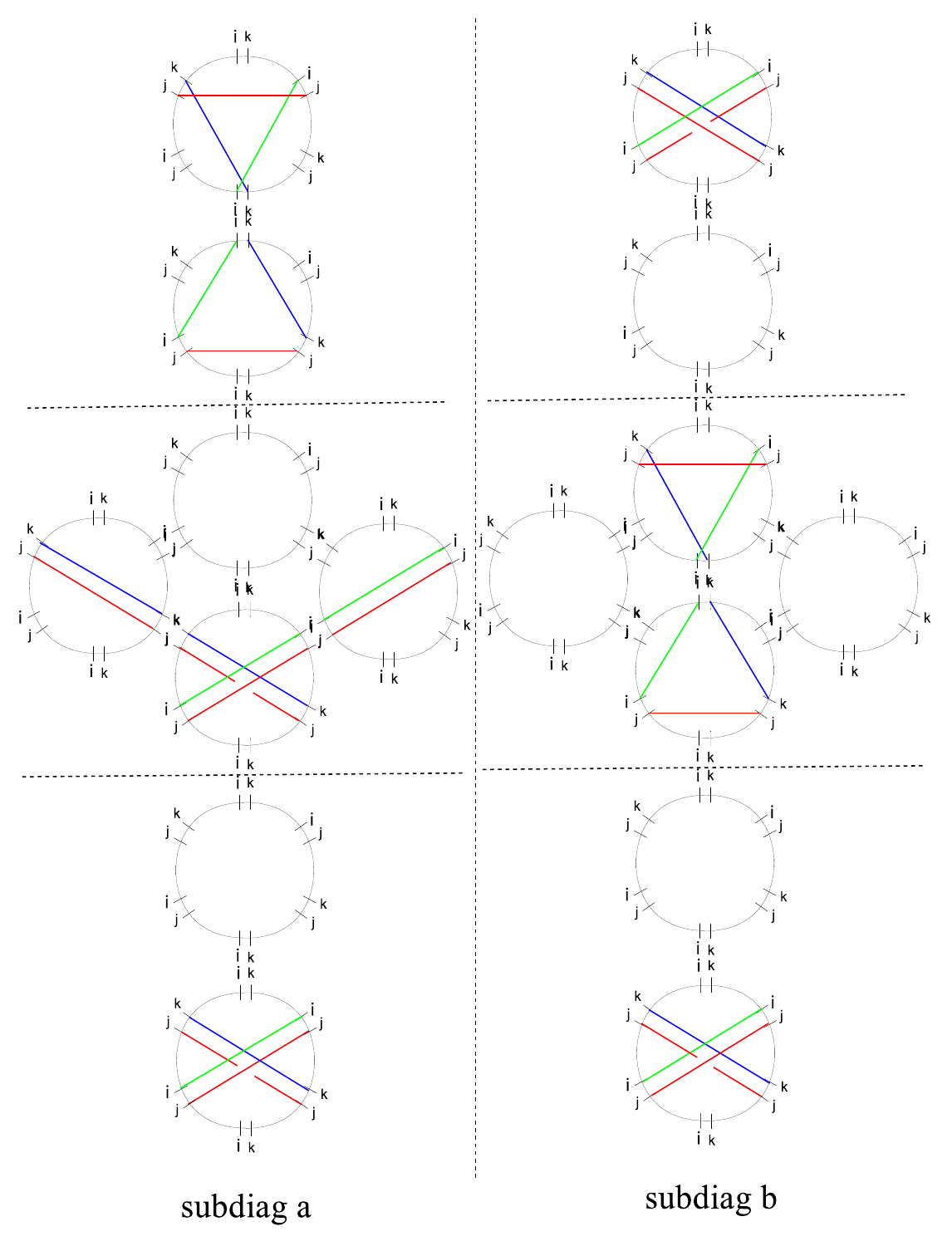}
\includegraphics[scale=0.42]{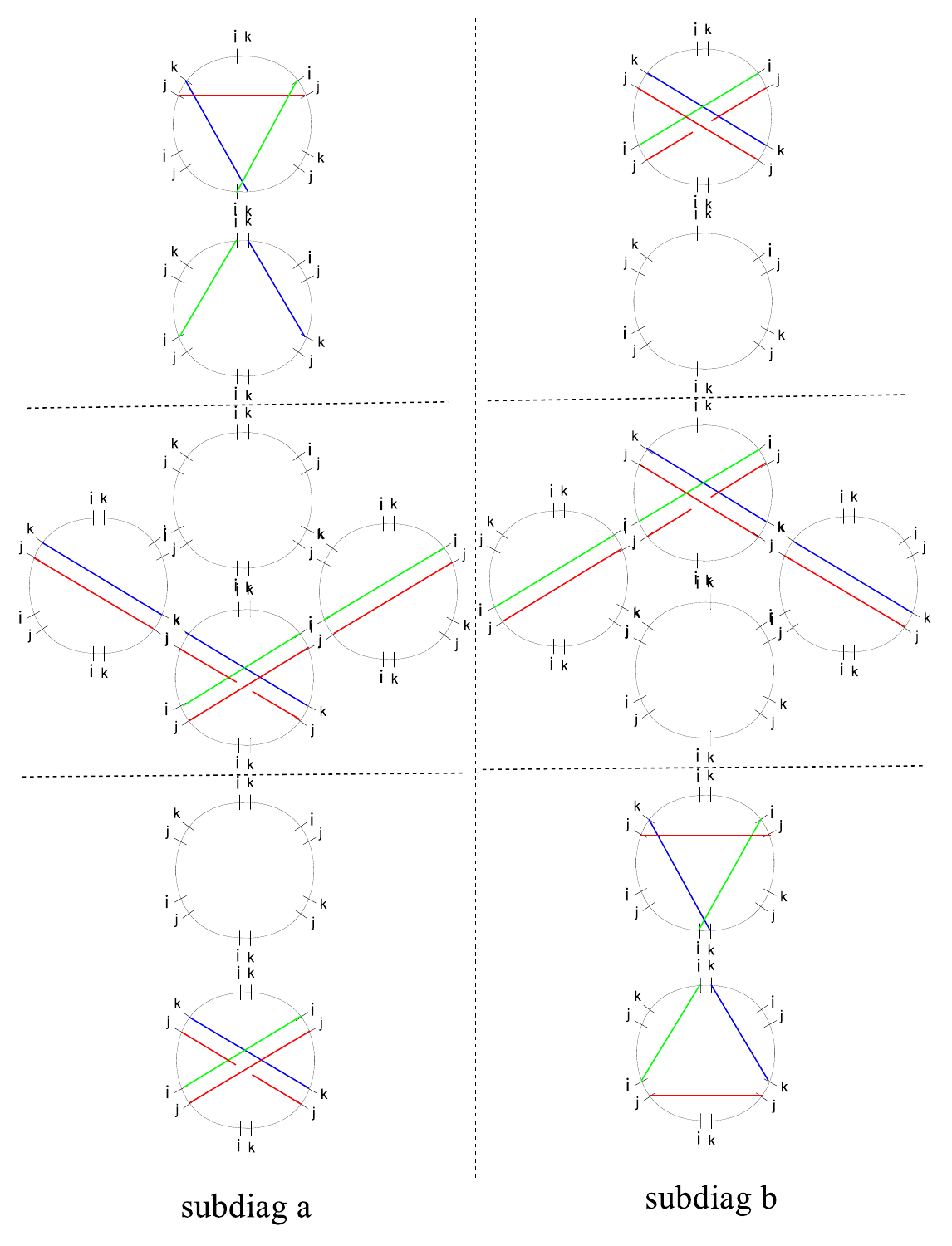}\includegraphics[scale=0.42]{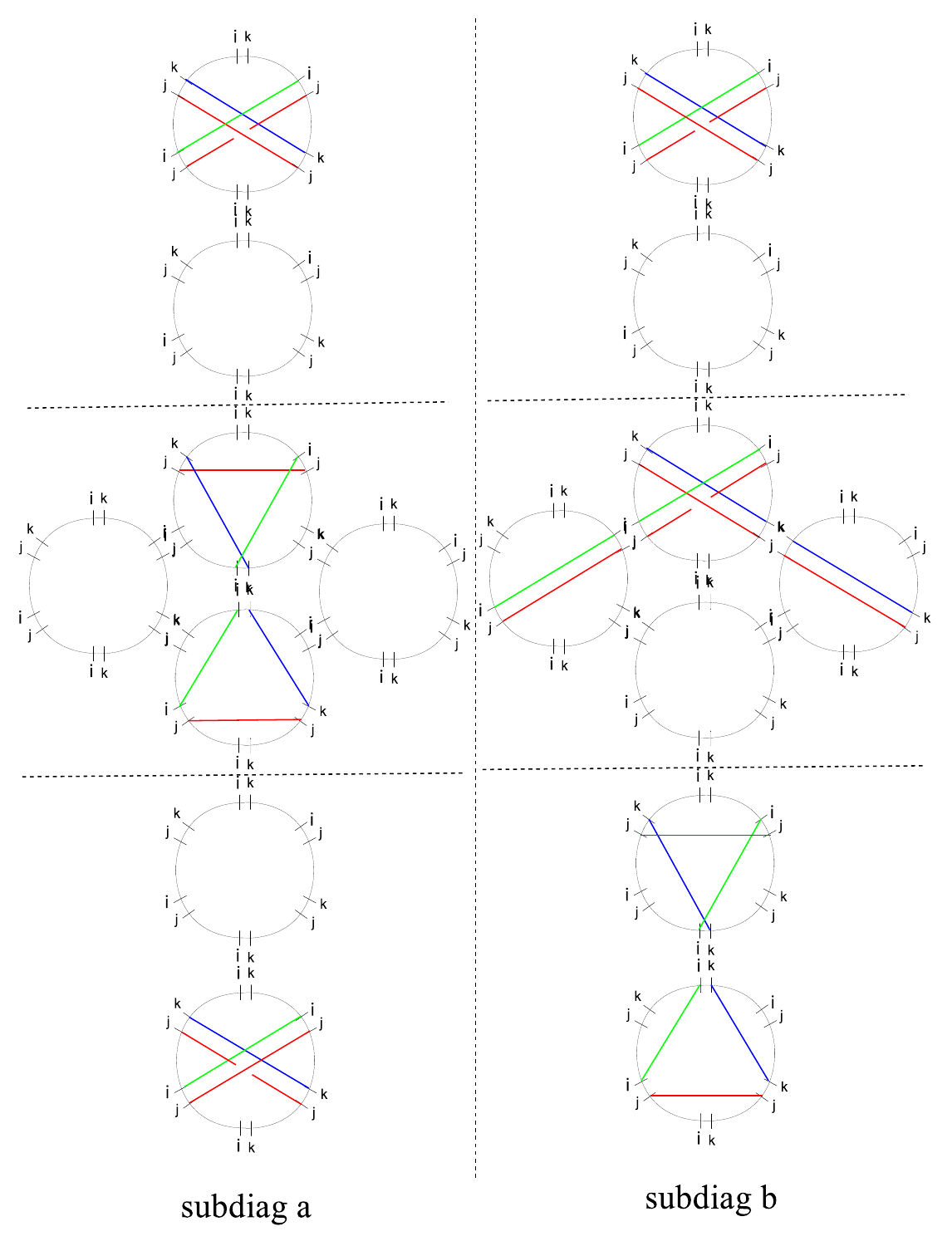}
\caption{Diagrams $\#$9-13 for the order two in 3-herd}
\end{center}
\end{figure}

\begin{tabular}{|c|ccccccccccccc|}
  \hline
  diag $\#$ &1&2&3&4&5&6&7&8&9&10&11&12&13\\
  \hline
  contribution  &1&$2 y^2$& $y^4$ & $y^4$ &$y^{-6}$&$y^6$&$y^{-4}$&$y^{-4}$&$2 y^{-2}$&1&$y^{-2}$&1&$y^2$\\
  \hline
\end{tabular}

Notice that diagrams of type $\#2$ and type $\#9$ come into play twice, so the total number of diagrams with multiplicities is 15. All the diagrams except $\#5$, 6 are independent of the order along what subdiagram one goes first, though for diagrams $\#5$, 6 this choice may switch the sign of the index to the opposite. The right choice in this case can be extracted from the order of crossing cuts what coincides with the order of simpleton generating functions in the final generating function expression.

\section{Vanilla states from refined charges}\label{app:L-r}


In section \ref{subsubsec:spec-lin-int} we motivated the definition of halo-saturated interfaces by noting that the framed wall-crossing of generic IR interfaces differs from that of IR line defects, due to the lack of a relation such as (\ref{eq:line-defect-property}) for generic interfaces. It is interesting to study the soliton combinatorics involved in contributions to $\langle \fa,L(\gamma)\rangle$ for a generic interface labeled by the regular homotopy class $\fa$ on $\Sigma^{*}$. In fact, since here we are interested in intersections of $\fa$ with closed homology classes, it will be sufficient to consider the relative homology class of $\fa$ on $\Sigma^{*}$, which will be denoted $a$.

\subsection{Refinement of halo charges}

In the classical K-wall formula of \cite{GMN4,GMN5}, the enhanced 2d-4d degeneracies $\omega(\gamma,a)$ keep track of the effects of the \emph{refinement} of 4d charges induced by the removal of points corresponding to the 2d vacua $z^{(i)},\, {z'}^{(j')}$ (the endpoints of $a$). In the context of spectral networks, this is identified with
\be
	\omega(\gamma,a) = \langle L(\gamma),a\rangle\,.
\ee
When positivity holds\footnote{In \cite{WWC} evidence was found, somewhat surprisingly, that positivity seems to hold for BPS boundstates corresponding to stable irreps of the $m$-Kronecker quiver.}, it tells us that $|\Omega(\gamma)|={\rm dim}({\mathfrak{h}}_{\gamma})$, in other words the BPS index really counts the number of oscillators generating the corresponding vanilla Fock sub-space. We conjecture that there exists a {unique} splitting
\be
	L(\gamma) = \sum_{r=1}^{|\Omega(\gamma)|}L_{r}(\gamma)\,,
\ee
with each term satisfying
\be
	L_{r}(\gamma) = \sum_{p\in\CW_{c}}\alpha_{r,\gamma}(p)\, p_{\Sigma},\qquad \partial L_{r}(\gamma)=0 \, , \qquad [L_{r}(\gamma)] = {\rm sgn}(\Omega(\gamma))\cdot \gamma \, ,
\ee
where $\alpha_{r,\gamma}(p)$ are integers determined by a set of rules which we will presently explain.
Heuristically, each $L_{r}(\gamma)$ should be associated with a $1$-particle vanilla BPS state in the multiplet $\fh_{\gamma}$, then $\omega(\gamma,a)$ counts the number of 4d vanilla as well as orbital oscillators contributing to the Fock space of \emph{framed} BPS states.

To present our construction of the $L_{r}$, we introduce a new homology lattice, naturally related to a classification of supersymmetric interfaces. Given a network at a critical phase $\CW(\vartheta_{c})$, consider the space $C\setminus\CW(\vartheta_{c})$, it will be a disconnected union of various components. Choose a point from each component, let $R$ be the set of these points. Then we define
\be
	C^{*}:=C\setminus R\qquad \Sigma^{*}:=\Sigma\setminus\pi^{-1}(R)\qquad \Gamma^{*}=H_{1}(\Sigma^{*},\IZ)
\ee
we will call $\Gamma^{*}$ the \emph{refined lattice}, while we denote by $H_{1}(\Sigma^{*},\IZ;\pi^{-1}(R))$ the $\Gamma^{*}$-torsor of relative homology classes on $\Sigma^{*}$ with endpoints in $\pi^{-1}(R)$.
We define $\Gamma^{*}_{c}$ to be the (not necessarily one-dimensional) sublattice which projects to $\Gamma_{c}\subset\Gamma$ upon filling the punctures at $\pi^{-1}(R)$. We also denote by $\gamma_{c}$ the generator of the one-dimensional lattice $\Gamma_{c}$ (the sign ambiguity is fixed by $\vartheta_{c}$).

Any IR interface labeled by $a\in H_{1}(\Sigma^{*},\IZ;\pi^{-1}(R))$, enjoys a well-defined pairing $\langle\tilde\gamma,a\rangle$ with any $\tilde\gamma\in\Gamma^{*}$. 
The $L_{r}$ -- so far defined as \emph{actual paths} -- can be clearly associated to homology classes of $\Gamma^{*}$, we define $\tgamma_{n,r}:=[L_{r,n\gamma_{c}}]_{\Gamma^{*}}\,{\rm sgn}(\Omega(n\gamma_{c}))$.

Correspondingly, we introduce a new set of formal variables $\tilde X$ associated with (relative) homology classes on $\Sigma^{*}$, satisfying
\be
	\tilde X_{\tgamma}\tilde X_{\tgamma'} = \tilde X_{\tgamma+\tgamma'}\qquad 	\tilde X_{a}\tilde X_{\tgamma'} = \tilde X_{a+\tgamma}\,.
\ee

Now, choose $a$ to be any relative homology class on $\Sigma^{*}$ with endpoints in $\pi^{-1}(R)$, and consider the generating function of its framed BPS states with halo charges in $\Gamma^{*}_{c}$ (cf. \ref{eq:restricted-framed-spin-functional}, where $a$ is played by $\wp^{(i)}$)
\be\label{eq:2d4d-interpretation}
	\sum_{\tgamma\in\Gamma_{c}^{*}}\tilde X_{a+\tgamma}
\ee
\begin{conj} the series (\ref{eq:2d4d-interpretation}) admits a factorization of the form
\be\label{eq:halo-interpretation}
	\tilde X_{a}\prod_{n,r}\Big( 1+\sigma(n\gamma_{c}) \tilde X_{\tgamma_{n,r}}\Big)^{\langle a, \tgamma_{n,r}\rangle}
\ee
where $\sigma(n\gamma_{c})={\rm sgn}(\Omega(n\gamma_{c}))$, $[\tgamma_{n,r}]_{\Gamma}=n\gamma_{c}$ and $r=1,\dots,|\Omega(n\gamma_{c})|$.
\end{conj}
Because of our choice of $\Gamma^{*}$, the refined homology classes $\tgamma^{*}$ uniquely determine the $L_{r}(\gamma)$ (i.e. the $\alpha_{r,\gamma}(p)$). This is our definition of the $L_{r}$, it relies on the conjectural factorization. We conclude by presenting some nontrivial evidence for the conjecture.

\subsection{The $3$-herd}
For the 3-herd, the BPS indices read
\be
	\Omega(\gamma_{c}) = 3,\quad \Omega(2\gamma_{c}) = -6,\quad \Omega(3\gamma_{c}) = 18,\quad \dots
\ee
We know that $L_{n,r}$ must run through each terminal street $n$ times, for all $r$, for homological reasons. Thus we expect for the halo generating function (\ref{eq:halo-interpretation}) of an interface crossing one terminal street (cf. fig.\ref{fig:3herd})
\be
\begin{split}
	\tilde X_{a} \,& (1+\tilde X_{\tgamma_{1,1}})(1+\tilde X_{\tgamma_{1,2}})(1+\tilde X_{\tgamma_{1,2}}) \\
	\times & (1-\tilde X_{\tgamma_{2,1}})^{-2}(1-\tilde X_{\tgamma_{2,2}})^{-2}(1-\tilde X_{\tgamma_{2,3}})^{-2}(1-\tilde X_{\tgamma_{2,4}})^{-2}(1-\tilde X_{\tgamma_{2,5}})^{-2}(1-\tilde X_{\tgamma_{2,6}})^{-2} \\
	\times & (1+\tilde X_{\tgamma_{3,1}})^{3}(1+\tilde X_{\tgamma_{3,2}})^{3}\ \cdots \ (1-X_{\tgamma_{3,18}})^{3} \\
	\times & \cdots
\end{split}
\ee
According to our conjecture, this \emph{predicts the following form for generating function of framed states (\ref{eq:2d4d-interpretation})}
\be
\begin{split}
	\tilde X_{a} \,& \Big[ 1+\Big(\tilde X_{\tgamma_{1,1}}+\tilde X_{\tgamma_{1,2}}+\tilde X_{\tgamma_{1,2}}\Big) \\
	+ & \Big(\tilde X_{\tgamma_{1,1}+\tgamma_{1,2}}+\tilde X_{\tgamma_{1,1}+\tgamma_{1,3}}+\tilde X_{\tgamma_{1,2}+\tgamma_{1,3}}+2\sum_{r=1}^{6}\tilde X_{\tgamma_{2,r}}\Big) \\
	+ &\Big(X_{\tgamma_{1,1}+\tgamma_{1,2}+\tgamma_{1,3}} + 2 \sum_{r=1}^{3}\sum_{r'=1}^{6} \tilde X_{\tgamma_{1,r}+\tgamma_{2,r'}}+ 3 \sum_{r=1}^{18}\tilde X_{\tgamma_{3,r}}\Big) \\
	+ & \cdots \Big]
\end{split}
\ee

Indeed, by studying the detours, \emph{we find exactly the predicted structure}, with the identifications (labels refer to the street map of figure \ref{fig:3herd}, the lifts of streets carry the orientations dictated by the WKB flow for each component of the lift)
\be
\begin{split}
	L_{1,1}=&\pi^{-1}( 	\gamma _1+\delta _1+\alpha _2+\beta _2+\delta _2+\delta _3+\alpha _4+\beta _4+\delta _4+\alpha _6+\beta _6 )\\
	L_{1,2}=&\pi^{-1}( \alpha _1+\beta _1+\delta _1+\gamma _2+\delta _2+\delta _3+\delta _4+\alpha _6+\beta _6 )\\
	L_{1,3}=& \pi^{-1}(  \alpha _1+\beta _1+\delta _1+\delta _2+\alpha _3+\beta _3+\gamma _3+\delta _3+\delta _4+\alpha _5+\beta _5 ) \\
\end{split}	
\ee
	\begin{figure}[h!]
	\begin{center}
	\includegraphics[width=.30\textwidth]{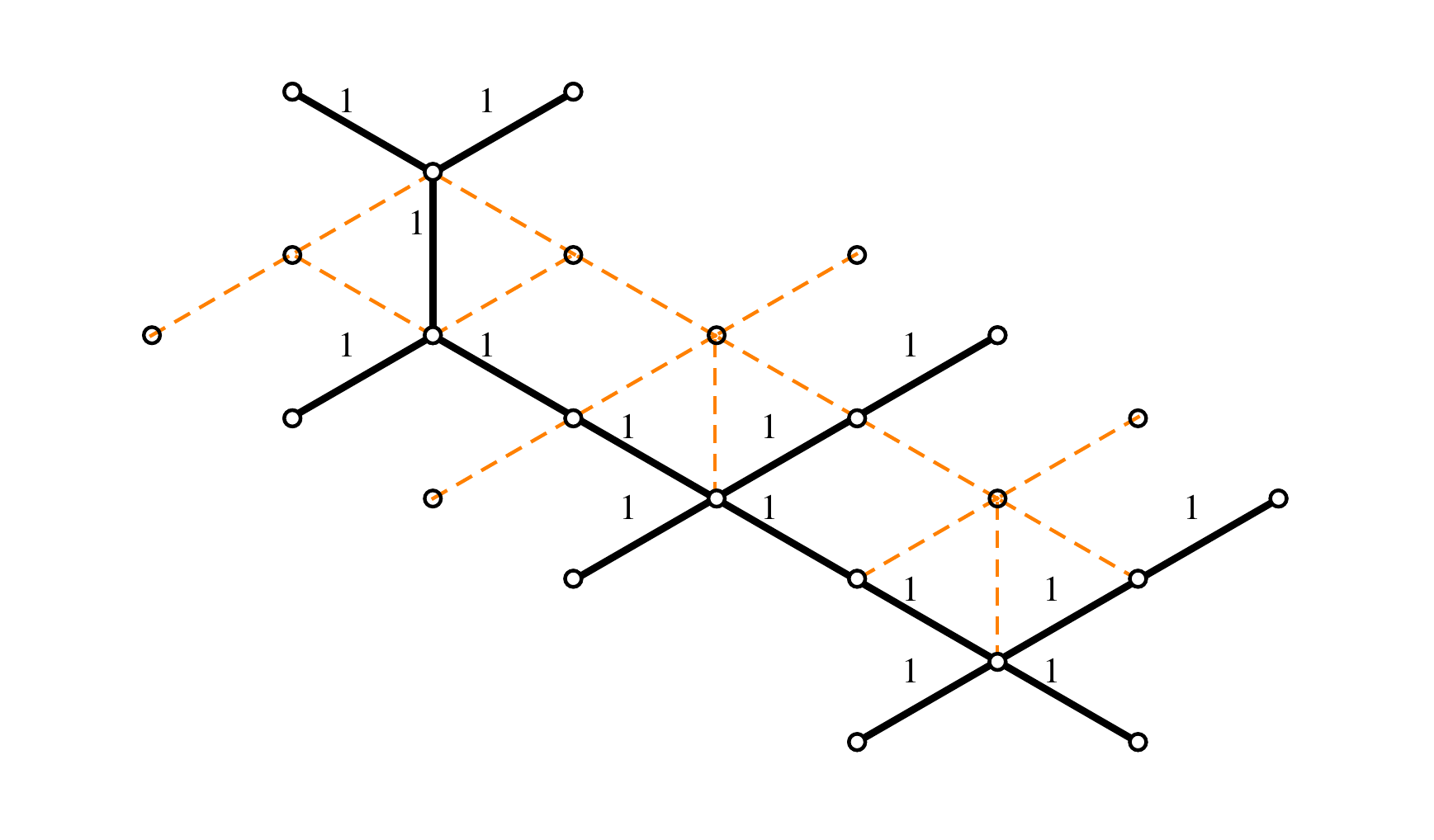}\includegraphics[width=.30\textwidth]{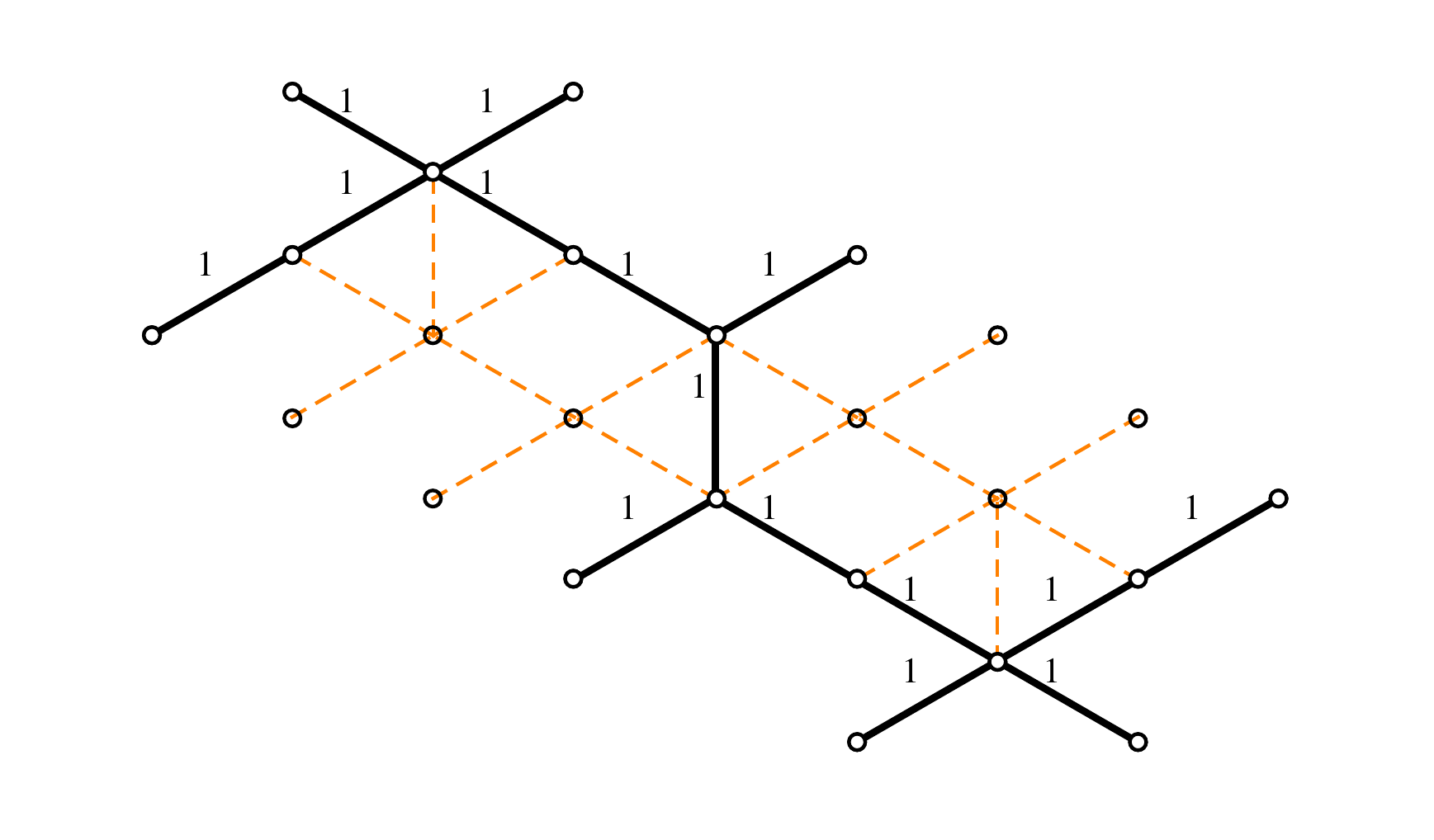}\includegraphics[width=.30\textwidth]{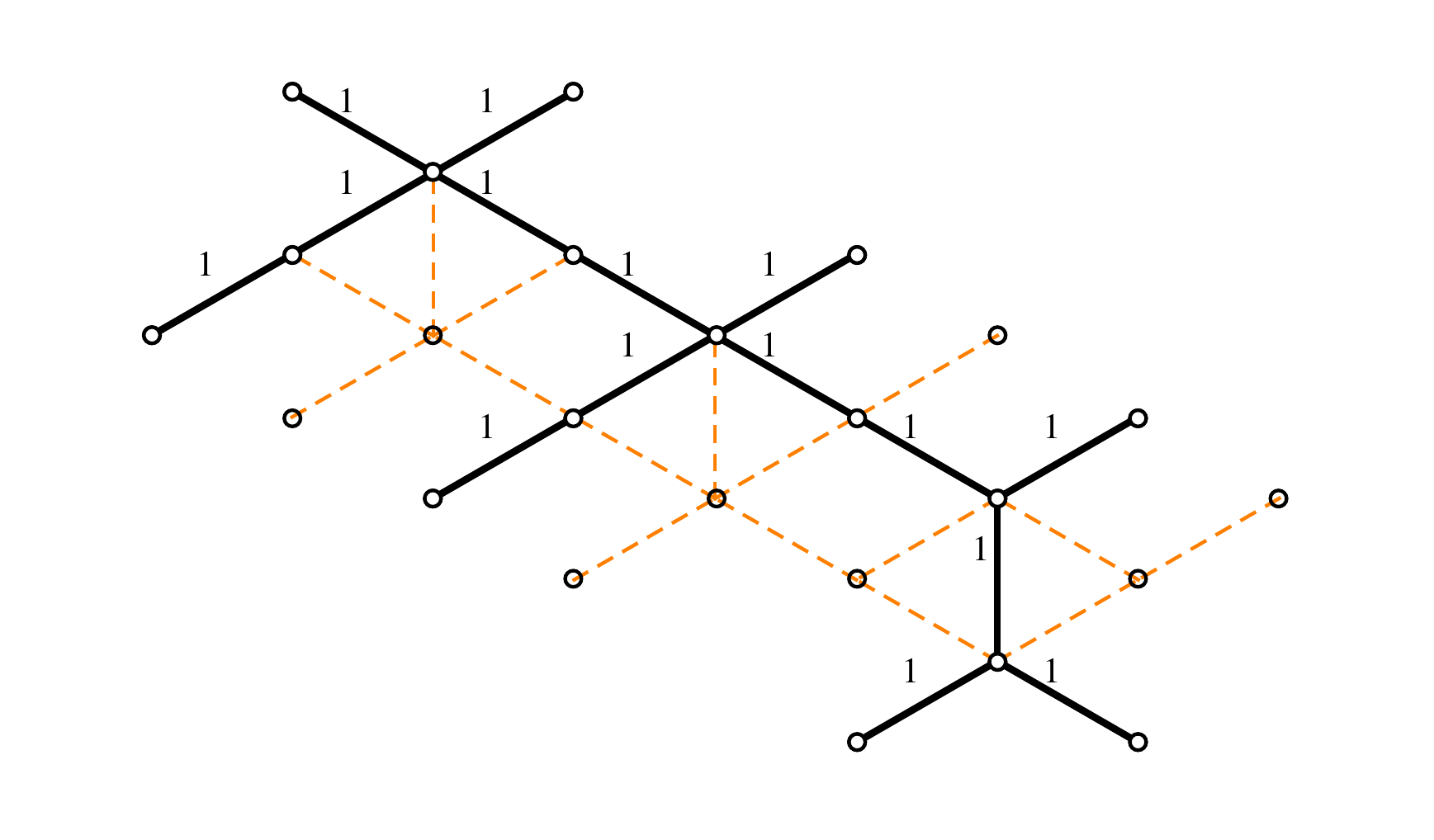}
	\caption{The set of $L_{1,r}$. Values of the $\alpha_{r}(p)$ are displayed.}
	\label{default}
	\end{center}
	\end{figure}
\be
\begin{split}
	L_{2,1}=& \pi^{-1}( 2(\alpha _1+\beta _1+\delta _1+\gamma _2+\delta _2+\delta _3+\delta _4+\alpha _6+\beta _6))\\
	L_{2,2}=&\pi^{-1}( \alpha _1+\beta _1+\gamma _1+2 \delta _1+\alpha _2+\beta _2+\gamma _2+2 \delta _2+2 \delta _3+\alpha _4+\beta _4+2 \delta _4+2 \alpha _6+2 \beta _6)\\
	L_{2,3}=&\pi^{-1}(  2 \alpha _1+2 \beta _1+2 \delta _1+\gamma _2+2 \delta _2+\alpha _3+\beta _3+\gamma _3+2 \delta _3+2 \delta _4+\alpha _5+\beta _5+\alpha _6+\beta _6   )\\
	L_{2,4}=&\pi^{-1}( 2 (\gamma _1+\delta _1+\alpha _2+\beta _2+\delta _2+\delta _3+\alpha _4+\beta _4+\delta _4+\alpha _6+\beta _6))\\
	L_{2,5}=& \pi^{-1}(   \alpha _1+\beta _1+\gamma _1+2 \delta _1+\alpha _2+\beta _2+2 \delta _2+\alpha _3+\beta _3+\gamma _3+2 \delta _3+\alpha _4+\beta _4+2 \delta _4\\
	& \ \ \,+ \alpha _5+\beta _5+\alpha _6+\beta _6   )\\
	L_{2,6}=&\pi^{-1}( 2 (   \alpha _1+\beta _1+\delta _1+\delta _2+\alpha _3+\beta _3+\gamma _3+\delta _3+\delta _4+\alpha _5+\beta _5    ) )\\
\end{split}	
\ee
	\begin{figure}[h!]
	\begin{center}
	\includegraphics[width=.30\textwidth]{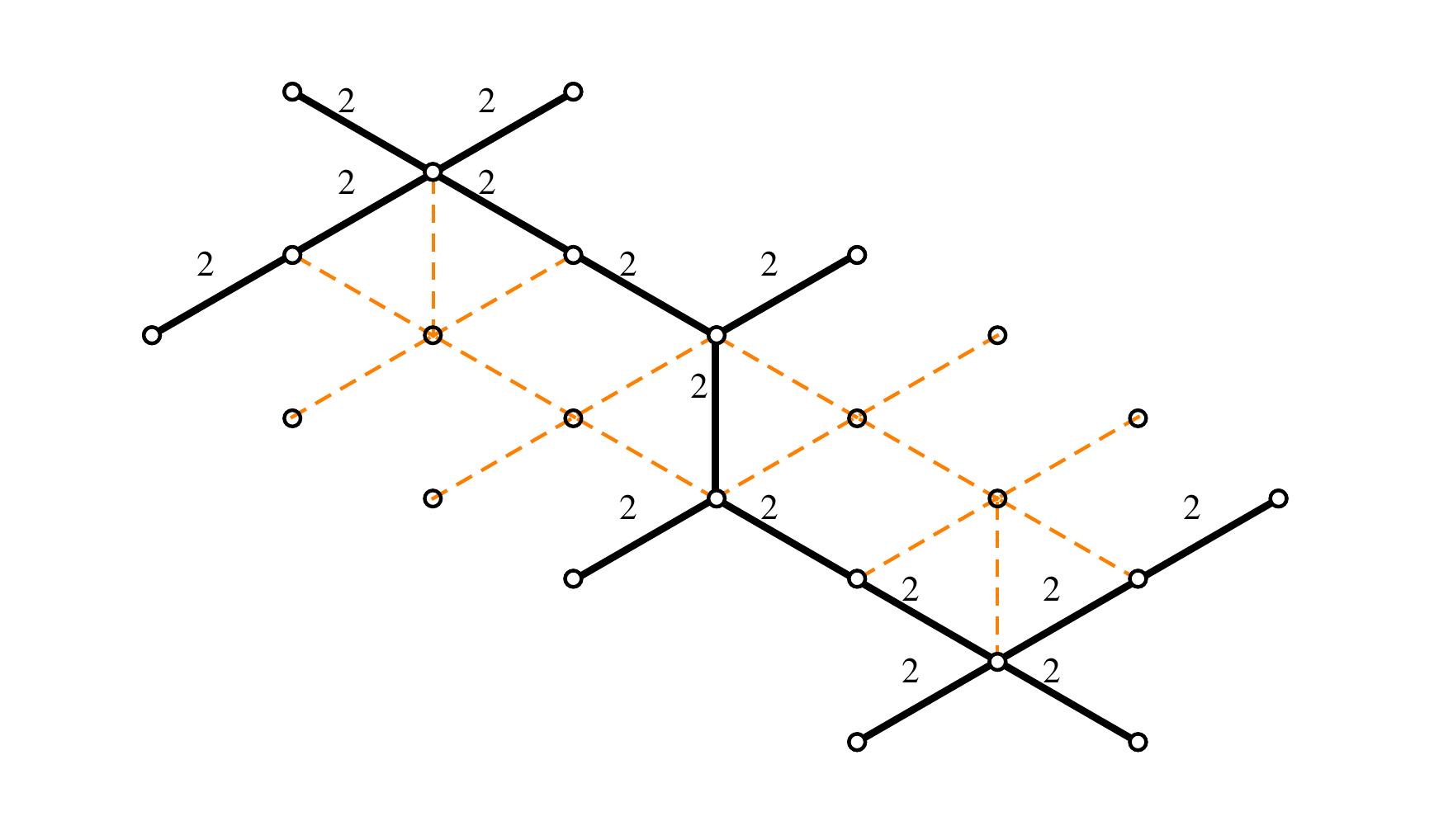}\includegraphics[width=.30\textwidth]{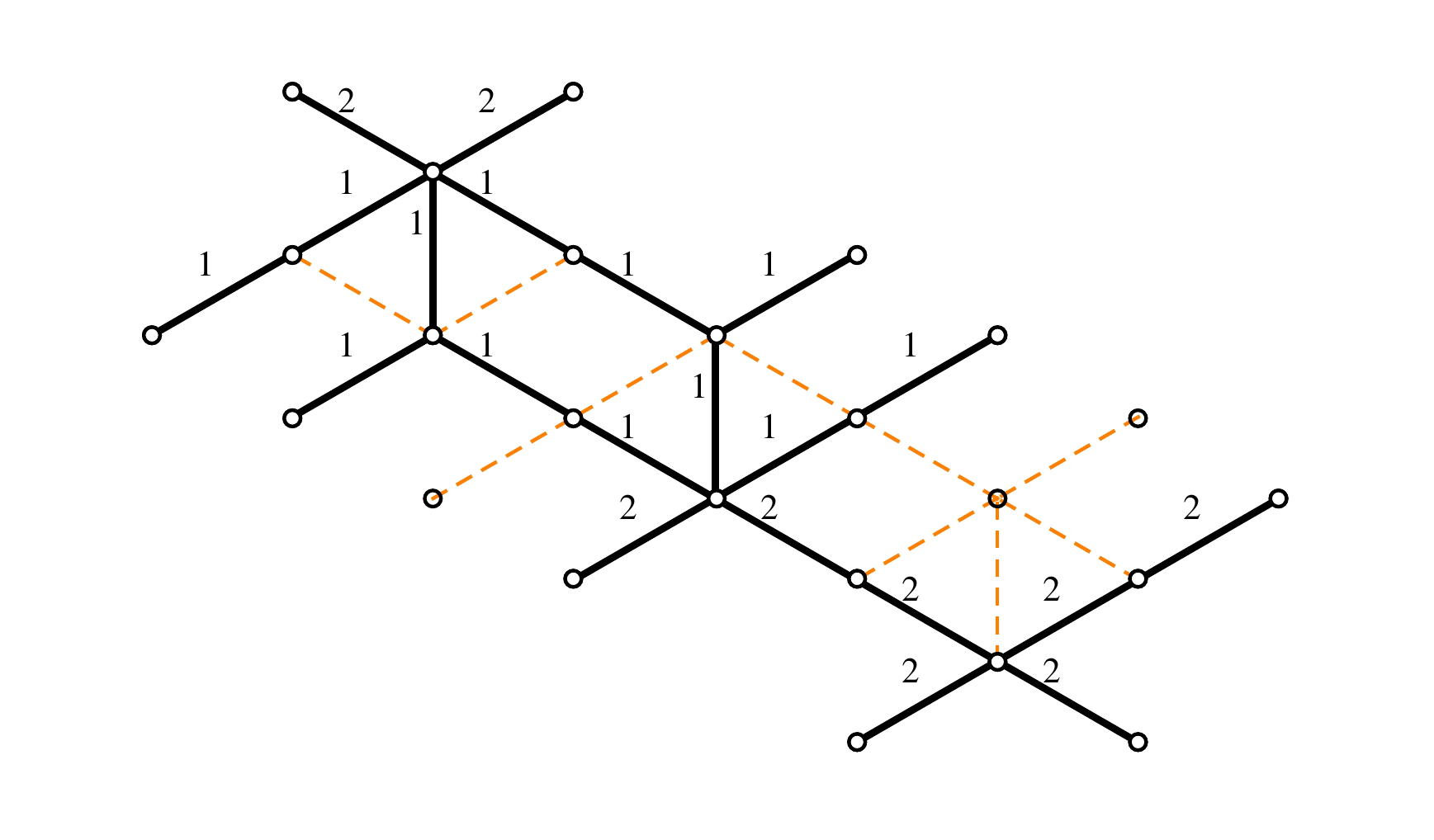}\includegraphics[width=.30\textwidth]{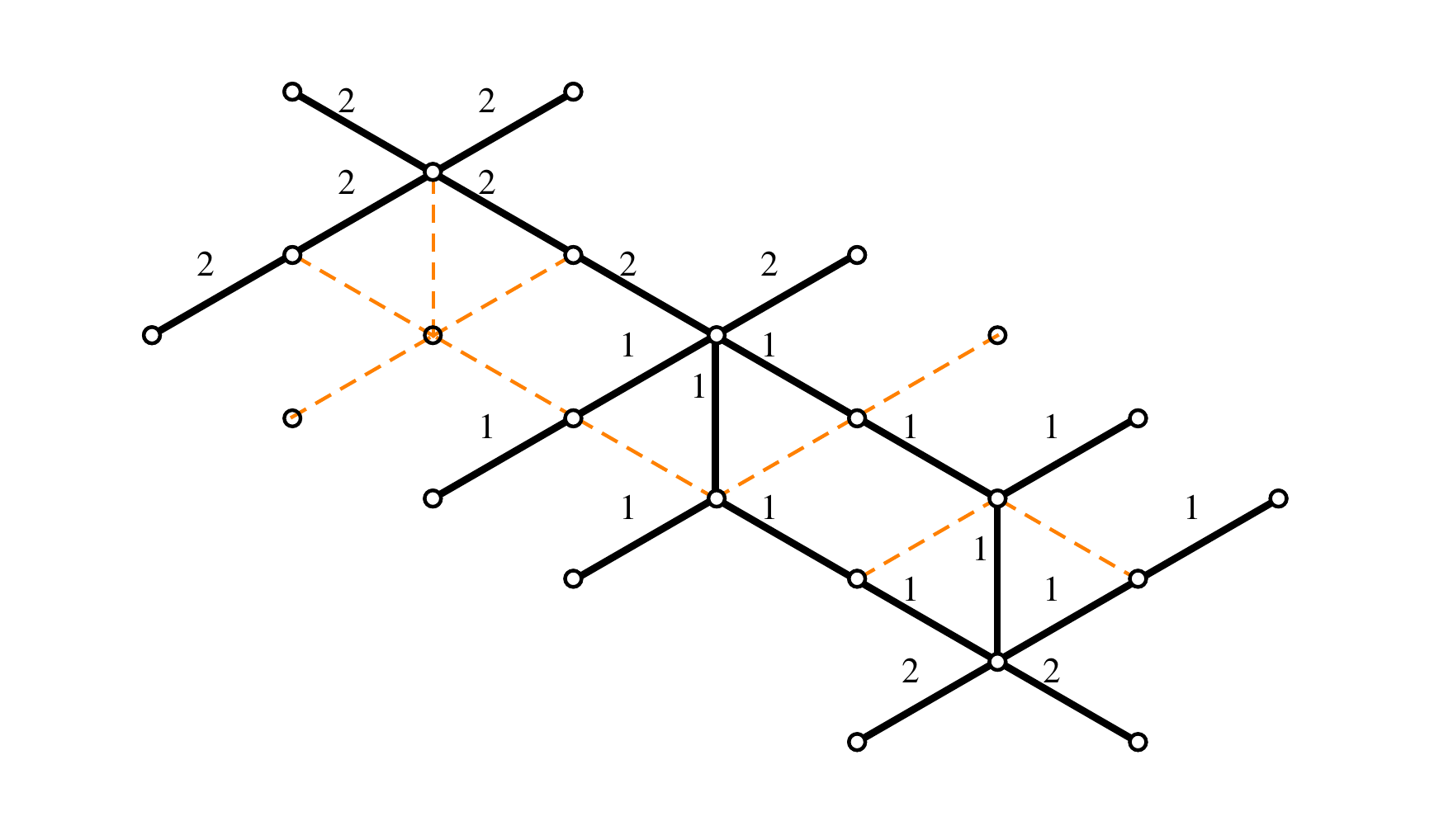} \\
	\includegraphics[width=.30\textwidth]{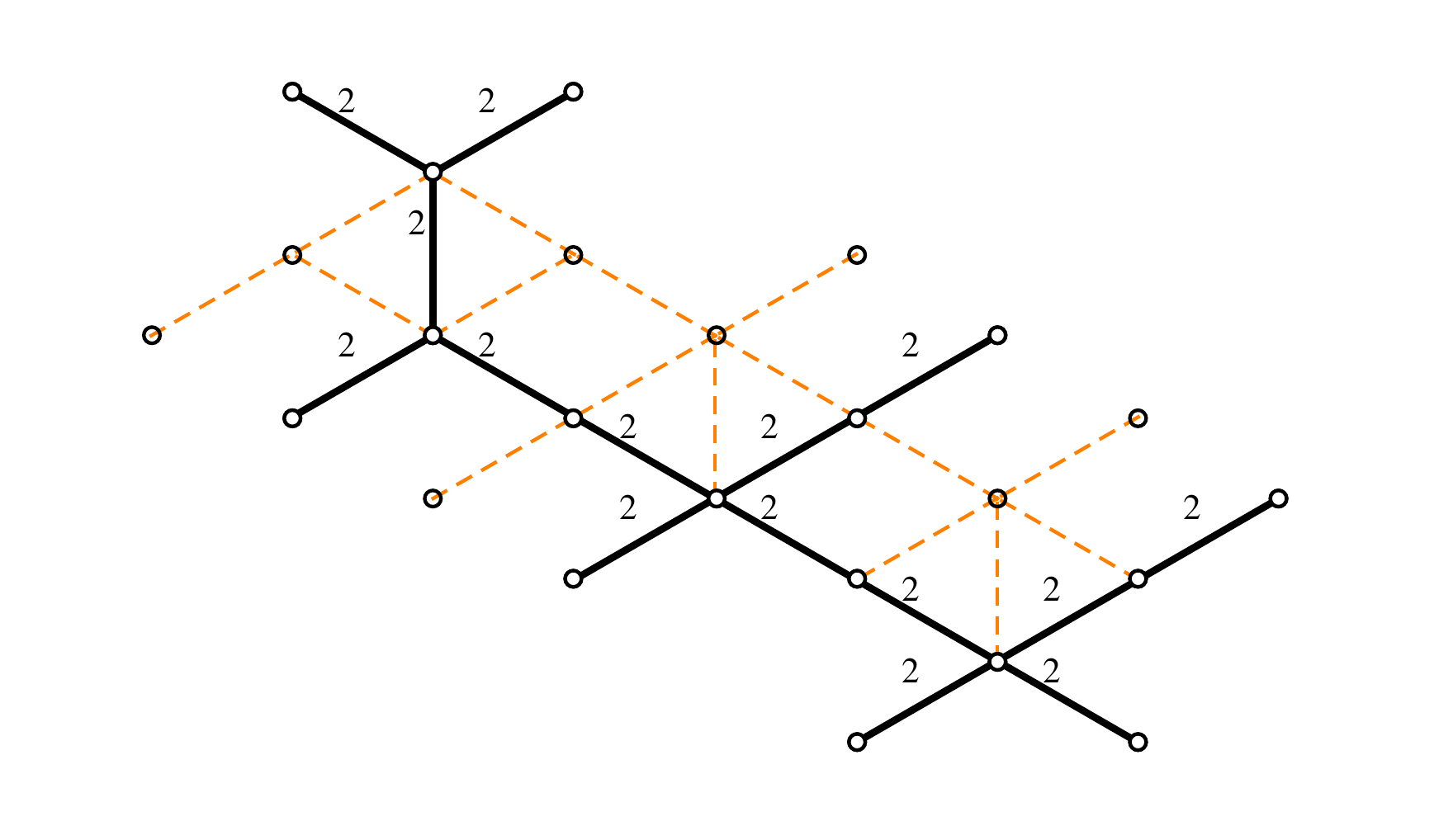}\includegraphics[width=.30\textwidth]{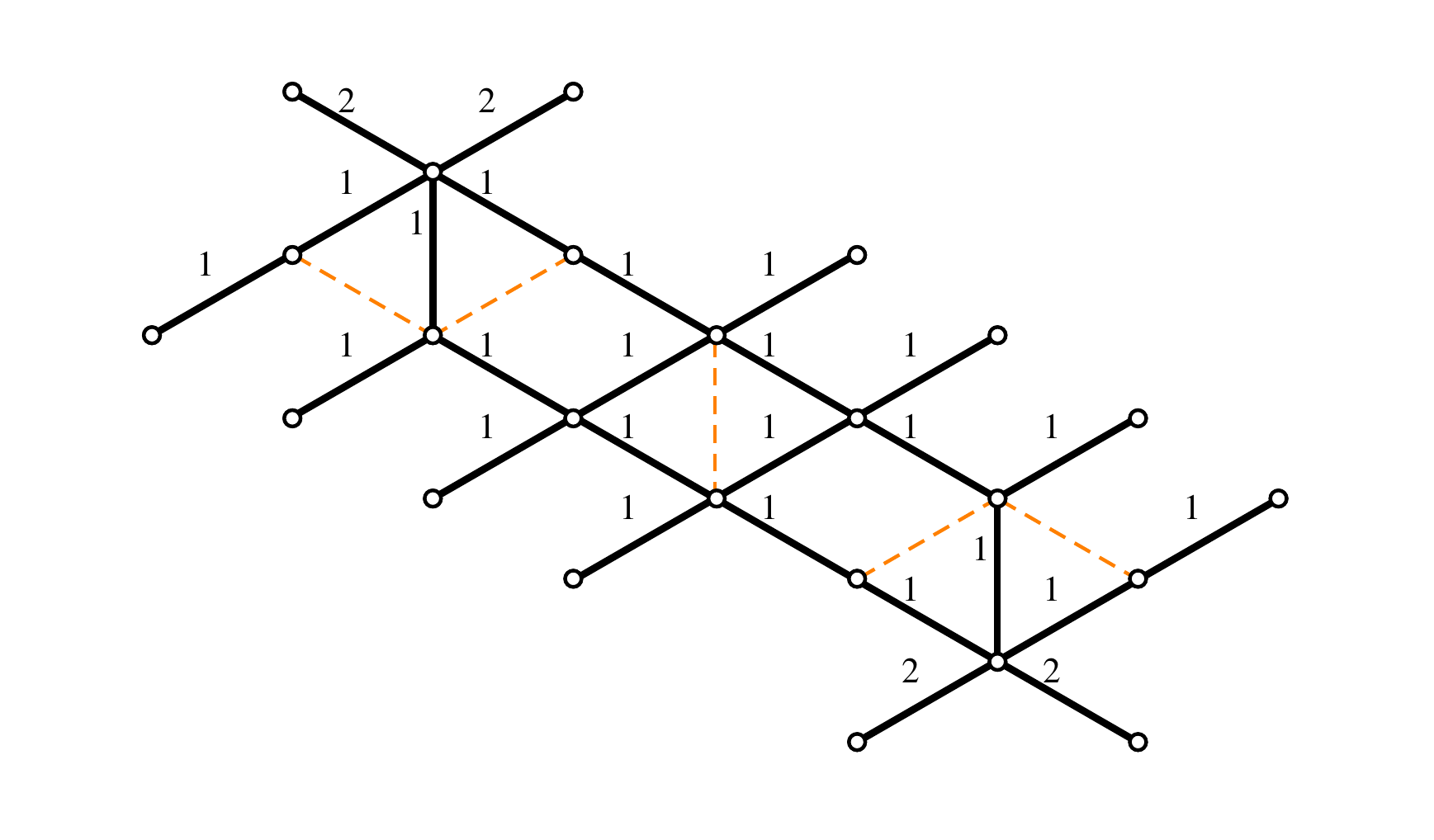}\includegraphics[width=.30\textwidth]{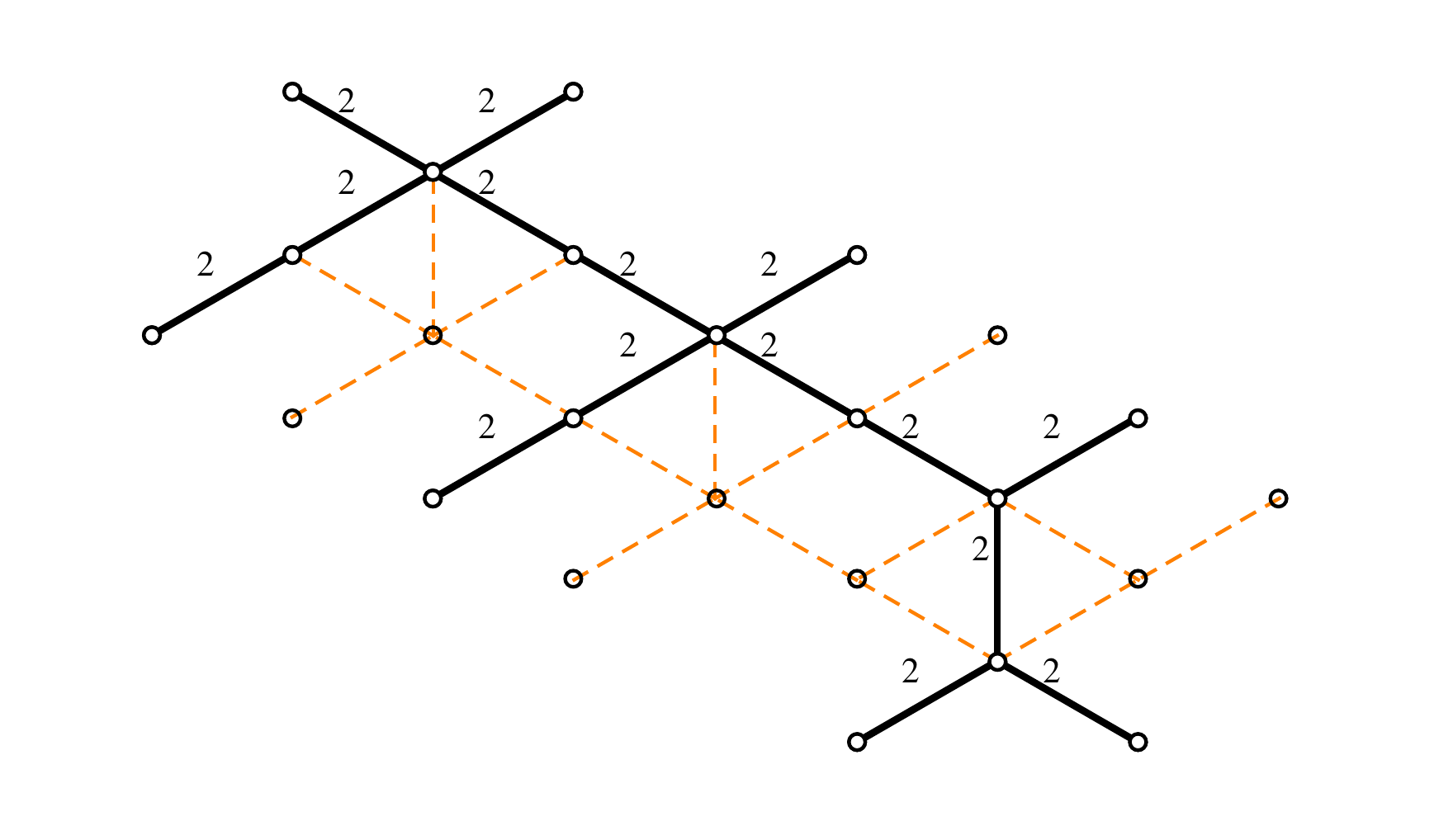}
	\caption{The set of $L_{2,r}$. Values of the $\alpha_{r}(p)$ are displayed.}
	\label{default}
	\end{center}
	\end{figure}
\be
\begin{split}
 	L_{3,1}&=\pi^{-1}( 3 (\alpha _1+\beta _1+\delta _1+\gamma _2+\delta _2+\delta _3+\delta _4+\alpha _6+\beta _6) )\\
  	L_{3,2}=L_{3,3}&=\pi^{-1}(   2 \alpha _1+2 \beta _1+\gamma _1+3 \delta _1+\alpha _2+\beta _2+2 \gamma _2+3 \delta _2+3 \delta _3+\alpha _4+\beta _4+3 \delta _4+3 \alpha _6+3 \beta _6   )\\
  	L_{3,4}=L_{3,5}&=\pi^{-1}(   3 \alpha _1+3 \beta _1+3 \delta _1+2 \gamma _2+3 \delta _2+\alpha _3+\beta _3+\gamma _3+3 \delta _3+3 \delta _4+\alpha _5+\beta _5+2 \alpha _6+2 \beta _6   )\\
  	L_{3,6}=L_{3,7}=L_{3,8}&=\pi^{-1}(  2 \alpha _1+2 \beta _1+\gamma _1+3 \delta _1+\alpha _2+\beta _2+\gamma _2+3 \delta _2+\alpha _3+\beta _3+\gamma _3+3 \delta _3+\alpha _4+\beta _4 \\
	& +3 \delta _4+\alpha _5+\beta _5+2 \alpha _6+2 \beta _6    )\\
 	L_{3,9}=L_{3,10}&= \pi^{-1}( \alpha _1+\beta _1+2 \gamma _1+3 \delta _1+2 \alpha _2+2 \beta _2+\gamma _2+3 \delta _2+3 \delta _3+2 \alpha _4+2 \beta _4+3 \delta _4+3 \alpha _6+3 \beta _6)\\
  	L_{3,11}=L_{3,12}&=\pi^{-1}(   3 \alpha _1+3 \beta _1+3 \delta _1+\gamma _2+3 \delta _2+2 \alpha _3+2 \beta _3+2 \gamma _3+3 \delta _3+3 \delta _4+2 \alpha _5+2 \beta _5+\alpha _6+\beta _6  )\\
 	L_{3,13}=L_{3,14}&=\pi^{-1}(   2 \alpha _1+2 \beta _1+\gamma _1+3 \delta _1+\alpha _2+\beta _2+3 \delta _2+2 \alpha _3+2 \beta _3+2 \gamma _3+3 \delta _3+\alpha _4+\beta _4   \\
	&  +3 \delta _4+2 \alpha _5+2 \beta _5+\alpha _6+\beta _6    )\\
 	L_{3,15}=L_{3,16}&=\pi^{-1}(  \alpha _1+\beta _1+2 \gamma _1+3 \delta _1+2 \alpha _2+2 \beta _2+3 \delta _2+\alpha _3+\beta _3+\gamma _3+3 \delta _3+2 \alpha _4+2 \beta _4  \\
	&   +3 \delta _4+\alpha _5+\beta _5+2 \alpha _6+2 \beta _6  ) \\
 	L_{3,17}&=\pi^{-1}(  3 ( \gamma _1+\delta _1+\alpha _2+\beta _2+\delta _2+\delta _3+\alpha _4+\beta _4+\delta _4+\alpha _6+\beta _6   ) )\\
 	L_{3,18}&= \pi^{-1}( 3 (  \alpha _1+\beta _1+\delta _1+\delta _2+\alpha _3+\beta _3+\gamma _3+\delta _3+\delta _4+\alpha _5+\beta _5  ) )
\end{split}	
\ee
	\begin{figure}[h!]
	\begin{center}
	\includegraphics[width=.30\textwidth]{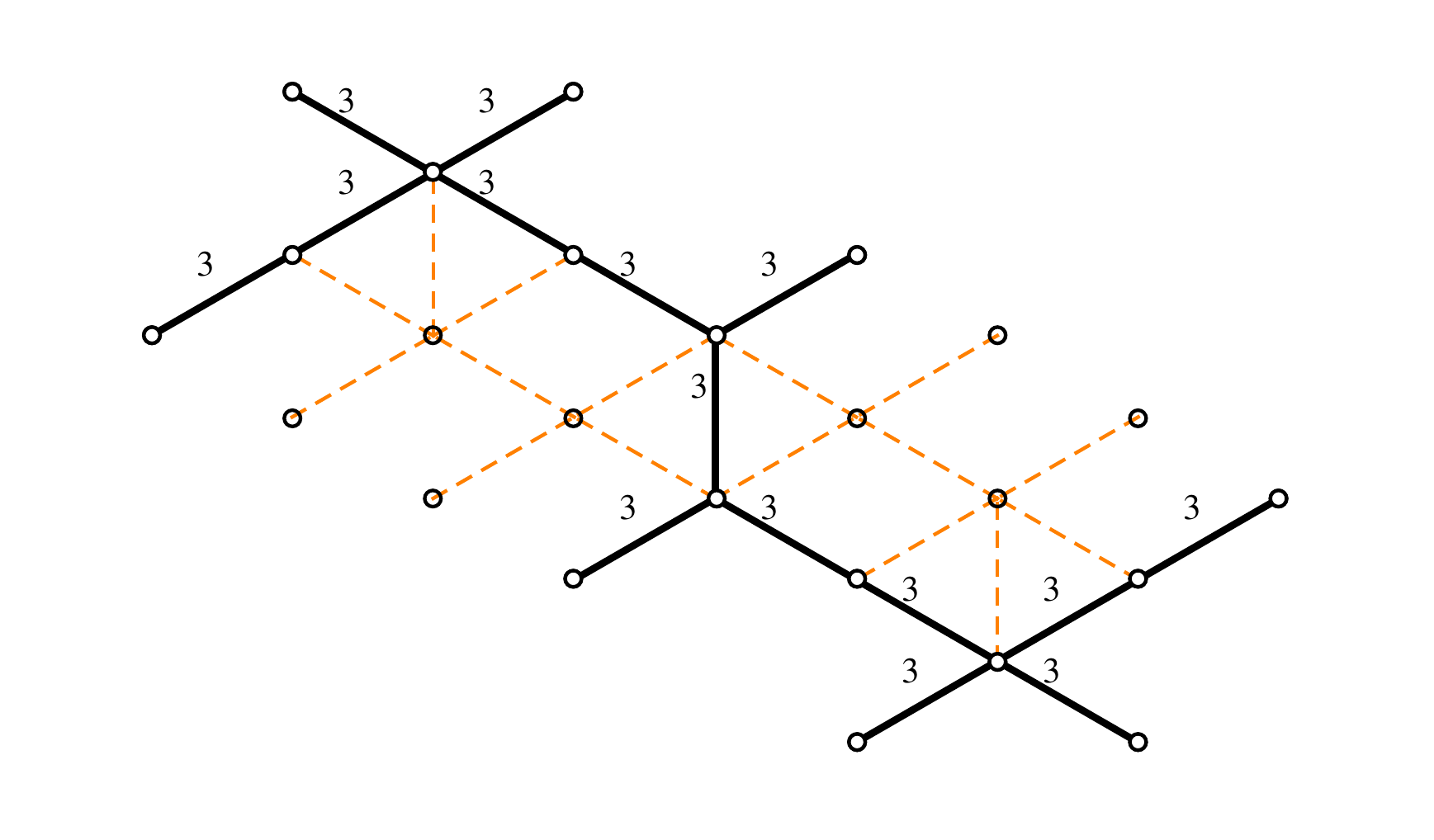}\includegraphics[width=.30\textwidth]{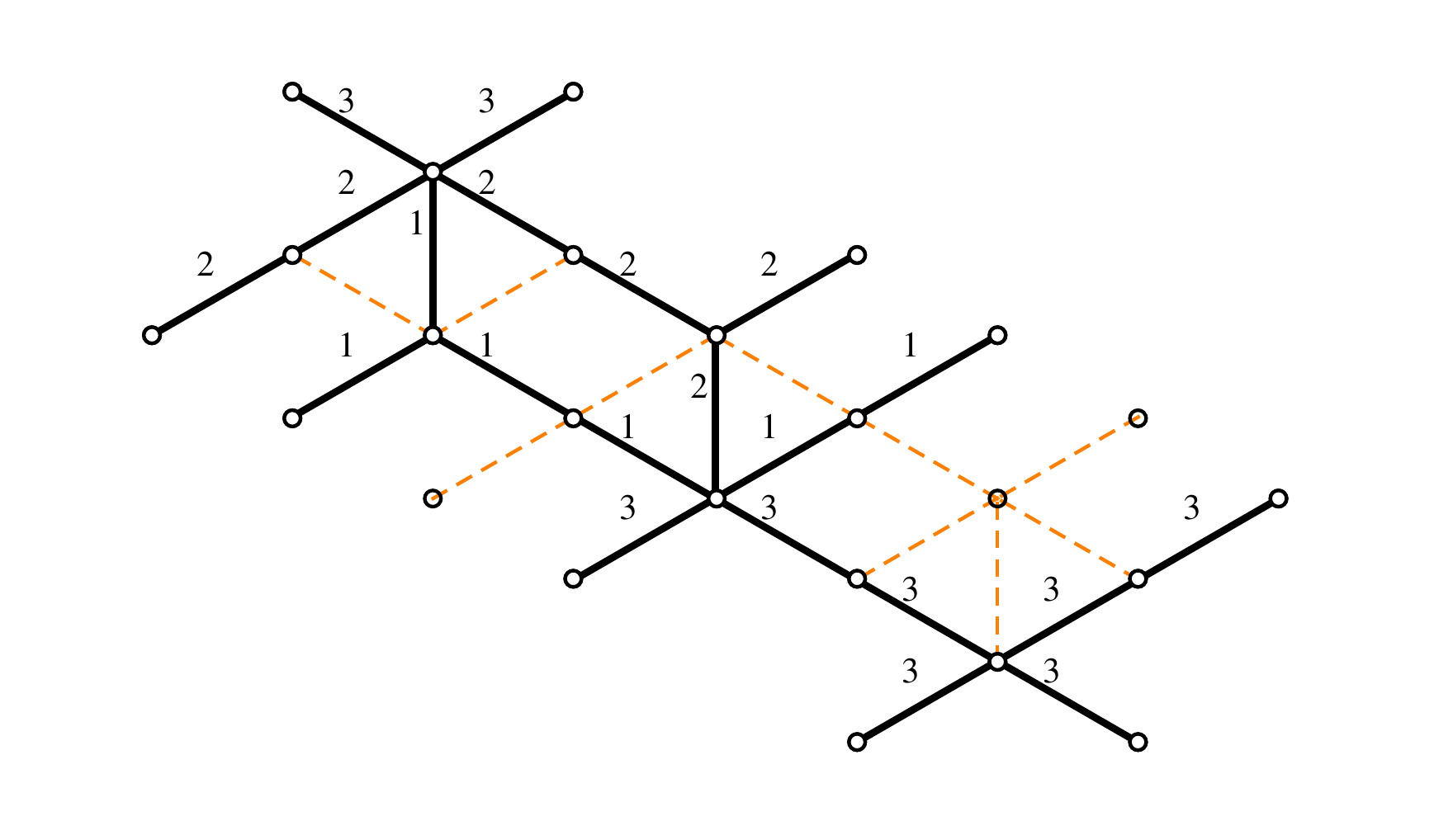}\includegraphics[width=.30\textwidth]{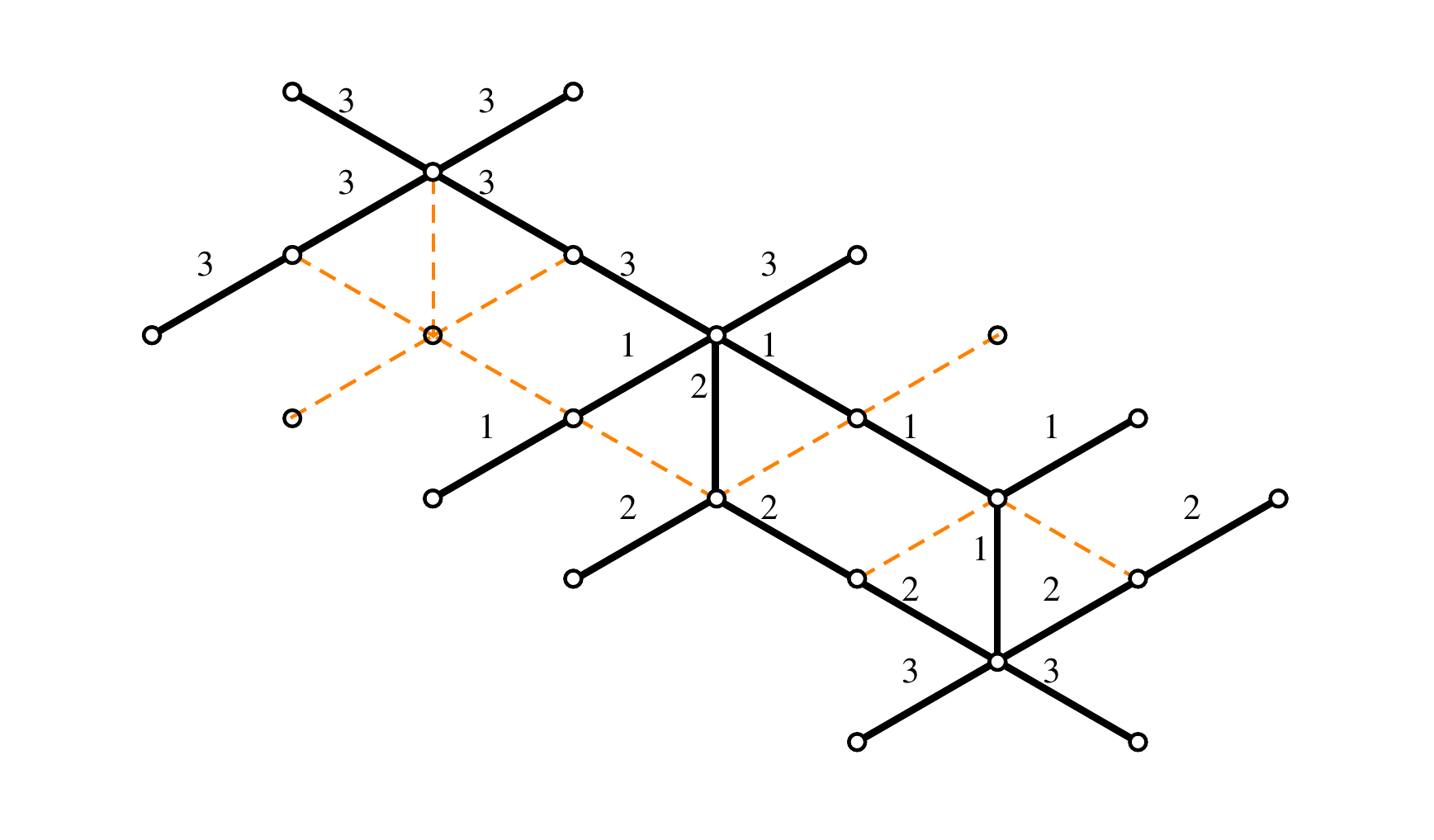} \\
	\includegraphics[width=.30\textwidth]{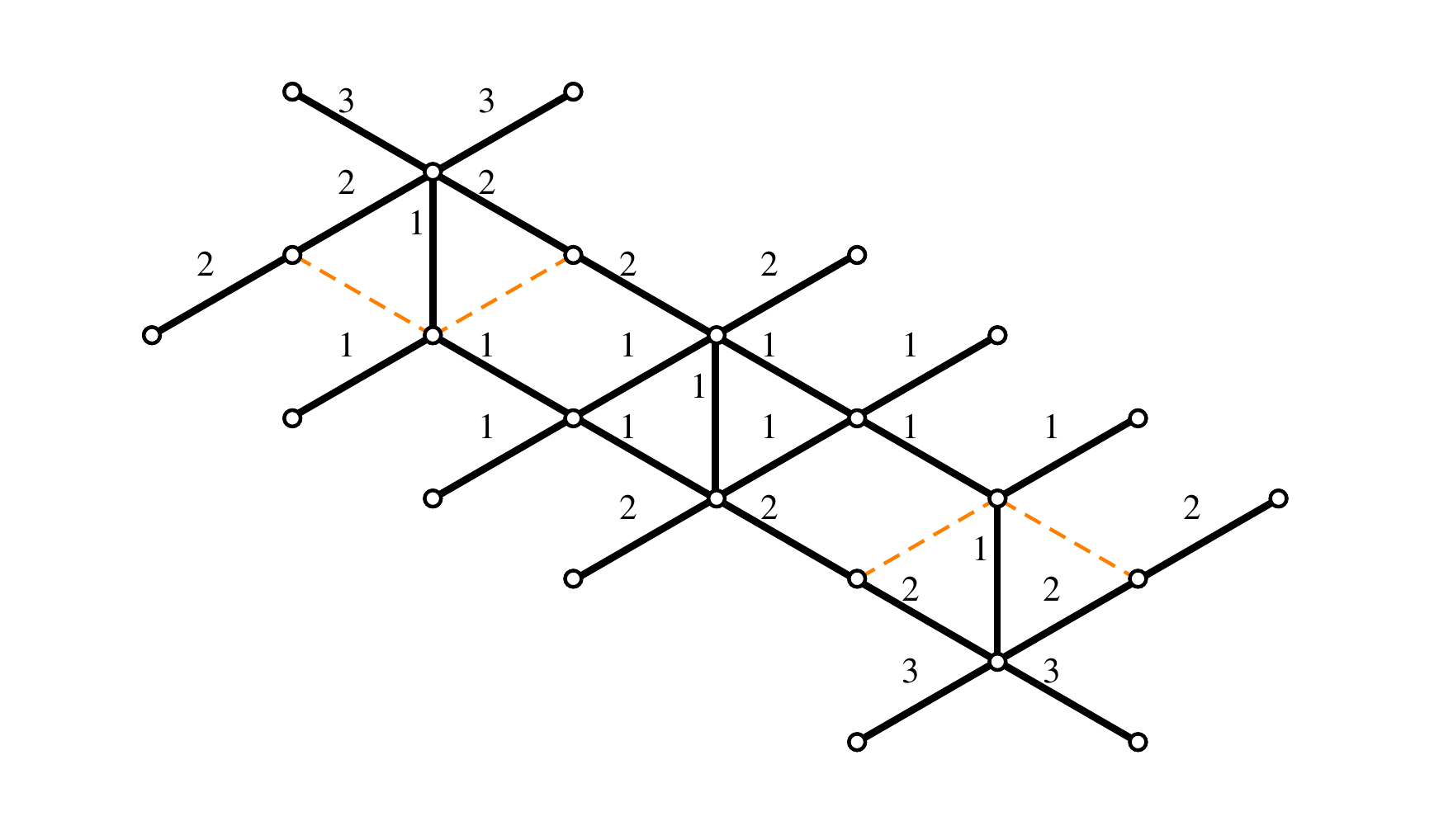}\includegraphics[width=.30\textwidth]{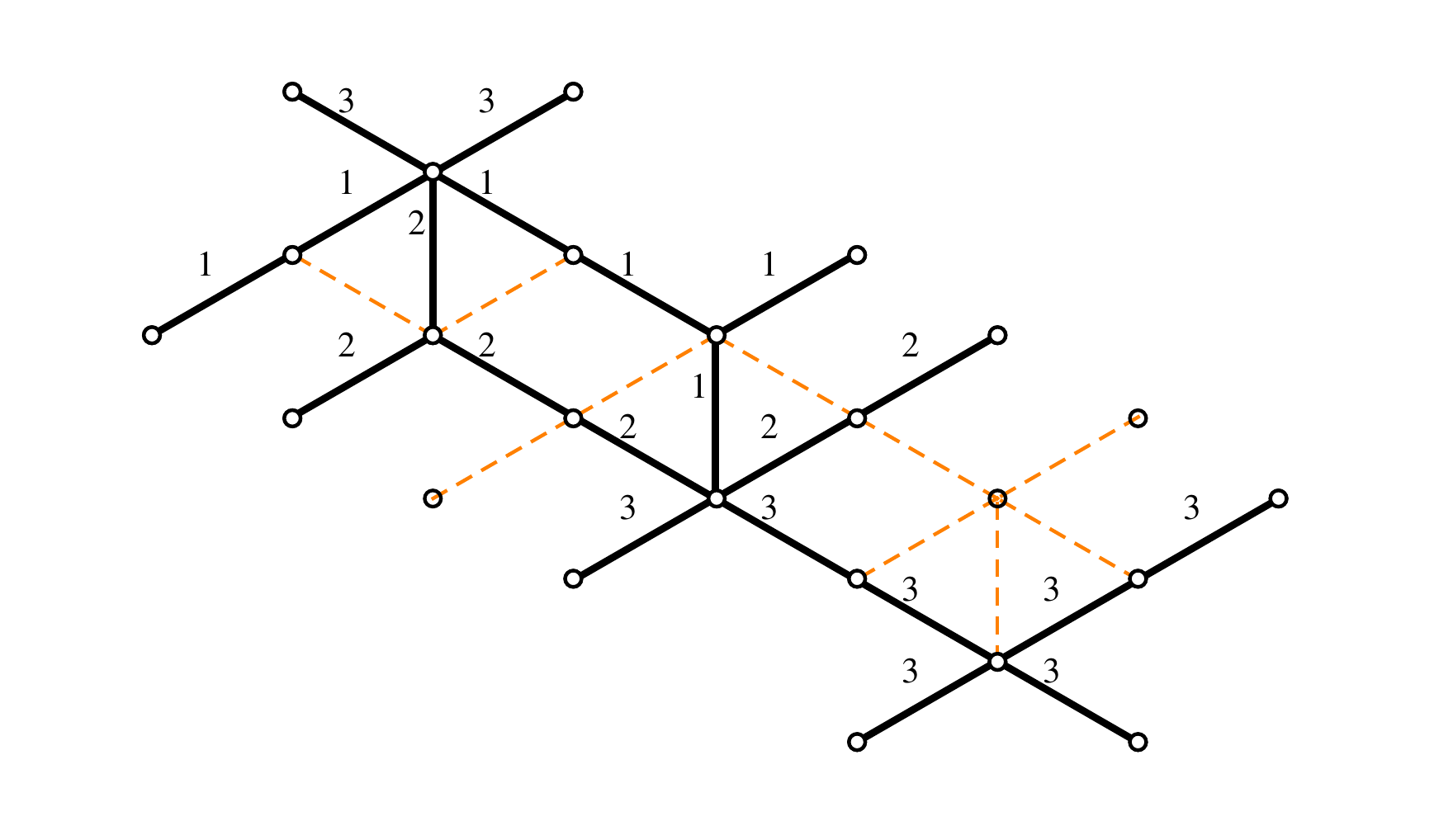}\includegraphics[width=.30\textwidth]{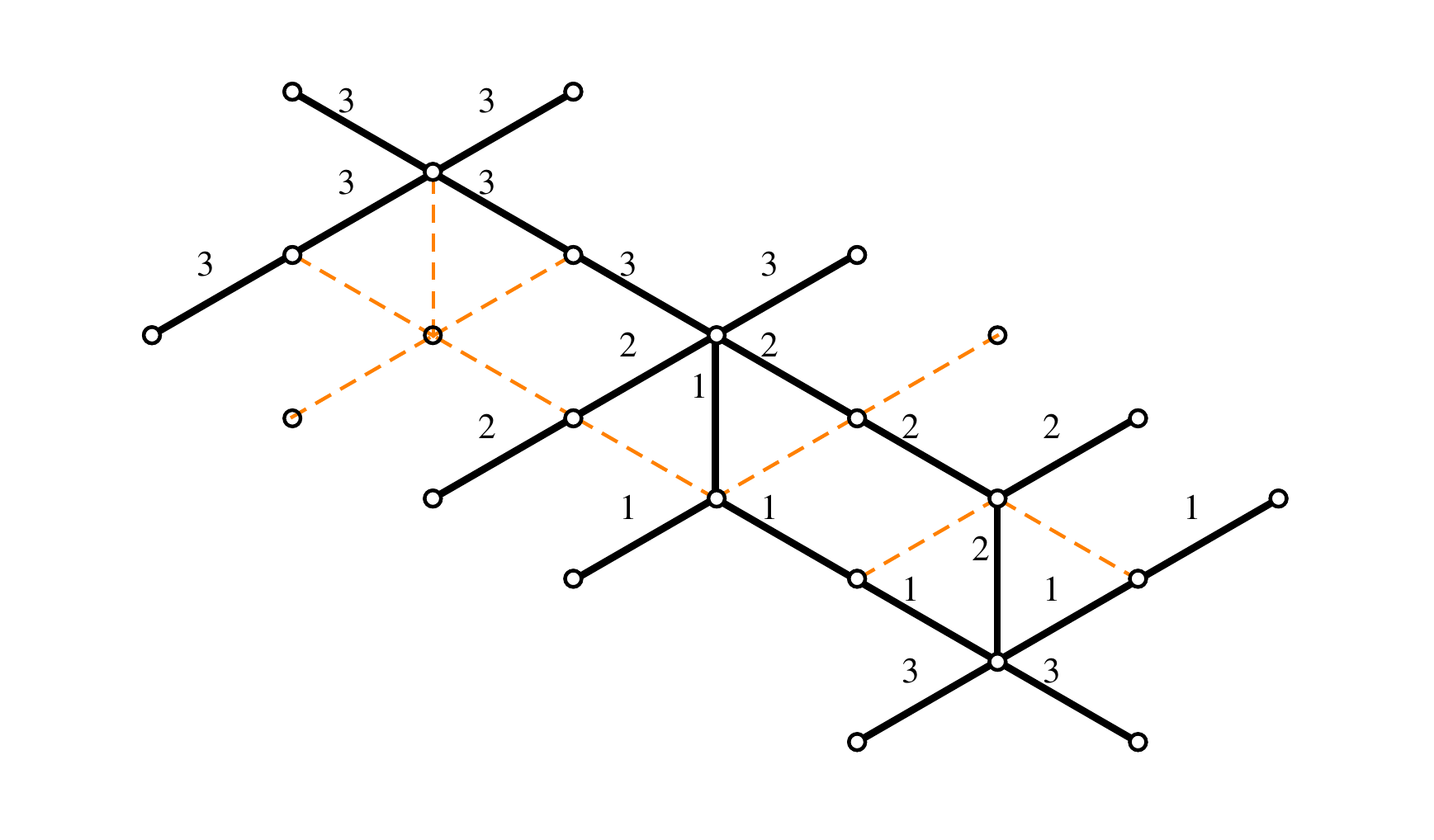} \\
	\includegraphics[width=.30\textwidth]{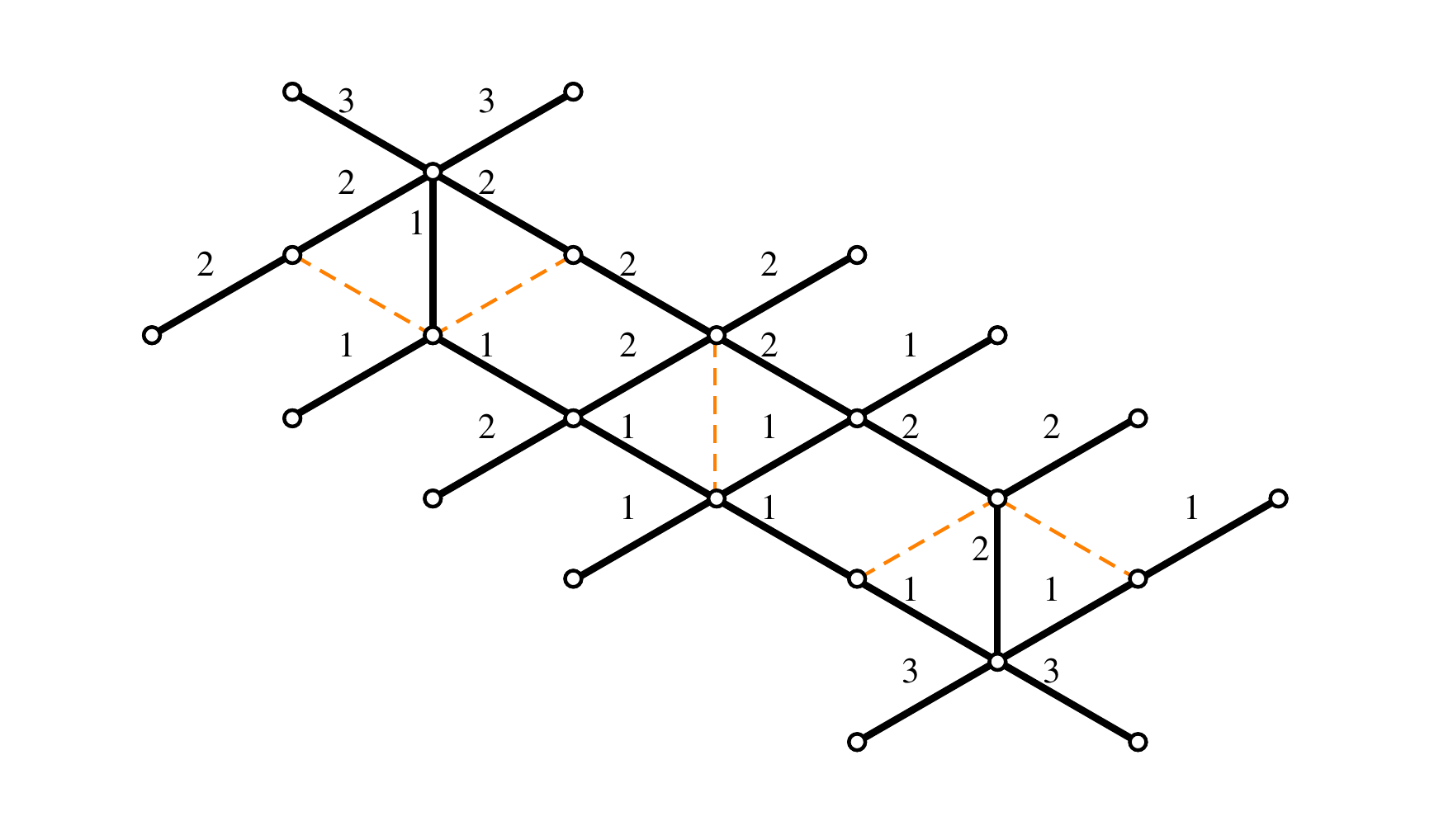}\includegraphics[width=.30\textwidth]{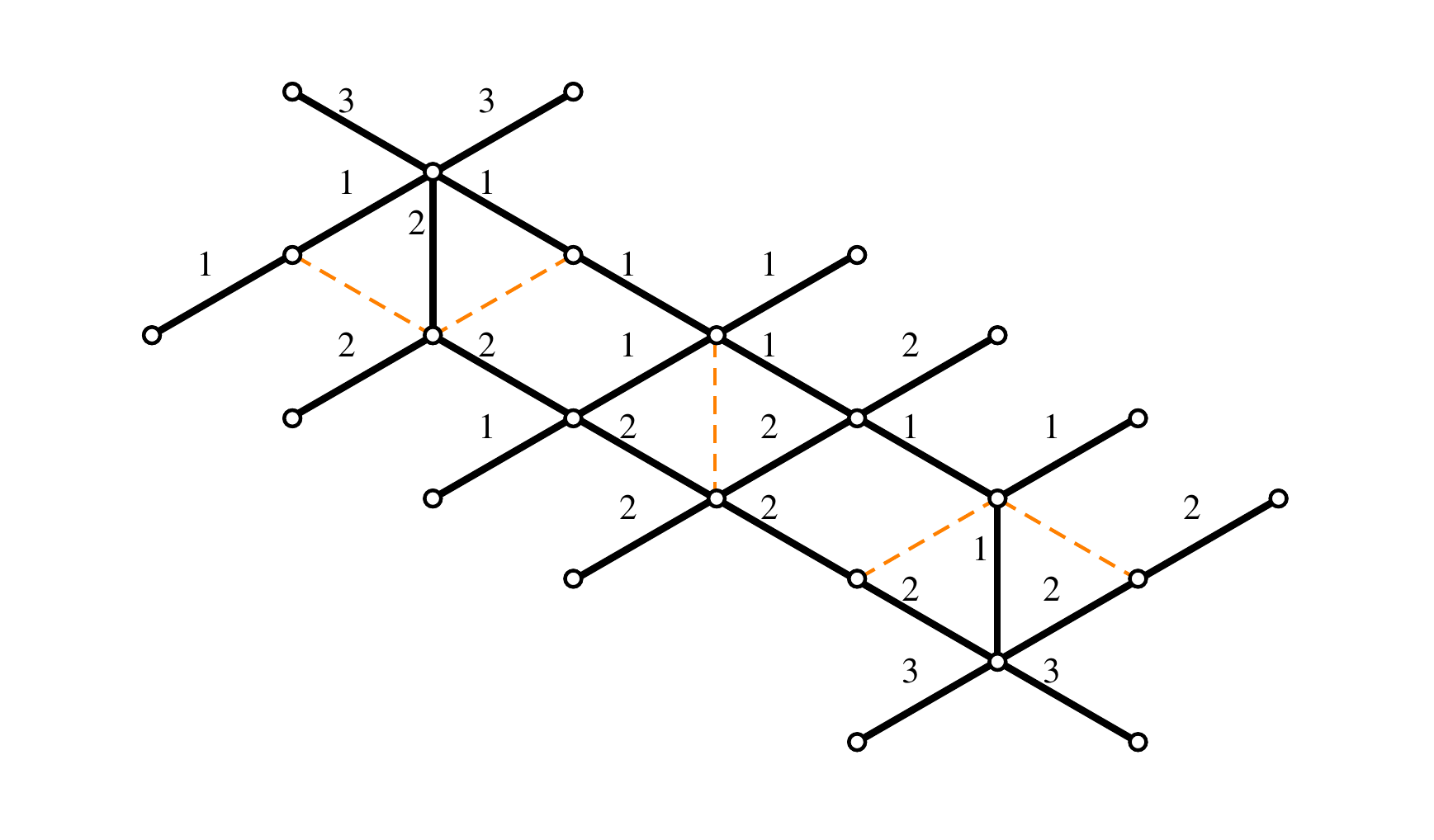}\includegraphics[width=.30\textwidth]{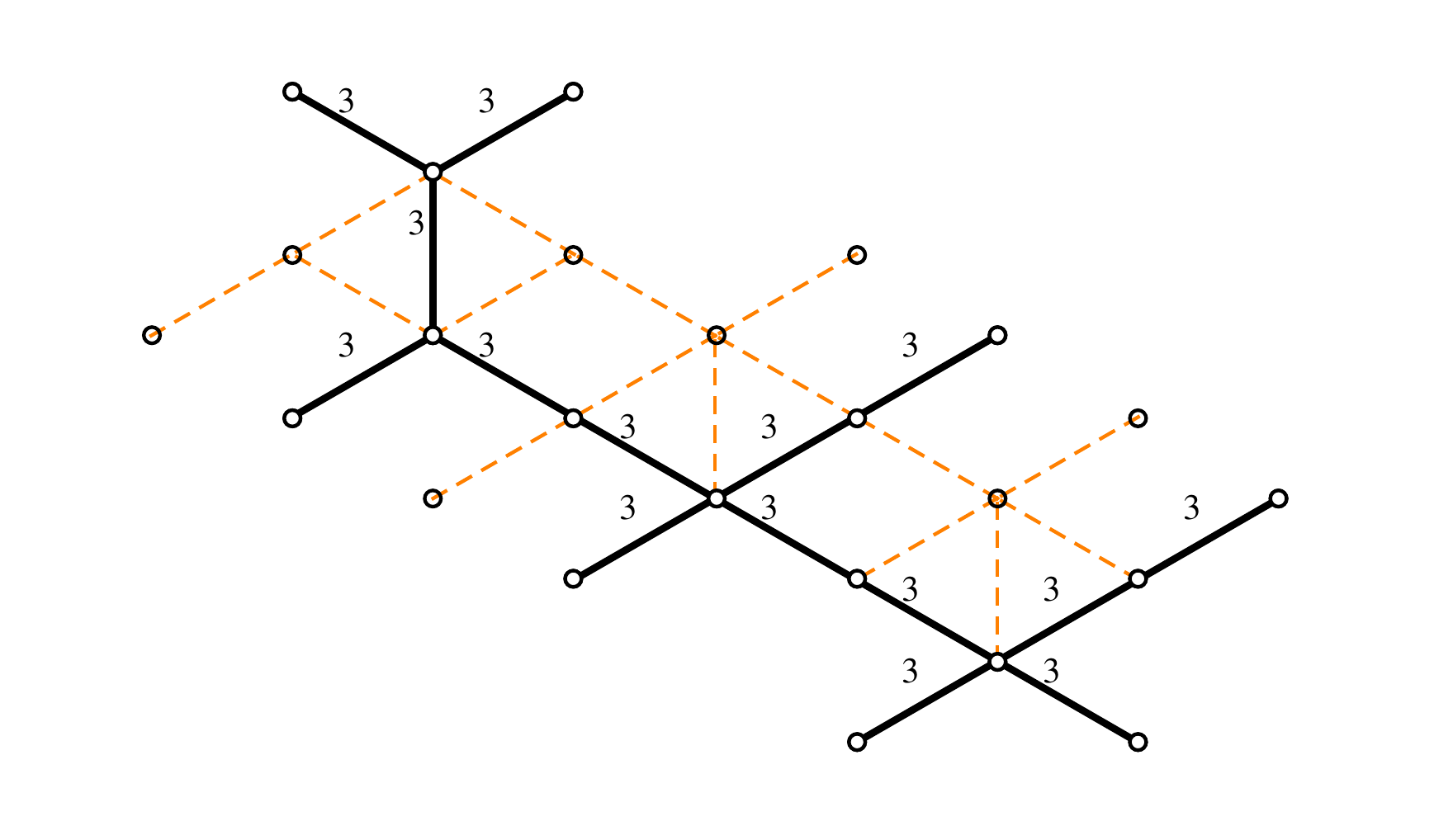} \\
	\includegraphics[width=.30\textwidth]{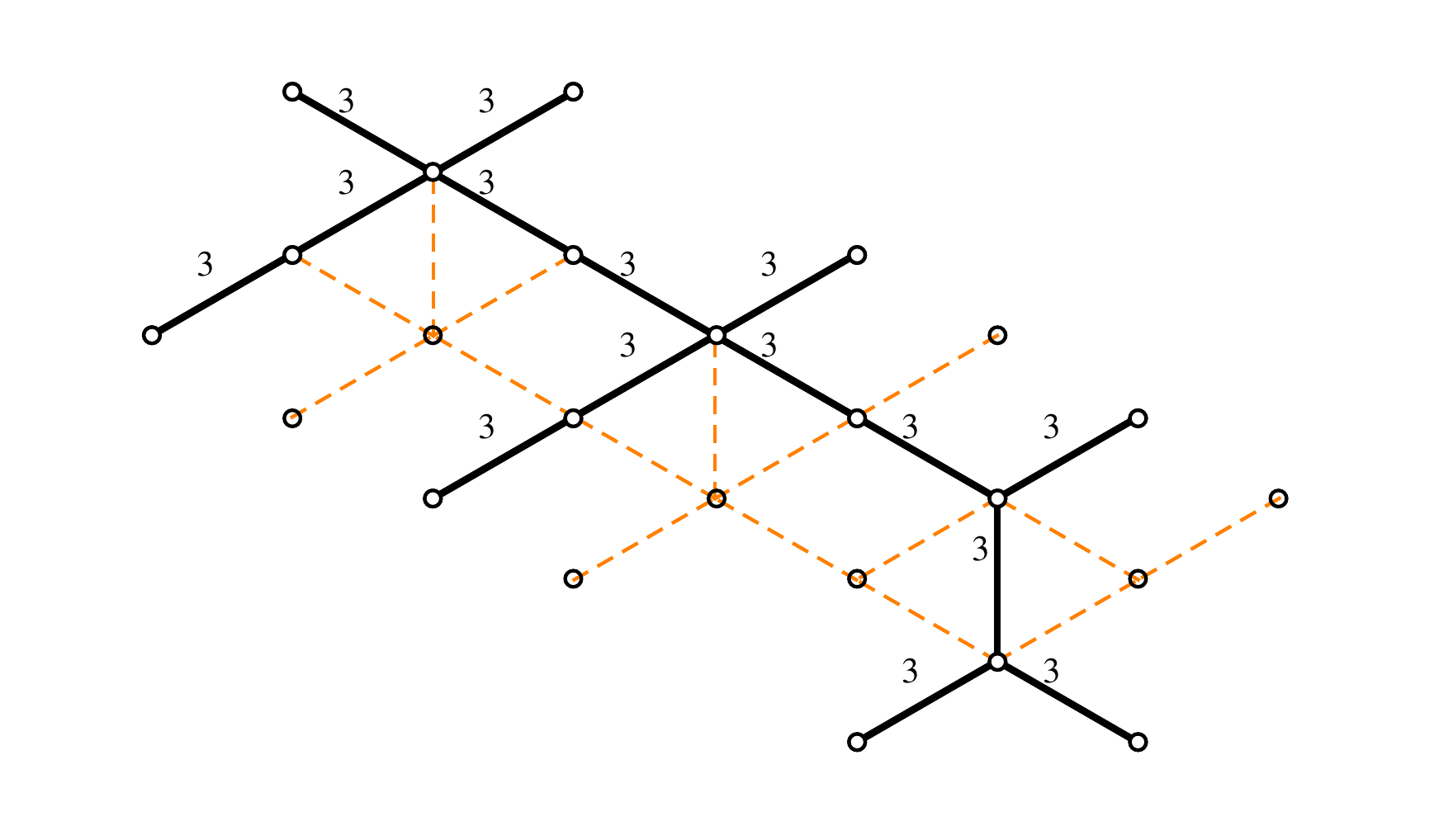}
	\caption{The set of $L_{3,r}$. Values of the $\alpha_{r}(p)$ are displayed.}
	\label{default}
	\end{center}
	\end{figure}
	
	The check can be extended to detours on {all streets of the critical network} (considering a short interface crosing only a single 2-way street). We checked that the correspondence between (\ref{eq:halo-interpretation}) and (\ref{eq:2d4d-interpretation}) holds: taking an interface crossing the lift of a single 2-way street, we find that the halo generating function gets contributions only from those $L_{n,r}$ which contain the street itself.

\section{A technical equivalence relation}\label{app:dirtytrick}

\begin{figure}[h!]
\begin{center}
\includegraphics[width=0.4\textwidth]{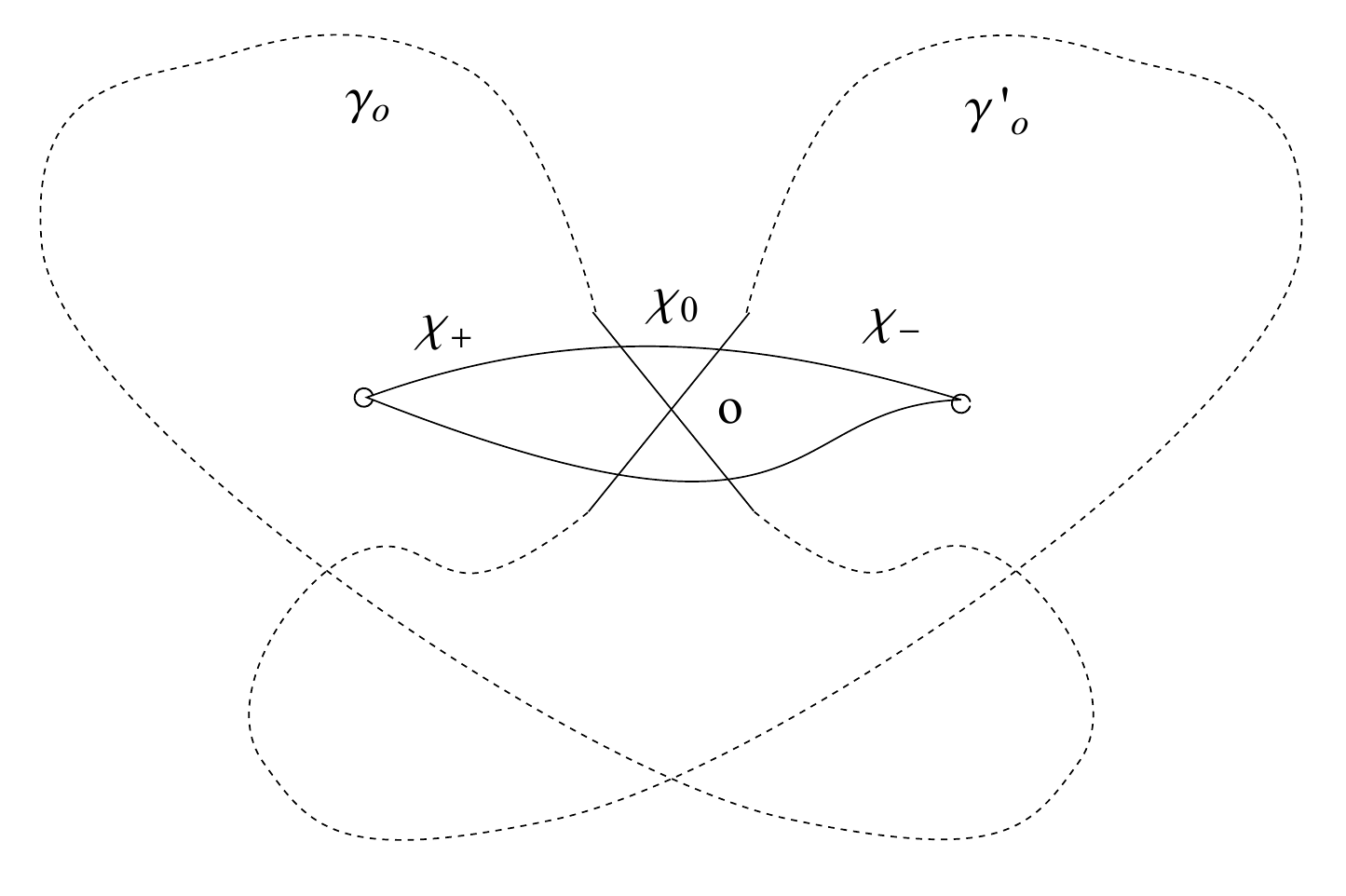}
\end{center}
\caption{The writhe as the origin of the quadratic refinement \label{fig:qref}}
\end{figure}

First we would like to stress that the writhe implements a quadratic refinement function for the intersection pairing in the homology basis. Indeed the both quantities are of the same nature: they both can be interpreted as certain signed sums over intersections or self-intersections of actual paths.

Suppose we consider two paths $\gamma_o$ and $\gamma_o'$ intersecting in some point $o$ and suppose we are able to concatenate them (somehow) then the writhe (a signed sum over all self-intersections of the concatenation) is expected to have the following three contributions:
\begin{enumerate}
\item The sum over self-intersections of $\gamma_o$
\item The sum over self-intersections of $\gamma_o'$
\item The sum over mutual intersections of $\gamma_o$ and $\gamma_o'$
\end{enumerate}

The problem is that according to our rules we are not able to concatenate two closed paths if they do not have a common tangent vector in the intersection point. Thus we add a small refinement: we consider an auxiliary path $\chi$ intersecting both $\gamma_o$ and $\gamma_o'$ near the point $o$ as it is depicted in fig.\ref{fig:qref}. There are two possibilities to choose $\chi$ going above or below the point $o$.

First consider the choice of $\chi$ going above and consider paths $\gamma_o$ and $\gamma_o'$ as detours then the writhe of the resulting path reads
\be\label{eq:auxil}
\wr(\aux_+ \gamma_o \aux_0\gamma_o' \aux_-)=\wr(\aux_+ \gamma_o \aux_0 \aux_-)+\wr(\aux_+ \aux_0\gamma_o' \aux_-)+\langle [\gamma_o],[\gamma_o']\rangle
\ee
where $\langle \star, \star\rangle$ is the intersection pairing of homology classes $[\star]$ on $\Sigma^*$\footnote{Incidentally, since $\Sigma^{*}$ differs from $\Sigma$ only by punctures, the pairing coincides with the intersection pairing of homology classes on $\Sigma$.}. Then we \emph{shrink} the auxiliary path to zero and rewrite this relation as
 \be
 \wr(\gamma_o \gamma_o')=\wr( \gamma_o )+\wr(\gamma_o')+\langle [\gamma_o],[\gamma_o']\rangle
 \ee
In this form the writhe represents a \emph{quadratic refinement} of the intersection pairing on cycles, and we imply a smooth gluing of the paths via an auxiliary path as in (\ref{eq:auxil}).

Notice that for the lower choice of $\chi$ we take a detour along $\gamma_o'$ first and then along $\gamma_o$, this we describe as
\be
\wr(\gamma_o' \gamma_o)=\wr( \gamma_o' )+\wr(\gamma_o)+\langle [\gamma_o'],[\gamma_o]\rangle
\ee

And the difference reads
\be
\wr(\gamma_o \gamma_o')-\wr(\gamma_o' \gamma_o )=2\langle [\gamma_o],[\gamma_o']\rangle
\ee

This relation can be continued to the $\rho$-projections of the algebraic variables
\be
\begin{split}
	\rho\Big( \hU_{\gamma_{o}}\hU_{\gamma_{o}'} \Big) = \rho\Big( \hU_{\gamma_{o}\gamma_{o}'} \Big)
	& = y^{2 \left\langle [\gamma_o],[\gamma_o']\right\rangle } \rho\Big( \hU_{\gamma_{o}'\gamma_{o}} \Big)\\
	&  = y^{2 \left\langle [\gamma_o],[\gamma_o']\right\rangle } \rho\Big( \hU_{\gamma_{o}}'\hU_{\gamma_{o}} \Big)
\end{split}
\ee
The punchline is that, formally, $\hU_{\gamma_o}\hU_{\gamma_o'}$ and $y^{2\left\langle [\gamma_o],[\gamma_o']\right\rangle}\hU_{\gamma_o'}\hU_{\gamma_o}$ give the same contribution when projected under $\rho$. We must stress that this by no means implies something like an algebra rule for the $\hU$ (nor the $\hY$) variables!

To lighten computations in the main body of the paper (most notably section \ref{sec:herds}) we will sometimes employ the following \emph{equivalence relation}
\be\label{eq:KS-rem}
\hU_{\gamma_o}\hU_{\gamma_o'}{\dot =}y^{2\left\langle [\gamma_o],[\gamma_o']\right\rangle}\hU_{\gamma_o'}\hU_{\gamma_o}
\ee
\emph{for the purpose of eventually projecting} through the map $\rho$ from $\hU$ variables to $\hY$ variables. The symbol $\dot =$ is meant to warn the reader that \emph{this is not an identity} regarding the algebra of $\hU$ variables.

\section{Off-diagonal herds}\label{app:off-diag-herds}

In section \ref{subsec:off-diag-herd} we encountered a particular type of wild critical network, which is actually part of a larger family of ``off-diagonal'' herds ${\cal H}_p^{(m,n)}$\footnote{T.~Mainiero has
independently come to the picture of the off-diagonal herds and is currently
studying them.}. We call these networks $p\,\mhyphen\,(m,n)$ herds, in this section we describe schematically their structure.

\begin{figure}[h!]
\begin{center}
\includegraphics[width=.45\textwidth]{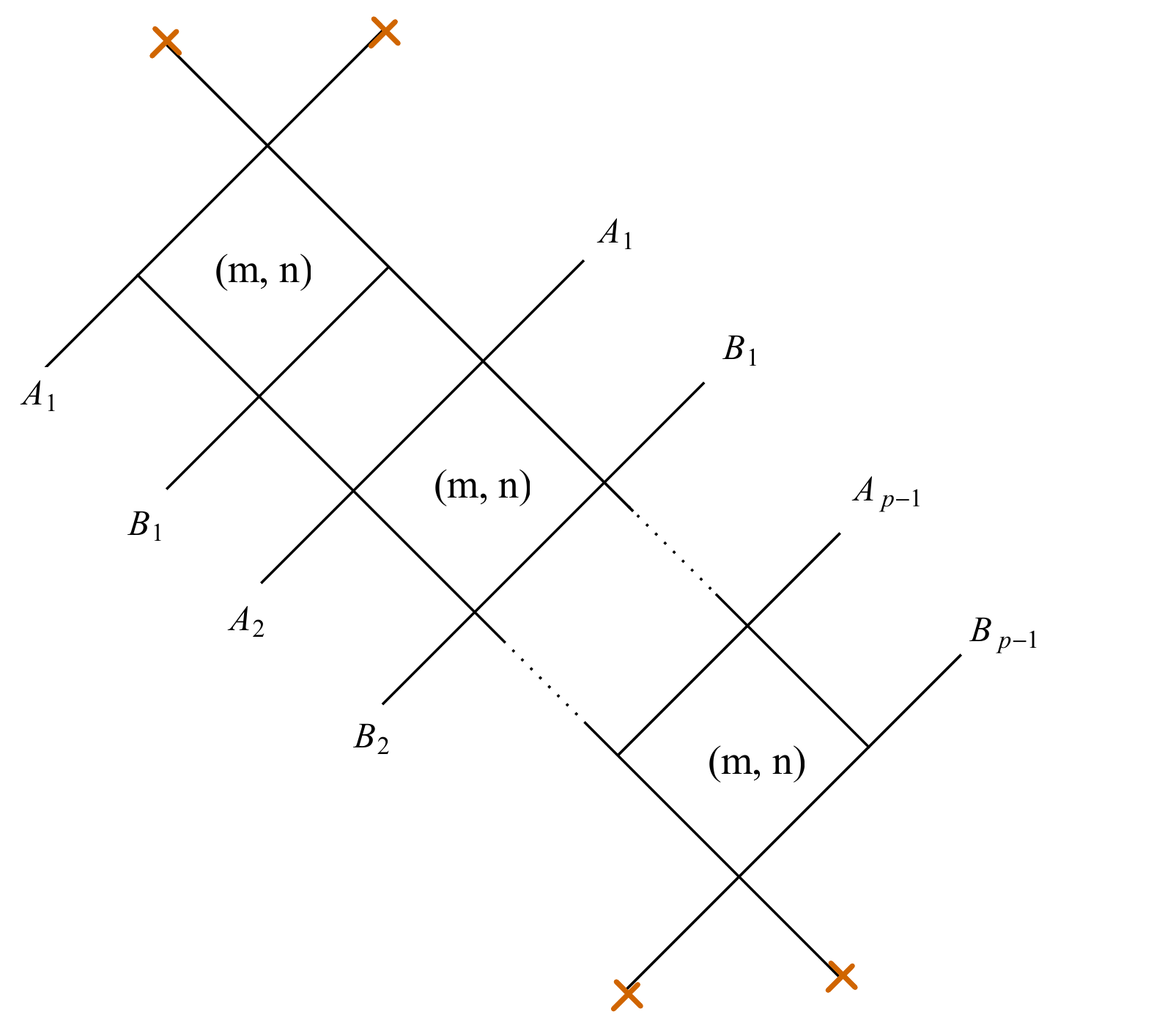}
\caption{A $p$-herd is a collection of ``fat'' horses glued together with appropriate boundary conditions.}
\label{fig:off-herd}
\end{center}
\end{figure}

The structure of off-diagonal herds is closely related to that of the usual ``diagonal'' herds: the general structure of the network consists of $p$ blocks glued together, as shown in figure \ref{fig:off-herd}

The gluing conditions are similar to those implemented in diagonal herds: blocks are glued to each other in the natural way, throughout the herd which wraps the cylinder of $\cal C$. Terminal blocks are connected to branch points.

\begin{figure}[h!]
\begin{center}
\includegraphics[width=.65\textwidth]{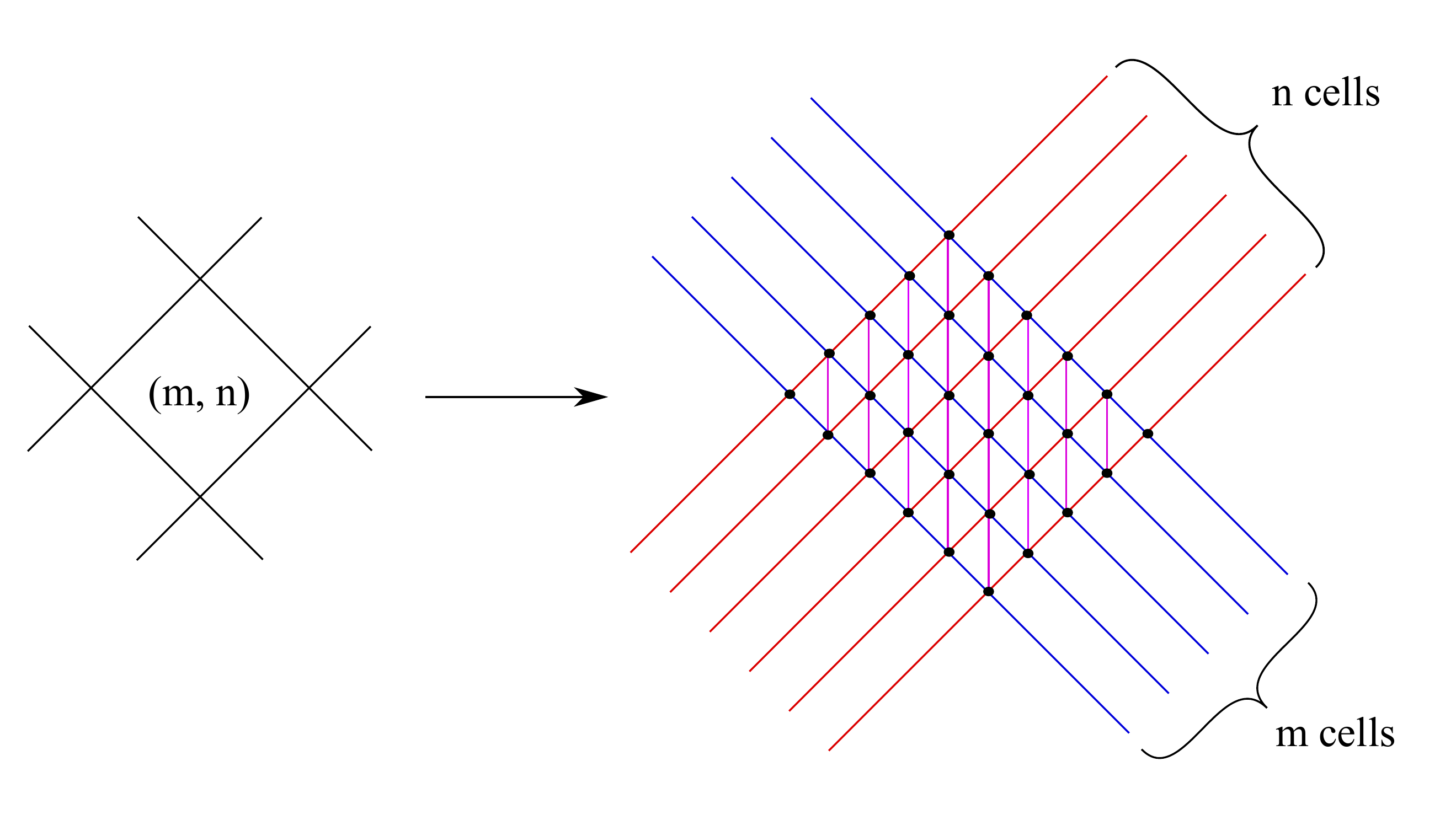}
\caption{A ``fat'' $(m,n)$-horse}
\label{fig:fat-horse}
\end{center}
\end{figure}

The novel feature of these types of networks in comparison to diagonal herds is that the single $(m,n)$-block is now an $m\times n$ array of elementary horses glued
together. This is displayed in figure \ref{fig:fat-horse},  street types are analogous to those of section \ref{sec:Qinside}: $ij, \,jk, \,ik$-types are marked in blue, red and purple respectively.

Physically $p\,\mhyphen\,(m,n)$ - herds encode the protected spin characters of bound states of particles of charges $\gamma$ and $\gamma'$ with 
$\langle\gamma,\gamma'\rangle=p$, with the ratio of particles of the first and the second types being $m:n$.

Put differently, these networks can be associated to ``slope-$m/n$'' $p$-Kronecker quiver representations \cite{WWC}
\be
{\cal K}_p^{(m,n)}: \quad \IC^{\alpha m} \mathop{\longrightarrow}\lm^p \IC^{\alpha n}, \quad \alpha\in \IN\,.
\ee

From this perspective, the Poincar\'e polynomial of the corresponding quiver variety is expected to coincide with the protected spin character calculated from the network.
The computation of the PSC from the network can be carried out applying the techniques  discussed above in section \ref{subsec:off-diag-herd}.

To give an example, consider the following generating function for the Euler characteristics of the moduli spaces
\be
Q({\cal K}_p^{(m,n)}):=\prod\lm_{\alpha\in\IN} (1+(-1)^{\alpha m n}z^{\alpha})^{\alpha\, \chi({\cal M}_p^{(\alpha n,\alpha m)})}\,.
\ee
From the network's side, we may associate soliton generating functions $\nu,\tau$ (solitons going into/out of the joint respectively) to every joint of ${\cal H}_{p}^{(m,n)}$. We also introduce the notation $\nu_k^{(i,j)}[g]$, $\tau_k^{(i,j)}[g]$, where $k=1,\ldots, p$ labels the fat horse within the herd, while $i=0,\ldots,m$ and $j=0,\ldots,n$ label the joint within the fat herd, and $g=1,\ldots, 6$ labels the six streets connected to the joint. The enumeration of the streets at a generic 6-way joint goes clockwise starting from ``noon''. With this notation, we define the following generating function:
\be
Q({\cal H}_p^{(m,n)}):=\sqrt[n]{1+\tau_p^{(m,n)}[3]\nu_p^{(m,n)}[3]}=\sqrt[m]{1+\tau_p^{(m,n)}[5]\nu_p^{(m,n)}[5]}
\ee

Some examples of the network-quiver correspondence are
\be
\begin{split}
Q({\cal K}_3^{(3,1)})&=Q({\cal H}_3^{(3,1)})=1+z\,,\\
Q({\cal K}_3^{(3,2)})&=Q({\cal H}_3^{(3,2)})=1 + 13 z + 1034 z^2 + 115395 z^3+ O(z^4)\,,\\
Q({\cal K}_3^{(4,3)})&=Q({\cal H}_3^{(4,3)})=1 + 68 z + 66378 z^2+O(z^3)\,,
\end{split}
\ee
in agreement with the general expectation
\be
Q({\cal K}_p^{(m,n)})=Q({\cal H}_p^{(m,n)})\,.
\ee

It appears to be a challenging problem to generalize equation (\ref{eq:fnceqn}) ( or (\ref{eq:P})  ) to the
case of   $p\,\mhyphen\,(m,n)$  herds, even in the classical case. One might expect a system of equations
for a well-chosen set of generating functions,  but finding a manageable such system is a problem we leave
for the future.

\section{Generic interfaces and the halo picture}\label{app:halo-factorization}
This section is devoted to showing how the factorization property deriving from the halo picture fails to capture the $\CK$-wall jump (\ref{eq:3-herd-variant}). More precisely, the $\CK$-wall jump cannot be written as a conjugation by dilogarithms \emph{unless} some extra techical assumptions are introduced (see immediately above \ref{eq:q-dilog-factorized}) about the algebra of formal variables. It is sufficient to consider the truncated expression

\be\label{eq:non-factorizable}
\begin{split}
	F_{jj}(\wp,\vartheta_{c}^{+};y) & = \hY_{\wp^{(j)}}  +  y^{2}\, \hY_{\wp^{(j)}+\tgamma_{1}}+ (y+ y^5)\hY_{\wp^{(j)}+\tgamma_{1}+\tgamma_{2}}+2 y^4\hY_{\wp^{(j)}+2\tgamma_{1}} +\cdots\\
	& = \hY_{\wp^{(j)}} \Big( 1 +  y^{3}\, \hY_{\tgamma_{1}}+ (y^{2}+ y^6)\hY_{\tgamma_{1}+\tgamma_{2}}+2 y^6\hY_{2\tgamma_{1}} +\cdots\Big)\,. 
\end{split}
\ee
In order to assess whether it admits a factorization similar to (\ref{eq:q-dilog-factorized}), involving quantum dilogs, consider the following identity
\be\label{eq:finite-quantum}
\begin{split}
	\Phi((-y)^{m}\,\hY_{\tgamma})^{k}\, \hY_{\wp^{(j)}}\, \Phi((-y)^{m}\,\hY_{\tgamma})^{-k} & = \hY_{\wp^{(j)}}\, \Phi_{-\langle\wp^{(j)},\tgamma\rangle}((-y)^{m}y^{-2 \langle \wp^{(j)},\tgamma\rangle}\,\hY_{\tgamma})^{-k\,{\rm sgn}(\langle\wp^{(j)},\tgamma\rangle)} \\
	& = 1+ k\, \frac{1-y^{2\langle \wp^{(j)},\tgamma\rangle}}{1-y^{-2}}  \, y^{-1}\, (-y)^{m}y^{-2 \langle \wp^{(j)},\tgamma\rangle}\,\hY_{\tgamma} + O(\hY_{2\tgamma})\,,
\end{split}
\ee
where in the last line we expanded in powers of $\hY_{\tgamma}$ and used the fact cycles $\tgamma$ appearing in the expression of interest all satisfy $\langle\wp^{(j)},\tgamma\rangle<0$. 
From this, taking $\tgamma=\tgamma_{1}$ and comparing with the above we find only one possibility compatible with a dilog factorization: $\Phi((-y)^{2}\,\hY_{\tgamma_{1}})$. Note that this would contribute a factor of $\Phi_{1}((-y)^{4}\,\hY_{\tgamma_{1}})$ when switching to finite-type dilogs, which is equal to $1+ y^{3}\,\hY_{\tgamma_{1}}$. Therefore this dilog would not contribute to any other term in parentheses on the RHS of (\ref{eq:non-factorizable}), hence we may use (\ref{eq:finite-quantum}) directly on the other terms as well. Thus considering the term in $\hY_{2\tgamma_{1}}$, if it were coming from a dilog expansion we would expect the following pre-factor
\be
	k\,y^{3}\,(-y)^{m}\,\frac{1-y^{-4}}{1-y^{-2}} = k\,(-y)^{m}(y+y^{3})
\ee
which clearly cannot match $2 y^{6}$. This establishes that (\ref{eq:3-herd-variant}) cannot be cast into the form of conjugation by quantum dilogarithms.

\nocite{*}
\bibliographystyle{utphys}
\bibliography{biblio}

\end{document}